\documentclass[reprint,amsmath,amssymb,aps,pra]{revtex4-2}
\usepackage[utf8]{inputenc}
\usepackage{indentfirst}
\usepackage{amsmath}
\usepackage{amssymb}
\usepackage{bm}
\usepackage{braket}
\usepackage{dcolumn}
\usepackage{graphicx}
\usepackage{tabularx}
\usepackage{ltablex}
\usepackage{multirow}
\usepackage{tikz}
\usepackage{hyperref}

\newcommand{\od}[2]{\frac{d #1}{d #2}}

\newcommand{\eq}[1]{Eq.\,(\ref{#1})}
\newcommand{\eqnm}[1]{(\ref{#1})}

\newcommand{\smbr}[1]{\left( #1 \right)}
\newcommand{\cubr}[1]{\left\{ #1 \right\}}
\newcommand{\dbrc}[1]{\left\langle #1 \right\rangle}

\newcommand{\sqbr}[1]{\left[#1\right]}
\newcommand{\abs}[1]{\left|#1\right|}

\newcommand{\mfk}[1]{\mathfrak{#1}}
\newcommand*\mathinhead[2]{\texorpdfstring{$\boldsymbol{#1}$}{#2}}

\usetikzlibrary{shapes.geometric, arrows}
\tikzstyle{startstop} = [rectangle, rounded corners, minimum width=3cm, minimum height=1cm,text centered, draw=black, fill=red!30]
\tikzstyle{io} = [trapezium, trapezium left angle=70, trapezium right angle=110, minimum width=3cm, minimum height=1cm, text centered, draw=black, fill=blue!30]
\tikzstyle{input} = [rectangle, rounded corners, minimum width=3cm, minimum width=0.2cm, minimum height=0.2cm, text centered, draw=black, fill=cyan!30, opacity=1.0]
\tikzstyle{process} = [rectangle,text centered, draw=black, fill=orange!45]
\tikzstyle{decision} = [diamond, text centered, draw=black, fill=green!30]
\tikzstyle{inpoint}=[circle,minimum size =0.02cm, draw=black, thin, fill=blue!20,text=black]

\tikzstyle{arrow} = [thick,->,>=stealth]

\DeclareFontFamily{OT1}{pzc}{}
\DeclareFontShape{OT1}{pzc}{m}{it}{<-> s * [1.3500] pzcmi7t}{}
\DeclareMathAlphabet{\mpzc}{OT1}{pzc}{m}{it}

\begin{document}

\title{Supersymmetric Quantum Mechanics of Hypergeometric-like Differential Operators}

\author{Tianchun Zhou}
\email{tczhou@hit.edu.cn}
\affiliation{
 School of Physics $\&$ Laboratory for Space Environment and Physical Science, Harbin Institute of Technology, Harbin, 150001 P. R. China}

\date{\today}

\begin{abstract}
 Systematic iterative algorithms dictated purely by the principles of supersymmetric quantum mechanics (SUSYQM) for solving the discrete eigenvalue problem of the principal hypergeometric-like differential operator and for generating the associated hypergeometric-like differential equation itself as well its eigen-solutions are developed, without recourse to any traditional method or any result out of it. These SUSYQM algorithms are initiated by devising \emph{two} distinct types of {\it active supersymmetrization} transformations and momentum operator maps, and applying them to quickly transforming the \emph{same} hypergeometric-like differential equation yet in its two trivial asymmetric factorizations into two different supersymmetrically factorized Schr\"odinger equations. The rest iteration flows are completely controlled by two repeated operations -- intertwining action and incorporating some generalized commutator relations -- for renormalizing the superpartner equation of the present iteration level eigen-equation into the eigen-equation of next level. These algorithms therefore give us a simple SUSYQM answer to the question regarding why, for the {\it same} hypergeometric-like differential operator, there exist simultaneously a series of the principal as well as the associated eigenfunctions in their canonical forms, which boils down to the two basic observations: 1) {\it two} distinct types of quantum momentum kinetic energy operators and two superpotentials  are rooted in this  operator; 2) each initial superpotential can proliferate into a hierarchy of descendant superpotentials in \emph{a shape-invariant fashion}. The two active supersymmetrizations and their accompanying momentum operator maps establish the isomorphisms between the \emph{nonstandard} as well as \emph{standard} coordinate representations of the SUSYQM algorithms of generic level for the principal hypergeometric-like operator and its associate, so these algorithms can be constructed in either coordinate representation with equal efficiency. In comparison with the widely used {\it passive differential} method for passing to a Schr\"odinger type equation, the method of {\it algebraic} supersymmetrization is much simpler and more efficient in building the SUSYQMs  not only for the hypergeometric-like operators and their associates but also for the more complicated Sturm-Liouville-type differential operators. Due to their relatively high efficiency, algebraic elementariness and logical independence, the iterative SUSYQM algorithms developed in this paper could become the hopefuls for supplanting some traditional  methods for solving the eigenvalue problems of the hypergeometic-like differential operators and their associated cousins.

 \end{abstract}
% \pacs{}
% \keywords{supersymmetrization}

\maketitle

\section{Introduction}
Hypergeometric-type differential operators appear widely in mathematics, physics and engineering. One famous prototypical eigen-equation of such type of operators is Euler's hypergeometric equation \cite{Abramowitz_Handbook_Math_functions_1965}
\begin{eqnarray}\label{Eulerian_HDE}
 -\cubr{x(x-1)\frac{d^2}{d x^2} +\sqbr{(a+b+1)x-c}\od{}{x}}\Phi = ab \Phi
\end{eqnarray}
where $a$, $b$ and $c$ are constants. Its solution (often denoted by $_2F_1(a, b; c; x)$ in literature) can be expressed as a hypergeometric series, which earns this equation  such a name. 
Lots of linear differential equations of second order either directly belong or can be converted to this equation. For instance, the equation \cite{Szego_OrthoPolynomials_1939}
\begin{eqnarray}\label{Jacobi_DE}
 &&\cubr{y^2-1)\frac{d^2}{d y^2} + \sqbr{(\alpha+\beta+2)y+\alpha-\beta}\od{}{y}}\Phi \\
 &=& n\smbr{n+\alpha+ \beta +1} \Phi, \nonumber
\end{eqnarray}
whose solution defines the Jacobi polynomials $P^{(\alpha,\beta)}_n$ for integer $n$ and constants $\alpha$ and $\beta$, can be transformed into \eq{Eulerian_HDE}, by setting $y=1-2x$, $a=-n$, $b=n+\alpha+\beta+1$ and $c=\alpha+1$. The Legendre, Gegenbauer, radial Zernike and Chebyshev polynomials are the special cases of $P^{(\alpha,\beta)}_n$.  In terms of this type of Jacobi polynomials, the eigenfunctions of the Schr\"odinger equations can be expressed  for several analytically solvable potentials, such as trigonometric and hyperbolic Rosen-Morse, trigonometric and hyperbolic Scarf,  Eckart and generalized P\"oschl-Teller potentials (see these in \cite{Cooper_SUSY_and_QM_1995} and the references therein).

The other prototypical and widely encountered hypergeometric-type differential equation is the confluent hypergeometric differential equation \cite{Abramowitz_Handbook_Math_functions_1965}
\begin{eqnarray}\label{Confluent_HDE}
 \sqbr{x\frac{d^2}{d x^2} +\smbr{m-x}\od{}{x}}\Phi = n \Phi
\end{eqnarray}
for constants $m$ and $n$, which is a limiting case of \eq{Eulerian_HDE} in the sense that two regular singular points of \eq{Eulerian_HDE} are confluent into one. The famous associated (generalized) Laguerre differential equation is of this type, whose solution -- the associated Laguerre polynomials -- are utilized to analytically express the radial wave functions of Coulomb potential for an electron without the spin effect (this can be seen almost in any text of quantum mechanics of medium or advanced level) and the (radial) wave functions of Morse potential \cite{Landau_QM_1977, Dong_Factorization_methods_in_QM_2007}, the potential of 3-D isotropic harmonic oscillator \cite{Cooper_SUSY_and_QM_1995} and the pseudo-harmonic oscillator \cite{Dong_Factorization_methods_in_QM_2007}.
The Hermite polynomials and even the Bessel-type functions are also closely related to the solutions of \eq{Confluent_HDE}.

The modest common generalization of Eqs. \eqnm{Eulerian_HDE} and \eqnm{Confluent_HDE} is the following {\it hypergeometric-like} differential equation
\begin{eqnarray}\label{eigen_equ}
 H_0 \Phi_l =\lambda_l \Phi_l
\end{eqnarray}
for the hypergeometric-like differential operator
\begin{eqnarray}\label{H_0_operator}
H_0 \equiv- p(x) \frac{d^2 }{dx^2} - q(x) \od{}{x},
\end{eqnarray}
where $p(x)= (p''/2) x^2+ p'_0 x+ p_0$ is at most a binomial in $x$ and $q(x)=q' x + q_0$ is at most linear in $x$, $p_0\equiv p(x=0)$, $p'_0 \equiv p'(x=0)$, $q_0\equiv q(x=0)$ and the prime denotes the derivative with respect to $x$. Usually \eq{eigen_equ} has a series of discrete eigen-solutions with $\lambda_l$  the eigenvalues and $\Phi_l$ the  eigen-functions, labeled by an integer $l$.

For every $H_0(x)$, besides \eq{eigen_equ} that is termed the \emph{base} or \emph{principal} equation, there exists also the {\it associated} differential equation. One famous example is that, for the special Legendre differential operator $H_0(x)$ for which $p(x)= 1-x^2$ and $q(x)=-2 x$, there exist the base as well as the associated Legendre differential equations; traditionally the latter can be derived from the former by sequentially $m$-fold differentiating the former, multiplying with $\sqrt{p}^m$ and moving this multiplier next to the $m$-fold differential operator. In much the same fashion, the associated hypergeometric-like differential equation of the generic $H_0(x)$ in \eq{H_0_operator} was derived from \eq{eigen_equ} in \cite{jafarizadeh_SUSY_and_diff_equs_1997, Cotfas_SI_raising_lowering_hypergeometric_equations_2002}.  In this paper, a pure \emph{algebraic} algorithm of deriving the associated hypergeometric-like differential equation and its eigen-solution associated with $H_0(x)$ will be provided, which are motivated by the principles of supersymmetric quantum mechanics (SUSYQM).

In view that the hypergeometric-like differential equation \eqnm{eigen_equ} and its associated counterpart appear so widely in various realms of mathematics, physics and engineering, a natural question is how to solve it efficiently. Historically, the two types of equations \eqnm{Eulerian_HDE} and \eqnm{Confluent_HDE} are solved prevailingly by the {\it analytical} Frobenius method, i.e. the eigen-functions are obtained as various types of special functions in the forms of power series. These series can terminate at some finite orders into polynomials under special conditions. 
Despite of its prevalence and power, many people think that such power series method can not be counted as either an elegant or simple solving technique. 

People then attempted to search for {\it algebraic} ways of solving this type of differential equations. A monumental development along this course is the discovery of the \emph{factorization methods} that can be applied to many eigen-differential equations of second order, appearing in mathematics, physics and engineering. In such algebraic approach each differential equation is equivalently split into a pair of differential equations of first order that involve the eigen-functions of two nearest neighboring indices. Many of these methods were summarized in the classical paper \cite{Infeld_factorization_methods_1951}.

The factorization methods have also been directly applied to solving the bound-state problems of the Schr\"odinger equation for a large class of potentials in quantum mechanics. The earliest such example probably dates back to applying the ladder operator method to obtaining the spectrum of 1-D quantum harmonic oscillator by Dirac and Schr\"odinger. Indeed, the various types of methods of factorizing the Sturm-Liouville type of eigen differential equations (see Appendix \ref{generalized_DO}) addressed in \cite{Infeld_factorization_methods_1951} were not directly applied to these equations themselves but to their converted Schr\"odinger-type equations. A more recent reference on the applications of factorization methods in solving various potential problems of quantum mechanics is \cite{Dong_Factorization_methods_in_QM_2007}, which also includes quite comprehensive references and emphasizes the implementations of the dynamic Lie algebras $su(1,1)$ and $su(2)$ in terms of the ladder operators for various Hamiltonians and the applications of these factorizations in quantum control. 

For the purpose of being self-contained, the main idea of the of factorization method that is applied to quantum mechanics problems and leads to the concept of supersymmetric quantum mechanics is outlined here. Many linear eigenvalue problems of higher dimensions can be reduced to the eigenvalue problem of a $1$-D Schr\"odinger equation with Hamiltonian $H_l(y)\equiv -d^2/d y^2 + V_l(y)$ (represented in $y$-coordinate and in the units $\hbar=2M=1$) in its factorized form
\begin{eqnarray}\label{Schrodinger_equ_introduction}
 H_l \psi_l(y) &=& A^\dagger_l A_l \psi_l(y)= E_l \psi_l(y), \\
 A_l&\equiv& \od{}{y}+ W_l(y),\quad A^\dagger_l \equiv -\od{}{y}+ W_l(y),\nonumber 
\end{eqnarray}
where the Hamiltonian $H_l$, potential $V_l(y)$, eigenvalue $E_l$, eigen-function $\psi_l(y)$, ladder operators $A_l$ and $A^\dagger_l$, and the so-called superpotential $W_l(y)$ are labeled by a discrete integer $l$; $W_l(y)$ is related to the potential $V_l(y)$ through the Ricatti equation $V_l(y)=-d W_l/dy+ W^2_l$. {\it Intertwiningly acting} \eq{Schrodinger_equ_introduction} with $A_l$ ( a formally {\it leftward braiding} in the operator product) yields the \emph{superpartner} eigen-equation
\begin{eqnarray}\label{partner_Schrodinger_equ_introduction}
 H^s_l \psi^s_l(y) &\equiv& A_l A^\dagger_l  \psi^s_l(y)= E_l \psi^s_l(y), \\
 \psi^s_l(y)&\propto& A_l \psi_l(y) ,\nonumber 
\end{eqnarray}
where the superpartner Hamiltonian $H^s_l \equiv-d^2/dy^2+ V^s_l(y)$ with the superpartner potential $V^s_l(y)\equiv d W_l/dy+ W^2_l$. 

One key feature is that the Hamiltonian pair $H_l$ and $H^s_l$ in Eqs. \eqnm{Schrodinger_equ_introduction} and \eqnm{Schrodinger_equ_introduction} have  common eigenvalue $E_l$. This degeneracy allows us %Witten \cite{Witten_SUSY_breaking_1981} 
to pack Eqs. \eqnm{Schrodinger_equ_introduction} and \eqnm{partner_Schrodinger_equ_introduction}  into one $2$-D matrix eigen-equation ${\bf H}_l {\boldsymbol\psi}_l(y)= E_l {\boldsymbol\psi}_l(y)$ for the diagonal matrix Hamiltonian ${\bf H}_l \equiv diag\{A^\dagger_l A_l,\,A_l A^\dagger_l \}$ and the Pauli spinor ${\boldsymbol \psi}_l(y)\equiv \smbr{\psi_l(y),\,\psi^s_l(y)}^T$.  It is easy  to check that 
${\bf H}_l=Q^+_l Q^-_l + Q^-_l Q^+_l \equiv \cubr{Q^+_l,\,Q^-_l}$ for the two supercharges $Q^+_l \equiv A^\dagger_l \sigma^+$ and $Q^-_l \equiv A_l \sigma^-$ , and $\sigma^\pm\equiv \smbr{\sigma_1 \pm i \sigma_2}/2$ for the Pauli matrices $\sigma_{1,2}$. Furthermore, ${\bf H}_l$ and $Q^\pm_l$  form the ${\cal N}=2$ supersymmetric Lie algebra $su(1,\,1)$ \cite{Witten_SUSY_breaking_1981, Cooper_SUSY_and_QM_1995}, as they satisfy the commutator/anticommutator relations $\sqbr{Q^{\pm}_l,\, {\bf H}_l}=0$ and $\cubr{Q^+_l,\,Q^+_l}=\cubr{Q^-_l,\,Q^-_l} = 0$. Because of such supersymmetry (SUSY for short), this quantum mechanics is termed supersymmetric quantum mechanics and the prefix {\it super} frequently appears in the relevant terminologies.

Gendenshte\"in recognized \cite{Gendenshtein_SUSY_Schr_Equ_1983} that, for a broad class of potentials in quantum mechanics, the superpartner potential $V^s_l$ and the original potential $V_l$ belong to the {same type} of function series. Particularly,  $V^s_l$ differs from the nearest neighboring potential, say $V_{l+1}$, merely by a constant $\delta_l$, i.e. $V^s_l = V_{l+1} - \delta_l$ -- a specific symmetry among the potential series $\{V_l\}$. In applications we find out that it is more convenient to translate this type of potential relation into the {\it generalized commutator} relation $A^\dagger_{l+1} A_{l+1} - A_l A^\dagger_l=\delta_l$ (we will come to concrete examples in later sections and will see that this relation is the nontrivial generalization of the commutator relation for the 1-D quantum harmonic oscillator). Such commutator relation allows us to rearrange the supepartner eigen-equation \eqnm{partner_Schrodinger_equ_introduction} into the {\it normally ordered form} 
\begin{align}
&A^\dagger_{l+1} A_{l+1} \psi_{l+1}(x)= E_{l+1} \psi_{l+1}(x),\\
&\psi_{l+1}(x) \propto A_l \psi_l(y),\quad E_{l+1}= E_l + \delta_l,
\end{align}
which is just the $(l+1)$-version of the original eigen-equation  \eqnm{Schrodinger_equ_introduction}. Here by the normally ordered form we mean that in such a form the daggered operator sits on the immediate left of the undaggered one and all subscripts are the same. It is for this reason that such a symmetry among the potential series $\{ V_l\}$ is called shape (pattern) invariance symmetry (SISY for short). Since the eigenvalue renormalization accompanies this rearrangement, such a process is also called \emph{renormalization}.

It is quite amazing that breaking a system of differential equations of {\it second} order equivalently into another system of coupled differential equations of {\it first} order, often brings to light more underlying symmetries carried by this system. One such famous paradigm is that the symmetry of symplectomorphism in classic mechanics carried by the first-order Hamiltonian formulation becomes manifest, when passing from the second-order Lagrangian or Newtonian formulation. In SUSYQMs we see another such paradigm, i.e. appropriately splitting a family of second-order differential equations equivalently into a series of raising and lowering recurrence relation pair $\Psi_{l+1}(x) \propto A_l \Psi_l(x)$ and $\Psi_{l-1}(x) \propto A^\dagger_l \Psi_l(x)$, which are just coupled differential equations of first order if $A_l$ and $A^\dagger_l$ are viewed respectively as the raising and lowering operators, actually brings to the fore the underlying SUSY and SISY.

The level-crossing SISY together with the SUSY for each fixed $l$-th level allow us to build a hierarchy of SUSYQMs and to exhaust the spectra of series of Hamiltonians iteratively by {\it repeatedly performing the leftward braiding action and renormalization}, and solving a trivial first-order differential equation, which could be either the annihilation condition $A_0 \psi_0(x)=0$ of the initial level that corresponds to $E_0=0$, or the annihilation condition $A_{l_{max}} \psi_{l_{max}}(x)=0$ of the top level that corresponds to $E_{l_{max}+1}=0$. This algorithm of solving eigenvalue problems of Schr\"odinger equations and other types of differential equations of second order that can be cast into Schr\"odinger equations is termed {\it SUSYQM algorithm}, and it will be applied to solving the hypergeometric-like differential equation and its associated counterpart in this paper.

At the advent of SUSYQM, a few people have tried to exhibit the SUSY and SISY hidden in the hypergeometric-like differential operator $H_0$ and its associated counterpart by constructing the factorizing schemes of these operators \cite{jafarizadeh_SUSY_and_diff_equs_1997, Cotfas_SI_raising_lowering_hypergeometric_equations_2002}. However, since they mainly aimed at looking for some successful factorizations  of these operators rather than developing the algorithms of solving these eigenvalue problems that are based {\it purely} on SUSYQMs, at various points either certain results out of the traditional methods were referred to as the key inputs for going forward  or some traditional methods themselves were directly made use of. 
For instance, in \cite{jafarizadeh_SUSY_and_diff_equs_1997, Cotfas_SI_raising_lowering_hypergeometric_equations_2002}, the general Rodriguez formula for classical polynomials as the eigen-solutions, which is supposed to come out of some traditional (such as the Frobenius and the generating function) methods rather than the factorization algorithms themselves, is directly referred to either make some key arguments or derive the recurrence relations for factorizing $H_0$ or its associated cousin. Hence, logically their factorization methods can not be counted as the genuine SUSYQM algorithms, that are completely independent from the traditional ones, for finding the eigen-solutions of these operators.

In \cite{jafarizadeh_SUSY_and_diff_equs_1997, Cotfas_SI_raising_lowering_hypergeometric_equations_2002}, the associated hypergeometric-like differential equation for $H_0$ and its eigenfunction were first  derived from the base equation \eqnm{eigen_equ} in the traditional way -- acting \eq{eigen_equ} with the differential operator $\smbr{d/dx}^m$ for $m$ being a positive integer and moving the multiplier $\sqrt{p}^m$ next to the operator $\smbr{d/dx}^m$. The authors \cite{jafarizadeh_SUSY_and_diff_equs_1997, Cotfas_SI_raising_lowering_hypergeometric_equations_2002} then searched for the methods of directly factorizing the derived associated hypergeometric-like equation with generic association level. This {\it two-stage} approach of factorizing the associated hypergeometric-like differential operator is fairly cumbersome because this equation derived by the traditional {\it differential} method often does not come out in such a well-organized form that the underlying SUSY and SISY are manifest. It is then highly desired to construct a certain {\it one-stage} yet quite simple algorithm for solving the eigenvalue problem of the associated hypergeometric-like differential operator. Such an ideal algorithm is expected to come out with the following four features: 1) a well-organized associated differential equation of $H_0$ with generic association level should comes out step by step {\it iteratively}, starting from factorizing the much simpler principal equation \eqnm{eigen_equ}; in other words, we do not need to derive the messy associated hypergeometric-like differential equation of generic association level from \eq{eigen_equ} in the traditional way, let alone directly factorize it; 2) to move to each higher association level iteratively, the associated differential equation itself and its eigen-solution follow from those of the previous association level, in such an efficient and {\it well-organized} algebraic manner that not only the two symmetries -- SUSY and SISY -- are manifest, but also they should work as the {\it two guiding principles} for constructing this algorithm; 3) the construction of this algorithm will not consult any traditional method or resort to any result out of it at all, but only to SUSYQM; 4) the algorithm should be both algebraically and conceptually elementary enough so that it has the chance to compete with or even to supplant the traditional solving techniques. It is also highly desired to have a systematic algorithm for solving the eigenvalue problem of the base differential equation (\ref{eigen_equ}),  that also comes out with these features.  These algorithms,  if any, can therefore be counted as pure {\it SUSYQM algorithms} and truly work as the \emph{independent} alternatives to or even the substitutes for the traditional methods. We will see that just because we lift the SUSY and SISY to the height of two guiding principles, such types of independent SUSYQM algorithms are successfully obtained.

There is an another consideration which justifies the worth of pursuing such type of SUSYQM algorithms. Although {\it mathematically} it is well-known that the associated Legendre equation can be derived from its base counterpart by $m$-fold differentiating the latter, and the method has also been applied to deriving the more general associated hypergeometric-like differential equation from \eq{eigen_equ}, one may still have the impression that this method seems to be motivated from nowhere, and such {\it differential} connection between the two types of hypergeometric-like equations is far from being obvious. Even from today's perspective after many generations of education on the theories of the base and associated Legendre functions, as the eigen-functions of the Legendre differential operator, the recognition of the differential connection between these two types of special functions still requires amazing mathematical intuition and imagination. A more natural and elementary {\it algebraic} approach of establishing this type of connection, if any, would be a blessing. 

Furthermore, even though  there is already such a pure mathematical proof of this differential connection or  there may be some algebraic ways of establishing such connection, one may still ask the innocent questions, e.g., why do there exist in general two such distinct types of eigenfunctions for the {\it same} hypergeometric-like differential operator $H_0$? Is there any more straightforward and shorter answer to this question other than the traditional \emph{high-tech} mathematical proof? It turns out that such simple answer does exist, that is full of {\it SUSYQM flavor}. 

The above three aspects of considerations motivate us to develop some systematic SUSYQM algorithms that not just works to demonstrate the intrinsic symmetry properties (SUSY and SISY) owned by the base as well as the associated hypergeometric-like differential operators, but more importantly to serve as the economical approaches to solving the discrete eigenvalue problems of these two types of operators. Such algorithms are expected not only to be completely removed from the traditional solving methods or from any derived result out of them, but also to be both algebraically and conceptually elementary enough in order to compete with the traditional methods, hence to be palatable to freshmen or even to high school students, once they have been exposed to the elementary calculus involved. For this latter purpose and the clarity, the presentation of this paper will try to be as pedagogic as possible. The pre-knowledges on the SUSYQM other than those mentioned in this introduction are not assumed for understanding the material in this paper.

The rest of this paper mainly falls into four sections. 

In Section \ref{susyqm_H_like_functions}, a systematic SUSYQM algorithm of solving the base \eq{eigen_equ} is developed. This section consists of five subsections. In Subsection \ref{base_kinetic_energy_operator}, aiming at preparing the squared form $-(p d/dx)^2$, an algebraic supersymmetrization transformation ${\cal T}$ is devised (in four ways) to recast the operator $p H_0(x)$ into its supersymmetric factorization. Via this transformation, the superpotential $W_0(x)$, the momentum operator $-i p\, d/dx$ and the rescaling factor of the eigen-function all emerge simultaneously in a simple algebraic way. In Subsection \ref{Standard_Hermitian_factorization}, the same transformation ${\cal T}$ is then applied to the Hamiltonian ${\mathbb{H}}_l(x)$ and its standard form of Schr\"odinger equation is reached in $y$-coordinate defined by the momentum operator map $-i p\, d/dx=-i d/dy$. Also the formal supersymmetric factorizations of this Schr\"odinger equation and its superpartner equation are given. In Subsections \ref{minus_route_iteration} and \ref{upward_SIC}, we show that, corresponding to both the downward and the upward connected shape invariance conditions  for the potential series $\{V_l(y)\}$, there actually exist {\it two} routes of supersymmetric factorizations of the same ${\mathbb{H}}_l(x)$, which are slightly different from and closely related to each other. The algorithms of pinning down (in one go as well as in some bottom-up procedures of iteration) the details of the supersymemtric factorizations, such as the energy levels $E^\mp_l$, the superpotentials $W^\mp_l$, the eigenvalues $\lambda^\mp_l$ and the constants $\delta^\mp_l$ appearing in the shape invariance conditions, are also provided in these two subsections. In Subsection \ref{non_standard_factorization}, we list the main steps of constructing the SUSYQM algorithm for solving the eigenvalue problem of ${\mathbb{H}}_l(x)$  directly within the original $x$-coordinate, and the one-to-one correspondences between the representation of this algorithm in $y$-coordinate and that in $x$-coordinate are pointed out. Two other eigen-equations, that involve the more general hypergeometric-like differential operator $H_l(x)$, and are  equivalent to the base hypergeometic-like equation \eqnm{eigen_equ}, are derived in Subsection \ref{equivalent_eigen_equ}.

In Section \ref{associated_hypergeometric_functions}, a systematic iterative SUSYQM algorithm for simultaneously building a hierarchy of associated hypergeometric-like differential equations and their eigen-solutions is developed. This section is made up of five subsections. In Subsection \ref{2nd_type_momentum_op}, parallel to ${\cal T}$ introduced in Subsection \ref{base_kinetic_energy_operator}, another algebraic active supersymmetrization transformation ${\cal S}$ is devised to directly supersymmetrize the asymmetric factorization of $H_0$ into another supersymmetric factorization with $W_0^a$ being the superpotential for the lowest association level, aiming at preparing $-i \sqrt{p}d/dx$ -- another type of momentum operator. In addition, the one-to-one correspondence between the nonstandard $x$-coordinate representation and the standard $z$-coordinate representation of this supersymmetric factorization of $H_0$ is explicated.  Respectively in Subsections \ref{positive_association_levels} for positive association level and \ref{negative_association_levels} for negative association level, within the non-standard {\it x}-coordinate representation, a hierarchy of associated hypergeometric-like equations and their eigen-solutions are built simultaneously in a bottom-up iterative approach, starting from the two slightly different initial supersymmetric factorization of the base differential equation \eqnm{eigen_equ}, and proceeding with the requirement that the SUSY is manifested at each iteration step and SISY is ensured from one association level to the next one. The main steps of passing from the $x$-coordinate representation to the standard $z$-coordinate representation of the SUSYQM algorithms for building the associated hypergeometric-like functions via the combination of the momentum operator map and the active supersymmetrization transformation of the second type is then outlined in Subsection \ref{associated_standard_reps}. An eigen-equation that is of the \emph{principal} hypergeometric type while equivalent to the usual \emph{associated} hypergeometric-like differential equation is derived in Subsection \ref{equivalent_associated_eigen_equ} by a certain partially asymmetrizing transformation. Another type of factorization of the associated hypergeometric-like differential equation is achieved in Subsection \ref{another_factorization_associated_eigen_equ}, in the same spirit of factorizing the principal hypergeometric-like differential equation, i.e. aiming at preparing the kinetic energy operator $-\smbr{p d/dx}^2$. This section is closed by making some remarks in Subsection \ref{some_remarks_associated_eigen_equ}.

The degenerate case in which the associated hypergeometric-like differential operators collapse into the corresponding principal counterparts, which are of Hermite type, is addressed in Section \ref{degenerate_cases}. In Subsection \ref{operator_algebra_dg}, under special conditions, it is demonstrated that how the ladder operators of the associated hypergeomtric-like operators and their principal counterparts collapse into the same types, so that the shape invariance conditions and the eigen-solutions of these two types of operators also coincide. In Subsection \ref{eigen_solution_dg}, the eigen-solutions of these Hermite types of operators are obtained by running the systematic SUSYQM algorithm.

Summary and conclusions are given in Section \ref{summary}. Particularly, two flow charts of running the same standard SUSYQM algorithm for generating the principal hypergeometric-like equation of generic level as well as its associated counterpart are provided, which give us a unified picture on why there exist the base as well as the associated hypergeometric-like functions for the same hypergeometric-like differential operator $H_0$. In Appendix \ref{solution_Wl_and_El}, the calculation of various coefficients and eigenvalues for the SUSYQM of the base hypergeometric-like functions by directly matching the two expressions of the same binomial potential $V_l(x)$ is offered. In Appendix \ref{generalized_DO}, our method of active supersymmetrization is applied to turn the generic Sturm-Liouville eigen-equations into two types of standard Schr\"odinger equations in their partial or full supersymmetric factorizations. 

\section{Iteratively Building the SUSYQM Related to the Base Hypergeometric-like Functions}\label{susyqm_H_like_functions}
In this section, we are going to find out the regular {\it base} solution of \eq{eigen_equ} in a purely algebraic way, i.e., by developing its SUSYQM algorithm.

\subsection{Preparing the Kinetic Energy Operator  \mathinhead{\hat{p}^2}{} and Disguising the First-order Differentiation via Active Supersymmetrization}\label{base_kinetic_energy_operator}
In order to disclose the close affinity between the hypergeometric-like differential equation \eqnm{eigen_equ} and its solutions and the underlying supersymmetric quantum mechanics, the natural thing above all to do is to construct certain Hamiltonians from the operator $H_0$ given in \eqnm{H_0_operator}, by performing some simple algebraic manipulations and transformations on $H_0$ or \eq{eigen_equ}. To this end, the first step is to directly construct certain operators from $H_0$ that contain the squared forms  $\hat{p}^2$ for some differential operators $\hat{p}$ of first-order, so that these squared forms can be identified with the quantum kinetic energy operators, once these $\hat{p}$ are mapped to the standard quantum momentum operators in some new coordinates, just like the squared form $-\smbr{d/dy}^2$ in the Hamiltonian $H_l$ in \eq{Schrodinger_equ_introduction} is the standard kinetic energy operator and $-i d/dy$ the standard momentum operator for $i$ being the unit imaginary. Meanwhile, the reminder consisting of pure function and constant inside the constructed operators can be identified with the potential minus the eigenvalue. In doing so, the standard Schr\"odinger equations equivalent to \eq{eigen_equ} can be reached. Based on these Schr\"odinger equations, the structures of SUSY and SISY can be further explored.

The squared form $\hat{p}^2$ is a product of {\it two identical} differential operators of first order. This {highly symmetric} double-layer product reminds that, in order to prepare the kinetic energy operator, one can {\it symmetrize certain asymmetric product of two first-order differential operators} that can be easily constructed from the operator $H_0$. Note that
\begin{eqnarray}
  H_0=- \od{}{x} p \od{}{x} +\smbr{p'-q}\od{}{x},
\end{eqnarray}
where the first term is made of an asymmetric product of $-d/dx$ and $p d/x$. This form of $H_0$ reminds us to simply multiply it with $p$ from left hand side for preparing the (symmetric) squared form $\hat{p}^2 \equiv-\smbr{p d/dx}^2$ and construct the trivial asymmetric factorization 
\begin{eqnarray}\label{pH0}
 pH_0 &=& \sqbr{-p\od{}{x}+ 2 W_0(x)} \smbr{p\od{}{x}}, \label{simple_factorization} \\
      &=& w^{-1}\smbr{- p\od{}{x}} w \smbr{p\od{}{x}},\label{pH0_times_selfadjoint}
\end{eqnarray}
in which each factor comes with the common piece $p\, d/dx$,
\begin{eqnarray}\label{W_0}
 W_0(x)\equiv\frac{1}{2}(p'-q)
\end{eqnarray}
and 
\begin{eqnarray}\label{w_integration_factor}
 w(x) \equiv \exp\smbr{-\int d x \frac{2 W_0}{p}},
\end{eqnarray}
which is the integration factor of the first differential operator inside the square brackets in \eq{simple_factorization}. Accordingly, we will deal with the new equivalent differential equation
\begin{eqnarray}\label{eigen_equ_with_multiplier}
 {\mathbb{H}}_l(x) \Phi_l\equiv\sqbr{p(x)H_0 - \lambda_l p(x)+E_l}\Phi_l = E_l \Phi_l,
\end{eqnarray}
which is obtained by multiplying \eq{eigen_equ} with $p$ and adding the term proportional to the new eigenvalue $E_l$ on both sides of it.

In what immediately follows, we will discuss how to symmetrize the asymmetric form ${\mathbb{H}}_0(x) \equiv pH_0$ in \eq{simple_factorization} or \eq{pH0_times_selfadjoint}, i.e. the $\lambda_l=E_l=0$ case of ${\mathbb{H}}_l(x)$ in \eq{eigen_equ_with_multiplier}, as a preparation for factorizing the more general operator ${\mathbb{H}}_l(x)$ in the next subsection. Note that the second term  in \eq{simple_factorization} -- an unpleasant differential operator of first order -- should be swept under the rug somehow.  

By some trivial and straightforward  manipulations, the asymmetric factorization \ref{pH0_times_selfadjoint} is turned into
\begin{eqnarray}
pH_0 &=&\sqrt{w}^{-1}\sqbr{\sqrt{w}^{-1}\smbr{- p\od{}{x}} \sqrt{w}}  \label{proliferated_pH0} \\
&&\cdot\sqbr{\sqrt{w}\smbr{p\od{}{x}}\sqrt{w}^{-1}}\sqrt{w}. \nonumber
\end{eqnarray}
This form of $pH_0$ reminds us to introduce the {\it wrapper} or bilateral transformations
\begin{equation}\label{T_transformation}
 {\cal T}\smbr{-} \equiv \sqrt{w}\smbr{-}\sqrt{w}^{-1},\quad
 {\cal T}^{-1}\smbr{-} \equiv \sqrt{w}^{-1}\smbr{-}\sqrt{w},
\end{equation}
that are used to wrap up $\pm p d/x$ and to denote \eq{proliferated_pH0} as 
\begin{eqnarray} \label{pH01}
pH_0 &=&{\cal T}^{-1}\cubr{\sqbr{{\cal T}^{-1}\smbr{-p\od{}{x}}}\sqbr{{\cal T}\smbr{p\od{}{x}}}},
\end{eqnarray}
or, equivalently,
\begin{eqnarray}\label{pH0_supersymmetrized0}
{\cal T}\smbr{pH_0} &\equiv& \sqbr{{\cal T}^{-1}\smbr{-p\od{}{x}}}\sqbr{{\cal T}\smbr{p\od{}{x}}}. 
\end{eqnarray}

With $w$ given in \eq{w_integration_factor}, it is easy to check that
\begin{equation}\label{annihilation_condition0}
 \begin{aligned}
 \sqbr{p\od{}{x}+ W_0}\sqrt{w}=0,\;  \sqbr{-p\od{}{x}+ W_0}\sqrt{w}^{-1}=0,
 \end{aligned}
\end{equation}
where $W_0(x) = -p \smbr{\log \sqrt{w}}'$. Consequently,
\begin{eqnarray}\label{T_inverse_minus}
{\cal T}^{-1}\smbr{-p\od{}{x}}=- p\od{}{x}+ W_0(x)
\end{eqnarray}
and
\begin{eqnarray}\label{T_plus}
{\cal T} \smbr{p\od{}{x}}= p\od{}{x} + W_0(x).
\end{eqnarray}
 Eqs. \eqnm{T_inverse_minus} and \eqnm{T_plus} essentially say that $\sqrt{w}^{\pm 1}$, that form the double-sided transformations ${\cal T}$ and ${\cal T}^{-1}$, are the integration factors of the two two-term linear operators on their right-side hands. Formally, ${\cal T}^{-1}$ works to shift $-pd/dx$ and ${\cal T}$ to shift $pd/dx$ by {\it the same amount}  $W_0(x)$. 
 
 \eq{pH0_supersymmetrized0} is then rewritten as the following {\it supersymmetrically factorized} form
\begin{eqnarray}
 {\cal T} \smbr{pH_0}&=&\sqbr{- p\od{}{x}+ W_0(x)}\sqbr{p\od{}{x}+ W_0(x)}, \label{pH0_supersymmetrized}\\
                     &=&-\smbr{p\od{}{x}}^2+ V_0(x),\label{pre_Hamiltonian_H0}
\end{eqnarray}
 where $W_0(x)$ is the so-called {\it superpotential} for the potential 
\begin{eqnarray}\label{Riccati_equ_V0}
 V_0(x)\equiv - p\od{W_0(x)}{x}+W^2_0(x),
\end{eqnarray}
 which is in general a {\it binomial} in $x$ in view that $W_0(x)$ is {\it linear} in $x$. In \eq{pH0_supersymmetrized}, the two factors, each of which is made of two-term first-order differential operator, look almost the same, particularly they share the superpotenital $W_0(x)$, except that inside the first factor the pure differential operator carries an extra minus sign. This particular factorization guarantees that the first-order derivatives cancels automatically after expansion. As it turns the asymmetric factorization \eqnm{simple_factorization} of $p H_0$ into the supersymmetric one \eqnm{pH0_supersymmetrized}, the transformation ${\cal T}$ is termed the {\it supersymmetrization} transformation. %, which is the so-called Ricatti equation that relates $W_0(x)$ with $V_0(x)$.

We have shown that when $-pd/dx$ is wrapped up with the transformation ${\cal T}^{-1}$ and simultaneously $pd/dx$ with ${\cal T}$, as shown on the right hand side of \eq{pH0_supersymmetrized0}, their product is the supersymmetric factorization \eqnm{pH0_supersymmetrized}. There is a more straightforward reason why ${\cal T}$ alone can directly  supersymmetrize the asymmetric factorization \eqnm{simple_factorization} of $p H_0$. Note that the action of $\cal T$ is distributive over each factor of an operator  product and does nothing on any ordinary function. The property \eqnm{T_plus}  implies that the action of $\cal T$ on $-pd/dx$ amounts to shifting it by $-W_0(x)$. Consequently,
\begin{eqnarray}
{\cal T}\smbr{- p\od{}{x}+ 2 W_0(x)}&=&- p\od{}{x} + W_0(x),\\
                                    &=&{\cal T}^{-1}\smbr{-p\od{}{x}}. \nonumber
\end{eqnarray}
Therefore, when ${\cal T}$ acts on $p H_0$ in its asymmetric factorized form \eqnm{simple_factorization},  the previous {\it unevenly} distributed term $2W_0(x)$ inside the first factor is now {\it evenly} distributed over the two factors, by subtracting $W_0$ from the first factor and adding the same amount to the second one. This way, the supersymmetric factorization \eqnm{pH0_supersymmetrized} is naturally reached, by the direct action of ${\cal T}$ on $p H_0$ in its trivially asymmetric factorization \eqnm{simple_factorization}.

Under the operator map
\begin{eqnarray}\label{momentum_map_type1}
  -i p\od{}{x}  = -i \od{}{y} \equiv \hat{p}_y
\end{eqnarray}
 with $i$ the imaginary unit and $\hat{p}_y$ the quantum momentum operator in the units $\hbar=2M=1$ ($\hbar$ is the Planck constant and $M$ the point mass) in some $y$-coordinate, 
 the squared term in \eq{pre_Hamiltonian_H0} becomes the standard quantum kinetic energy operator of a particle, i.e.
 \begin{eqnarray}\label{KE_map_type1}
 -\smbr{p\od{}{x}}^2=\hat{p}^2_y=-\frac{d^2}{dy^2}.
 \end{eqnarray}
The operator identifications \ref{momentum_map_type1} and \ref{KE_map_type1} are respectively referred to as {\it momentum} and {\it kinetic energy} operator maps. Note that the other operator map $i\, p\,d/dx = \hat{p}_y $ parallel to \eq{momentum_map_type1} also serves for this purpose, but this sign choice does not stick to the definition of the momentum operator in quantum mechanics.
 
 The operator map \ref{momentum_map_type1} is equivalent to the differential equation $dy/dx= p^{-1}(x)$, whose solution 
 \begin{eqnarray}\label{y_x_coord_trans}
 y(x)= \int^x \frac{d\tilde{x}}{p\smbr{\tilde{x}}}
 \end{eqnarray}
defines the coordinate transformation between $x$ and $y$-coordinates. 

Now under the momentum map \ref{momentum_map_type1} or the inverse of the coordinate transformation \ref{y_x_coord_trans}, the supersymmetric factorization \ref{pH0_supersymmetrized} represented in $x$-coordinate is mapped to the {\it standard} supersymmetric factorization in $y$-coordinate in the form
\begin{eqnarray}\label{cal_H_0}
  {\cal H}_0(y)&=&\sqbr{-\od{}{y}+{\cal W}_0(y)} \sqbr{\od{}{y}+{\cal W}_0(y)}, \\
              &=&-\frac{d^2}{d y^2}+ {\cal V}_0(y)
 \end{eqnarray}
 with the standard Hamiltonian ${\cal H}_0(y)\equiv {\mathbb  H}_0(x(y)) = {\cal T} \smbr{pH_0}$, the potential ${\cal V}_0(y)\equiv V_0(x(y))$  and the corresponding superpotential ${\cal W}_0(y) \equiv W_0(x(y))$.

The supersymmetric factorization \eqnm{pH0_supersymmetrized} can also be reached in the following straightforward manner. Recall that $pH_0$ given in \eqnm{pH0} is just a binomial of the {\it operator} $p d/dx$ without the zeroth order term. In order to transform this $pH_0$ into some standard quantum Hamiltonian operator, the unwanted first order differentiation (the second term) can be {\it disguised} through the trick of {\it completing the square} of this binomial, regarding this first-order differential operator as a cross term, i.e.
\begin{eqnarray}\label{pH0_symmetrized}
 p(x)H_0 &=& -\sqbr{-p\od{}{x} + W_0(x)}^2 - p \od{W_0(x)}{x}  + W^2_0(x), \nonumber \\
\end{eqnarray}
where the last two terms on the right hand side have been added as the compensating ones, which just constitute the potential $V_0(x)$ defined in \eq{Riccati_equ_V0}. Furthermore, upon finding the integration factor of the linear differential operator (i.e. making use of \eq{T_inverse_minus}) inside the square brackets and noticing that the action of ${\cal T}^{-1}$  is distributive over an operator product, we have
\begin{eqnarray}\label{squared_form}
 \sqbr{-p\od{}{x} + W_0(x)}^2&=&{\cal T}^{-1} \sqbr{\smbr{-p \od{}{x}}^2}.
\end{eqnarray}
Thus, applying the transformation ${\cal T}$ to $p H_0$ in \eq{pH0_symmetrized} directly yields \eq{pH0_supersymmetrized}.

Since ${\cal T}\smbr{p H_0}=p {\cal T}\smbr{H_0}$, \eq{pH0_supersymmetrized} can also be interpreted as the result of symmetrizing  a certain asymmetric double -layer product inside the operator ${\cal T}H_0 $ by the left-multiplier $p$. In other word, there must be the asymmetric double-layer structure $-{\cal T}^{-1} \sqbr{\smbr{d/dx} p \smbr{d/dx} }$ inside $H_0$. In fact, one can directly check that $H_0$ admits the splitting
\begin{eqnarray}\label{second_double_layer_splitting}
 H_0 &=&- \frac{1}{\sqrt{w}} \od{}{x} p \od{}{x} \sqrt{w} + \frac{\smbr{p\sqrt{w}'}'}{\sqrt{w}},\nonumber \\
     &=&-{\cal T}^{-1}\smbr{\od{}{x} p \od{}{x}} + \frac{\smbr{p\sqrt{w}'}'}{\sqrt{w}}.
\end{eqnarray}
Applying the combined transformation $p{\cal T}$ to the above form of $H_0$ also leads to \eq{pre_Hamiltonian_H0} with
\begin{eqnarray}\label{V01}
 V_0(x)=p \frac{\smbr{p\sqrt{w}'}'}{\sqrt{w}}.
\end{eqnarray}

Eqs. \eqnm{W_0} and \eqnm{w_integration_factor} together yield 
\begin{eqnarray}\label{w_definition1}
q=\frac{(w\,p)'}{w}.
\end{eqnarray}
Upon making this substitution and factorizing $d/dx$ out from the right side, $H_0$ in \eq{H_0_operator}  is then recast into the familiar self-adjoint form 
\begin{eqnarray}\label{self_adjoint_form}
H_0=-p\sqbr{\od{}{x}+\frac{\smbr{w p}'}{w p}}\od{}{x}=-\frac{1}{w} \od{}{x} w p \od{}{x},
\end{eqnarray}
where $w(x)$ is assumed to be well behaved so that it can work as the weight function to define the inner product in the Hilbert space of $H_0$, with respect to which the Hermitian adjoint can be defined. This self-adjoint form  of $H_0$ is another {\it asymmetric double-layer factorization} rooted in $H_0$, in addition to the first term in \eq{second_double_layer_splitting}. 

By left-multiplying the self-adjoint form of $H_0$ with $p$,  one also reaches the asymmetric factorization  \eqnm{pH0_times_selfadjoint}. This tells us that one can directly start with this self-adjoint form of $H_0$ and then follow the same rest route as in the first approach to reach the supersymmetric factorization \ref{pH0_supersymmetrized}. Again multiplication of $p$ ensures that the squared form $-\smbr{p d/dx}^2$ as the kinetic energy operator is prepared. 

If we set $\lambda_0=E_0=0$, then the $l=0$ case of \eq{eigen_equ_with_multiplier} is 
\begin{eqnarray}\label{initial_eigen_equ_assymmetric}
pH_0 \Phi_0(x)=\sqbr{- p\od{}{x}+ 2 W_0(x)} \sqbr{p\od{}{x}}\Phi_0(x)=0.  
\end{eqnarray}
Left-multiplying with $\sqrt{w}$ and making use of \eq{pH0_supersymmetrized}, this equation is then supersymmetrized into the form
\begin{eqnarray}\label{initial_eigen_equ_minus}
\sqbr{- p\od{}{x}+ W_0(x)}\sqbr{p\od{}{x}+ W_0(x)}\smbr{\sqrt{w}\Phi_0(x)}=0.
\end{eqnarray}
We see that in this process of supersymmetrization the eigenfunction has to be rescaled to a new one by the integration factor $\sqrt{w}$ ( the left/right side factor of $\cal T$/${\cal T}^{-1}$). One sufficient condition for this equation to hold is the following annihilation condition
\begin{eqnarray}\label{ground_state_minus_route}
 \sqbr{p\od{}{x}+ W_0(x)}\smbr{\sqrt{w}\Phi_0(x)}=\sqrt{w}p \od{\Phi_0(x)}{x}=0,
\end{eqnarray}
which has the solution $\Phi_0(x)=\text{const.}$ and is essentially the first equation in \eqnm{annihilation_condition0}. 
Later on, we will see that \eq{initial_eigen_equ_minus} will work as the starting (ground-state) eigen-equation for proliferating the generic eigen-equation \eqnm{eigen_equ_with_multiplier} by some systematic SUSYQM algorithm. Meanwhile, the solution $\Phi_0(x)$ will give birth to the solution to \eq{eigen_equ_with_multiplier}.

In the above we  have introduced a {\it pure algebraic} method, which is termed {\it active supersymmetrization}, to obtain the standard Hamiltonian ${\cal H}_0(y)$ that comes  in its supersymmetrically factorized form, from the hypergeometric-like operator $H_0(x)$, by firstly aiming at preparing the kinetic energy operator  $-\smbr{p d/dx}^2$ and the trivial asymmetric factorization \eqnm{simple_factorization} of $p H_0(x)$, and then appropriately pairing the simple algebraic transformation ${\cal T}$ that sandwiches $pd/dx$ and the inverse transformation ${\cal T}^{-1}$ that sandwiches $-p d/dx$ in the particular form \eqnm{pH0_supersymmetrized0}. These two bilateral transformations simply arise when lumping the two operators  $p d/dx \pm W_0$ via their integration factors. In such a simple algebraic manner, not only is the unpleasant first-order differential operator in \eq{pH0} disguised and the kinetic energy operator identified as in \eq{KE_map_type1}, but also the {\it superpotential} $W_0$, hence the supersymmetric factorization of $p H_0(x)$ are {\it actively} reached simultaneously. This is in sharp contrast to the traditional method that will be illustrated shortly, in which the first order differential operator is eliminated in a `hard' way to arrive at a certain Schr\"odinger equation and  the superpotential is then obtained by solving a Ricatti equation like \eq{Riccati_equ_V0} for given $V_0(x)$.
For this reason, this procedure is termed {\it active supersymmetrization}  or the {\it active supersymmetric factorization}. One more advantage of this method lies in that the rescaling factor of the eigenfunction, that is needed for transforming the eigen-equation of $p H_0$ into a certain standard Schr\"odinger equation in some new coordinate, is just the integration factor $\sqrt{w}$ (the one forming the transformation ${\cal T}$), so unlike in traditional approach, one does need to solve a differential equation satisfied by it. Such light algebraic method well fits in with the tone of quickly establishing the SUSYQM of some linear differential operator of second order, and will be made use of again to establish the SUSYQM of the associated hypergeometric-like functions of $H_0(x)$ in Section \ref{associated_hypergeometric_functions}, and to quickly transform a general Sturm-Liouville eigen-equation into some standard Schr\"odinger equations in Appendix \ref{generalized_DO}. Soon it will become clear that this active supersymmetric factorization of the Hamiltonian ${\cal H}_0(y)$ will give birth to the supersymmetric factorization of the generic operator ${\mathbb{H}}_l(x)$ as well as its eigenfunction in \eq{eigen_equ_with_multiplier} by running some iterative SUSYQM algorithm.

Traditionally, in order to find out the standard Hamiltonian ${\cal H}_0(y)$ corresponding to $pH_0$ in \eq{pH0} and to turn the eigen-equation $pH_0 \Phi_0(x)=0$ into a standard Schr\"odinger equation, the {\it ‘hard’} way of eliminating the first-order derivative is often followed (e.g., see \cite{Courant_and_Hilbert_1931, *Courant_and_Hilbert_1989} for transforming a generic Sturm-Liouville-type eigen-equation into a Schr\"odinger-type one, which is referred to as Courant-Hilbert differential method in this paper). Firstly, one assumes a coordinate transformation $y=y(x)$ and utilizes the chain rules $d/dx =(dy/dx)d/dy$ and $d^2/dx^2= \smbr{d^2y/dx^2} d/dy + \smbr{dy/dx}^2 d^2/dy^2$ to rewrite $p H_0$ as a differential operator of second order with respect to $y$-coordinate. For instance, in order to simplify the algebra and to make the superpotential $W_0$ to be more easily identified, instead of traditionally starting with the form of $p H_0$ with $H_0$ in its  \emph{expanded form} given in \eq{H_0_operator}, one could start with the \emph{factorized form} of $p H_0$ in \eq{simple_factorization} and rewrite it as $pH_0=\smbr{-{\tilde p} d/dy+2W_0}\smbr{{\tilde p} d/dy}$ with $\tilde{p}\equiv p dy/dx$. Secondly, one rescales the eigenfunction by $\Phi_0(x)=\mu(x)\Psi_0(y)$ and forces the overall coefficient of the first-order derivative $d\Psi_0(y)/dy$ in the $y$-coordinate differential equation $pH_0\smbr{\mu(x)\Psi_0(y)}=0$ to be vanishing to obtain the differential equation of second order for $y(x)$ and first order for $\mu$ in the form
\begin{eqnarray}\label{coeff+1st_order_derivative}
 \tilde{p}\sqbr{\smbr{2W_0-\od{\tilde{p}}{y}} \mu- 2\tilde{p} \od{\mu}{y}}=0.
\end{eqnarray}
The survived terms are then rearranged as the Schr\"odinger equation $-d^2\Psi_0/dy^2+ {\cal V}_0(y)\Psi_0=0$ in $y$-coordinate, with the collection of the coefficients of zeroth derivatives being the potential
\begin{eqnarray}\label{V02}
 {\cal V}_0(y)\equiv \frac{1}{\tilde{p} \mu}\sqbr{2 W_0 \od{\mu}{y}- \od{}{y}\tilde{p}\od{\mu}{y}}.
\end{eqnarray}

In \eq{coeff+1st_order_derivative}, we can try the simplest situation in which $\tilde{p}=1$, which means that $dy/dx =1/p(x)$ and its solution is just \eqnm{y_x_coord_trans}, then \eq{coeff+1st_order_derivative} reduces to the first-order differential equation $d\log\mu/dy = W_0$ for the rescaling factor $\mu(x)$, which has the solution $\mu(x)= \exp\smbr{\int^x d\tilde{x} W_0(\tilde{x})/p(\tilde{x})}=w^{-1/2}(x)$. As a consequence, the potential \eqnm{V02} is simplified into the form \eqnm{Riccati_equ_V0}. 

In this traditional {\it differential} method, the linear differential equation of second order is {\it passively } converted into a standard Schr\"odinger equation in some new coordinate by eliminating the first-order derivative {\it bona fide} and solving a linear differential equation of second order for this coordinate transformation and for the rescaling factor of the eigenfunction, though not so complicated in this particular example, is full of {\it mathematical} flavor and not so straightforward or algebraically so simple for reaching the supersymmetric factorization of the more general operator ${\mathbb{H}}_l(x)$, in contrast to the {\it active} supersymmetrization -- a pure {\it algebraic} method proposed above by us -- that is full of SUSYQM flavor. Recall that the `key trick' in our \emph{algebraic} approach, if can be so called, is to find out the integration factor of some two-term differential operators of first order, such as $-pd/dx+2W_0$ and $pd/dx \pm W_0$, which is at the elementary calculus level and trivial, and the coordinate transformation $y(x)$, the superpotential $W_0$ and the rescaling factor of the eigenfunction just come out naturally {\it in one go}.  In contrast, in many cases, applying the Courant-Hilbert differential method to transform an eigen-equation of some linear differential operator of second-order, one obtains the corresponding Schr\"odinger equation in its {\it unfactorized} form with some complicated resultant potential, hence one still has to solve separately for the relevant superpotential from the Recatti equation in the form like \eq{Riccati_equ_V0}, in order to reach the supersymmetric factorization of the Schr\"odinger equation. For the above reasons, we prefer our algebraic method to supersymmetrize the linear differential operators of second order like $p H_0$, and to solve their eigenvalue problems with this supersymmetric factorization method. We think that this method deserves more amplification.

\subsection{The Standard Schr\"odinger Equation of Generic Level and Its SUSY Factorization in {\it y}-coordinate}\label{Standard_Hermitian_factorization}

We are now in a position to address the issue of factorizing \eq{eigen_equ_with_multiplier} with generic level $l$. We have shown that the first-order differential operator contained in $p H_0$ in this equation can be disguised by applying the transformation ${\cal T}$, according to \eq{pre_Hamiltonian_H0}. To prepare this transformation and apply it to \eq{eigen_equ_with_multiplier}, $\sqrt{w}$  is  multiplied  from the left of ${\mathbb H}_l(x)$ and the unit operator $\sqrt{w}^{-1}\sqrt{w}$ is inserted between ${\mathbb{H}}_l(x)$ and $\Phi_l$. Introduce the notations 
\begin{eqnarray}\label{transformed_Hl}
{\cal  H}_l(y) \equiv {\cal T}\sqbr{{\mathbb H}_l(x)}_{x=x(y)},\; \Psi_l (y)\equiv \sqbr{\sqrt{w}\Phi_l(x)}_{x=x(y)},
\end{eqnarray}
as the transformed Hamiltonian and the rescaled eigenfunction, respectively, where $x=x(y)$ is understood as the inverse function of the coordinate transformation \eqnm{y_x_coord_trans}. Note that $\sqrt{w}\Phi_0(x)$ in \eq{ground_state_minus_route} is just the $l=0$ case of this $\Psi_l (y)$.  \eq{eigen_equ_with_multiplier} is then  transformed into the following standard Schr\"odinger equation in $y$-coordinate,
\begin{eqnarray} \label{Schr_equ_y_coord}
 {\cal H}_l(y) \Psi_l(y) = E_l \Psi_l(y),
\end{eqnarray}
where, by making use of Eqs. \eqnm{pre_Hamiltonian_H0} and \eqnm{KE_map_type1}, ${\cal H}_l(y)$ in \eq{transformed_Hl} can be expressed as
\begin{eqnarray}\label{Hamiltonian_y_coordinate}
 {\cal  H}_l(y) &\equiv& -  \frac{d^2}{d y^2}  + {\cal V}_l\smbr{y} 
\end{eqnarray}
with ${\cal V}_l(y)\equiv V_l(x(y))$ and
\begin{eqnarray}\label{Vl_potential}
V_l(x) &\equiv&  V_0(x)- \lambda_l p + E_l
\end{eqnarray}
with $V_0(x)$ given in \eq{Riccati_equ_V0}. Since both $V_0(x)$ and $p(x)$ are in general binomials in $x$, $V_l(x)$ is a {\it binomial} in $x$ as well. Although the potential ${\cal V}_l(y)$ may look drastically different from a binomial in $y$, it can be recast into the binomial $V_l(x)\equiv {\cal V}_l(y(x))$ under the coordinate transformation \eqnm{y_x_coord_trans}. We say such type of potentials ${\cal V}_l(y)$ are  {\it binomializable}.

In the above, the kinetic energy operator map \eqnm{KE_map_type1} and the equivalent coordinate transformation \eqnm{y_x_coord_trans} have been made use of to pass from \eq{eigen_equ_with_multiplier} (the $x$-coordinate (nonstandard) representation) to \eq{Schr_equ_y_coord} (the $y$-coordinate (standard) representation).  Later on, for convenience we will switch frequently between these two different coordinate representations of the operators and eigenfunctions, by keeping in mind such operator map and coordinate transformation. Notice that, in $x$-coordinate representation,  the kinetic energy operator of the operator ${\cal T}\sqbr{{\mathbb H}_l(x)}$ is buried in the part ${\cal T}\sqbr{p(x) H_0(x)}$, hence it still takes the form $-\smbr{p d/dx}^2$, the same as that in the eigen-equation \eqnm{initial_eigen_equ_minus} for the case $l=0$.

We see that, the supersymmetrization transformation ${\cal T}$, introduced in the previous subsection for supersymmetrizing $pH_0$, in fact efficiently turns the eigen-equation \eqnm{eigen_equ_with_multiplier} into the standard Schr\"odinger equation \eqnm{Schr_equ_y_coord}. The remaining task is to find out the superpotential ${\cal W}_l(y)\equiv W_l(x(y))$ for ${\cal V}(y)$ and to construct the SUSYQM for ${\cal H}_l(y)$. We expect that the superpotential $W_0(x)$ for $V_0(x)$ is just {\it part} of $W_l(x)$.

Note that in \eq{Vl_potential}, the piece $-\lambda_l p + E_l$ is of the same type {\it binomial} in $x$ as $V_0(x)$ and $W^2_0(x)$. Such feature allows us to absorb this piece into $V_0(x)$ and introduce the overall \emph{linear} superpotential $W_l(x)$ for $V_l(x)$, i.e.
\begin{eqnarray}
 W_l(x) &=& \alpha_l x+\beta_l, \label{Wl_superpotential},
\end{eqnarray}
where the constant $\alpha_l$ and $\beta_l$  will be determined when additional conditions are imposed.
As usual, in $x$-coordinate we have 
\begin{eqnarray}\label{potential_x_coord}
 V_l(x)=- p \od{W_l(x)}{x} + W^2_l(x)
\end{eqnarray}
and in $y$-coordinate  
\begin{eqnarray}\label{potential_y_coord}
{\cal V}_l(y)&=&-\od {{\cal W}_l(y)}{y} + {\cal W}^2_l(y).
\end{eqnarray}

  This $W_l(x)$ is just the shift of $W_0(x)$ by another linear function $\Delta W_l(x)$, i.e.
\begin{eqnarray}\label{DeltaWl}
\Delta W_l(x) &\equiv& W_l(x)- W_0(x), \nonumber \\
              &=& \smbr{\alpha_l-\alpha_0} x +\smbr{\beta_l -\beta_0}.
\end{eqnarray}

In view of Eqs. \eqnm{Hamiltonian_y_coordinate}, \eqnm{potential_x_coord} and \eqnm{potential_y_coord}, very much similar to the factorization $x^2_1-x^2_2=\smbr{x_1+x_2}\smbr{x_1-x_2}$ for two numbers $x_1$ and $x_2$, 
the Hamiltonian operator ${\cal H}_l$ in \eq{Hamiltonian_y_coordinate} can be factorized as the product 
\begin{eqnarray}\label{Hamiltonian_SHF}
{\cal H}_l(y) &=& {\cal A}^\dagger_l(y) {\cal A}_l(y)
\end{eqnarray}
of the Hermitian conjugate pair
\begin{equation}\label{conjugate_pair}
\begin{aligned}
{\cal A}_l(y)&\equiv \od{}{y}+{\cal W}_l(y)=p\od{}{x}+W_l(x), \\
{\cal A}^\dagger_l(y)&\equiv -\od{}{y}+{\cal W}_l(y)=-p\od{}{x}+W_l(x).
\end{aligned}
\end{equation}

With the factorization \eqnm{Hamiltonian_SHF} in hand, \eq{Schr_equ_y_coord}  implies that  $E_l =\int_{D(y)} dy \abs{{\cal A}_l(y)\Psi_l(y)}^2/\int_{D(y)} dy \abs{\Psi_l(y)}^2 \ge 0$ ($D(y)$ is the domain of $y$), i.e. $E_l$ is a non-negative number.

As usual in SUSYQM, the superpartner Schr\"odinger equation of \eq{Schr_equ_y_coord} is obtained by acting this equation with the right-situated factor ${\cal A}_l(y)$ in ${\cal H}_l$ (which is often referred to as the intertwining multiplication or leftward braiding), that is,
\begin{eqnarray}\label{superpartner_Schr_equ}
 {\cal H}^s_l(y) \Psi^s_l\smbr{y} = E_l \Psi^s_l\smbr{y}, 
\end{eqnarray}
where the superpartner Hamiltonian 
\begin{eqnarray}\label{superpartner_Hamiltonian}
{\cal H}^s_l(y)\equiv{\cal A}_l(y) {\cal A}^\dagger_l(y) = - \frac{d^2}{d y^2} + {\cal V}^s_l\smbr{y}, 
\end{eqnarray}
the superpartner potential
\begin{eqnarray}\label{superpartner_Vl}
 {\cal V}^s_l\smbr{y}\equiv \od{{\cal W}_l}{y} +{\cal  W}^2_l(y) = p \od{W_l(x)}{x} + W^2_l(x)\equiv V^s_l\smbr{x}\nonumber \\
\end{eqnarray}
and the eigenfunction 
\begin{eqnarray}\label{recursion_relation_unspecified}
 \Psi^s_l\smbr{y} \propto {\cal A}_l(y)\Psi_l\smbr{y}.
\end{eqnarray}

One key feature of the pair of Hamiltonians ${\cal H}_l(y)$ and ${\cal H}^s_l(y)$ is that they have the common eigenvalue $E_l$.  As has been pointed out in the introduction section, this feature allows one to adopt the supercharge-plus-spinor formulation of the pair of eigen-equations, in which the underlying $sl(1,1)$ supersymmetry becomes manifest \cite{Witten_SUSY_breaking_1981, Gendenshtein_SUSY_Schr_Equ_1983, Cooper_SUSY_and_QM_1995}. However, to proceed, we will not used this matrix formulation.

The other key feature is that the potential pair $V_l$ and $V^s_l$ share the superpotential $W_l$. This makes the roles of these two potentials interchangeable, i.e. if the original superpotential $W_l$ is replaced by $-W_l$, \eq{superpartner_Vl} and \eq{potential_y_coord} are interchanged. 

Subtracting \eq{potential_x_coord} from \eq{superpartner_Vl} yields
\begin{eqnarray}\label{potential_difference}
 V^s_l(x)-V_l(x)= 2p(x) \od{W_l(x)}{x}=2 \alpha_l p.
\end{eqnarray}
Later on, we will demonstrate that this equation together with some shape invariance symmetry, which essentially says that $V^s_l(x)$ and $V_l(x)$ are similar, will completely determine $\alpha_l$, $\beta_l$, $E_l$ and $\Psi_l(y)$.

\subsection{The Downward-connected Shape Invariant Potential Series and the Related Hierarchy of SUSYQMs}\label{minus_route_iteration}
By design, the superpartner Hamiltonian ${\cal H}^s_l$ and the original one ${\cal H}_l$ are expected to belong to the same hierarchy of Hamiltonians, which requires that the superpartner potential ${\cal V}^s_l$ and the original potential ${\cal V}_l$ be of the same type of functions and be different from each other solely by a certain constant shift. For instance, we can assume that ${\cal V}^s_l$ differs  from ${\cal V}_{l-1}$, the {\it lower} nearest neighboring potential of ${\cal V}_l$, only by a constant $\delta_l$, i.e. 
\begin{eqnarray}\label{SI_potential_minus}
{\cal V}^{s-}_l(y)={\cal V}^-_{l-1}(y) + \delta^-_l.
\end{eqnarray}
From hereon the superscript {\it minus} is attached to each involved quantity and operator to indicate that the superpartner potential is {\it downard} connected to a neighbor in the potential series $\{ {\cal V}^-_l(y)\}$. Correspondingly,  \eq{Schr_equ_y_coord}  is relabeled as
\begin{eqnarray}\label{Schr_equ_y_coord_minus}
{\cal H}^-_l(y) \Psi_l(y) = E^-_l \Psi_l(y) 
\end{eqnarray}
with
\begin{eqnarray}\label{factorized_Hamiltonian_minus}
 {\cal H}^-_l={\cal A}^{-\dagger}_l{\cal A}^-_l =-  \frac{d^2}{d y^2}  + {\cal V}_l\smbr{y},
\end{eqnarray}
\begin{equation}\label{Wl_minus}
\begin{aligned}
 {\cal A}^-_l(y)&\equiv \od{}{y}+{\cal W}^-_l(y)=p \od{}{x} + W^-_l(x), \\
 {\cal A}^{-\dagger}_l(y) &\equiv - \od{}{y}+{\cal W}^-_l(y)= -p \od{}{x} + W^-_l(x), \\
 W^-_l(x)&=\alpha^-_l x +\beta^-_l, \\
 {\cal V}_l\smbr{y} &= V_l(x) \equiv  V_0(x)- \lambda^-_l p + E^-_l,
 \end{aligned}
\end{equation}
 \eq{superpartner_Hamiltonian} as
\begin{eqnarray}\label{superpartner_Hamiltonian_minus}
 {\cal H}^{s-}_l={\cal A}^-_l{\cal A}^{-\dagger}_l = - \frac{d^2}{d y^2} + {\cal V}^{s-}_l\smbr{y}
\end{eqnarray}
and \eq{superpartner_Schr_equ} as
\begin{eqnarray}\label{superpartner_Schr_equ_minus}
 {\cal H}^{s-}_l(y) \Psi^{s-}_l\smbr{y} = E^-_l \Psi^{s-}_l\smbr{y}.
\end{eqnarray}

In \eq{Schr_equ_y_coord_minus}, we do not attach a minus sign to $\Psi_l\smbr{y}$ as its superscript. The reason for this will become clear later on.

Upon  adding $-d^2/dy^2$ on both sides of the relation \eqnm{SI_potential_minus} and making use of \eq{superpartner_Hamiltonian_minus} and the $(l-1)$ version of \eq{factorized_Hamiltonian_minus}, the relation \eqnm{SI_potential_minus} is equivalent to the following generalized commutator relation 
\begin{eqnarray}\label{SI_operator_product_minus}
{\cal A}^-_l{\cal A}^{-\dagger}_l -{\cal A}^{-\dagger}_{l-1}{\cal A}^-_{l-1}=\delta^-_l.
\end{eqnarray}

Substituting for ${\cal A}^-_l{\cal A}^{-\dagger}_l$ via \eq{SI_operator_product_minus},  \eq{superpartner_Schr_equ_minus} is then turned into 
\begin{eqnarray}\label{superpartner_Schr_equ_lowering_after_rnmlz}
 {\cal H}^-_{l-1} \Psi_{l-1} \smbr{y}= E^-_{l-1} \Psi_{l-1}\smbr{y}
\end{eqnarray}
where 
\begin{eqnarray}\label{prelowering_relation}
\Psi_{l-1} \smbr{y}=\Psi^{s-}_l \smbr{y} \propto {\cal A}^-_l(y)\Psi_l\smbr{y},
\end{eqnarray}
\begin{eqnarray}\label{E_down_shift}
 E^-_{l-1} = E^-_l -\delta^-_l
\end{eqnarray}
and 
\begin{eqnarray}
{\cal H}^-_{l-1} = {\cal A}^{-\dagger}_{l-1}(y){\cal A}^-_{l-1}(y).
\end{eqnarray}

\eq{superpartner_Schr_equ_lowering_after_rnmlz} assumes the \emph{same pattern} as the original \eq{Schr_equ_y_coord_minus}, except that the label $l$ is substituted with $l-1$ and this type of order is termed {\it normal order} in this paper, and for such a reason the condition \eqnm{SI_potential_minus} of the potential series is usually called the pattern (shape) invariance condition, which is a sort of generalized commutator relation between two pairs of raising and lowering operators with neighboring indices. Later on, we will see that this is indeed a a generalization of the level-index-independent commutator for the more familiar quantum harmonic oscillator. For obvious reason, the condition \eqnm{SI_potential_minus} is termed the {\it downward-connected} pattern invariance condition. The procedure of rearranging the superpartner eigen-equation \eqnm{superpartner_Schr_equ_minus} of \eq{Schr_equ_y_coord_minus} with label $l$ into the same type of eigen-equation \eqnm{superpartner_Schr_equ_lowering_after_rnmlz} but with label $l-1$, by making use of the generalized commutator relation \eqnm{SI_operator_product_minus} is termed \emph{renormalization} or \emph{normally ordering} in this paper.

By fixing the particular proportionality constant, the relation \eqnm{prelowering_relation}  becomes the {\it lowering} recurrence relation
\begin{eqnarray}\label{lowering_recursion_relation_Hermitian}
\Psi_{l-1}\smbr{y}\equiv \frac{{\cal A}^-_l(y)}{\sqrt{E^-_l}} \Psi_l\smbr{y}
\end{eqnarray}
for $l\ge 1$; that is, ${\cal A}^-_l(y)$ works as the lowering ladder operator. The combination of Eqs. \eqnm{Schr_equ_y_coord_minus}, \eqnm{factorized_Hamiltonian_minus}  and \eqnm{lowering_recursion_relation_Hermitian} yields the {\it raising} recurrence relation
\begin{eqnarray}\label{raising_recursion_relation_Hermitian}
 \Psi_l = \frac{{\cal A}^{-\dagger}_l}{\sqrt{E^-_l}} \Psi_{l-1},
\end{eqnarray}
that is, ${\cal A}^{-\dagger}_l(y)$ works as the raising ladder operator accordingly.

The constants $\alpha^-_l$, $\beta^-_l$, $\lambda^-_l$ and $E^-_l$in \eq{Wl_minus}, and the eigenvalue shift $\delta^-_l$ appearing in \eq{E_down_shift}  will be determined iteratively in Subsection \ref{Minus_route_factorization}. 

As has been demonstrated in \eq{ground_state_minus_route} and will be demonstrated again in \ref{Minus_route_factorization}, the ground state is determined as $\Psi_0(y)\propto w^{1/2}$. 

By repeatedly making use of the raising recurrence relation \eqnm{raising_recursion_relation_Hermitian}, the generic eigenfunction $\Psi_j$ ($j>0$) is then obtained as
\begin{eqnarray}\label{Psi_j_continuous_raising}
 \Psi_j = \frac{{\cal A}^{-\dagger}_j}{\sqrt{E^-_j}} \frac{{\cal A}^{-\dagger}_{j-1}}{\sqrt{E^-_{j-1}}}\cdots \frac{{\cal A}^{-\dagger}_2}{\sqrt{E^-_2}} \frac{{\cal A}^{-\dagger}_1}{\sqrt{E^-_1}}\Psi_0.
\end{eqnarray}

As will be demonstrated in Subsection \ref{positive_association_levels},  in the expression $\Psi_l(y)=w^{1/2}(x)\Phi_l(x)$, $\Phi_l(x)$ can be given by the compact Rodriguez formula \eqnm{Rodriguez_formula} for any $l>0$. Then, complementary to the bottom-up result \eqnm{Psi_j_continuous_raising}, for all $0\le j<l$, there is the following formula for $\Psi_j(y)$,
\begin{eqnarray}\label{Psi_j_continuous_lowering}
 \Psi_j \smbr{y} = \frac{{\cal A}^-_{j+1}(y)}{\sqrt{E^-_{j+1}}}\frac{{\cal A}^-_{j+2}(y)}{\sqrt{E^-_{j+2}}} \cdots \frac{{\cal A}^-_{l-1}(y)}{\sqrt{E^-_{l-1}}}\frac{{\cal A}^-_l(y)}{\sqrt{E^-_l}} \Psi_l\smbr{y},\nonumber \\
\end{eqnarray}
which is constructed in a top-down approach by consecutively applying the lowering recurrence relation \eqnm{lowering_recursion_relation_Hermitian} to some top-level eigenfunction $\Psi_l(y)=w^{1/2}(x)\Phi_l(x)$. 

\subsubsection{Determining \mathinhead{\alpha^-_l}{}, \mathinhead{\beta^-_l}{}, \mathinhead{\delta^-_l}{},  \mathinhead{\lambda^-_l}{} and \mathinhead{E^-_l}{}} \label{Minus_route_factorization}

Although conceptually the standard SUSYQM are formulated in $y$-coordinate, the algebras for finding the factorization parameters $\alpha^-_l$, $\beta^-_l$, $E^-_l$, $\lambda^-_l$ and $\delta^-$ are much simpler in $x$-coordinate, just because the simple {\it linear} character of the superpotential $W^-_l(x)$ and the simple {\it binomial} character of the corresponding potential $V_l(x)$ (see \eq{Vl_potential}) exist only in $x$-coordinate. 

As demonstrated in Appendix \ref{solution_Wl_and_El}, the coefficients $\alpha^-_l$ and $\beta^-_l$ in $W^-_l(x)=\alpha^-_l x +\beta^-_l$ and the eigenvalue $E^-_l$ can be determined from those algebraic equations obtained by directly matching the coefficients of the two distinct expressions (Eqs. \eqnm{Vl_potential} and \eqnm{potential_x_coord}) of the binomial $V^-_l(x)$. 
Here, we provide another  {\it iterative} approach of determining these quantities, which is even algebraically simpler. This approach is motivated by the following three considerations. Firstly, the shape invariance condition \eqnm{SI_potential_minus}  itself is a recurrence relation between the two neighboring superpotentials $W^-_l$ and $W^-_{l-1}$. Although this is a linear differential recurrence relation, it simply reduces to some {\it algebraic} recurrence relations for $\alpha^-_l$, $\beta^-_l$, $\alpha^-_{l-1}$ and $\beta^-_{l-1}$, in view of the simple linear function $W^-_l(x)=\alpha^-_l x +\beta^-_l$. Solving the generic expressions for these coefficients from those algebraic recurrence relations involves merely elementary algebras. Secondly, the initial data  such as $\alpha^-_0$, $\beta^-_0$ and $E^-_0$, that are required for kicking off this iterative algorithm, can be trivially read off from the $l=0$ case of $V^-_l(x)$ in \eq{Vl_potential}. Thirdly,  these initial data and the recurrence relations allow us to build all levels of eigen-equations and their solutions \emph{explicitly level by level} by running a systematic iterative SUSYQM algorithm, which will be provided in \ref{proliferation_minus_route}.

We first specify the initial data that are needed for the iteration. With $W_0(x)$ defined in \eq{W_0} and the choice $\Phi_0(x)=1$, if we make the identifications
\begin{equation}\label{W0_minus}
\begin{aligned}
 W^-_0(x)&\equiv W_0(x) =  \frac{1}{2}\smbr{p''-q'}x + \frac{1}{2}\smbr{p_0-q_0}, \\
 \Psi^-_0(y)&\equiv\Psi_0(y)=\sqrt{w},
 \end{aligned}
\end{equation}
then the $l=0$ version of \eq{Schr_equ_y_coord_minus} reads
\begin{eqnarray}\label{initial_eigen_equ_minus_branch}
 {\cal H}^-_0\Psi_0= {\cal A}^{-\dagger}_0 {\cal A}^-_0 \Psi_0 = 0,
\end{eqnarray}
which is just \eq{initial_eigen_equ_minus} expressed in $y$-coordinate. This equation holds when
\begin{eqnarray}\label{annihilation_psi_0_minus}
 {\cal A}^-_0\Psi^-_0=0,
\end{eqnarray}
which is just the annihilation condition \eqnm{ground_state_minus_route} expressed in $y$-coordinate.

The corresponding eigenvalues are
\begin{equation}\label{E0_minus}
\begin{aligned}
 E^-_0&=E_0=0, \\
 \lambda^-_0&=\lambda_0=0,
\end{aligned} 
\end{equation}
where the first equation can also be derived from $V^-_0=V_0$ for
\begin{eqnarray}\label{V_0_and_W_0_minus}
 V^-_0(x) =- p \od{W^-_0(x)}{x} + W^{-2}_0(x)
\end{eqnarray}
and
\begin{eqnarray}\label{V_0_and_E_0}
 V_0 = - p \od{W_0}{x}  + W^2_0 + E_0,
\end{eqnarray}
where the latter is the $l=0$ case of \eq{Vl_potential}.

Comparing
\begin{eqnarray}\label{Wminus0expansion}
 W^-_0 (x) = \alpha^-_0 x+\beta^-_0
\end{eqnarray}
with the upper one in \eqnm{W0_minus}, we obtain 
\begin{equation}\label{alpha_0_beta_0_minus}
\begin{aligned}
\alpha^-_0 &=\frac{1}{2}\smbr{p''-q'},\\
\beta^-_0 &= \frac{1}{2}\smbr{p'_0-q_0}.
\end{aligned}
\end{equation}

These initial data $\alpha^-_0$, $\beta^-_0$, $E^-_0$ and $\lambda^-_0$, together with certain algebraic recurrence relations that we are going to work out immediately,  completely determine $\alpha^-_l$, $\beta^-_l$, $\delta^-_l$, $\lambda^-_l$ and $E^-_l$.

The combination of condition \eqnm{SI_potential_minus} and \eq{potential_difference} for $V^-_l$ leads to the recurrence relation 
\begin{eqnarray}
V^-_l-V^-_{l-1}=-2\alpha^-_l p +\delta^-_l 
\end{eqnarray}
for the potential series $\{V^-_i\}$. On the other hand, making use of the definition \eqnm{Vl_potential} for $V^-_l(x)$ yields 
\begin{eqnarray}
V^-_l-V^-_{l-1}=\smbr{\lambda^-_{l-1}-\lambda^-_l} p+ E^-_l-E^-_{l-1}. 
\end{eqnarray}
Equating these two expressions for $V^-_l-V^-_{l-1}$ then gives
\begin{equation}\label{alpha_l_minus}
 \begin{aligned}
 \alpha^-_l &=\frac{1}{2}\smbr{\lambda^-_l-\lambda^-_{l-1}}, \\
 \delta^-_l&=E^-_l-E^-_{l-1}.
 \end{aligned}
\end{equation}

Inserting $V^s_l(x)$ and $V_{l-1}(x)$ given by Eqs. \eqnm{superpartner_Vl} and \eqnm{potential_x_coord} respectively,  the downward-connected shape invariance condition \eqnm{SI_potential_minus} becomes the following differential recurrence relation in terms of the two consecutive superpotentials in the series $\{W^-_l(x)\}$
\begin{eqnarray}\label{recursion_RL_W_minus}
 p \od{}{x}\sqbr{W^-_l(x)+W^-_{l-1}(x)} + W^{-2}_l(x)-W^{-2}_{l-1}(x)=\delta^-_l.\nonumber \\
\end{eqnarray}
Inserting $p(x)=(p''/2)x^2+p'_0 x + p_0$, $W^-_l(x)=\alpha^-_l x+ \beta^-_l$ and $W^-_{l-1}(x)=\alpha^-_{l-1} x+ \beta^-_{l-1}$ then yields the equation
\begin{eqnarray}
 &&\sqbr{\frac{p''}{2} \smbr{\alpha^-_l+\alpha^-_{l-1}} + \smbr{ \alpha^{-2}_l - \alpha^{-2}_{l-1}}} x^2 \nonumber \\
 && + \sqbr{p'_0 \smbr{\alpha^-_l+\alpha^-_{l-1}} + 2 \smbr{\alpha^-_l \beta^-_l-\alpha^-_{l-1} \beta^-_{l-1}}}x  \nonumber \\
 &&+ p_0\smbr{\alpha^-_l+\alpha^-_{l-1}}  +\smbr{\beta^{-2}_l-\beta^{-2}_{l-1}}=\delta^-_l
\end{eqnarray}
that holds for any $x$ in its definition domain. Thus, the overall coefficients of $x^2$, $x$ and $x^0$ of the above equation have to be vanishing separately, which then give us the following algebraic recurrence relations,
\begin{eqnarray}\label{coefficient_x^2}
   \alpha^-_l - \alpha^-_{l-1}=-\frac{p''}{2},
\end{eqnarray}
\begin{eqnarray}\label{coefficient_x}
 \alpha^-_l \beta^-_l-\alpha^-_{l-1} \beta^-_{l-1} &=&-\frac{p'_0} {2}\smbr{\alpha^-_l+\alpha^-_{l-1}}
\end{eqnarray}
and
\begin{eqnarray}\label{coefficient_x^0}
\delta^-_l = p_0\smbr{\alpha^-_l+\alpha^-_{l-1}}  +\smbr{\beta^{-2}_l-\beta^{-2}_{l-1}}.
\end{eqnarray}

The two-term recurrence relation \eqnm{coefficient_x^2}  defines the arithmetical series $\{\alpha^-_l\}$. With the initial $\alpha^-_0$ given by the upper equation in \eqnm{alpha_0_beta_0_minus}, $\alpha^-_l$ is solved as  
\begin{eqnarray}\label{alpha_l_minus_recursion}
 \alpha^-_l =\alpha^-_0 - \frac{l p''}{2}=-\frac{1}{2}\sqbr{\smbr{l-1}p''+q'}\equiv -c_{l-1},
\end{eqnarray}
which is the same as that given in \eq{alpha_l_plus_minus_third_way} by directly matching.

Combining \eq{alpha_l_minus_recursion} and the upper equation in \eqnm{alpha_l_minus} yields
\begin{eqnarray}\label{difference_lambdal_minus}
 \lambda^-_l -\lambda^-_{l-1} = -\smbr{l-1} p''- q'.
\end{eqnarray}
Thus,
\begin{eqnarray}\label{lambda_minu_eigenvalue}
 \lambda^-_l = \lambda^-_0 + \sum^l_{j=1}\smbr{\lambda^-_j -\lambda^-_{j-1}} = -l q' -\frac{l\smbr{l-1}}{2} p''.
\end{eqnarray}

Here we see another advantage of this iteration algorithm: the eigenvalue $\lambda^-_l$ of $H_0$ comes out of this algorithm \emph{itself}. In contrast, the direct match method shown in Appendix \ref{solution_Wl_and_El} requires this eigenvalue to be known beforehand via other approaches. In \cite{jafarizadeh_SUSY_and_diff_equs_1997}, this $\lambda^-_l$ is obtained by directly matching the coefficients of the highest order ($2l$-th) terms on both sides of the original \eq{eigen_equ} before factorizing $H_0$, assuming that $\Phi_l(x)$ takes the polynomial form \eqnm{Rodriguez_formula}. This latter assumption, as an input, can only be understood as a result out of other traditional methods.

Making the substitution for $\alpha^-_j \beta^-_j-\alpha^-_{j-1} \beta^-_{j-1}$ via \eq{coefficient_x} and making use of Eqs. \eqnm{coefficient_x^2} and \eqnm{alpha_l_minus_recursion}, we obtain
\begin{eqnarray}\label{alphal_betal_product}
 \alpha^-_l \beta^-_l &=&\alpha^-_0 \beta^-_0 + \sum^{l}_{j=1}\smbr{\alpha^-_j \beta^-_j-\alpha^-_{j-1} \beta^-_{j-1}}, \nonumber \\
                      &=&\alpha^-_0 \beta^-_0- l p'_0\smbr{\alpha^-_l +\frac{l}{4} p''}.
\end{eqnarray}
Upon further making use of \eq{alpha_l_minus_recursion} and the upper equation in \eqnm{alpha_0_beta_0_minus}, we obtain 
\begin{eqnarray}\label{betal_relative}
 \beta^-_l = -\frac{l}{2}  p'_0+   \frac{\alpha^-_0}{\alpha^-_l} \smbr{ \beta^-_0 -\frac{l}{2}  p'_0}.
\end{eqnarray}

Eqs. \eqnm{alpha_l_minus_recursion} and \eqnm{betal_relative} together give us the minus branch of $\Delta W_l(x)$, defined in \eq{DeltaWl}, in the form
\begin{eqnarray}
 \Delta W^-_l(x) = - \frac{l p''}{2} x +   \frac{l p''}{2 \alpha^-_l} \smbr{ \beta^-_0 -\frac{l}{2}  p'_0}  -l  p'_0.
\end{eqnarray}

$\beta^-_l$ can also be expressed in terms of the elements in the series $\{c_l\equiv \smbr{l p''+q'}/2\}$  (see the definition of $c_{l-1}$ in \eq{alpha_l_minus_recursion}) and $\{d_l \equiv l p'_0 + q_0\}$. First we notice that $p'' =2\smbr{c_l-c_{l-1}}$, $q'=2\smbr{l-1}c_l+ 2\,l\,c_{l-1}$,  $p'_0=d_l -d_{l-1}$ and $q_0=l\,d_{l-1}-(l-1)d_l$. After these substitutions, \eq{alphal_betal_product} is rewritten as
\begin{eqnarray}\label{alpha_beta_product}
 \alpha^-_l \beta^-_l=\frac{1}{2}\sqbr{\smbr{c_{l-1}-l c_l} d_{l-1} + l c_{l-1} d_l}\equiv g_l,
\end{eqnarray}
and \eq{betal_relative} as
\begin{eqnarray}\label{beta_minus_l}
\beta^-_l = -\frac{g_l}{c_{l-1}} =-\frac{1}{2c_{l-1}}\sqbr{\smbr{c_{l-1}-l c_l} d_{l-1} + l c_{l-1} d_l}.
\end{eqnarray}
This form of $\beta^-_l$ is the same as that given in \eq{beta_l_minus_other_way} via direct match.

Substituting for $\alpha^-_l$ via \eq{alpha_l_minus_recursion} and $\beta^-_l$ via \eq{beta_minus_l}, the expression \eqnm{coefficient_x^0} for $\delta^-_l$ is rewritten as
\begin{eqnarray}\label{delta_l_minus}
\delta^-_l = \smbr{\frac{g_l}{c_{l-1}}}^2-\smbr{\frac{g_{l-1}}{c_{l-2}}}^2 -p_0\smbr{c_{l-1}+c_{l-2}}.
\end{eqnarray}

With $E^-_0=0$ and the lower equation in \eqnm{alpha_l_minus}, $E^-_l$ is obtained as $E^-_l = E^-_0+ \sum^l_{j=1}\smbr{E^-_j-E^-_{j-1}}=\sum^l_{j=1} \delta^-_j$. Further substituting for $\delta^-_j$ via \eq{delta_l_minus} yields
\begin{eqnarray}
 E^-_l &=& -p_0 \sum^l_{j=1} \smbr{c_{l-1}+c_{l-2}} + \smbr{\frac{g_l}{c_{l-1}}}^2-\smbr{\frac{g_0}{c_{-1}}}^2,\nonumber  \\
       &=&l\sqbr{\frac{d_{l-1}}{4 c^2_{l-1}}  \sqbr{2 c_{l-1} d_l - \smbr{c_{l-1}+ c_l } d_{l-1}} -  p_0}\nonumber \\
       &&\cdot \sqbr{(l+2) c_{l-1} - l c_l }. \label{E_minus_eigenvalue}
\end{eqnarray}

\subsubsection{Am \emph{Explicit} Bottom-up Iterative Flow of Proliferating a Hierarchy of Eigen-equations and Their Solutions -- the Minus Branch}\label{proliferation_minus_route}
Recall what we have essentially done up to this stage. We ran the standard SUSYQM algorithm for solving the eigen-equation \eqnm{eigen_equ_with_multiplier} or \eqnm{Schr_equ_y_coord} for the generic level $l$, whose exact forms are {\it already known} to us, and obtained their SUSYQM factorization \eqnm{superpartner_Schr_equ_lowering_after_rnmlz} and the general eigenfunction \ref{Psi_j_continuous_raising}, and the eigenvalue $\lambda^-_l$ in \eq{lambda_minu_eigenvalue} or $E^-_l$ in \eq{E_minus_eigenvalue}. 

However, there is another interpretation of this seeming {\it one-step} factorization algorithm. The recurrence relations and the initial data for the series $\{\alpha^-_l\}$, $\{\beta^-_l\}$, $\{\delta^-_l\}$, $\{\lambda^-_l\}$ and $\{E^-_l\}$ given in \ref{Minus_route_factorization} tell us that these series can be determined \emph{level by level} in a bottom-up \emph{iterative} way.
We can then regard the supersymmetrically factorized initial eigen-equation \eqnm{initial_eigen_equ_minus} as the {\it only known} eigen-equation to us, and the eigen-equation of generic level is what we are seeking for. Then, following the iterative SUSYQM algorithm, that could be called standard one, in which the two key operations: 1) intertwiningly multiplying the right-situated factorial operator for deriving the superpartner eigen-equation, 2) renormalizing this superpartner eigen-equation into the normally ordered form the original eigen-equation takes, 
are consecutively performed in repeated way (which of these two operations goes first depends on concrete situation), this initial equation will proliferate into all the eigen-equations of higher levels that automatically come with their supersymmetrically factorized forms. Along with this procedure, the eigen-solutions of these derived equations will also come out in a fairly neat way. For one of the two key operations to be successfully performed, a hierarchy of generalized commutator relations between any two neighboring levels need to be carefully designed, each of these relations takes the form  \ref{SI_operator_product_minus}, by requiring that, for the $l$-th level $\alpha^-_l$ marches according to the recurrence relation \ref{coefficient_x^2}, $\beta^-_l$ to  \ref{coefficient_x}, and $\lambda^-_l$ and $E^-_l$ according to the two recurrence relations in \eq{alpha_l_minus}. The first several levels of proliferating eigen-equations and their solutions are listed as below.

 For $l=0$, setting $\lambda^-_0=0$ and  $E^-_0=0$, the initial eigen-equation is still \eq{initial_eigen_equ_minus} or \eq{initial_eigen_equ_minus_branch}, i.e.
\begin{eqnarray}\label{base_eigen_equ_minus}
 {\cal A}^{-\dagger}_0 {\cal A}^-_0\Psi_0(y)=0,
\end{eqnarray}
where  $W^-_0(y)=W_0(x(y))=\alpha^-_0 x+\beta^-_0$ appearing in ${\cal A}^-_0$, with $\alpha^-_0$ and $\beta^-_0$ given in \eq{alpha_0_beta_0_minus}.
This equation is interpreted as the annihilation condition ${\cal A}^-_0\Psi_0(y)=0$ (\eq{ground_state_minus_route} or \eq{annihilation_psi_0_minus}), which gives the ground state $\Psi_0(y)\propto w^{1/2}(x)$.

Note that in \eq{base_eigen_equ_minus} the annihilation (lowering) operator ${\cal A}^-_0$ is situated in a \emph{wrong} position, in the sense that one can  get the superpartner eigen-equation of the $l=1$ level by {\it directly} intertwiningly acting on \eq{base_eigen_equ_minus} with this ${\cal A}^-_0$. Instead, we need a certain factorized eigen-equation in which some creation (raising) operator is situated right on the immediate left of the eigenfunction $\Psi_0(y)$ so that after performing the intertwining action with this raising  operator the superpartner eigen-equation of the $l=1$ level can be reached. To this end, ${\cal A}^{-\dagger}_0 {\cal A}^-_0$ is substituted for via the shape invariance condition 
\begin{eqnarray}\label{SI_operator_product_minus_l1}
{\cal A}^{-\dagger}_0 {\cal A}^-_0={\cal A}^-_1 {\cal A}^{-\dagger}_1 -\delta^-_1, 
\end{eqnarray}
which is just the $l=1$ case of \eqnm{SI_operator_product_minus}. Here we have the designs ${\cal A}^-_1=d/dy+{\cal W}^{-}_1(y)$, ${\cal A}^{-\dagger}_1=-d/dy+{\cal W}^{-}_1(y)$ and ${\cal W}^{-}_1(y)=W^-_1(x)=\alpha^-_1 x + \beta^-_1$, and $\alpha^-_1$, $\beta^-_1$ and $\delta^-_1$ are determined from  $\alpha^-_0$ and $\beta^-_0$ according to the recurrence relations \eqnm{coefficient_x^2}-\eqnm{coefficient_x^0} for $l=1$, respectively. \eq{base_eigen_equ_minus} is then rewritten as
\begin{eqnarray}\label{base_eigen_equ_raising_op_right_situated}
  {\cal A}^-_1 {\cal A}^{-\dagger}_1 \Psi_0(y)=E^-_1 \Psi_0(y)
\end{eqnarray}
where $E^-_1\equiv E^-_0+ \delta^-_1=\delta^-_1$. 

Now in \eq{base_eigen_equ_raising_op_right_situated} the raising operator ${\cal A}^{-\dagger}_1$  is  situated in the {\it correct} position.
Acting this equation with ${\cal A}^{-\dagger}_1$ yields its superpartner equation
\begin{eqnarray}\label{EE_standard_form_minus_1}
 {\cal A}^{-\dagger}_1 {\cal A}^-_1 \Psi_1(y)= E^-_1 \Psi_1(y)
\end{eqnarray}
for $\Psi_1(y) = \smbr{{\cal A}^{-\dagger}_1/\sqrt{E^-_1}} \Psi_0(y)$. Here we see that, the shape invariance condition \eqnm{base_eigen_equ_raising_op_right_situated} is used for correctly positioning the operator that participates the intertwining action, and it is the intertwining action action that completes the normally ordering.

Before going to the $l=2$ level via intertwining action, the operator product ${\cal A}^{-\dagger}_1 {\cal A}^-_1$ in \eq{EE_standard_form_minus_1} has to be substituted for via the shape invariance condition 
\begin{eqnarray}
{\cal A}^{-\dagger}_1 {\cal A}^-_1={\cal A}^-_2 {\cal A}^{-\dagger}_2 -\delta^-_2
\end{eqnarray}
which is the $l=2$ case of \eqnm{SI_operator_product_minus}, where  ${\cal A}^-_2=d/dy+{\cal W}^{-}_2(y)$, ${\cal A}^{-\dagger}_2=-d/dy+{\cal W}^{-}_2(y)$, ${\cal W}^{-}_2(y)=W^-_2(x)=\alpha^-_2 x + \beta^-_2$, and $\alpha^-_2$, $\beta^-_2$ and $\delta^-_2$ are determined from $\alpha^-_1$ and $\beta^-_1$ according to the recurrence relations \eqnm{coefficient_x^2}-\eqnm{coefficient_x^0} with $l=2$, respectively. \eq{EE_standard_form_minus_1} is then rewritten as
\begin{eqnarray}\label{eigen_equ_1st_level_raising_op_right_situated}
{\cal A}^-_2 {\cal A}^{-\dagger}_2  \Psi_1(y)= E^-_2 \Psi_1(y)
\end{eqnarray}
with $E^-_2=E^-_1+\delta^-_2$. Then acting \eq{eigen_equ_1st_level_raising_op_right_situated} with ${\cal A}^{-\dagger}_2$ generates the normally ordered eigen-equation 
\begin{eqnarray}
 {\cal A}^{-\dagger}_2 {\cal A}^-_2   \Psi_2(y)= E^-_2 \Psi_2(y)
\end{eqnarray}
 of level $l=2$ as well as the solution $\Psi_2(y)=\smbr{{\cal A}^{-\dagger}_2/\sqrt{E^-_2}}  \Psi_1(y)$.

Keeping running this procedure to the generic $l$-th level ($l>2$), we obtain the eigen-equation \eqnm{Schr_equ_y_coord_minus}  and the raising recurrence relation \eqnm{raising_recursion_relation_Hermitian}. Thus,  once the factorized structure of the lowest level eigen-equation \eqnm{base_eigen_equ_minus} for the operator $pH_0$ is figured out, as has been demonstrated in Subsection \ref{base_kinetic_energy_operator}, and the superpotentials of high levels are appropriately designed, the SUSY and SISY together then {\it naturally give birth to} all Hamiltonians of higher levels and their eigen-solutions.  This bottom-up iterative approach highlights the fundamental roles played by constructing the kinetic energy operator $-\smbr{p d/dx}^2$  and the supersymmetric factorization of $p H_0$.  
That is, beginning with this ‘grandmother’ Hamiltonian ${\mathbb H}_0(x)\equiv p H_0$ and its SUSY factorization, and running the {\it standard} SUSYQM algorithm (the intertwining action of the right-situated operator in the product form of a Hamiltonian, plus the normal ordering), a hierarchy of descendant Hamiltonians $\{{\mathbb H}_l(x)={\cal H}_l(y)\}$ of higher levels, their SUSYQM factorization and their eigen-solutions, are proliferated iteratively. In Section \ref{associated_hypergeometric_functions}, we will see that this \emph{powerful} iterative procedure can also generate the associated hypergeometric-like differential equations of all association levels and their eigen-solutions, starting from another type of SUSY factorization of $H_0$. 

In this explicit iterative procedure, we see that the SUSYQM algorithm actually guides and even {\it completely dictates} the whole flow of proliferating the eigen-equations as well as their solutions for all higher levels, once we figure out the supersymmetric factorization \eqnm{pH0_supersymmetrized} or \eqnm{initial_eigen_equ_minus} of the initial Hamiltonian ${\cal T}(pH_0)$, that is constructed for the specific kinetic energy operator $-(p d/dx)^2$. Such flow chart of proliferation is plotted as the left column of Diagram \ref{diagram1}.

\subsection{The Upward-connected Shape Invariant Potential Series and the Related Supersymmetric Factorizations}\label{upward_SIC}
 The superpartner potential ${\cal V}^s_l$ in \eq{superpartner_Vl} can also be different from the upper nearest neighboring potential ${\cal V}_{l+1}$ by a constant. This type of relation is termed {\it upward-connected} shape invariance condition and the related functions, operators of factorizations based on this condition are superscripted with a plus sign, and the corresponding factorization is termed the plus branch or route. That is, parallel to the potential relation \eqnm{SI_potential_minus}, we also have
\begin{eqnarray}\label{SI_potential_plus}
 {\cal V}^{s+}_l = {\cal V}^+_{l+1} - \delta^+_l,
\end{eqnarray}
and the equivalent generalized commutator relation reads
\begin{eqnarray}\label{SI_operator_product_plus}
{\cal A}^+_l{\cal A}^{+\dagger}_l -{\cal A}^{+\dagger}_{l+1}{\cal A}^+_{l+1}=- \delta^+_l,
\end{eqnarray}
where
\begin{equation}\label{A_and_W_plus}
\begin{aligned}
{\cal A}^+_l(y) & \equiv \od{}{y} + {\cal W}^+_l(y) = p \od{}{x}+ W^+_l(x), \\
{\cal A}^{+\dagger}_l(y) &\equiv - \od{}{y} + {\cal W}^+_l(y) =-p\od{}{x} +W^+_l(x),\\
{\cal W}^+_l(y) &\equiv W^+_l(x) = \alpha^+_l x + \beta^+_l. 
\end{aligned}
\end{equation}

Now the {\it plus} version of \eq{Schr_equ_y_coord} reads
\begin{eqnarray}\label{eigeneua_plus_l}
 {\cal H}^+_l \Psi_l \smbr{y}= E^+_l \Psi_l\smbr{y}
\end{eqnarray}
where 
\begin{eqnarray}\label{H_plus_l}
{\cal H}^+_l={\cal A}^{+\dagger}_l(y){\cal A}^+_l(y). 
\end{eqnarray}

Intertwiningly acting \eq{eigeneua_plus_l} with  ${\cal A}^+_l(y)$ yields the superpartner eigen-equation
\begin{eqnarray}\label{superpartner_eigen_equ_plus}
 {\cal H}^{s+}_l \Psi^{s+}_l \smbr{y} \equiv {\cal A}^+_l(y) {\cal A}^{+\dagger}_l(y) \Psi^{s+}_l \smbr{y}= E^+_l \Psi^{s+}_l\smbr{y}, \nonumber \\
\end{eqnarray}
where
\begin{eqnarray}
\Psi^{s+}_l \smbr{y}\propto A^+_l \Psi_l(y).
\end{eqnarray}

The shape invariance condition \eqnm{SI_potential_plus} enables us to normally order \eq{superpartner_eigen_equ_plus} into 
\begin{eqnarray}\label{superpartner_Schr_equ_raising_after_rnmlz_plus}
 {\cal H}^+_{l+1} \Psi_{l+1} \smbr{y}= E^+_{l+1} \Psi_{l+1}\smbr{y}
\end{eqnarray}
that has the same pattern as \eq{eigeneua_plus_l}, with ${\cal H}^+_{l+1}$ being the $(l+1)$ version of \eqnm{H_plus_l}, 
\begin{eqnarray}
E^+_{l+1} = E^+_l + \delta^+_l
\end{eqnarray}
and 
\begin{eqnarray}\label{pre_raising_recurrence_plus}
\Psi_{l+1}(y)= \Psi^{s+}_l \smbr{y}\propto A^+_l \Psi_l(y).
\end{eqnarray}

In view of \eq{superpartner_Schr_equ_raising_after_rnmlz_plus} and by choosing a specific normalization,  the relation \eqnm{pre_raising_recurrence_plus} is then turned into the raising recurrence relation 
\begin{eqnarray}\label{raising_recursion_relation}
 \Psi_{l+1}\smbr{y} = \frac{{\cal A}^+_l(y)}{\sqrt{E^+_{l}}}\Psi_l\smbr{y},
\end{eqnarray}
in which $A^+_l(y)$ works as the raising ladder operator. Accompanying with is the lowering recurrence relation
\begin{eqnarray}\label{lowering_recursion_relation}
 \Psi_l\smbr{y}= \frac{{\cal A}^{+\dagger}_l(y)}{\sqrt{E^+_{l}}}\Psi_{l+1}\smbr{y}.
\end{eqnarray}
in which  ${\cal A}^{+\dagger}_l(y)$ works as the lowering ladder operator. Here the role of the daggered operator ${\cal A}^{+\dagger}_l(y)$ is parallel to that of the undaggered operator ${\cal A}^-_l(y)$ and ${\cal A}^{-\dagger}_l(y)$ in the minus route, whereas the role of ${\cal A}^+_l(y)$ is to that of the daggered ${\cal A}^{-\dagger}_l(y)$. 

Note that the notation of the eigenfunction for all $l\ge 0$ adopted here for the plus route is the same as for the minus one. This is why we do not add either plus or minus superscript to the eigenfunction. The consistency of this assumption will be further justified as we proceed. 

In this plus route of factorization, just because the potential series of shape invariance are upward-connected in the form \eqnm{SI_potential_plus}, the right-situated operator ${\cal A}^+_l(y)$ inside ${\cal H}^+_l={\cal A}^{+\dagger}_l(y){\cal A}^+_l(y)$ then works as the raising ladder operator, and it is the shape invariance condition \eqnm{SI_operator_product_plus} that plays the role of renormalizing the superpartner eigen-equation into the normally ordered one. This is in contrast to the minus route of factorization for the downward-connected potential series, as shown in \ref{minus_route_iteration}, where it is the intertwining operator multiplication that plays such a role.
 
\subsubsection{The Relations between the Two Routes of Factorizations}\label{Plus_route_factorization}
Starting with the upward-connected shape invariance condition \eqnm{SI_potential_plus} and following the steps that are similar to the factorization of the minus route demonstrated in Subsection \eqnm{Minus_route_factorization}, we can find out the algebraic recurrence relations for $\alpha^+_l$ and $\beta^+_l$ for the plus route factorization. 

According to \eq{superpartner_Vl},
\begin{eqnarray}\label{superpartner_Vl_plus}
 V^{s+}_l\smbr{x}= p \od{W^+_l(x)}{x} + W^{+2}_l(x),
\end{eqnarray}
and to \eq{potential_x_coord},
\begin{eqnarray}
V^+_{l+1}(x)=-p\, \od{W^+_{l+1}(x)}{x} + W^{+2}_{l+1}(x).
\end{eqnarray}
With these substitutions, \eq{SI_potential_plus} becomes the differential recurrence relation
\begin{eqnarray}\label{recursion_RL_W_plus}
p \od{}{x}\sqbr{W^+_l+ W^+_{l+1}}+ W^{+2}_l- W^{+2}_{l+1}=-\delta^+_l
\end{eqnarray}
for the series $\{W^+_l(x)\}$. Upon further inserting $W^+_l(x)=\alpha^+_l x +\beta^+_l$ and $W^+_{l+1}(x)=\alpha^+_{l+1} x +\beta^+_{l+1}$ and balancing the coefficients in front of $x^2$, $x$ and $x^0$, respectively, we obtain the following algebraic recurrence relations
\begin{eqnarray}\label{alpha_l_plus_recursion}
 \alpha^+_{l+1}-\alpha^+_l =\frac{p''}{2},
\end{eqnarray}
\begin{eqnarray}\label{beta_plus_recursion}
 \alpha^+_{l+1}\beta^+_{l+1} - \alpha^+_l \beta^+_l = \frac{p'_0}{2}\smbr{\alpha^+_{l+1} + \alpha^+_l}
\end{eqnarray}
and 
\begin{eqnarray}\label{delta_l_plus}
 \delta^+_l = - p_0 \smbr{\alpha^+_{l+1} + \alpha^+_l} +\smbr{\beta^{+2}_{l+1} - \beta^{+2}_l}.
\end{eqnarray}

Under $\alpha^+_l \rightarrow -\alpha^-_{l+1}$ and $\beta^+_l \rightarrow -\beta^-_{l+1}$, $W^+_l$ is transformed into $-W^-_{l+1}$, and \eq{recursion_RL_W_plus} into the $(l+2)$-version of \eq{recursion_RL_W_minus}, as long as $\delta^+_l = \delta^-_{l+2}$. In addition, Eqs. \eqnm{alpha_l_plus_recursion}, \eqnm{beta_plus_recursion} \eqnm{delta_l_plus} are transformed into the $(l+2)$-versions of Eqs. \eqnm{coefficient_x^2},  \eqnm{coefficient_x} and \eqnm{coefficient_x^0}, respectively.  These symmetries indicate that there exist the following relations between the plus and minus routes of factorizations,
\begin{equation}\label{symmetry1}
\begin{aligned}
\alpha^+_l &= -\alpha^-_{l+1},\\
\beta^+_l &= -\beta^-_{l+1},\\
W^+_l &=-W^-_{l+1},
\end{aligned}
\end{equation}
\begin{equation}\label{symmetry2}
 \begin{aligned}
 \delta^+_l &=\delta^-_{l+2},\\
 E^+_l &= E^-_{l+1},
 \end{aligned}
\end{equation}
which will be rigorously proved later on.  With these relations in hand, the expressions for $\alpha^+_l$, $\beta^+_l$, $\delta^+_l$ and $E^+_l$ in terms of the elements in $\{c_l\}$ and $\{d_l\}$ are obtained from those for the minus route of factorization.

The third equation in \eqnm{symmetry1} implies that
\begin{equation}\label{symmetry3}
 \begin{aligned} 
  {\cal A}^+_l(y) &=-{\cal A}^{-\dagger}_{l+1}(y),\\
  {\cal A}^{+\dagger}_l(y)&=-{\cal A}^-_{l+1}(y), 
 \end{aligned}
\end{equation}
which tell us explicitly why ${\cal A}^+_l(y)$ and ${\cal A}^-_l(y)$ (resp. ${\cal A}^{+\dagger}_l(y)$ and ${\cal A}^{-\dagger}_l(y)$), as ladder operators, play opposite roles in the two routes of factorization.

The combination of \eq{symmetry3}, the $(l+1)$-version of \eq{Schr_equ_y_coord_minus} and the condition \eqnm{SI_operator_product_plus} leads to   
\begin{eqnarray}
{\cal A}^{+\dagger}_{l+1}(y){\cal A}^+_{l+1}(y)\Psi_{l+1}(y)=\smbr{E^-_{l+1}+\delta^+_l}\Psi_{l+1}(y). 
\end{eqnarray}
Comparing this equation with \eq{superpartner_Schr_equ_raising_after_rnmlz_plus}, we have
\begin{eqnarray}\label{E_plus_minus_relation}
E^+_{l+1}=E^-_{l+1}+\delta^+_l. 
\end{eqnarray}

Substituting for $V^{s+}_l$ in the plus-version of \eq{potential_difference} via the condition \eqnm{SI_potential_plus} yields the recurrence relation 
\begin{eqnarray}\label{V_difference1}
 V^+_{l+1}-V^+_l=2\alpha^+_l p+\delta^+_l
\end{eqnarray}
for the potential series $\{V^+_i\}$.  On the other hand, the plus-version of \eq{Vl_potential}, which is
\begin{eqnarray}\label{Vl_potential_plus}
V^+_l(x) &\equiv&  V_0(x)- \lambda^+_l p + E^+_l,
\end{eqnarray}
yields
\begin{eqnarray}\label{V_difference2}
 V^+_{l+1}-V^+_l=\smbr{\lambda^+_l-\lambda^+_{l+1}} p + E^+_{l+1}-E^+_l.
\end{eqnarray}
Eqs \eqnm{V_difference1} and \eqnm{V_difference2} together give rise to
\begin{equation}\label{alpha_l_plus}
\begin{aligned}
 \alpha^+_l &=\frac{1}{2}\smbr{\lambda^+_l-\lambda^+_{l+1}},\\
 \delta^+_l &= E^+_{l+1}-E^+_l.
\end{aligned} 
\end{equation}

The second relation in \eqnm{symmetry2} is then proved by combining \eq{E_plus_minus_relation} with the second equation in \eqnm{alpha_l_plus}.

In the minus route of factorization, $l$ is originally restricted to be a non-negative integer. However, the relations \eqnm{symmetry1}-\eqnm{symmetry3} indicate that $l$ can start with $-1$ in the plus route of factorization.  Then, we can make the identification
\begin{eqnarray}\label{V_minus1_identification}
 V^+_{-1}(x) \equiv V_0(x).
\end{eqnarray}
Compare this equation with
\begin{eqnarray}\label{V_plus_minus1}
V^+_{-1}(x)\equiv V_0(x) - \lambda^+_{-1} p + E^+_{-1},
\end{eqnarray}
which is just the $l=-1$ case of the plus version of \eq{Vl_potential}, we have to choose
\begin{equation}\label{lambda_plus_minus1}
 \begin{aligned}
 \lambda^+_{-1}&=0,\\
 E^+_{-1}&=E_0=0.
 \end{aligned}
\end{equation}
On the other hand, compare \eq{V_minus1_identification} with \eq{V_0_and_E_0}, we have
\begin{eqnarray}\label{V_plus_minus2}
V^+_{-1} = p \od{W^+_{-1}}{x} +\smbr{W^+_{-1}}^2
\end{eqnarray}
with 
\begin{eqnarray}\label{W_plus_minus_one}
 W^+_{-1} \equiv -W_0=-W^-_0,
\end{eqnarray}
where
\begin{eqnarray}\label{Wplusminus1expansion}
 W^+_{-1}=\alpha^+_{-1} x+ \beta^+_{-1}.
\end{eqnarray}

The combination of Eqs. \eqnm{W_plus_minus_one}, \eqnm{Wplusminus1expansion}, \eqnm{W0_minus} and \eqnm{Wminus0expansion} then gives the initial data $\alpha^+_{-1}$ and $\beta^+_{-1}$  as 
\begin{eqnarray}
\alpha^+_{-1}=-\frac{1}{2}(p''-q') = -\alpha^-_0, \label{inital_alpha_plus}\\
\beta^+_{-1}=-\frac{1}{2}(p'_0-q_0)=-\beta^-_0, \label{initial_beta_plus}
\end{eqnarray}
which confirm the $l=-1$ case of the first two generic relations in \eqnm{symmetry1}. 

Now with $\alpha^+_{-1}$ in \eq{inital_alpha_plus}, the arithmetical recurrence relation \eqnm{alpha_l_plus_recursion} is solved as
\begin{eqnarray}\label{alpha_plus_l}
 \alpha^+_l = \frac{l+1}{2} p'' +\alpha^+_{-1} =\frac{1}{2} \smbr{l p'' + q'}\equiv c_l 
\end{eqnarray}
for $l\ge -1$, where the definition of $c_l$ is consistent with that of $c_{l-1}$ in \eq{alpha_l_minus_recursion}. \eq{alpha_plus_l} together with $\alpha^-_{l+1}=-c_l$ (the $(l+1)$-version of \eq{alpha_l_minus_recursion}), confirm the first relation in  \eq{symmetry1}.  

In the identity   $\alpha^+_l \beta^+_l = \alpha^+_{-1} \beta^+_{-1} + \sum^{l}_{j=0}\smbr{\alpha^+_j \beta^+_j-\alpha^+_{j-1} \beta^+_{j-1}}$, making substitutions for $\alpha^+_{-1}$ via \eq{inital_alpha_plus}, for $\beta^+_{-1}$ via \eq{initial_beta_plus} and for $\alpha^+_j \beta^+_j-\alpha^+_{j-1} \beta^+_{j-1}$ via \eq{beta_plus_recursion} then yields
\begin{eqnarray}\label{alpha_beta_l_plus}
&&\alpha^+_l \beta^+_l \nonumber \\
&=&\frac{p'_0}{2}\sum^l_{j=0}\smbr{2 \alpha^+_j-\frac{p''}{2}}+\frac{1}{4}\smbr{p''-q'}\smbr{p'_0-q_0},\nonumber \\
 &=&\frac{(l+1) p'_0}{4}\sqbr{(l-1)p''+2 q'} +\frac{1}{4}\smbr{p''-q'}\smbr{p'_0-q_0},\nonumber \\
 &=&g_{l+1},
\end{eqnarray}
where the definition of $c_l$ in \eq{alpha_plus_l} and that of $g_l$ in \eq{alpha_beta_product} have been made use of, to reach respectively the second and third equalities.

Substituting for $\alpha^+_l$ via the last equality in \eq{alpha_plus_l} and making use of \eq{alpha_beta_l_plus}, $\beta^+_l$ is solved as 
\begin{eqnarray}\label{beta_plus_l}
 \beta^+_l &=& \frac{\alpha^+_l \beta^+_l}{\alpha^+_l}=\frac{g_{l+1}}{c_l}.
\end{eqnarray}
Comparing this with \eq{beta_minus_l} for $\beta^-_l$, the second relation in \eq{symmetry1}) is confirmed.

In the identity $E^+_l=E^+_{-1} + \sum^l_{j=0}\smbr{E^+_j-E^+_{j-1}}$, making use of $E^+_j-E^+_{j-1}=\delta^+_{j-1}$ (the second equation in \eq{alpha_l_plus} with $l \rightarrow j-1$) and \eq{delta_l_plus} yields
\begin{eqnarray}\label{E_plus_l}
 E^+_l &=&-p_0\sum^j_{j=0}\smbr{\alpha^+_j + \alpha^+_{j-1}} + \beta^{+2}_l-\beta^{+2}_{-1},\nonumber\\
       &=&p_0\sum^{l+1}_{j=1}\smbr{\alpha^-_j + \alpha^-_{j-1}} + \beta^{-2}_{l+1}-\beta^{-2}_0,
\end{eqnarray}
where the second line has been reached by making use of the confirmed first two relations in \eqnm{symmetry1}.
On the other hand, making the substitution for $\delta^-_j$ via \eq{coefficient_x^0} in $E^-_l=E^-_0 + \sum^l_{j=1}\smbr{E^-_j-E^-_{j-1}}=\sum^l_{j=1}\delta^-_j$ gives
\begin{eqnarray}\label{E_minus_l}
 E^-_l = p_0\sum^{l}_{j=1}\smbr{\alpha^-_j + \alpha^-_{j-1}} + \beta^{-2}_l-\beta^{-2}_0.
\end{eqnarray}
Eqs. \eqnm{E_plus_l} and \eqnm{E_minus_l} together then confirm the second relation in \eq{symmetry2}.  

Combining the two lower relations in Eqs. \eqnm{alpha_l_plus} and \eqnm{alpha_l_minus}, the upper relation in \eq{symmetry2} is then confirmed.

Making use of the upper equation for $\lambda^+_{-1}$ in \eqnm{lambda_plus_minus1}, 
the $(j-1)$-version of the upper relation in \eq{alpha_l_plus} for $\lambda^+_j-\lambda^+_{j-1}$ and \eq{alpha_plus_l} for $\alpha^+_{j-1}$, the identity $\lambda^+_l =\lambda^+_{-1}+\sum^l_{j=0}\smbr{\lambda^+_j-\lambda^+_{j-1}}$ then reduces to
\begin{eqnarray}
 \lambda^+_l &=&-\smbr{l+1} q'-\frac{1}{2} \smbr{l-2}\smbr{l+1} p'',\nonumber\\
             &=& \lambda^-_{l+1}+ \smbr{l+1}p''=\lambda^-_l+ p''-q', \label{lambda_plus_l}
\end{eqnarray}
where to reach the second equality, \eq{lambda_minu_eigenvalue} has been made use of. We see that the relation between $\lambda^+_l$ and $\lambda^-_{l+1}$ or $\lambda^-_l$ is not so homogeneous as those for other parameters in \eqnm{symmetry1} and \eqnm{symmetry2}.
  
The combination of the three upper equations in \eqnm{alpha_l_minus}, \eqnm{alpha_l_plus} and  \eqnm{symmetry1} gives us
\begin{eqnarray}\label{2nd_relation_lambda_plus_l}
\lambda^+_{l+1} - \lambda^+_l =\lambda^-_{l+1} - \lambda^-_l.
\end{eqnarray}
It is easy to check that this relation can also be obtained by combining the relation \eqnm{lambda_plus_l} with \eq{difference_lambdal_minus}. 

The correctness of the above calculations for the constants $\alpha^+_l$, $\beta^+_l$, $\delta^+_l$, $\lambda^+_l$ and $E^+_l$ is further confirmed by directly matching presented in Appendix \ref{solution_Wl_and_El}.

It is tempting to make the guess that $\lambda^+_l=\lambda^-_l$ by looking at the relation \eqnm{2nd_relation_lambda_plus_l} alone. However, the inhomogeneous term in \eq{lambda_plus_l} defies such naive guess. It is worthwhile to further clarify this counterintuitive fact and to answer the natural question: what is the differential operator that has the eigenvalue $\lambda^+_l$ and whose eigen-equation is analogous to \eq{eigen_equ}? 

The fact that $E^+_{-1}=0$ (the lower equation in \eqnm{lambda_plus_minus1})  implies the $l=-1$ case of \eq{eigeneua_plus_l} reads 
\begin{eqnarray}\label{initial_eigeneq_plus_route}
{\cal H}^+_{-1} (y)\Psi^+_{-1}(y) \equiv {\cal A}^{+\dagger}_{-1}(y){\cal A}^+_{-1}(y) \Psi^+_{-1}(y)=0
\end{eqnarray}
for 
\begin{equation}
\begin{aligned}
{\cal A}^{+}_{-1}(y) &\equiv p \od{}{x}+ W^+_{-1}(x),\\
{\cal A}^{+\dagger}_{-1}(y)&\equiv-p \od{}{x}+ W^+_{-1}(x)
\end{aligned}
\end{equation}
and $\Psi^+_{-1}(y)$  being the ground state.

With \eq{V_minus1_identification}, \eq{Vl_potential_plus} is rewritten as 
\begin{eqnarray}
V^+_l = V^+_{-1} - \lambda^+_l p + E^+_l. 
\end{eqnarray}
Adding $-d^2/dy^2$ on both sides of the above equation and making use of Eqs. \eqnm{H_plus_l} and \eqnm{initial_eigeneq_plus_route} give rise to
\begin{eqnarray}
{\cal H}^+_l (y) = {\cal H}^+_{-1}(y)+ E^+_l - \lambda^+_l p.
\end{eqnarray}
Insert this expression for ${\cal H}^+_l (y)$ into \eq{eigeneua_plus_l} leads to 
\begin{eqnarray}\label{partner_eigen_equ_y_coord}
p^{-1}{\cal H}^+_{-1}(y)\Psi_l(y) = \lambda^+_l \Psi_l(y).
\end{eqnarray}

In view of the $l=-1$ case of the two relations in \eqnm{symmetry3} and \eq{initial_eigeneq_plus_route}, we have 
\begin{eqnarray}
{\cal H}^+_{-1}(y) = {\cal A}^-_0(y) {\cal A}^{-\dagger}_0(y). 
\end{eqnarray}
Applying the transformation ${\cal T}^{-1}\smbr{-}$ in \eq{T_transformation} to the operator $p^{-1}{\cal H}^+_{-1}(y)$ in \eq{partner_eigen_equ_y_coord}, we obtain
\begin{eqnarray}\label{H0_plus}
 H^+_0(x) = p^{-1} \mathsf{A}^-_0 \mathsf{B}^-_0 =\smbr{\od{}{x}}\smbr{-p \od{}{x} + 2W_0}, 
\end{eqnarray}
where
\begin{eqnarray}
 H^+_0(x)&\equiv& p^{-1} {\cal T}^{-1}\smbr{{\cal H}^+_{-1}(y)},\nonumber \\
 \mathsf{A}^-_0 &\equiv& {\cal T}^{-1}\smbr{{\cal A}^-_0(y)}=p \od{}{x},\label{A0_minus} \\
 \mathsf{B}^-_0 &\equiv& {\cal T}^{-1}\smbr{{\cal A}^{-\dagger}_0(y)}=-p \od{}{x} + 2W_0. \label{B0_minus}
\end{eqnarray}

Multiplying \eq{partner_eigen_equ_y_coord} with $\sqrt{w}^{-1}$ from the left on both sides and inserting the unit $\sqrt{w} \sqrt{w}^{-1}$ between $\Psi(y)$ and the rest on the left hand side for preparing the transformation ${\cal T}^{-1}$, and making use of \eq{H0_plus}, \eq{partner_eigen_equ_y_coord} is then recast into the following eigen-equation
\begin{eqnarray}\label{eigenequ_lambda_plusl}
H^+_0(x)\Phi_l(x)=p^{-1} \mathsf{A}^-_0 \mathsf{B}^-_0\Phi_l(x)=\lambda^+_l  \Phi_l(x)
\end{eqnarray}
for the eigenfunction $\Phi_l(x)\equiv \sqrt{w}^{-1}\Psi_l(y)$ and the eigenvalue $\lambda^+_l$. This is the sought-after eigen-equation that has the eigenvalue $\lambda^+_l$ and is analogous to \eq{eigen_equ}. 

Though the two operators $H^\pm_0(x)$ share the eigenfunction $\Phi_l(x)$, the {\it partner} eigen-equation \eqnm{eigenequ_lambda_plusl} is \emph{different} from the original eigen-equation \eqnm{eigen_equ}. The latter can be rewritten as
\begin{eqnarray}\label{eigenequ_lambda_minusl}
 H^-_0(x)\Phi_l(x)=p^{-1} \mathsf{B}^-_0 \mathsf{A}^-_0 \Phi_l(x)=\lambda^-_l \Phi_l(x)
\end{eqnarray}
for eigenvalue $\lambda^-_l$ and 
\begin{eqnarray}\label{H0_minus}
H^-_0(x)\equiv H_0(x)=\smbr{-\od{}{x}\; p  +2W_0}\smbr{\od{}{x}} 
\end{eqnarray}
which follows from \eq{simple_factorization}. In fact, the key difference between $H^+_0$ given in \eq{H0_plus} and $H^-_0$ in \eq{H0_minus} is order of $\mathsf{B}^-_0$ and $\mathsf{A}^-_0$ in the operator product, which explains why $\lambda^+_l\ne \lambda^-_l$,  and it is easy to check that
\begin{eqnarray}\label{H0_difference}
H^+_0 - H^-_0=p^{-1}\sqbr{\mathsf{A}^-_0, \mathsf{B}^-_0}=2 W'_0.
\end{eqnarray}
Eqs. \eqnm{H0_difference}, \eqnm{eigenequ_lambda_plusl} and \eqnm{eigenequ_lambda_minusl} together imply that 
\begin{eqnarray}
\lambda^+_l=\lambda^-_l+2 W'_0, 
\end{eqnarray}
which is just the relation \eqnm{lambda_plus_l} in view of \eq{W_0} for $W_0$. That is, the difference $p''-q'$  in \eq{lambda_plus_l} as the inhomogeneous piece results from swapping $\mathsf{A}^-_0$ and $\mathsf{B}^-_0$  when passing from \eq{eigenequ_lambda_plusl} to \eq{eigenequ_lambda_minusl}.

The plus route of factorization just established is actually a direction factorization of the eigen-equation
\begin{eqnarray}
 \sqbr{pH^+_0(x) -\lambda^+_l\, p + E^+_l}\Phi_l(x) = E^+_l \Phi_l(x)
\end{eqnarray}
that is equivalent to \eq{eigenequ_lambda_plusl}, with integers $l\ge -1$ and the initial eigenvalue $E^+_{-1} =0$.

\subsubsection{An Explicit Bottom-up Iterative Flow of Proliferating a Hierarchy of Eigen-equations and Their Solutions -- the Plus Branch}\label{proliferation_plus_route}
Again, in order to clearly see how the standard SUSQM algorithm
dictates the flow of simultaneously proliferating all higher levels of Hamiltonians ${\cal H}^+_l(y)$ and their eigen-solutions, from the initial (ground-state) Hamiltonian ${\cal H}^+_{-1}(y)$ in its SUSYQM factorization ${\cal H}^+_{-1}(y)={\cal A}^{+\dagger}_{-1}(y){\cal A}^+_{-1}(y)$ and its eigen-equation, quite similar to the minus route of factorization, the above plus route of SUSYQM factorization can also be explicitly turned into as a bottom-up iteration procedure. 

Recall that in the plus route, the undaggered operator ${\cal A}^+_{-1}$ works as a raising operator (which will not kill any state) and the daggered one ${\cal A}^{+\dagger}_{-1}$ as a lowering operator. Hence, it is inappropriate to assume that ${\cal A}^+_{-1}(y) \Psi^+_{-1}(y)=0$. Instead, it is reasonable to assume that
\begin{eqnarray}\label{lowest_raising_relation_plus}
{\cal A}^+_{-1}\Psi^+_{-1} \propto \Psi^+_0 \ne 0,
\end{eqnarray}
and \eq{initial_eigeneq_plus_route} is interpreted as the consequence of
\begin{eqnarray}\label{annihilation_psi_0_plus}
{\cal A}^{+\dagger}_{-1} \Psi^+_0=\sqbr{-p \od{}{x} + W^+_{-1}(x)} \Psi^+_0= 0.
\end{eqnarray}
Note that $W^+_{-1}(x)=-W_0(x)$ (\eq{W_plus_minus_one}). Thus, \eq{annihilation_psi_0_plus} is just \eq{ground_state_minus_route}, the annihilation condition \eqnm{annihilation_psi_0_minus} of the ground state for the {\it minus} route of factorization, and it has
the solution $\Psi^+_0 \propto \sqrt{w}$, i.e. $\Psi^+_0$ can be taken to be the same as $\Psi^-_0=\sqrt{w}$, the ground state of the minus route. For this reason, there is no need to make distinction between $\Psi^+_0$ and $\Psi^-_0$, and both of them are denoted as $\Psi_0$, as we have done for any level of eigenfunction. Then running the SUSYQM algorithm will generate $\Psi^+_l$ for all higher $l$, each of which, as will be seen, is also the same as that of the minus route of the same $l$-level. For this reason all the eigenfunctions of these two routes of factorizations are denoted by the same $\Psi_l$. 

With $\Psi^+_0 \propto \sqrt{w}$, the {\it auxiliary} eigenfunction $\Psi^+_{-1}$ of the lowest level can be solved from \eq{lowest_raising_relation_plus} as
\begin{eqnarray}\label{eigenfunction_minus1}
 \Psi^+_{-1} \propto \frac{1}{\sqrt{w}} \int^x d\bar{x} \frac{w(\bar{x})}{p(\bar{x})}.
\end{eqnarray}
Conversely, if we take $\Psi^+_0 \propto \sqrt{w}$, then \eq{annihilation_psi_0_plus} gives the solution $W^+_{-1}=\smbr{p d w/dx}/\smbr{2 w}= -\smbr{p'-q}/2=-W_0$, and $\alpha^+_{-1}= -\smbr{p''-q'}/2$ and $\beta^+_{-1}=-\smbr{p'_0-q_0}/2$. This is an alternative way of providing the initial data for solving $\alpha^+_l$, $\beta^+_l$ and $\delta^+_l$ for the plus route of factorization. 

In view that the raising operator ${\cal A}^+_{-1}(y)$ is situated in the {\it right} position in \eq{initial_eigeneq_plus_route}, directly acting on this equation or \eq{annihilation_psi_0_plus} with it then raises it into the eigen-equation of level $l=0$ in its pre-normally-ordered form, i.e. 
\begin{eqnarray}\label{initial_eigen_equ_plus}
{\cal A}^+_{-1}{\cal A}^{+\dagger}_{-1} \Psi_0=0,
\end{eqnarray}
which is essentially \eq{initial_eigen_equ_minus} or \eq{base_eigen_equ_minus}. 

Making use of 
\begin{eqnarray}
{\cal A}^+_{-1}{\cal A}^{+\dagger}_{-1}={\cal A}^{+\dagger}_0 {\cal A}^+_0-\delta^+_{-1}, 
\end{eqnarray}
which is the $l=0$ version of the upward-connected shape invariance condition \eq{SI_operator_product_plus}, \eq{initial_eigen_equ_plus} is then renormalized into the eigen-equation of the {\it first excited level} ($l=0$)
\begin{eqnarray}\label{eigen_equ_1st_level}
{\cal A}^{+\dagger}_0 {\cal A}^+_0 \Psi_0=E^+_0 \Psi_0,
\end{eqnarray}
where $E^+_0 \equiv \delta^+_{-1}$. 

Acting on \eq{eigen_equ_1st_level} with  ${\cal A}^+_0$ and renormalizing the resulting superpartner eigen-equation by making use of the  shape invariance condition 
\begin{eqnarray}
{\cal A}^+_0 {\cal A}^{+\dagger}_0={\cal A}^{+\dagger}_1 {\cal A}^+_1- \delta^+_0, 
\end{eqnarray}
which is the $l=1$ version of \eq{SI_operator_product_plus}, then yield the eigen-equation of {\it second excited level} ($l=1$)
\begin{eqnarray}\label{eigen_equ_2nd_level}
 {\cal A}^{+\dagger}_1 {\cal A}^+_1 \Psi_1 = E^+_1 \Psi_1
\end{eqnarray}
for $\Psi_1  = \smbr{{\cal A}^+_0/\sqrt{E^+_0}} \Psi_0$ and $E^+_1=E^+_0+ \delta^+_0$.

By keeping running this algorithm to higher and higher levels, eventually all levels of the eigen-equations, eigenfunctions and eigenvalues will be constructed deductively. In particular,  the generic two-term recurrence relations of the eigenfunction are obtained in the same form as in \eq{raising_recursion_relation}.  Throughout this procedure, we clearly see that the initial SUSYQM factorization \eqnm{initial_eigeneq_plus_route} or \eqnm{initial_eigen_equ_plus} plays the foundational role in initiating this iterative procedure, and that the rest of the process of proliferating all eigen-equations of higher levels and their eigen-solutions is completely controlled by the standard SUSYQM algorithm, especially by the two repeating operations: intertwining action and normally ordering, when appropriate shape invariance conditions are incorporated. We also see that, in each step of this plus route of factorization, it is the shape invariance condition that completes the renormalization of the superpartner eigen-equation into the normally ordered one. 

Comparing to the $l=0$ level SUSYQM factorization $pH_0={\cal H}^-_0(y)={\cal A}^{-\dagger}_0 {\cal A}^-_0$ that has been introduced in Subsection \ref{base_kinetic_energy_operator} as the initial SUSYQM factorization for proliferating all equations of higher levels in the minus route, the initial $l=-1$ SUSYQM factorization ${\cal H}^+_{-1}(y) = {\cal A}^{+\dagger}_{-1}(y){\cal A}^+_{-1}(y)$ in \eq{initial_eigeneq_plus_route} for this plus route stands out as a peculiarity in the sense that there is an {\it extra} $l=-1$ level of factorization, with the eigenfunction $\Psi^+_{-1}(y)$ and eigenvalue $\lambda^+_{-1}$ in this plus route. One may wonder why this extra level comes out and how it is related to the minus route solutions. Obviously, ${\cal H}^+_{-1}(y)$ is the superpartner Hamiltonian of ${\cal H}^-_0(y)$, as ${\cal A}^{+\dagger}_{-1}(y){\cal A}^+_{-1}(y)={\cal A}^-_0 {\cal A}^{-\dagger}_0$ according to the relations in \eqnm{symmetry3}. That is, \eq{initial_eigeneq_plus_route} can be obtained from \eq{base_eigen_equ_minus} by intertwiningly acting the latter with the \emph{lowering} operator ${\cal A}^-_0$ and by introducing $\Psi^-_{-1} \propto {\cal A}^-_0\Psi_0 \propto\Psi^+_{-1}$, which is just the definition \eqnm{lowest_raising_relation_plus}. In doing so, we head to the first {\it negative} level ($l=-1$) eigen-equation in the minus route. Traditionally people seek for the {\it polynomial} eigenfunctions of $H_0$ with nonnegative level index $l$, the non-polynomial solution $\Psi^+_{-1}$ given in \eq{eigenfunction_minus1} can then be excluded from the overall solution series if one aims to build all the solutions of nonnegative levels by starting with \eq{initial_eigeneq_plus_route}; or one can directly start with \eq{initial_eigen_equ_plus} as the initial factorized eigen-equation for $\Psi_0$ and run the SUSYQM algorithm for the plus route to obtain all the solutions of higher levels. But if we do not restrict ourselves to the nonnegative levels, all the solutions with negative levels can be reached by starting with \eq{base_eigen_equ_minus} for the minus route or with \eq{initial_eigeneq_plus_route} for the plus route and running the SUSYQM algorithm in the {\it negative} direction. \eq{initial_eigeneq_plus_route} is just the $(-1)$-level instance of the eigen-equations of negative levels in the minus route. From this perspective, the distinctions between the two routes of solutions and the procedures for obtaining them almost disappear because one route of solutions can cover the other's. Thus, adopting either route of these two factorization schemes will reach all the same series of eigen-solutions.  Running the logic for passing from \eq{initial_eigeneq_plus_route} to \eq{eigenequ_lambda_plusl} reversely, the reason why there exists such $l=-1$ level eigen-equation \eqnm{initial_eigeneq_plus_route} for the ground state of the plus route can also be traced back to the fact that $\lambda^+_l$ satisfies its own eigen-equation \eqnm{eigenequ_lambda_plusl} that is different from \eq{eigenequ_lambda_minusl} satisfied by $\lambda^-_l$. Since \eq{eigenequ_lambda_plusl} holds even for $l=-1$ and $\lambda^+_{-1}=0$, applying the transformation ${\cal T}$ to \eq{eigenequ_lambda_plusl} gives rise to \eq{initial_eigeneq_plus_route}.

\subsection{Constructing the SUSYQM of the Base Hypergeometric-like Functions Directly in the Non-standard {\it x}-coordinate Representation}\label{non_standard_factorization}
Up to now, the SUSYQM for the base hypergeometric-like functions has been successfully established in $y$-coordinate. Since the corresponding kinetic energy operator takes its standard form, this $y$-coordinate representation is referred to as the \emph{standard representation}. One natural question to ask is: can this type of SUSYQM be established or represented directly in the original $x$-coordinate? The answer is yes. To approach such representation, we first note that, in $x$-coordinate with supersymmetric  factorization \eqnm{conjugate_pair} and the notations in \eq{transformed_Hl}, \eq{Schr_equ_y_coord} reads
\begin{eqnarray}\label{Schr_equ_x_coord_factorized}
&&\sqbr{-p \od{}{x} + W_l(x)} \sqbr{p \od{}{x} + W_l(x)} \sqbr{\sqrt{w}\Phi_l(x)} \nonumber \\
&&= E_l \sqbr{\sqrt{w}\Phi_l(x)}.
\end{eqnarray}
The equation satisfied by $\Phi_l(x)$ as eigenfunction can be obtained by peeling off the rescaling factor $\sqrt{w}$. This peeling-off procedure is equivalent to applying the transformation ${\cal T}^{-1}(-) \equiv \sqrt{w}^{-1}\smbr{-}\sqrt{w}$ to \eq{Schr_equ_x_coord_factorized}.  Multiplying both sides of \eq{Schr_equ_x_coord_factorized} with $\sqrt{w}^{-1}$ and inserting the unit $\sqrt{w} \sqrt{w}^{-1}$ between the product of the two operators, we obtain the factorization of \eq{eigen_equ_with_multiplier} in the form
\begin{eqnarray}\label{eigen_equ_NHF}
{\mathbb{H}}_l(x) \Phi_l \equiv \cubr{{\cal T}^{-1} \sqbr{{\cal H }_l(y)}} \Phi_l= \mathsf{B}_l \mathsf{A}_l \Phi_l =E_l \Phi_l,
\end{eqnarray}
 where
\begin{eqnarray}\label{raising_ops_NHF}
 \mathsf{B}_l\smbr{x} \equiv  {\cal T}^{-1} {\cal A}^\dagger_l (y) =-\sqbr{p\od{}{x}- W_0(x)}+W_l(x) 
 \end{eqnarray}
 and
\begin{eqnarray}\label{lowering_ops_NHF}
\mathsf{A}_l\smbr{x} \equiv {\cal T}^{-1} {\cal A}_l (y) = \sqbr{p\od{}{x}- W_0(x)} +W_l(x),
\end{eqnarray}
where $W_0(x)$ and $\sqrt{w}$ are related to each other through $\sqbr{p d/dx + W_0(x)}\sqrt{w}=0$ (the annihilation condition \eqnm{ground_state_minus_route}). We see that the net effect of peeling off the factor $\sqrt{w}$ from the eigenfunction $\Psi_l(y)$ is to asymmetrize the supersymmetric factorization \eqnm{Hamiltonian_SHF} by shifting the two operators ${\cal A}^\dagger_l (y)$ and ${\cal A}_l (y)$ by the same amount but in opposite directions, i.e. 
\begin{equation}\label{xy_connection}
\begin{aligned}
 \mathsf{B}_l(x) &= {\cal A}^\dagger_l (y) + W_0(x),  \\
 \mathsf{A}_l(x) &= {\cal A}_l (y) - W_0(x).
\end{aligned} 
\end{equation}
Here the sign difference in $\pm W_0$ memorizes that in $\mp p\,d/dx$ inside ${\cal A}^\dagger_l (y)$ and ${\cal A}_l (y)$  under the same {\it asymmetrizing} transformation ${\cal T}^{-1}$. 

Conversely, the difference -- $W_0(x)$ in $\mathsf{B}_l(x)$ and $-W_0(x)$ in $\mathsf{A}_l(x)$ that causes the asymmetry in $\mathsf{B}_l(x)\mathsf{A}_l(x)$ -- can be hidden as an additional rescaling factor $\sqrt{w}$ of the eigen-function, so that \eq{eigen_equ_NHF} is then turned back into \eq{Schr_equ_x_coord_factorized}. This is formally implemented by applying ${\cal T}$ to \eq{eigen_equ_NHF}. 

For $l=0$, the factorization ${\mathbb{H}}_l=\mathsf{B}_l \mathsf{A}_l$ in \eq{eigen_equ_NHF} reduces to 
\begin{eqnarray}
p H_0 = \mathsf{B}_0 \mathsf{A}_0, 
\end{eqnarray}
which is \eq{simple_factorization}, where
\begin{eqnarray}\label{A0_minus1}
\mathsf{A}_0 = p\, \od{}{x}\equiv \mathsf{A}^-_0 
\end{eqnarray}
and 
\begin{eqnarray}\label{B0_minus1}
\mathsf{B}_0 = -p\, \od{}{x }+2 W_0\equiv \mathsf{B}^-_0,
\end{eqnarray}
where $\mathsf{A}^-_0$ and $\mathsf{B}^-_0$ have been introduced in Eqs. \eqnm{A0_minus} and \eqnm{B0_minus}, respectively.

There are three key differences in the $x$-coordinate factorization. Firstly, with respect to the inner product $\dbrc{\Psi_i, \smbr{\cdots}\Psi_j} \equiv \int dy \Psi_i(y) \smbr{\cdots} \Psi_j(y)$ in $y$-coordinate, ${\cal A}^+_l(y)$ and ${\cal A}_l(y)$ are Hermitian conjugate pair; however, the shifted operators $\mathsf{A}_l(x)$ and $\mathsf{B}_l(x)$ in general are not a Hermitian conjugate pair any more with respect to the usual $w$-weighted inner product $\dbrc{\Phi_i, \smbr{\cdots}\Phi_j}_w\equiv \int dx w(x) \Phi_i(x) \smbr{\cdots} \Phi_j(x)$ in $x$-coordinate.
In fact, $\mathsf{A}^\dagger_l=\mathsf{B}_l-p'\ne \mathsf{B}_l$ with respect to the latter inner product. 
Moreover, in the $x$-coordinate, the momentum operator $-i pd/dx$ and kinetic energy operator $-(pd/dx)^2$ appear in their {\it non-standard} forms.  The last one is that the factorization \eqnm{eigen_equ_NHF} is obviously {\it asymmetric} in that the non-derivative terms $W_0+W_l$ in $\mathsf{B}_l$ and $-W_0+W_l$ in $\mathsf{A}_l$ are not the same.   Thus, taking into account those three key differences, it is better to regard the above asymmetric factorization in $x$-coordinate as the {\it non-standard representation}, which is parallel to the previous standard representation in $y$-coordinate, of the {\it same SUSYQM}. Despite of these differences, there is a one-to-one correspondence of the two factorizations in $x$ and $y$-coordinates, embodied by the connections \eqnm{raising_ops_NHF} and \eqnm{lowering_ops_NHF}, the momentum operator map \eqnm{momentum_map_type1} and the eigenfunction rescaling relation in \eq{transformed_Hl}. We will see more one-to-one correspondences  between these two representations, such as the shape invariance conditions, many recurrence relations. This type of isomorphisms tell us that, if there exists a certain algebraic structure of SUSYQM in one representation, there must be a similar one in the other.

Since the non-derivative terms $W_0+W_l$ in $\mathsf{B}_l$ and $-W_0+W_l$ in $\mathsf{A}_l$ are not the same, neither of them can work as the superpotential of the factorization \eqnm{eigen_equ_NHF}. However, for any given hypergeometric-like differential operator $H_0$, singling out  respectively $\pm W_0$  from these two non-derivative terms in $\mathsf{B}_l(x)$ and $\mathsf{A}_l$ brings us {\it two advantages}: 1) the common remainder $W_l(x)$ is singled out and works as the superpotential in the $x$-coordinate representation, so that in order to fix this factorization we do not need to determine the two different non-derivative terms separately; 2) the connection of this nonstandard representation of the SUSYQM factorization to the standard one in $y$-coordinate is \emph{transparently} given by the two equations in \eqnm{xy_connection}.

In contrast, if only $\pm p\, d/dx$ are singled out (as the result of symmetrizing the double-layer structure $-p d^2/dx^2$ in $H_0$) in $\mathsf{A}_l$ and $\mathsf{B}_l$ for searching for the factorization \eqnm{eigen_equ_NHF}, the rest two pieces to be determined in both $\mathsf{A}_l$ and $\mathsf{B}_l$ have to be assumed to be two {\it different} linear functions in general, i.e.
\begin{equation}\label{involved_operator_forms}
 \begin{aligned}
 \mathsf{A}_l &=p \od{}{x}+a_{1l} x+a_{2l},\\
 \mathsf{B}_l &=-p \od{}{x}+b_{1l} x+ b_{2l}. 
 \end{aligned}
\end{equation}
Insert these two expressions into \eq{eigen_equ_NHF} and match the coefficients of the binomials in $x$ on both sides, the algebraic equations for the {\it five} constants $a_{1l}$, $a_{2l}$, $b_{1l}$, $b_{2l}$ and $E_l$ are then obtained. In this scheme, solving these algebraic equations becomes much more involved than the iterative procedure of solving for the {\it three} constants $\alpha_l$, $\beta_l$ and $E_l$, as demonstrated in Subsections \ref{minus_route_iteration} and \ref{Plus_route_factorization}, simply because the superpotential $W_l(x)$ is smeared out when mixed with $\pm W_0$ in both $\mathsf{A}_l \smbr{x}$ and  $\mathsf{B}_l \smbr{x}$, and the shape invariance conditions such as \eqnm{SIC_minus_NHF} are not invoked in the latter procedure, which makes this scheme less efficient.  

Now the procedure of looking for a non-standard SUSYQM factorization \eqnm{eigen_equ_NHF}, {\it directly} in $x$-coordinate and {\it without} explicitly referring to the SUSYQM factorization represented in $y$-coordinate, clearly surfaces out. The key steps involved are summarized as below. Firstly, tweak the original eigen-equation \eqnm{eigen_equ} into the equivalent \eq{eigen_equ_with_multiplier} by multiplying it with $p(x)$ to symmetrize the asymmetric double-layer structure $-pd^2/dx^2$ into the symmetric one $-\smbr{p d/dx}^2$, so that the latter is viewed as the kinetic energy operator (in $x$-coordinate). Secondly, in \eq{eigen_equ_with_multiplier} turn the binomial $p H_0$ of the operator $p d/dx$ into the form \eqnm{pH0_symmetrized} by preparing the complete squared form $-\sqbr{p d/dx-W_0(x)}^2$, in so doing the isolated  differential operator of first order is disguised and the non-derivative terms are collected as the new potential energy function $V_l(x)$, which takes the form in \eq{Vl_potential}.  \eq{pH0_symmetrized} can also be reached by applying the active supersymmetrization ${\cal T}^{-1}$ to \eq{pre_Hamiltonian_H0} and making use of \eq{squared_form}. Thirdly, upon further expressing $V_l(x)$ formally in term of the superpotential $W_l(x)$ (\eq{potential_x_coord}), the eigen-equation \eqnm{eigen_equ_with_multiplier} then takes the form
\begin{eqnarray}
 \cubr{-\sqbr{p\od{}{x} - W_0(x)}^2 -p \od{W_l(x)}{x} + W^2_l(x)}\Phi_l 
 =E_l\Phi_l. \nonumber \\
\end{eqnarray}
Up to this stage, the steps are almost the same as the first several steps adopted in pursuing the supersymmetric factorization in the $y$-coordinate representation. Fourthly, quite similar to factorizing the difference of two squared numbers, the operator inside the curly brackets is directly factorized into $\mathsf{B}_l\mathsf{A}_l$, with $\mathsf{B}_l$ and $\mathsf{A}_l$ defined respectively in Eqs. \eqnm{raising_ops_NHF} and \eqnm{lowering_ops_NHF}. Finally, the formal factorization \eqnm{eigen_equ_NHF} is reached. 

To completely determine the factorization and the eigen-solutions, \eq{eigen_equ_NHF} has to be combined with its superpartner equation as well as some shape invariance condition for proceeding to the neighboring level. The superpartner equation is obtained by intertwiningly acting \eq{eigen_equ_NHF} with $\mathsf{A}_l$, i.e.
\begin{eqnarray}\label{superpartner_eigen_equ_NHF}
\mathsf{A}_l \mathsf{B}_l  \Phi^s_l =E_l \Phi^s_l,
\end{eqnarray}
where $\Phi^s_l \propto \mathsf{A}_l \Phi_l$. This is the counterpart of \eq{superpartner_Schr_equ}, as it can be derived from \eq{superpartner_Schr_equ} by applying the transformation ${\cal T}^{-1}(-) \equiv \sqrt{w}^{-1}(-)\sqrt{w}$ and making use of Eqs. \eqnm{raising_ops_NHF} and \eqnm{lowering_ops_NHF}. 
 
To make sure that \eq{superpartner_eigen_equ_NHF} is also among the hierarchy of eigen-equations which take the same generic pattern as \eq{eigen_equ_NHF} but with the neighboring labels $l\pm 1$, certain shape invariance conditions are needed. As we have encountered before in the $y$-coordinate representation in this section, there exist the upward-connected as well as the downward-connected shape invariance conditions, which lead to two slightly different routes of factorizations. To proceed, the following downward-connected (minus route of) shape invariance condition can be imposed,
\begin{eqnarray}\label{SIC_minus_NHF}
 \mathsf{A}^-_l \mathsf{B}^-_l = \mathsf{B}^-_{l-1} \mathsf{A}^-_{l-1} +\delta^-_l,
\end{eqnarray}
which is the $x$-coordinate counterpart of condition \eqnm{SI_operator_product_minus}, and in fact can be derived from the latter by simply applying the transformation ${\cal T}^{-1}$ and making use of the relations \eqnm{raising_ops_NHF} and \eqnm{lowering_ops_NHF}. Consequently, \eq{superpartner_eigen_equ_NHF} is renormalized into
\begin{eqnarray}
\mathsf{B}^-_{l-1} \mathsf{A}^-_{l-1} \Phi_{l-1} = E^-_{l-1} \Phi_{l-1},
\end{eqnarray}
where $\quad E^-_{l-1} = E^-_l -\delta^-_l$ and $\Phi_{l-1}\propto \mathsf{A}^-_l\Phi_l$.

For this minus route of factorization, obviously $\mathsf{A}^-_l$ works as the lowering ladder operator and $\mathsf{B}^-_l$ as the raising one. Concretely, the raising and lowering recurrence relations are respectively given as
\begin{equation}\label{raising_lowering_RCR_NHF}
 \begin{aligned}
 \Phi_l &= \frac{\mathsf{B}^-_l}{\sqrt{E^-_l}} \Phi_{l-1},\\
 \Phi_{l-1} &= \frac{\mathsf{A}^-_l}{\sqrt{E^-_l}} \Phi_l. 
 \end{aligned}
\end{equation}
These two recurrence relations are the analogues of  \eqnm{raising_recursion_relation_Hermitian} and \eqnm{lowering_recursion_relation_Hermitian}. In fact, they can be derived from the latter relations by simply multiplying with $\sqrt{w}^{-1}$ and making use of Eqs. \eqnm{raising_ops_NHF} and \eqnm{lowering_ops_NHF}.

Consecutively applying the lowering recurrence relation in \eq{raising_lowering_RCR_NHF} yields
\begin{eqnarray}\label{base_hypergeometric_like_function_x_coord}
 \Phi_l = \frac{\mathsf{B}^-_l}{\sqrt{E^-_l}} \frac{\mathsf{B}^-_{l-1}}{\sqrt{E^-_{l-1}}}\cdots \frac{\mathsf{B}^-_2}{\sqrt{E^-_2}} \frac{\mathsf{B}^-_1}{\sqrt{E^-_1}} \Phi_0
\end{eqnarray}
where we can take $\Phi_0=1$. This can also be derived from \eqnm{Psi_j_continuous_raising}, by multiplying $\sqrt{w}^{-1}$ on both sides and making use of the operator relation \eqnm{raising_ops_NHF}.

The problem is then turned into determining the coefficients in $W^-_l =\alpha^-_l x+ \beta^-_l$, $\{E^-_l\}$ and $\{\lambda^-_l\}$. For this,  the shape invariance condition \eqnm{SIC_minus_NHF} is the pivotal equation that reduces to the differential recurrence relation  \eqnm{recursion_RL_W_minus} for series $\{W^-_l(x)\}$, which further gives rise to the algebraic recurrence relations \eqnm{coefficient_x^2}-\eqnm{coefficient_x^2} for determining the series  $\{\alpha^-_l\}$, $\{\beta^-_l\}$ and $\{\delta^-_l\}$, by further incorporating the initial data in \eq{alpha_0_beta_0_minus}. The relations in \eq{alpha_l_minus} and the initial data in \eq{E0_minus} are then used to determine $\{E^-_l\}$ and $\lambda^-_l$.  In other words, the procedure of iteratively determining $\alpha^-_l$, $\beta^-_l$, $\delta^-_l$, $\lambda^-_l$ and $E^-_l$ is exactly the same as that demonstrated in \ref{Minus_route_factorization}.

Adding/subtracting  \eq{lowering_ops_NHF} to/from the $(l+1)$-version of \eq{raising_ops_NHF} gives us 
\begin{equation}\label{Blminus1}
 \begin{aligned}
 &\mathsf{B}^-_{l+1} = W^-_{l+1} + W^-_l - \mathsf{A}^-_l,\\
 &\mathsf{B}^-_{l+1}  =-2 p \od{}{x} + W^-_{l+1} - W^-_l + 2 W_0 + \mathsf{A}^-_l.
 \end{aligned}
\end{equation}
On the other hand, the $(l+1)$-version of the upper equation in \eqnm{raising_lowering_RCR_NHF} reads $\sqrt{E^-_{l+1}}\Phi_{l+1}= \mathsf{B}^-_{l+1}\Phi_l$. Substituting for $\mathsf{B}^-_{l+1}$ in this equation via the upper equation and the lower one in \eq{Blminus1}, respectively, and further making use of the lower equation in \eqnm{raising_lowering_RCR_NHF},  we reach the following two \emph{three-term} recurrence relations
\begin{eqnarray}
 \sqrt{E^-_{l+1}}\Phi_{l+1}=\smbr{W^-_{l+1} + W^-_l}\Phi_l - \sqrt{E^-_l}\Phi_{l-1}
\end{eqnarray}
and 
\begin{eqnarray}
 \sqrt{E^-_{l+1}}\Phi_{l+1} &=&\smbr{-2 p \od{}{x} +W^-_{l+1} - W^-_l + 2 W_0}\Phi_l \nonumber \\
                            &&+ \sqrt{E^-_l}\Phi_{l-1}.
\end{eqnarray}

The minus route of the factorization in the form similar to \eqnm{involved_operator_forms} in the $x$-coordinate representation (the non-standard one) has already been found in a {\it drastically different} approach from ours by Jafarizadeh and Fakhri \cite{jafarizadeh_SUSY_and_diff_equs_1997}. They did not directly aim at constructing the kinetic energy operator $-(pd/dx)^2$ or pursue a factorization of the partially symmetrized eigen-equation \eqnm{eigen_equ_with_multiplier}.
Instead, they directly factorized the original eigen-equation \eqnm{eigen_equ} into $\mathit{B}(l) \mathit{A}(l)\Phi_l=E_l\Phi_l$, starting with the two linear differential operators in the most general asymmetric forms $\mathit{A}(l)=f_1(x)d /dx+g_1(x)$ and $\mathit{B}(l)=f_2(x)d/dx+g_2(x)$.  
To pin down the four functions $f_{1,2}$ and $g_{1,2}$ and eventually arrive at the forms given in \eqnm{involved_operator_forms}, the expanded form of $\mathit{B}(l) \mathit{A}(l)\Phi_l=E_l\Phi_l$ was compared with the original eigen-equation $H_0 \Phi_l=\lambda_l \Phi_l$ to obtain the equations satisfied by $f_{1,2}$, $g_{1,2}$ and $E_l$. In their approach, neither the simple linear function $W^-_l=\alpha^-_l x+ \beta^-_l$ was taken as the input (in fact the concept of superpotential does not appear at all in \cite{jafarizadeh_SUSY_and_diff_equs_1997}, let alone in operational level), nor the explicit shape invariance conditions such as \eq{SI_operator_product_minus} and \eq{SI_operator_product_plus} were directly made use of. In this sense their algorithm of determining the {\it five} constants similar to $a_{1l}$, $a_{2l}$, $b_{1l}$, $b_{2l}$ and $E_l$ in \eq{involved_operator_forms} does not take the full advantages of SUSYQM, hence is not so light or elementary as that provided in this paper.  

In comparison, our approaches (both the nonstandard and standard ones) at least have the following advantages. Firstly, the intimate connection between the nonstandard and standard representations of the same SUSYQM is established by the two trivial algebraic manipulations: the double-sided supersymmetrization transformations ${\cal T}$ and ${\cal T}^{-1}$, plus the momentum operator map \eqnm{momentum_map_type1}; as the results of this connection, all the constituents in both $\mathsf{A}_l$ and $\mathsf{B}_l$, as given in Eqs. \eqnm{raising_ops_NHF} and \eqnm{lowering_ops_NHF}, appear not merely as a mathematical coincidence, instead they have the clear {\it SUSYQM origin}. In particular, the appearance of $\pm p\,d/dx$ comes out as the requirement of preparing the quantum kinetic energy operator $-\smbr{p\, d/dx}^2$, a necessary ingredient of the quantum Hamiltonian in $x$-coordinate; the linear shifts $\pm W_0=\pm (p'-q)/2$ of $\mp p\,d/dx$, singled out delicately in Eqs. \eqnm{raising_ops_NHF}, \eqnm{lowering_ops_NHF} and \eqnm{xy_connection}, source the double-sided transformations ${\cal T}$ and ${\cal T}^{-1}$, in the sense that these two transformations emerge as lumping the two-term operators $\mp p\, d/dx\pm W_0$ into the forms in which $\mp p\,d/dx$ are sandwiched by the integration factors $\sqrt{w}^{\pm 1}$; the third \emph{common} piece $W_l$ then comes as the {\it superpotential} required for establishing the corresponding SUSYQM in both $x$ and $y$-coordinate representations, and it is the only piece that needs to be calculated in our approaches. Secondly, by imposing the shape invariance conditions \eqnm{SIC_minus_NHF} and \eqnm{SI_potential_minus} and introducing the simple linear superpotential  $W^-_l=\alpha^-_l x+ \beta^-_l$, the factorization can be determined level by level iteratively, and this involves merely such high school algebras as solving certain simple algebraic recurrence relations for series $\{\alpha^-_l\}$, $\{\beta^-_l\}$ and $\{E^-_l\}$.  Thirdly, instead of resorting to some methods or arguments external to the factorization algorithm, the eigenvalue $\lambda^-_l$ is determined {\it within} this SUSYQM algorithm.

\subsection{Two Other Equivalent Eigen-equations}\label{equivalent_eigen_equ}
An important issue facing us in quantum mechanics is to solve a Schr\"odinger equation for various types of potentials. One may try to convert such equation into the principal hypergeometric-like differential equation \eqnm{eigen_equ} for some types of potentials, before applying the algorithms, developed in Subsections \ref{base_kinetic_energy_operator}-\ref{non_standard_factorization}, to solving this type of equation. Of course, the method of conversion usually has to be developed case by case, as there is no unique one. Nevertheless, we have already developed some techniques that can be applied to this conversion. For instance, in some cases, the type of potential ${\cal V}_l(y)$ expressed in $y$-coordinate inside the Schr\"odinger \eq{Schr_equ_y_coord}  can be \emph{binomialized}  into $V_l(x)$ in some $x$-coordinate, and accordingly the linear superpotential $W_l(x)$ is easy to figure out.  Meanwhile, this binomialization defines $p(x)$ and the inverse momentum operator map in the form \eqnm{momentum_map_type1}. If $p(x)$ is at most a binomial in $x$, then \eq{Schr_equ_y_coord} is successfully converted into \eq{Schr_equ_x_coord_factorized} or \eq{eigen_equ_NHF}, and eventually into \eq{eigen_equ}. However, this is just a quite limiting case in that the inverse supersymmetrization and the inverse momentum operator map are made use of \emph{simultaneously} in the particularly combined fashion shown in \eq{Schr_equ_x_coord_factorized}. In many cases, \emph{partial supersymmetrization} (whose meaning will become clear shortly) and inverse momentum operator map have to be utilized \emph{at different stages} of a  conversion process. Furthermore, other operations other than these two techniques, such as multiplying some function to turn a fractional coefficient of the eigen-equation into a binomial, may have to be invoked as well. The various combinations of these conversion tricks may explain why those Schr\"odinger equations with many types of potentials can be converted into hypergeometric-like differential equations. In this subsection, we will provide other forms of eigen-equations that are equivalent to the standard principal hypergeometric-like differential equation \eqnm{eigen_equ}, and that can be targeted as the intermediate eigen-equations for ultimately passing to the standard \eq{eigen_equ}. At the same time, the technique of \emph{partial supersymmtrization} and its inverse transformation, that turn out to be useful for the purpose of this conversion, will be exemplified.

For the minus route of factorization, Eqs. \eqnm{raising_ops_NHF}, \eqnm{lowering_ops_NHF}, \eqnm{A0_minus1} and \eqnm{B0_minus1} together tell us that 
\begin{eqnarray}\label{Bl_B0_relation1}
 \mathsf{B}^-_l(x)= \mathsf{B}^-_0(x) + \Delta W^-_l(x)= {\cal T}^{-1}_\Delta \sqbr{\mathsf{B}^-_0(x)}, 
\end{eqnarray}
\begin{eqnarray}\label{Al_A0_relation1}
  \mathsf{A}^-_l(x) = \mathsf{A}^-_0(x) + \Delta W^-_l(x) = {\cal T}_\Delta \sqbr{\mathsf{A}^-_0(x)},
\end{eqnarray}
where the second equality in \eq{Bl_B0_relation1} and that in \eq{Al_A0_relation1} have been reached by absorbing $\Delta W^-_l(x)\equiv W^-_l(x)-W_0(x)$ into the integration factors $u^{\pm 1}_l(x)\equiv e^{\mp \int d x \smbr{\Delta W^-_l/p}}$ and defining the two bilateral transformations ${\cal T}_\Delta \sqbr{-}\equiv u_l(x)\sqbr{-}u^{-1}_l(x)$ and ${\cal T}^{-1}_\Delta \sqbr{-}\equiv u^{-1}_l(x)\sqbr{-}u_l(x)$. 

Applying ${\cal T}^{-1}_\Delta$ to both \eq{Bl_B0_relation1} and \eq{Al_A0_relation1} and then forming their product yield
\begin{eqnarray}\label{partially_asymmetrized_Hl}
 {\cal T}^{-1}_\Delta \sqbr{\mathsf{B}^-_l(x)\mathsf{A}^-_l(x) } &=& u^{-2}_l \sqbr{\mathsf{B}^-_0(x)} u^2_l \,\mathsf{A}^-_0(x), \nonumber \\
                                                         &=&\sqbr{-p \od{}{x} + 2 W^-_l(x)}\sqbr{p \od{}{x}}, \nonumber \\
                                                         &=& p H_l, 
\end{eqnarray}
where 
\begin{eqnarray}\label{Hl}
 H_l(x) \equiv -p \frac{d^2}{d x^2} + \sqbr{2 W^-_l(x) - p'} \od{}{x},
\end{eqnarray}
which is obtained by factorizing out $p$ from the differential operator in the second line. This $H_l(x)$ is a \emph{hypergeometric-like} differential operator, which subsumes $H_0(x)$ defined in \eq{H_0_operator} as its special $l=0$ case, noticing that for $l=0$, $2 W_0-p'=-q$. 
Multiplying  with $u^{-1}_l(x)$ from the left, denoting $\varphi_l (x)\equiv u^{-1}_l(x) \Phi_l(x)$  and making use of \eq{partially_asymmetrized_Hl}, \eq{eigen_equ_NHF} is then turned into
\begin{eqnarray}\label{asymmetrized_eigen_equ_general_l0}
 p H_l \varphi_l= E^-_l \varphi_l,
\end{eqnarray}
which can be viewed as a nontrivial generalization of \eq{initial_eigen_equ_assymmetric}, noticing that, for $l=0$, $u^{-1}(x)=1$, $\varphi_0(x)=\Phi_0(x)$, $E^-_0 =0$, hence \eq{asymmetrized_eigen_equ_general_l0} reduces to \eq{initial_eigen_equ_assymmetric}.

\eq{asymmetrized_eigen_equ_general_l0} has the generic eigenvalue $E^-_l$ in common with the operator $\mathsf{B}^-_l \mathsf{A}^-_l$. Of course, this is not the direct eigen-equation for the hypergeometric-like differential operator $H_l$ except for $p$ being a constant. In what follows, we will demonstrate that there actually exists another eigen-equation that is equivalent to \eq{asymmetrized_eigen_equ_general_l0}, and comes with the interesting properties in that its operator consist of $H_l$ as its only differential parts and another pure function, and its shares the eigenfunction $\varphi_l$ with $p H_l$ while shares the eigenvalue $\lambda^-_l$ with the operator $H_0$.
Note that $2 W_0-p'=-q$ and $\Delta W^-_l=W^-_l- W_0$. \eq{Hl} and \eq{H_0_operator} together then imply that 
\begin{eqnarray}\label{Relation_H0_Hl}
 H_0 =H_l - 2\Delta W^-_l \od{}{x},
\end{eqnarray}
by which $H_0$ in \eq{eigen_equ} will be substituted in favor of $H_l$. The first-order differential operator $- 2\Delta W^-_l d/dx$ can be turned into some \emph{non-differential} terms by a certain {partial supersymmetrization}. Note that 
\begin{eqnarray}
 -p \frac{d^2}{d x^2} - 2\Delta W^-_l \od{}{x}&=&p {\cal T}^{2}_\Delta\sqbr{-\od{}{x}}\;\sqbr{\od{}{x}}.
\end{eqnarray}
Applying ${\cal T}^{-1}_\Delta$ to the above equation then yields 
\begin{eqnarray}\label{partial_SUSY_trans}
 &&{\cal T}^{-1}_\Delta\sqbr{-p \frac{d^2}{d x^2} - 2\Delta W^-_l \od{}{x}} = p {\cal T}_\Delta \sqbr{-\od{}{x}} {\cal T}^{-1}_\Delta \sqbr{\od{}{x}}, \nonumber \\
\quad &&=-p \frac{d^2}{d x^2} + p\sqbr{\smbr{\frac{\Delta W^-_l}{p}}' + \smbr{\frac{\Delta W^-_l}{p}}^2},
\end{eqnarray}
where the second equality has been reached by making use of the properties 
\begin{equation}\label{SUSY_Tran_Delta}
 \begin{aligned}
 {\cal T}_\Delta\sqbr{-\od{}{x}} &=-\od{}{x} - \frac{\Delta W^-_l}{p},\ \\
 {\cal T}^{-1}_\Delta\sqbr{\od{}{x}} &=\od{}{x} - \frac{\Delta W^-_l}{p},
 \end{aligned}
\end{equation}
which follow directly from the definitions of ${\cal T}_\Delta$, ${\cal T}^-_\Delta$ and $u^{\pm 1}_l(x)$, and are analogs of Eqs. \eqnm{T_inverse_minus} and \eqnm{T_plus}.

At the same time, according to the second property in \eq{SUSY_Tran_Delta}, applying the same partial supersymmetrization transformation ${\cal T}^{-1}_\Delta$ to $\sqbr{2 W^-_l(x)-p' }d/dx$ -- the first-order differential operator inside $H_l$ -- will shifts by certain non-differential term in the form
\begin{eqnarray}\label{shift_1st_derivative}
  {\cal T}^{-1}_\Delta\sqbr{\smbr{2W^-_l -p'}\od{}{x}} = \smbr{2W^-_l-p'}\sqbr{\od{}{x}- \frac{\Delta W^-_l}{p}}. \nonumber \\
\end{eqnarray}
 Because only the second first-order differential operator on the right hand side of \eq{Relation_H0_Hl} is removed by the supersymmetrization transformation ${\cal T}^{-1}_\Delta$, and the one inside $H_l$ is preserved, we call such transformation \emph{partial supersymmetrization}.

Adding Eqs.  \eqnm{partial_SUSY_trans} and \eqnm{shift_1st_derivative} up and making use of \eq{potential_x_coord} for $V_l$ and \eq{Riccati_equ_V0} for $V_0$, we arrive at
\begin{eqnarray}
 {\cal T}^{-1}_\Delta \sqbr{H_0}  = H_l -\frac{V_l - V_0}{p},
\end{eqnarray}
which is not a hypergeometic-type any more but a Sturm-Liouville-type. Then, applying the transformation ${\cal T}^{-1}_\Delta$ to \eq{eigen_equ} yields
\begin{eqnarray}\label{2nd_equivalent_eigen_equ}
 \sqbr{H_l -\frac{V_l - V_0}{p}}\varphi_l =\lambda^-_l \varphi_l.
\end{eqnarray}
This is another eigen-equation that is equivalent to \eq{eigen_equ}. It also follows simply from the combination of \eq{asymmetrized_eigen_equ_general_l0} and $\lambda^-_l p + V_l-V_0=E^-_l$ (the last equation in \eqnm{Wl_minus}). The important point is that, although the overall operator in \eq{2nd_equivalent_eigen_equ} is more than a hypergeometic-like one due the extra term $(V_0-V_l)/p$, its eigen-equation can be turned into the standard hypergeometric-like differential \eq{eigen_equ}, by applying the \emph{partial asymmetrization} transformation ${\cal T}_\Delta$ to \eq{2nd_equivalent_eigen_equ} to eliminating this extra piece and restoring $-2\Delta W^-_l d/dx$  and going back to \eq{Relation_H0_Hl}.

Introducing the definition $w_l\equiv e^{-\int dx 2 W^-_l(x)/p(x)}$, we have $w_0 =w$ for $w$ given in \eq{w_integration_factor} and $\sqrt{w_l}  =\sqrt{w} u_l$. The overall active supersymmetrization transformation ${\cal T}_l(-) \equiv \sqrt{w_l}(-)\sqrt{w_l}^{-1}$ then has the decomposition ${\cal T}_l = {\cal T} \cdot {\cal T}_\Delta = {\cal T}_\Delta \cdot  {\cal T}$ for ${\cal T}$ defined in \eq{T_transformation}. The asymmetrically factorized \eq{asymmetrized_eigen_equ_general_l0} is then supersymmetrized into \eq{Schr_equ_x_coord_factorized} by applying ${\cal T}_l$. 

\section{Building the Associated Hypergeometric-like Equations and Their Solutions via the Iterative SUSYQM Algorithm}\label{associated_hypergeometric_functions}
In this section, we will give the answers to the following two questions from the viewpoints of SUSYQM: 1) why do there exist a hierarchy of associated hypergeometric-like differential operators and their eigen-solutions for the base hypergeometic-like differential operator $H_0$?  2) how can these associated hypergeometric-like differential equations and their solutions be built \emph{level by level (iteratively)} from the base level equation \eqnm{eigen_equ} via the SUSYQM algorithm that is \emph{completely removed} from  the traditional differential method?  In particular, we expect that the derivation of the associated hypergeometric-like differential equation and its eigen-solution of generic association level comes with the four pleasant features as promised in the introduction section. 

\subsection{Another Type of Momentum Operator Buried in \mathinhead{H_0}{} and the Related Active Supersymmetrization}\label{2nd_type_momentum_op}
The story of building the associated hypergeometric-like differential equations and their solutions via the SUSYQM algorithm can be initiated by the basic observation that another squared form $-\smbr{\sqrt{p} d/dx}^2$ is buried in $H_0$, which is the parallel of the squared form $-(p d/dx)^2$ singled out from $p H_0$ in \eq{pH0}.
Accordingly, when viewing this squared form as a quantum kinetic energy operator, there exist another type of potential energy function that can be constructed from $H_0$ and $H_0$ can be regarded as a Hamiltonian of initial level and \eq{eigen_equ} as a Schr\"odinger equation represented in $x$-coordinate. More importantly, we will see that $H_0$ itself admits a certain supersymmetric factorization. This situation is parallel to the operator $p H_0$ discussed in Subsection \ref{base_kinetic_energy_operator}, that has the kinetic energy operator $-\smbr{p d/dx}^2$ and has the explicit supersymmetric factorization \eqnm{pH0_supersymmetrized}. Therefore, applying the {\it same} sort of SUSYQM algorithm, that has already been successfully applied to building the principal hypergeometric-like eigen-functions, to certain supersymmetrically factorized form of \eq{eigen_equ} with $-\smbr{\sqrt{p} d/dx}^2$ as the kinetic energy operator would proliferate a hierarchy of associated hypergeometric-like eigen-equations and their solutions. 

In fact, $-\smbr{\sqrt{p} d/dx}^2$ can be singled out from the first term in $H_0$ in \eq{H_0_operator},  by rewriting it as 
\begin{eqnarray}
-p \frac{d^2}{d x^2} =-\smbr{\sqrt{p} \od{}{x}}^2+\sqrt{p}'\smbr{\sqrt{p}\od{}{x}}. 
\end{eqnarray}
Adding on the second term in $H_0$ with the form $-q\, d/dx= -\smbr{q/\sqrt{p}}\smbr{\sqrt{p}d/dx}$, $H_0$ has the natural asymmetric factorization
\begin{eqnarray} \label{assymetric_factorization_H0II}
H_0 &=& \sqbr{-\sqrt{p} \od{}{x}+ 2 W^a_0(x)}\sqbr{\sqrt{p} \od{}{x}}
\end{eqnarray}
for
\begin{eqnarray}\label{superpotential_W0a}
 W^a_0(x) \equiv \frac{1}{2}\smbr{-\frac{q}{\sqrt{p}} + \sqrt{p}'}
          =-\frac{\sqrt{p}}{2}\sqbr{\log \smbr{w \sqrt{p}}}',
\end{eqnarray}
where the substitution $q=\smbr{pw}'/w$ (\eq{w_definition1}) has been made use of to reach the second equality. 

Introduce the notations
\begin{eqnarray}
\mfk{h}_0 &\equiv& -\sqrt{p} \od{}{x} +2 W^a_0(x), \label{mfkh_00} \\
            &=& - \frac{1}{w \sqrt{p} } \smbr{\sqrt{p} \od{}{x}} w  \sqrt{p}, \label{mfkh_01} \\
\mfk{h}^\dagger_0 &\equiv& \sqrt{p} \od{}{x},\label{mfkh0_dagger}
\end{eqnarray}
where the two terms in \eq{mfkh_00} are lumped into the form in \eq{mfkh_01} in terms of the integration factors $\smbr{\sqrt{p}w}^{\pm 1}$ of the differential operator $\mfk{h}_0$.
Then the factorization \eqnm{assymetric_factorization_H0II} is rewritten as
\begin{eqnarray}\label{H0_Hermitian_factorization}
H_0 = \mfk{h}_0 \mfk{h}^\dagger_0
    = - \frac{1}{w \sqrt{p} } \smbr{\sqrt{p} \od{}{x}} \smbr{w  \sqrt{p}} \smbr{\sqrt{p} \od{}{x}}.
\end{eqnarray}
This factorization is nothing but the familiar self-adjoint form \eqnm{self_adjoint_form}, another asymmetric product of two first-order differential operators that is parallel to the first term on the right hand side of \eq{second_double_layer_splitting} for the \emph{same} operator $H_0$. In this factorization, the definitions of $\mfk{h}_0$ and $\mfk{h}^\dagger_0$ is not unique. Here the minus sign is deliberately attributed to $\mfk{h}_0$ for later purpose. It is easy to check that $\mfk{h}_0$ and $\mfk{h}^\dagger_0$ are the Hermitian conjugate pair with respect to the $w$-weighted inner product. Again, $w(x)$ is assumed to be well behaved in the definition domain of this inner product so that the latter can be well defined. 

The asymmetric factorization of $H_0$ in \eq{assymetric_factorization_H0II} is an analog of the factorization of $p H_0$ in \eq{simple_factorization}, i.e. $\sqrt{p}\, d/dx $ is parallel to $p\,d/dx$ and $W^a_0(x)$ to $W_0(x)$. According to our experience gained in Subsection \ref{base_kinetic_energy_operator}, this asymmetric factorization  is expected to be supersymmetrized into some form that is analogous to \eqnm{pH0_supersymmetrized} for $p\,H_0$, by some active supersymmetrization transformation, so that $-i\sqrt{q}\,d/dx $ is convincingly identified with another type of momentum operator and $W^a_0(x)$ with another type of superpotential of the initial level for the Hamiltonian $H_0(x)$ represented $x$-coordinate.
Such transformation can be reached by performing some trivial identical manipulations on the self-adjoint form \eqnm{H0_Hermitian_factorization} and following the procedure that is similar to constructing the transformation ${\cal T}$ in \eq{T_transformation}, as demonstrated in detail in Subsection \ref{base_kinetic_energy_operator}. 
Concretely,  we first partition evenly the middle factor $ w \sqrt{p}$ in \eqnm{H0_Hermitian_factorization} over its left and right neighboring factor operators by the amount $\smbr{w \sqrt{p} }^{1/2}$, and use this function as the building block to construct the {\it bilateral} transformation
\begin{eqnarray}\label{S_transformation}
 {\cal S}\smbr{-} &\equiv & \smbr{w \sqrt{p}}^{\frac{1}{2}}\smbr{-}\smbr{w \sqrt{p}}^{-\frac{1}{2}} 
\end{eqnarray}
to wrap up $-\sqrt{p}d/x$, 
and its inverse
\begin{eqnarray}\label{S_inverse_transformation}
 {\cal S}^{-1}\smbr{-}&\equiv&\smbr{w \sqrt{p}}^{-\frac{1}{2}}\smbr{-}\smbr{w \sqrt{p}}^{\frac{1}{2}}
\end{eqnarray}
to wrap up $\sqrt{p}d/dx$. In doing so, \eq{H0_Hermitian_factorization} is then rewritten as
\begin{eqnarray}\label{pre_supersymmetrized_H00}
 H_0 = {\cal S}^{-1} \cubr{\sqbr{{\cal S}^{-1}\smbr{-\sqrt{p}\od{}{x}}} \sqbr{{\cal S}{\smbr{\sqrt{p}\od{}{x}}}}},
\end{eqnarray}
or equivalently,
\begin{eqnarray}\label{supersymmetrized_H00}
{\cal S}\smbr{H_0} &=&\sqbr{{\cal S}^{-1}\smbr{-\sqrt{p}\od{}{x}}} \sqbr{{\cal S}{\smbr{\sqrt{p}\od{}{x}}}}.
\end{eqnarray}

Eqs.\eqnm{pre_supersymmetrized_H00} and \eqnm{supersymmetrized_H00} take respectively the similar forms as Eqs. \eqnm{pH01} and \eqnm{pH0_supersymmetrized0}, and we have already known that ${\cal T}$ can supersymmetrize the asymmetric factorization \eqnm{pH0} of $p H_0$. So we expect that ${\cal S}$ can supersymmetrize the asymmetric factorization \eqnm{assymetric_factorization_H0II} or \eqnm{H0_Hermitian_factorization}). It is easy to check that
\begin{eqnarray}\label{S_inverse_transformed_h}
 {\cal S}^{-1}\smbr{-\sqrt{p}\od{}{x}} = -\sqrt{p} \od{}{x}+W^a_0(x)
\end{eqnarray}
and 
\begin{eqnarray}\label{S_transformed_h}
 {\cal S}\smbr{\sqrt{p}\od{}{x}}={\cal S}\smbr{\mfk{h}^\dagger_0}= \sqrt{p}\od{}{x} +W^a_0(x)
\end{eqnarray}
with $W^a_0(x)$ given in \eq{superpotential_W0a}. Looking at these two above equations from right to left, they tell us that $\smbr{w \sqrt{p}}^{\pm 1/2}$ are just the integration factors of the two first-order differential operators on the right-hand sides. Then \eq{supersymmetrized_H00} is rewritten as
\begin{eqnarray}\label{supersymmetrized_H0}
{\cal S}\smbr{H_0} &=&\sqbr{-\sqrt{p} \od{}{x}+W^a_0(x)}\sqbr{\sqrt{p} \od{}{x}+W^a_0(x)},\\
                   &=&\smbr{-\sqrt{p} \od{}{x}}^2 + V^a_0(x),
\end{eqnarray}
where $W^a_0(x)$ is identified as the superpotential for the potential 
\begin{eqnarray}\label{initial_potential_association}
 V^a_0(x)=-\sqrt{p}\od{W^a_0(x)}{x} + \smbr{W^a_0(x)}^2.
\end{eqnarray}
Here we see that, parallel to ${\cal T}$ that sueprsymmetrizes $p H_0$, the \emph{wrapper} transformation ${\cal S}$ indeed works as the desired {\it active} supersymmetrization transformation that supersymmetrizes $H_0$, or its the asymmetric factorization \eqnm{assymetric_factorization_H0II} or \eqnm{H0_Hermitian_factorization} in a simple algebraic way.

Why can ${\cal S}$ directly supersymmetrize the asymmetric factorization \eqnm{assymetric_factorization_H0II} of $H_0$ into its supersymmetric form \eqnm{supersymmetrized_H0}? \eq{S_transformed_h} says that ${\cal S}$, when applying to $\pm \sqrt{p}d/dx$, just shifts them by $\pm W^a_0(x)$, respectively. Consequently, 
\begin{eqnarray}\label{down_shift_S}
 {\cal S}\smbr{\mfk{h}_0}=\mfk{h}_0-W^a_0={\cal S}^{-1}\smbr{\mfk{h}^\dagger_0}.
\end{eqnarray}
${\cal S}$ acts distributively on $H_0$ in its factorized form \eqnm{assymetric_factorization_H0II}, i.e. ${\cal S}\smbr{\mfk{h}_0 \mfk{h}^\dagger_0}={\cal S}\smbr{\mfk{h}_0 } {\cal S}\smbr{\mfk{h}^\dagger_0}$. ${\cal S}$ shifts $\mfk{h}_0$ {\it down} by $-W^a_0$, as clearly seen from \eq{down_shift_S}, and $\mfk{h}^\dagger_0$ {\it up} by the opposite amount $W^a_0$ (see \eq{S_transformed_h}). This way, the {\it unevenly} distributed term $2W^a_0$  in the first factor of asymmetric factorization (\ref{assymetric_factorization_H0II}) is then {\it evenly} distributed between $\mfk{h}_0$ and $\mfk{h}^\dagger_0$ in the form (\ref{supersymmetrized_H0}).

Now, similar to the momentum operator map \eqnm{momentum_map_type1}, we introduce another momentum operator map 
\begin{eqnarray}\label{momentum_map_type2}
 -i \sqrt{p} \od{}{x} \equiv \hat{p}_z=-i \od{}{z},
\end{eqnarray}
which gives rise to the new coordinate $z$ that is related to $x$ by the coordinate transformation 
\begin{eqnarray}\label{z_coord_transformation}
 z(x)=\int^x \frac{d \tilde{x}}{\sqrt{p(\tilde{x})}}.
\end{eqnarray}
 Then the squared form $\smbr{-i\sqrt{p} d/dx}^2$ in ${\cal S} \smbr{H_0(x)}$ in the form \eqnm{supersymmetrized_H0} or in $H_0(x)$ in the form \eqnm{H0_Hermitian_factorization} is naturally identified with the standard quantum kinetic energy operator $-d^2/d z^2$ in $z$-coordinate. Notice that the map $i \sqrt{p} d/dx \equiv \hat{p}_z$ also works for this purpose. 

Under the map \eqnm{momentum_map_type2} or the coordinate transformation \eqnm{z_coord_transformation}, the factorization \eqnm{supersymmetrized_H0} is rewritten as
\begin{eqnarray}\label{standardized_factorization_base_level}
 \mathsf{H}_0(z)&\equiv& {\cal S}\smbr{H_0(x)}|_{x=x\smbr{z}} ={\cal S}\smbr{\mfk {h}_0(x)}{\cal S}\smbr{\mfk {h}^\dagger_0(x)}|_{x=x\smbr{z}} \nonumber \\
                &=& \mathsf{ h}_0(z) \mathsf{ h}^\dagger_0(z), 
\end{eqnarray}
where, according to Eqs. \eqnm{S_inverse_transformed_h} and \eqnm{S_transformed_h},
\begin{eqnarray} \label{standard_h0}
 \mathsf{h}_0 (z) &\equiv &{\cal S}^{-1}\smbr{-\mfk{h}^\dagger_0(x)} = -\od{}{z}+\mathsf{W}^a_0(z),\\
                  &=&-\left.{\smbr{ w\sqrt{p}}^{-\frac{1}{2}}\od{}{z}\smbr{ w \sqrt{p}}^{\frac{1}{2}}}\right|_{x=x(z)} \nonumber
\end{eqnarray}  
and
\begin{eqnarray} \label{standard_h0_dagger}
\mathsf{h}^\dagger_0(z) &\equiv& {\cal S}\smbr{\mfk {h}^\dagger_0(x)} = \od{}{z}+\mathsf{W}^a_0(z),\\
                        &=&\left.{\smbr{w \sqrt{p}}^{\frac{1}{2}}\od{}{z}\smbr{w \sqrt{p}}^{-\frac{1}{2}}}\right|_{x=x(z)},
\end{eqnarray}
for
\begin{eqnarray}\label{Wa0_z_coord}                        
\mathsf{W}^a_0(z) \equiv W^a_0\sqbr{x(z)} = -\left.{\od{}{z}\log \smbr{w \sqrt{p}}^{\frac{1}{2}}}\right|_{x=x(z)}. 
\end{eqnarray}
That is, in $z$-coordinate $\mathsf{H}_0(z)$ appears as the {\it standard} Hamiltonian that has the supersymmetric factorization \eqnm{standardized_factorization_base_level}, with the superpotential  $\mathsf{W}^a_0(z)$ for the potential 
\begin{eqnarray}
 V^a_0(z) = -\od{\mathsf{W}^a_0(z)}{z}+\smbr{\mathsf{W}^a_0(z)}^2.
\end{eqnarray}
\eq{Wa0_z_coord} says that in $z$-coordinate $\log \smbr{w \sqrt{p}}^{\frac{1}{2}}$ is the `potential' of the superpotential $\mathsf{W}^a_0(z)$.

The original $H_0(x)$ and its supersymmetrized counterpart ${\cal S}\smbr{H_0(x)}$ in \eq{supersymmetrized_H0}  can be regarded as the Hamiltonian operators in their $x$-coordinate representations (the {\it non-standard} ones), for the latter $-i\sqrt{p}d/dx$ is regarded as the momentum operator.

Multiplying \eq{eigen_equ} with $\smbr{\sqrt{p} w }^{1/2}$, which  is the left factor of the bilateral transformation ${\cal S}$, for preparing ${\cal S}$ and rescaling the original eigenfunction $\Phi_l\smbr{x}$ into the new eigenfunction  $ \mathsf{\Phi}_l(z) \equiv \smbr{w \sqrt{p}}^{1/2}\Phi_l\smbr{x}|_{x=x(z)}$, and making use of the supersymmetric factorization \eqnm{standardized_factorization_base_level} for $\mathsf{H}_0(z)$, \eq{eigen_equ} is then turned into the following  Schr\"odinger equation in its standard supersymmetrically factorized form, 
\begin{eqnarray}\label{Schrodinger_equ_base_level}
\mathsf{H}_0(z) \mathsf{\Phi}_l(z) &=&\mathsf{h}_0(z) \mathsf{h}^\dagger_0(z) \mathsf{\Phi}_l(z) = \lambda_l \mathsf{\Phi}_l(z).
\end{eqnarray}
This is the parallel of \eq{Schr_equ_y_coord}, another incarnation of the original eigen-equation \eqnm{eigen_equ}.
Soon it will become clear that the very existence of such supersymmetric factorized eigen-equation is the foundation of the existence of a hierarchy of {\it associated} hypergeometric-like differential equations related to $H_0$.
 
Here again, instead of following the traditional ‘{\it differential}' way  (see, e.g., \cite{Courant_and_Hilbert_1931, *Courant_and_Hilbert_1989}) of eliminating the first-order derivative in \eq{eigen_equ} to obtain the Schr\"odinger equation \eqnm{Schrodinger_equ_base_level}, we have devised a simple {\it algebraic} supersymmetrization transformation ${\cal S}$, which directly turns the asymmetrical factorization of \eq{eigen_equ}, i.e.
\begin{eqnarray}\label{asymmetric_eigen_equ_initial_association_level}
H_0(x)\Phi_l(x)=\mfk{h}_0(x) \mfk{h}^\dagger_0(x) \Phi_l(x) = \lambda_l \Phi_l(x) 
\end{eqnarray}
into the supersymmetrically factorized Schr\"odinger equation \eqnm{Schrodinger_equ_base_level}. Via such an algebraic approach, the new coordinate transformation \eqnm{z_coord_transformation} is simply determined by the momentum operator identification \eqnm{momentum_map_type2}, the rescaling factor of the eigenfunction, $\smbr{w \sqrt{p}}^{1/2}$, just comes out naturally as the integration factor that makes up the transformation ${\cal S}$ itself, \emph{the superpotential $W^a_0$ of the sought-after supersymmetric factorization is even identified before the supersymmetrization transformation is performed}, as $W^a_0$ already appears in the first factor of the asymmetric factorization \eqnm{assymetric_factorization_H0II}. Most importantly, all these ingredients for finding out the supersymmetric factorized \eq{Schrodinger_equ_base_level} just come out quickly as a whole {\it in one go} in this algebraical approach.

Through passing to the $z$-coordinate, we have been explicitly convinced that the {\it asymmetric} factorization $H_0={\mfk h}_0 {\mfk h}^\dagger_0$ in $x$-coordinate (which is just the variant of the usual self-adjoint form of $H_0$)  has the corresponding {\it standard} SUSYQM factorization $\mathsf{H}_0(z) = \mathsf{ h}_0(z) \mathsf{ h}^\dagger_0(z)$ in $z$-coordinate.
% , formally the latter is of course more {\it symmetric}. 
In other words, these two factorizations can be viewed as \emph{two equivalent representations} of the {\it same} SUSYQM associated with $H_0$ in (two different) $x$ and $z$-coordinate systems, and these two representations are simply connected to each other by the transformation ${\cal S}$ and the momentum operator map \eqnm{momentum_map_type2}, i.e. Eqs. \eqnm{down_shift_S} and \eqnm{standard_h0} together give us the relation
\begin{eqnarray}\label{h0_relation}
 \mathsf{h}_0(z)={\cal S}\smbr{\mfk{h}_0(x)} = \mfk{h}_0(x)-W^a_0
\end{eqnarray}
and Eqs. \eqnm{S_transformed_h} and \eqnm{standard_h0_dagger} together give us the relation
\begin{eqnarray}\label{h0_dagger_relation}
 \mathsf{h}^\dagger_0(z)={\cal S}\smbr{\mfk{h}^\dagger_0(x)} = \mfk{h}^\dagger_0(x) + W^a_0.
\end{eqnarray}
Later on, we will see that this type of isomorphism, not only preserves the SUSYQM factorizations of the startup level of the Hamiltonians $H_0$ and $\mathsf{H}_0(z)$, but also the structures of the shape invariance conditions  that will be derived in the next subsection and will be used to pass to all higher association levels, hence preserves the SUSYQM of the Hamiltonians of all  association levels, too. Because of this isomorphism, for the purpose of expounding this type of SUSYQMs of $H_0$ and its cousins of higher association levels and constructing the associated eigen-functions, one can choose either $x$ or $z$-coordinate representation at his/her own disposal.

One reminder is in order. $\mathsf{H}_0(z)$ is not completely removed from ${\cal H}_l(y)$ in \eq{transformed_Hl}. 
With the notation ${\cal P}\smbr{-}\equiv p^{1/4}\smbr{-}p^{-1/4}$ and the wrapper transformation ${\cal T}\smbr{-}$  in \eq{T_transformation}), we have 
\begin{eqnarray}\label{S_PT_combined}
{\cal S}\smbr{-} = {\cal P}{\cal T}\smbr{-} ={\cal T} {\cal P}\smbr{-}.  
\end{eqnarray}
By combining Eqs. \eqnm{transformed_Hl}, \eqnm{standardized_factorization_base_level} and \eqnm{S_PT_combined}, it is easy to prove that $\mathsf{H}_0(z)$  and  ${\cal H}_l(y)$ are connected to each other through the relation
\begin{eqnarray}\label{Standard_Hermitian_relation}
{\cal H}_l(y) =  {\cal P}^{-1}\sqbr{p \mathsf{H}_0(z)} - p \lambda_l + E_l.
\end{eqnarray}
In particular, for the minus route of the factorization for the base hypergeometric-like function, we have ${\cal H}^-_0(y) =  {\cal P}^{-1}\sqbr{p \mathsf{H}_0(z)}$, as $\lambda^-_0=0$ and $E^-_0=0$.

\subsection{Iteratively Building the Associated Hypergeometric-like Eigen-equation Series and Their Solutions with {\it Positive} Association Levels in the Non-standard {\it x}-coordinate Representation}\label{positive_association_levels} %in Ascending Order}

Just like the principal hypergeometric-like differential equations of generic higher levels and their eigen-functions can be proliferated iteratively from those of the lowest level either in $x$-coordinate or in $y$-coordinate representation, the associated hypergeometric-like differential equations and their solutions of higher levels can also be proliferated, either from the supersymmetrized Schr\"odinger equation \eq{Schrodinger_equ_base_level} of the lowest level  represented in $z$-coordinate or from the asymmetrically factorized \eq{asymmetric_eigen_equ_initial_association_level} represented in $x$-coordinate, just because both the supersymmetry and the shape invariance symmetry represented in these two coordinates are isomorphic to each other, as have been pointed out before. In Subsections \ref{Standard_Hermitian_factorization}-\ref{upward_SIC}, we have demonstrated in detail how to iteratively build the principal hypergeometric-like differential equations and their eigenfunctions of higher levels in the standard $y$-coordinate representation, whereas only outlined the key steps of accomplishing this in the nonstandard $x$-coordinate representation in Subsection \ref{non_standard_factorization}. Complementary to this, in this and the next two subsections, we will deliberately exemplify in detail that the SUSYQM of the associated Hypergeometric-like differential equations of higher levels and their eigen-functions can be iteratively constructed directly in the nonstandard $x$-coordinate presentation.

Instead of performing a direct factorization of the already known associated hypergeometric-like differential equation of generic association level (which can be derived from the base level eigen-equation \eqnm{eigen_equ} in traditional ways) and conducting a posterior check of SUSY and shape invariance symmetry owned by such factorization, now we assume that this equation is \emph{unknown or blind} to us and the {\it only known inputs} that are needed are the base eigen-equation \eqnm{eigen_equ} and its solutions $\Phi_l(x)$ and $\lambda^\pm_l$ that have already been worked out via the pure SUSYQM algorithms in Section \ref{susyqm_H_like_functions}. Actually, we will see that the concrete form of $\Phi_l(x)$ is not relevant.  As an example of illustration, \eq{asymmetric_eigen_equ_initial_association_level},  its eigenvalue $\lambda^-_l$ out of the minus route of factorizing \eq{eigen_equ} and its solution $\Phi_l$ are chosen to start with, and, for convenience, these are relabeled as
\begin{eqnarray}\label{eigen_equ_base_level}
 H^a_0\Phi_{l0} &=&\mfk{h}_0\mfk{h}^\dagger_0 \Phi_{l0}= \lambda_{l0} \Phi_{l0}, \\
 H^a_0 &\equiv& H_0,\;\Phi_{l0} \equiv \Phi_l,\; \lambda_{l0} \equiv \lambda^-_l. \nonumber 
\end{eqnarray}
We will demonstrate that \eq{eigen_equ_base_level} will iteratively proliferate into the associated hypergeometric-like differential equations of all higher levels and their eigen-solutions through the {\it same} SUSYQM algorithm we have used to build the principal hypergeometric-like differential equations of generic level and their eigen-solutions.

Intertwiningly acting \eq{eigen_equ_base_level} with the right-situated operator $\mfk{h}^\dagger_0$, one obtains the superpartner eigen-equation 
\begin{eqnarray}\label{eigen_equ_level_1_before_rnmlz}
 H^{as}_0 \Phi_{l1} \equiv \mfk{h}^\dagger_0\mfk{h}_0 \Phi_{l1}= \lambda_{l0}  \Phi_{l1}
\end{eqnarray}
with the degenerate eigenvalue $\lambda_{l0}$, where $H^{as}_0\equiv \mfk{h}^\dagger_0\mfk{h}_0$ is the superpartner Hamiltonian of $H^a_0=\mfk{h}_0\mfk{h}^\dagger_0$, which is obtained by simply switching the order of the two factors in $H^a_0$, and $\Phi_{l1}$ is proportional to $\mfk{h}^\dagger_0 \Phi_{l0}$. Assume that $\Phi_{l0}$ has already been normalized to unit, then according to \eq{eigen_equ_base_level},  we can take the normalized $\Phi_{l1}$ as
\begin{eqnarray}\label{base_raising}
\Phi_{l1}\equiv \frac{\mfk{h}^\dagger_0}{\lambda^{\frac{1}{2}}_{l0}} \Phi_{l0} = \frac{\sqrt{p}}{\lambda^{\frac{1}{2}}_{l0}} \Phi^{(1)}_{l0},
\end{eqnarray}
where the second equality has been reached by making use of the definition \eqnm{mfkh0_dagger} for $\mfk{h}^\dagger_0$, the superscript $(1)$ denotes the derivative of first order with respect to $x$, and $\Phi_{l1}$ is referred to as the associated hypergeometric-like function of first level.  We see that $\mfk{h}^\dagger_0$ works as the raising ladder operator and $\mfk{h}_0$ in turn as the lowering one. Although $\Phi_{l1}$ introduced this way coincides with that out of the traditional differentiation method -- taking the $m=1$ version of the traditional differential action $(d/dx)^m$ and multiplication with $\sqrt{p}^{m}$,  here $\Phi_{l1}$ is a natural result of the intertwining action -- a trivial operation in the SUSYQM algorithm.
 
 In order to allow this procedure to go ahead to generate an associated hypergeometric-like differential operator $H^a_1 =\mfk{h}_1 \mfk{h}^\dagger_1$ of $m=1$ association level,  with its eigen-solution $\Phi_{l1}$ introduced in \eq{base_raising}, it would be sufficient to devise the generalized commutator
 \begin{eqnarray}\label{SI_level_1}
\Delta^+_1=\mfk{h}^\dagger_0\mfk{h}_0- \mfk{h}_1 \mfk{h}^\dagger_1
\end{eqnarray}
for some first-order differential operators $\mfk{h}_1$ and $\mfk{h}^\dagger_1$,  and constant $\Delta^+_1$ as yet to be determined, in order
to substitute for $H^{as}_0\equiv \mfk{h}^\dagger_0\mfk{h}_0 $ in \eq{eigen_equ_level_1_before_rnmlz} and to renormalize it into the following  one 
 \begin{equation}\label{eigen_equ_level_1_after_rnmlz}
  \begin{aligned}
   H^a_1 \Phi_{l1} &\equiv \mfk{h}_1 \mfk{h}^\dagger_1 \Phi_{l1}=\lambda_{l1}\Phi_{l1},\\
 \lambda_{l1} &\equiv \lambda_{l0}-\Delta^+_1,  
  \end{aligned}
 \end{equation}
that is, this eigen-equation recovers the pattern of \eq{eigen_equ_base_level} by simply replacing the subscript $0$ by $1$, and it works as the renewed start-up eigen-equation for marching to the eigen-equations of even higher association levels. In many cases, the existence of such pattern invariance condition like \eqnm{SI_level_1} is the permit for successfully building a hierarchy of self-similar eigen-equations and their solutions via the SUSYQM algorithm. 

Then the issue is how to design $\mfk{h}_1$ and $\mfk{h}^\dagger_1$ , which ought to work respectively as the lowering and raising operators in connecting $\Phi_{l1}$ with its nearest neighbors $\Phi_{l0}$ and $\Phi_{l2}$, where $\Phi_{l2}$ should be as yet appropriately introduced. Here the situation that faces us is slightly different from that in Section \ref{susyqm_H_like_functions}, in that there the ladder operators for all $l$ are known to us  once the eigen-equation with generic $l$ has been obtained. In contrast, here the form of the eigen-equation for each high association level is assumed to be unknown to us and this eigen-equation itself and the corresponding ladder operator pair are needed to be constructed somehow.

The definition \eqnm{base_raising} says that $\mfk{h}^\dagger_0$ serves to raise the association level of the eigenfunction by $1$. This gives us the hint that acting this relation with $\mathfrak{h}^\dagger_0$ one more time may result in some equation that can be used to introduce the definitions of $\mfk{h}^\dagger_1$ and $\Phi_{l2}$. In fact,
\begin{eqnarray}\label{dagger_h0_action_once}
 \mfk{h}^\dagger_0 \Phi_{l1} = \sqbr{\sqrt{p} \od{}{x}}^2 \frac{\Phi_{l0}}{\lambda^{\frac{1}{2}}_{l0}}= \sqrt{p}'\Phi_{l1} + \frac{\sqrt{p}^2}{\lambda^{\frac{1}{2}}_{l0}} \Phi^{(2)}_{l0},
\end{eqnarray}
where the superscript $(2)$ denotes the second-order $x$-derivative. Grouping the two terms involving $\Phi_{l1}$,  this result can be rearranged as
\begin{eqnarray}\label{level_1_raising_unnormalized}
 \mfk{h}^\dagger_1 \Phi_{l1} = \frac{\sqrt{p}^2}{\lambda^{\frac{1}{2}}_{l0}} \Phi^{(2)}_{l0}
\end{eqnarray}
for
\begin{eqnarray}\label{raising_op_association_level_1}
  \mfk{h}^\dagger_1\equiv \mfk{h}^\dagger_0-\sqrt{p}'.
\end{eqnarray}
As a consequence, 
\begin{eqnarray}\label{lowering_op_association_level_1}
  \mfk{h}_1\equiv \mfk{h}_0-\sqrt{p}'. 
\end{eqnarray}

\eq{level_1_raising_unnormalized} together  with the eigen-equation in \eqnm{eigen_equ_level_1_after_rnmlz} allow us to introduce the normalized $\Phi_{l2}$ in the form 
\begin{eqnarray}\label{level_1_raising_normalized}
 \Phi_{l2} \equiv \frac{\mfk{h}^\dagger_1}{\lambda^{\frac{1}{2}}_{l1}} \Phi_{l1}= \frac{\sqrt{p}^2}{\lambda^{\frac{1}{2}}_{l1} \lambda^{\frac{1}{2}}_{l0}} \Phi^{(2)}_{l0},
\end{eqnarray}
which is analogous to \eq{base_raising}. Of course, $\Phi_{l2}$ introduced this way should be further justified by the eigen-equation it satisfies.

With Eqs. \eqnm{raising_op_association_level_1} and \eqnm{lowering_op_association_level_1} in hand, we have 
\begin{eqnarray}\label{quasi_commutator_positive_first0}
\mfk{h}^\dagger_0\mfk{h}_0- \mfk{h}_1 \mfk{h}^\dagger_1
&=&[\mfk{h}^\dagger_0, \mathfrak{h}_0]+ \mfk{h}_0 \sqrt{p}'+ \sqrt{p}' \mfk{h}^\dagger_0-\smbr{\sqrt{p}'}^2. \nonumber \\
\end{eqnarray}
It follows from the definitions \eqnm{mfkh_00} for  $\mathfrak{h}_0$ and \eqnm{mfkh0_dagger} for $\mfk{h}^\dagger_0$ that
\begin{eqnarray}\label{commutator}
[\mfk{h}^\dagger_0, \mathfrak{h}_0]= 2 \sqrt{p}\smbr{W^a_0}' 
\end{eqnarray}
and
\begin{eqnarray}\label{anti_commutator1}
\mfk{h}_0 \sqrt{p}'+ \sqrt{p}' \mfk{h}^\dagger_0 =- \sqrt{p} \sqrt{p}''+  2\sqrt{p}'W^a_0.
\end{eqnarray}
\eq{quasi_commutator_positive_first0} then reduces to  
\begin{eqnarray}\label{quasi_commutator_positive_first}
\mfk{h}^\dagger_0\mfk{h}_0- \mfk{h}_1 \mfk{h}^\dagger_1 &=& \sqbr{\sqrt{p}\smbr{2W^a_0-\sqrt{p}'}}' =- q'
\end{eqnarray}
where the second equality has been reached by making use of the definition \eqnm{superpotential_W0a} of $W^a_0$. We see that the shape invariance condition \eqnm{SI_level_1} indeed holds for the constant $\Delta^+_1=- q'=\lambda^-_1$, referring to \eq{lambda_minu_eigenvalue} for $\lambda^-_l$. Consequently, the lower equation in \eqnm{eigen_equ_level_1_after_rnmlz} is rewritten as
\begin{eqnarray}
 \lambda_{l1}=\lambda^-_l -\lambda^-_1.
\end{eqnarray}

Now, we proceed to construct $\mathfrak{h}_2$ and $\mathfrak{h}^\dagger_2$, hence $H^a_2\equiv \mathfrak{h}_2 \mathfrak{h}^\dagger_2$ in a similar way as done for $\mathfrak{h}_1$, $\mathfrak{h}^\dagger_1$ and $H^a_1$. By treating \eq{eigen_equ_level_1_after_rnmlz} as the {\it renewed} startup eigen-equation and intertwiningly acting it with $\mfk{h}^\dagger_1$, the new superpartner eigen-equation for association level $m=2$ is obtained in the form
\begin{equation}\label{eigen_equ_level_2_before_rnmlz}
 \begin{aligned}
  H^{as}_1 \Phi_{l2} &\equiv \mfk{h}^\dagger_1 \mfk{h}_1 \Phi_{l2}=\lambda_{l1} \Phi_{l2}, 
 \end{aligned}
\end{equation}
where the normalized eigenfunction  that follows from this construction is $\Phi_{l2} \equiv \mfk{h}^\dagger_1 \Phi_{l1}/\lambda^{1/2}_{l1}$, which is exactly the same as that given in \eq{level_1_raising_normalized}. 

We will show that there exist the operator pairs $\mfk{h}_2$ and $\mfk{h}^\dagger_2$ and some constant $\Delta^+_2$ such that the following generalized commutator relation
\begin{eqnarray}\label{SI_level_2}
\Delta^+_2 \equiv \mfk{h}^\dagger_1\mfk{h}_1- \mfk{h}_2 \mfk{h}^\dagger_2
\end{eqnarray}
holds. This relation allows us to substitute for $H^{as}_1\equiv \mfk{h}^\dagger_1\mfk{h}_1$ in favor of $\mfk{h}_2 \mfk{h}^\dagger_2$ in \eq{eigen_equ_level_2_before_rnmlz} and to renormalize it into the form
\begin{eqnarray}\label{eigen_equ_level_2_after_rnmlz}
 H^a_2 \Phi_{l2} \equiv \mfk{h}_2 \mfk{h}^\dagger_2 \Phi_{l2} =\lambda_{l2} \Phi_{l2}
\end{eqnarray}
with
\begin{eqnarray}\label{lambda_l20}
\lambda_{l2} \equiv \lambda_{l1}-\Delta^+_2=\lambda_{l0}-\smbr{\Delta^+_1+\Delta^+_2}.
\end{eqnarray}
We see that the tentatively defined $\Phi_{l2}$ in \eq{level_1_raising_normalized} is indeed an eigenfunction of $H^a_2$.

On the other hand, acting \eq{level_1_raising_normalized} with $\mfk{h}^\dagger_1$ defined in \eq{raising_op_association_level_1} yields
\begin{eqnarray}\label{level_2_raising_unnormalized}
 \mfk{h}^\dagger_1 \Phi_{l2}= \sqrt{p}' \Phi_{l2}+ \frac{\sqrt{p}^3}{\lambda^{\frac{1}{2}}_{l1} \lambda^{\frac{1}{2}}_{l0}} \Phi^{(3)}_l.
\end{eqnarray}
Upon dividing by $\lambda^{1/2}_{l2}$ for the normalization consideration, regrouping the two terms involving $\Phi_{l2}$ and introducing
\begin{eqnarray}\label{h2_dagger}
 \mfk{h}^\dagger_2 \equiv \mfk{h}^\dagger_1-\sqrt{p}'=\mfk{h}^\dagger_0-2 \sqrt{p}',
\end{eqnarray}
\eq{level_2_raising_unnormalized} is rearranged as the raising relation
\begin{eqnarray}
 \Phi_{l3}\equiv \frac{\mfk{h}^\dagger_2}{\sqrt{\lambda_{l2}}} \Phi_{l2}= \frac{\sqrt{p}^3}{\lambda^{\frac{1}{2}}_{l2} \lambda^{\frac{1}{2}}_{l1} \lambda^{\frac{1}{2}}_{l0}} \Phi^{(3)}_l.
\end{eqnarray}
Here it is delightful to see that the repetitive pattern exhibited in the raising relations \eqnm{base_raising} and \eqnm{level_1_raising_normalized} emerges once again.

The Hermitian conjugate of \eq{h2_dagger} reads 
\begin{eqnarray}\label{h2}
\mfk{h}_2 =\mfk{h}_1-\sqrt{p}'=\mfk{h}_0-2 \sqrt{p}'. 
\end{eqnarray}

In the expansion
\begin{eqnarray}\label{generalized_commutator20}
 \mfk{h}^\dagger_1\mfk{h}_1- \mfk{h}_2 \mfk{h}^\dagger_2 &=& [\mfk{h}^\dagger_0, \mathfrak{h}_0]+ 2\smbr{\mfk{h}_0 \sqrt{p}'+ \sqrt{p}' \mfk{h}^\dagger_0} \\
 && -\smbr{\sqrt{p}' \mfk{h}_0 +  \mfk{h}^\dagger_0 \sqrt{p}'}-3\smbr{\sqrt{p}'}^2,\nonumber 
\end{eqnarray}
making use of the fact that 
\begin{eqnarray}\label{anti_commutator2}
\mfk{h}^\dagger_0 \sqrt{p}'+ \sqrt{p}' \mfk{h}_0=\sqrt{p} \sqrt{p}''+ 2 W^a_0 \sqrt{p}',
\end{eqnarray}
and Eqs. \eqnm{commutator}-\eqnm{quasi_commutator_positive_first}, \eq{generalized_commutator20} is then simplified into
\begin{eqnarray}\label{generalized_commutator21}
 \mfk{h}^\dagger_1\mfk{h}_1- \mfk{h}_2 \mfk{h}^\dagger_2 &=& -q'-p''.
\end{eqnarray}
That is, the pattern invariance condition \eqnm{SI_level_2} indeed holds for the constant $\Delta^+_2 = -q'-p''$. Then it follows from this $\Delta^+_2$, $\Delta^+_1 = -q'$ and \eq{lambda_minu_eigenvalue} for $\lambda^-_l$ that $\Delta^+_1+\Delta^+_2=\lambda^-_2$. Consequently, \eq{lambda_l20} is rewritten as 
\begin{eqnarray}\label{lambda_l21}
\lambda_{l2} =\lambda^-_l- \lambda^-_2.
\end{eqnarray}

Repeating this iterative procedure and keeping going to the $m$-th level for $2<m\le l$, we would obtain the normally ordered eigen-equation
\begin{eqnarray}\label{eigen_equ_level_m_after_rnmlz}
 H^a_m \Phi_{lm} =\lambda_{lm} \Phi_{lm},
\end{eqnarray}
for
\begin{eqnarray}
 H^a_m &\equiv& \mfk{h}_m \mfk{h}^\dagger_m,\label{H_plus_m} \\ 
 \mfk{h}_m &\equiv& \mfk{h}_0- m \sqrt{p}', \label{lowering_ops_mth} \\           
 \mfk{h}^\dagger_m &\equiv& \mfk{h}^\dagger_0- m \sqrt{p}', \label{raising_ops_mth}\\
 \Phi_{lm} &=& \frac{\mfk{h}^\dagger_{m-1}}{\lambda^{\frac{1}{2}}_{l,m-1}} \Phi_{l,m-1}, \label{raising_recursion_relation_plus_m} \\
 \lambda_{lm} &\equiv& \lambda_{l0}-\sum^m_{n=1} \Delta^+_n, \label{lambda_lm00},
\end{eqnarray}
and the generic shape invariance condition 
\begin{eqnarray}\label{SI_associated_ascending}
\Delta^+_m \equiv \mathfrak{h}^\dagger_{m-1} \mathfrak{h}_{m-1}- \mfk{h}_m \mfk{h}^\dagger_m.
\end{eqnarray}
Again, combining the definitions \eqnm{lowering_ops_mth} and \eqnm{raising_ops_mth}, Eqs. \eqnm{commutator}, \eqnm{anti_commutator1}, \eqnm{quasi_commutator_positive_first} and \eqnm{anti_commutator2} yields the constant
\begin{eqnarray}\label{Delta_plus_n}
\Delta^+_n =- q' -\smbr{n-1}p''.  
\end{eqnarray}
Consequently, 
\begin{eqnarray}\label{lambda_m_def}
\sum^m_{n=1} \Delta^+_n=-m q'- \frac{1}{2} m(m-1) p''\equiv \lambda^-_m, 
\end{eqnarray}
and \eq{lambda_lm00} becomes
\begin{eqnarray}\label{lambda_lm0}
\lambda_{lm} = \lambda_l^- - \lambda^-_m.%=l\smbr{l+1}-m\smbr{m+1}.
\end{eqnarray}

Note that  \eq{lambda_m_def} actually works as another definition of $\lambda^-_m$, which is \emph{independent of} but consistent with that of $\lambda^-_l$ in \eq{lambda_minu_eigenvalue}.

It is easy to check that if Eqs. \eqnm{eigen_equ_level_m_after_rnmlz}-\eqnm{lambda_lm0} hold, then they hold for $m\rightarrow m+1$ ($m+1\le l$) as well.

For $1 \le m\le l$, upon repeatedly making use of the raising recurrence relation \eqnm{raising_recursion_relation_plus_m}, $\Phi_{lm}$ is then related to $\Phi_l$ in the following bottom-up way,
\begin{eqnarray}\label{Phi_lm_bottom_up_approach}
 \Phi_{lm} (x) &=& \frac{\mfk{h}^\dagger_{m-1}}{\lambda^{\frac{1}{2}}_{l,m-1}} \;\frac{\mfk{h}^\dagger_{m-2}}{\lambda^{\frac{1}{2}}_{l,m-2}}\; \cdots\; \frac{\mfk{h}^\dagger_1}{\lambda^{\frac{1}{2}}_{l1}}\; \frac{\mfk{h}^\dagger_0}{\lambda^{\frac{1}{2}}_{l0}} \;\Phi_l(x),\nonumber \\
               &=&\smbr{\displaystyle \prod^{m-1}_{j=0} \lambda^{-\frac{1}{2}}_{l,j}} \sqrt{p}^m\smbr{\od{}{x}}^m \Phi_l(x),
\end{eqnarray}
where the lumped form 
\begin{eqnarray}\label{h_j_dagger_lumped}
 \mfk{h}^\dagger_j =\sqrt{p}^{j+1} \od{}{x} \sqrt{p}^{-j},
\end{eqnarray}
which follows from the combination of Eqs \eqnm{mfkh0_dagger} and \eqnm{raising_ops_mth}, has been made use of to contract the operator product and reach the second equality in \eq{Phi_lm_bottom_up_approach}. 
We see that the canonical form of associated hypergeometric-like function of the generic association level $m$ (in terms of the base function $\Phi_l(x)$) is recovered, as the direct outcome of the SUSYQM algorithm. The structure on the right hand side of \eq{Phi_lm_bottom_up_approach} shows why $\Phi_{lm} (x)$ and the associated hypergeometric-like differential equation satisfied by it can be obtained in the traditional way -- first taking  $m$-fold differentiating the principal hypergeometic-like differential equation \eqnm{eigen_equ} satisfied by $\Phi_l(x)$ and then multiplying with $\sqrt{p}^m$ and moving this factor next to the operator $\smbr{d/dx}^m$. 

From \eq{lambda_lm0}, it is evident that $m=l$ is the top association level, for which 
\begin{eqnarray}
\lambda_{ll}\equiv\lambda_{l0} -\lambda^-_l =0.
\end{eqnarray}
This is another way of pinning down the eigenvalue of the principal hypergeometric-like operator $H_0$, i.e.  $\lambda_l=\lambda_{l0}=\lambda^-_l$, with $\lambda^-_l$ defined in \eq{lambda_m_def} for $m=l$.
\eq{eigen_equ_level_m_after_rnmlz} then implies the annihilation condition
\begin{eqnarray}\label{annihilation_condition_top_level}
 \mfk{h}^\dagger_l \Phi_{ll}= \sqbr{\sqrt{p}\od{}{x} - l \sqrt{p}'}\Phi_{ll}=0.
\end{eqnarray}
This equation is easy to be integrated and the solution of the top association level is
\begin{eqnarray}\label{Phi_ll}
\Phi_{ll}(x)= \sqbr{\Phi^{(l)}_l \displaystyle \prod^{l-1}_{j=0}\lambda^{-\frac{1}{2}}_{lj} } \sqrt{p(x)}^l,
\end{eqnarray}
where the constant inside the square brackets has been attached in order to be consistent with $\Phi_{ll}$ given by \eq{Phi_lm_bottom_up_approach} for $m=l$ ($\Phi^{(l)}_l$ is a constant because $\Phi_l$ is a polynomial of $l$-th order in $x$). 

On the other hand, the combination the $\smbr{m+1}$-version of the raising recurrence relation \eqnm{raising_recursion_relation_plus_m} and \eq{eigen_equ_level_m_after_rnmlz} yields the lowering recurrence relation
\begin{eqnarray}\label{lowering_recursion_relation_plus_m}
 \Phi_{lm}=\frac{\mfk{h}_m}{\lambda^{\frac{1}{2}}_{lm}} \Phi_{lm+1}.
\end{eqnarray}

Starting with the solution (\ref{Phi_ll}) and applying the relation (\ref{lowering_recursion_relation_plus_m}) (with $m\rightarrow j$) consecutively for $j=l-1,\; l-2,\;\cdots,\; m$  will give rise to the generic solution $\Phi_{lm}$ for $0\le m<l$ in the form
\begin{eqnarray}\label{Phi_lm_top_down_approach}
 \Phi_{lm}&=&\frac{\mfk{h}_m}{\lambda^{\frac{1}{2}}_{lm}} \frac{\mfk{h}_{m+1}}{\lambda^{\frac{1}{2}}_{l,m+1}} \cdots \frac{\mfk{h}_{l-2}}{\lambda^{\frac{1}{2}}_{l,l-2}} \frac{\mfk{h}_{l-1}}{\lambda^{\frac{1}{2}}_{l,l-1}} \Phi_{ll}, \nonumber \\
        &=&\sqbr{\Phi^{(l)}_l \displaystyle \prod^{l-1}_{j=m} \lambda^{-\frac{1}{2}}_{lj}} \frac{\smbr{-1}^{l-m}}{w \sqrt{p}^{m}} \smbr{\od{}{x}}^{l-m} w p^l, \nonumber \\
\end{eqnarray}
where the operator product has been contracted into the compact form by making use of the lumped form 
\begin{eqnarray}\label{h_j_lumped}
 \mfk{h}_j = -\sqrt{p}^{-j} w^{-1} \od{}{x} \sqrt{p}^{j+1}w, 
\end{eqnarray}
which results from the combination of Eqs \eqnm{mfkh_01} and \eqnm{lowering_ops_mth}. In fact, this provides the \emph{top-down} approach of building all $\Phi_{lm}$ with the association levels lower than $l$, besides the \emph{bottom-up} one given in \eq{Phi_lm_bottom_up_approach}. In particular, for $m=0$ \eq{Phi_lm_top_down_approach} reduces to the base hypergeometric-like polynomial $\Phi_l$ in the compact form
\begin{eqnarray}\label{Rodriguez_formula}
\Phi_l \equiv \Phi_{l0}=\sqbr{\Phi^{(l)}_l \displaystyle \prod^{l-1}_{j=0} \lambda^{-\frac{1}{2}}_{lj}} \frac{1}{w} \smbr{\od{}{x}}^l (-p)^l w, 
\end{eqnarray}
which is called the generalized Rodriguez' formula. 

The above procedure of obtaining $\Phi_{lm}$ for $m\ge0$ (particularly the  top-down subroutine) has two advantages in comparison with other approaches: 1) the top level eigen-function $\Phi_{ll}$ is easily solved from the annihilation condition \eqnm{annihilation_condition_top_level}, so that $\Phi_{lm}$ is determined in the simple compact form \eqnm{Phi_lm_top_down_approach}; 2) the starting level (base) eigenfunction $\Phi_l$ and eigenvalue $\lambda_l^-$ also come out {\it within} this procedure in a \emph{more efficient way} than those SUSYQM algorithms demonstrated in Section \ref{susyqm_H_like_functions}. Thus, for the purpose of building the eigen-solution $\Phi_l$ and $\lambda^-_l$ of $H_0$ alone, there is no need to resort to other separate algorithms, such as the latter ones (though these are theoretically important in revealing the SUSYQM reason for why there exist two distinct types of special functions for the same hypergeometric-like operator $H_0$, and they have their own advantages), once we follow the above SUSYQM algorithm for building the associated hypergeometric-like differential equation and its eigen-solution. To my best knowledge, such SUSYQM algorithm demonstrated in this subsection provides the quickest way of building the eigen-solutions of \eq{eigen_equ}, as well as the associated hypergeometric-like equation of generic level and its eigen-solutions. We will see in the next subsection that the case in which the association level is a negative integer can be accomplished in a similar fashion. 

We can make use of the \emph{two-term} raising and lowering recurrence relations \eqnm{raising_recursion_relation_plus_m} and \eqnm{lowering_recursion_relation_plus_m} to construct some \emph{three-term} recurrence relations. The $(m+1)$-version of  relation \eqnm{raising_recursion_relation_plus_m} reads $\lambda^{1/2}_{lm} \Phi_{lm+1}=\mfk{h}^\dagger_m \Phi_{lm}$.
The combination of Eqs. \eqnm{lowering_ops_mth} and \eqnm{raising_ops_mth} implies that $\mfk{h}^\dagger_m + \mfk{h}_{m-1} = 2 W^a_0 -(2m-1)\sqrt{p}'$ and $\mfk{h}^\dagger_m - \mfk{h}_{m-1} = 2 \sqrt{p} d/dx- 2W^a_0 - \sqrt{p}'$. Furthermore, the $(m-1)$-version of relation \eqnm{lowering_recursion_relation_plus_m} reads $\mfk{h}_{m-1} \Phi_{lm} = \lambda^{1/2}_{lm-1}\Phi_{lm-1}$. These three aspects of consideration together lead to the following 
three-term recurrence relations 
\begin{eqnarray}\label{3-term_recrr_rel1}
\lambda^{\frac{1}{2}}_{lm} \Phi_{lm+1} &=& - \sqbr{2\smbr{m-1}\sqrt{p}'+\frac{q}{\sqrt{p}}}\Phi_{lm} \\
 && - \lambda^{\frac{1}{2}}_{lm-1} \Phi_{lm-1} \nonumber
\end{eqnarray}
and 
\begin{eqnarray}\label{3-term_recrr_rel2}
\lambda^{\frac{1}{2}}_{lm} \Phi_{lm+1} &=& \smbr{2\sqrt{p} \od{}{x} -2\sqrt{p}' + \frac{q}{\sqrt{p}} }\Phi_{lm}  \\
                              && + \lambda^{\frac{1}{2}}_{lm-1} \Phi_{lm-1},\nonumber
\end{eqnarray}
where $W^a_0(x)$ has been substituted for via \eq{superpotential_W0a}.

In many applications, one is directly faced with the issue of solving the associated hypergeometric-like differential equation \eq{eigen_equ_level_m_after_rnmlz} in its prefactorized (expanded) form in $x$-coordinate, for which both $H^a_m(x)$ and $\lambda_{lm}$ take some ‘smeared’ forms, and these two even mix with each other in the form $H^a_m(x)-\lambda_{lm}$. In order to directly make use of the formula \eqnm{Phi_lm_bottom_up_approach} or \eqnm{Phi_lm_top_down_approach}  to quickly determine the eigen-function, upon utilizing Eqs. \eqnm{lowering_ops_mth}, \eqnm{raising_ops_mth} and \eqnm{anti_commutator1}, $H^a_m(x)$ should be identified as
\begin{eqnarray}\label{H_m_expansion}
 H^a_m(x)=H_0(x)+ \frac{m}{2}\sqbr{ p''+\frac{q-p'}{p}p'}+ \smbr{\frac{m}{2}}^2 \frac{p'^2}{p}, \nonumber \\
\end{eqnarray}
which is just the expansion of $\mfk{h}_m \mfk{h}^\dagger_m$. The $m$-independent part $H_0$ defined in \eq{H_0_operator} is easily picked up from a given operator $H^a_m -\lambda_{lm}$, as these two are the pure differentiation terms of second and first orders. Consequently, the functions $p(x)$ and $q(x)$ are pinned down. In consequence,  all the non-differential terms in \eq{H_m_expansion} should be identified with the last three $m$-dependent terms (up to some constant terms like $m\,p''/2$, which can be mixed with $-\lambda_{lm}$), which essentially are rational fractions with the numerators and denominators being at most binomials in $x$.
In doing so, the integer-valued $m$ as the association level can be correctly identified.  
According to \eq{lambda_lm0}, the eigenvalue $\lambda_{lm}$ is separated from the mixed form of $H^a_m-\lambda_{lm}$ in the expanded form
\begin{eqnarray}\label{lambda_lm}
 \lambda_{lm} =-\smbr{l-m} q'-\sqbr{l(l-1)-m(m-1)}\frac{p''}{2}
\end{eqnarray}
and the base level $l$ is then correctly fixed. Or, in view that the above iterative SUSYQM algorithm of deriving the generic associated hypergeometric-like differential equations and their solutions is not complex at all,  after $H_0$ is readily identified from $H^a_m-\lambda_{lm}$, one can repeat this whole deductive procedure to get the generic solution in the forms \eqnm{Phi_lm_bottom_up_approach}, \eqnm{Phi_lm_top_down_approach} and \eqnm{lambda_lm0}. 

\subsection{Iteratively Building the Associated Hypergeometric-like Eigen-equation Series and Their Solutions with {\it Negative} Association Levels in the Non-standard {\it x}-coordinate Representation}\label{negative_association_levels}

We have demonstrated in the previous subsection how to build a tower of associated hypergeometric-like differential equations and their eigen-solutions with positive association levels in \emph{ascending} order by applying the \emph{raising} ladder operators. In the following we will show that a tower of associated hypergeometric-like differential equations and their eigen-solutions for {\it negative} association levels can be constructed in a similar fashion but in \emph{descending} (\emph{negatively} ascending) order by applying some \emph{lowering} ladder operators. The famous example of this category is the associated Legendre differential equation and its eigenfunctions with negative association levels. 

As in Subsection \ref{positive_association_levels}, we still assume that the associated hypergeometric-like differential equation for generic negative association level is unknown to us and the only known one to us is \eq{eigen_equ} or its supersymmetrically factorized version \eq{asymmetric_eigen_equ_initial_association_level}, although one can make a guess on how it looks like from the form of $H^a_m$ in \eq{H_m_expansion} for positive $m$ by some parity symmetry arguments. Note that \eq{asymmetric_eigen_equ_initial_association_level} is not immediately suitable as the startup eigen-equation for performing the intertwining manipulation to generate the superpartner eigen-equation of association level $-1$, because the lowering operator is not situated on the right side inside the product $H^a_0 =\mfk{h}_0\mfk{h}^\dagger_0$. A slight twist of this can give us a factorization of $H^a_0$, in which a certain $\mpzc{h}_0$  is situated on the right side and works as the lowering operator, and its Hermitian conjugate $\mpzc{h}^\dagger_0$ is situated on the left side. Recall that in the factorization $H^a_0=\mfk{h}_0 \mfk{h}^\dagger_0$, the minus sign has been deliberately assigned to the inner layer and define $\mfk{h}_0$ (see \eq{mfkh_01}). Alternatively, we have the freedom to assign the minus sign to the outer layer of $H^a_0$ and define the slightly different factorization
\begin{eqnarray}\label{base_level_factorization_negative}
H^a_0= \mpzc{h}^\dagger_0 \mpzc{h}_0
\end{eqnarray}
with
\begin{eqnarray}\label{operator_relations_lowest}
\mpzc{h}^\dagger_0 \equiv \frac{1}{w} \od{}{x}w \sqrt{p} = - \mfk{h}_0,\;   \mpzc{h}_0 \equiv - \sqrt{p}\od{}{x}=-\mfk{h}^\dagger_0. %\nonumber \\
\end{eqnarray}
The base eigen-equation \eqnm{asymmetric_eigen_equ_initial_association_level} is then rewritten as
\begin{eqnarray}\label{base_eigen_equ1}
  \mpzc{h}^\dagger_0 \mpzc{h}_0 \Phi_{l0}=\lambda_{l0}\Phi_{l0}.
\end{eqnarray}
We will see that this $\mpzc{h}_0$ in fact works as a lowering operator 
% and eventually result in $\mpzc{h}_{-m}=\mpzc{h}_0 + m \sqrt{p}'$ for the negative $m$-th association level. That is, 
and it plays the role analogous to the previous raising operator $\mfk{h}_0^+=\sqrt{p}d /dx$ but marches in the \emph{negative} direction, which suits for building the descending series $\{\Phi_{l,-m}\}$ ($m$ is a positive integer) and their eigen-equations.

Acting \eq{base_eigen_equ1} with $\mpzc{h}_0$ yields the superpartner eigen-equation 
\begin{eqnarray}\label{eigen_equ_m1st_level_before_rnmlz}
 \mpzc{h}_0 \mpzc{h}^\dagger_0 \Phi_{l,-1}=\lambda_{l0} \Phi_{l,-1},
\end{eqnarray}
where 
\begin{eqnarray}\label{level_minus_1_lowering}
 \Phi_{l,-1} \equiv \frac{\mpzc{h}_0}{\lambda^{\frac{1}{2}}_{l0}} \Phi_{l0}=-\Phi_{l1}
\end{eqnarray}
and the second equation in \eq{operator_relations_lowest} and \eq{base_raising} have been made use of to reach the second equality above.

This eigen-equation is expected to be renormalized into the following one
\begin{eqnarray}\label{eigen_equ_m1st_level_after_rnmlz}
 H^a_{-1}\Phi_{l,-1} \equiv \mpzc{h}^\dagger_{-1} \mpzc{h}_{-1}  \Phi_{l,-1}=\lambda_{l,-1} \Phi_{l,-1}
\end{eqnarray}
for $\lambda_{l,-1} \equiv \lambda_{l0}-\Delta^-_1$, that has the pattern in common with \eq{base_eigen_equ1}, and the operators $\mpzc{h}^\dagger_{-1}$ and $\mpzc{h}_{-1}$ as yet have to be appropriately defined so that the pattern invariance condition
\begin{eqnarray}\label{SIC_minus1}
 \Delta^-_1\equiv \mpzc{h}_0 \mpzc{h}^\dagger_0 - \mpzc{h}^\dagger_{-1} \mpzc{h}_{-1}
\end{eqnarray}
holds for some constant $\Delta^-_1$.

On the other hand, the definition \eqnm{level_minus_1_lowering} indicates that $\mpzc{h}_0$ works as the lowering operator that lowers $\Phi_{l0}$ down to $\Phi_{l,-1}$. Acting on this relation with $\mpzc{h}_0$ one more time yields
\begin{eqnarray}\label{lowering_relation_minus_1}
  \mpzc{h}_0\Phi_{l,-1}  &=&- \sqrt{p}'\Phi_{l,-1}+ \smbr{-1}^2 \frac{\sqrt{p}^2}{\lambda^{\frac{1}{2}}_{l0}} \Phi^{(2)}_{l0}.
\end{eqnarray}
Upon dividing by $\lambda^{1/2}_{l,-1}$ for normalization consideration and grouping the two terms involving $\Phi_{l,-1}$, the above equation can be rearranged as
\begin{eqnarray}\label{tentative_def_level_minus2}
 \Phi_{l,-2} = \frac{\mpzc{h}_{-1}}{\lambda^{\frac{1}{2}}_{l,-1}}\Phi_{l,-1},
\end{eqnarray}
where
\begin{eqnarray}\label{h_minus_1}
\mpzc{h}_{-1}\equiv \mpzc{h}_0 + \sqrt{p}'=-\mfk{h}^\dagger_1
\end{eqnarray}
and 
\begin{eqnarray}\label{Phil_minus_2}
\Phi_{l,-2}\equiv \smbr{-1}^2 \frac{\sqrt{p}^2}{\lambda^{\frac{1}{2}}_{l,-1}\lambda^{\frac{1}{2}}_{l0}} \Phi^{(2)}_{l0}=\Phi_{l2}.
\end{eqnarray}
As a consequence of \eq{h_minus_1},
\begin{eqnarray}\label{h_minus_1_dagger}
\mpzc{h}^\dagger_{-1} \equiv \mpzc{h}^\dagger_0 + \sqrt{p}'= -\mfk{h}_1.
\end{eqnarray}
Note that, as the renewed lowering relation from association level $-1$ to $-2$, the relation \eqnm{tentative_def_level_minus2} follows the same pattern as \eqnm{level_minus_1_lowering}. 

In view of the relations \eqnm{operator_relations_lowest}, \eqnm{h_minus_1} and \eqnm{h_minus_1_dagger}, the commutator relation \eqnm{quasi_commutator_positive_first} then implies that there holds the generalized commutator relation 
\begin{eqnarray}
 \mpzc{h}_0 \mpzc{h}^\dagger_0 - \mpzc{h}^\dagger_{-1} \mpzc{h}_{-1} =-q'
\end{eqnarray}
That is, the pattern invariance condition \eqnm{SIC_minus1} indeed holds for $\Delta^-_1=-q'=\Delta^+_1$, and consequently, we have $\lambda_{l,-1}=\lambda_{l1}$. 

Intertwiningly acting \eq{eigen_equ_m1st_level_after_rnmlz} with $\mpzc{h}_{-1}$ and dividing it by $\lambda^{1/2}_{l,-1}$ yield
\begin{eqnarray}\label{eigen_equ_m2nd_level_before_rnmlz}
 \mpzc{h}_{-1} \mpzc{h}^\dagger_{-1} \Phi_{l,-2}=\lambda_{l,-1} \Phi_{l,-2},
\end{eqnarray}
where $\Phi_{l,-2}\equiv \mpzc{h}_{-1}  \Phi_{l,-1}/\lambda^{1/2}_{l,-1}$, which is the same as the lowering relation \eqnm{tentative_def_level_minus2}. Therefore, the tentative definition of $\Phi_{l,-2}$ in \eq{Phil_minus_2} is justified.

\eq{eigen_equ_m2nd_level_before_rnmlz} is expected to be renormalized into the following one 
\begin{eqnarray}\label{eigen_equ_m2nd_level_after_rnmlz}
H^a_{-2} \Phi_{l,-2}\equiv \mpzc{h}^\dagger_{-2} \mpzc{h}_{-2}  \Phi_{l,-2} = \lambda_{l,-2} \Phi_{l,-2}
\end{eqnarray}
that has the same pattern as Eqs. \eqnm{base_eigen_equ1} and \eqnm{eigen_equ_m1st_level_after_rnmlz}, for 
\begin{eqnarray}\label{lambdal_minus20}
\lambda_{l,-2}\equiv \lambda_{l,-1}-\Delta^-_2=\lambda_{l0}-\smbr{\Delta^-_1+\Delta^-_2}
\end{eqnarray}
and some operator $\mpzc{h}_{-2}$ as yet to be appropriately defined so that the pattern invariance condition
\begin{eqnarray}\label{SIC_minus2}
 \Delta^-_2 \equiv \mpzc{h}_{-1} \mpzc{h}^\dagger_{-1} -\mpzc{h}^\dagger_{-2} \mpzc{h}_{-2}
\end{eqnarray}
holds for some constant $\Delta^-_2$.

On the other hand, acting $\Phi_{l,-2}$ defined by \eq{Phil_minus_2} with $\mpzc{h}_0$ yields
\begin{eqnarray}
 \mpzc{h}_0\Phi_{l,-2}= -2 \sqrt{p}'\Phi_{l,-2}- \frac{\sqrt{p}^3}{\lambda^{\frac{1}{2}}_{l,-1}\lambda^{\frac{1}{2}}_{l0}} \Phi^{(3)}_{l0}.
\end{eqnarray}
Dividing $\lambda^{1/2}_{l,-2}$ on both sides and grouping the two terms involving $\Phi_{l,-2}$, the above equation can be reorganized as the renewed lowering relation
\begin{eqnarray}
 \Phi_{l,-3} = \mpzc{h}_{-2}\Phi_{l,-2}
\end{eqnarray}
for 
\begin{eqnarray}\label{mzch_minus_2}
\mpzc{h}_{-2}\equiv \mpzc{h}_0 + 2 \sqrt{p}'=-\mfk{h}^\dagger_2
\end{eqnarray}
and 
\begin{eqnarray}
 \Phi_{l,-3} \equiv \smbr{-1}^3 \frac{\sqrt{p}^3}{\lambda^{\frac{1}{2}}_{l,-2}\lambda^{\frac{1}{2}}_{l,-1}\lambda^{\frac{1}{2}}_{l0}}\Phi^{(3)}_{l0}=-\Phi_{l3}.
\end{eqnarray}

In view of the relation \eqnm{mzch_minus_2} between $\mpzc{h}_{-2}$ and $\mfk{h}^\dagger_2$ and \eqnm{h_minus_1} between $\mpzc{h}_{-1}$ and $\mfk{h}^\dagger_1$, the generalized commutator relation \eqnm{generalized_commutator21} then implies that
\begin{eqnarray}\label{SIC_minus21}
 \mpzc{h}_{-1} \mpzc{h}^\dagger_{-1} -\mpzc{h}^\dagger_{-2} \mpzc{h}_{-2} &=& -q'-p''.
\end{eqnarray}
Thus, the shape invariance condition \eqnm{SIC_minus2} indeed holds for constant $\Delta^-_2 =-q'-p''$. 

\eq{lambdal_minus20} can be rewritten as
\begin{eqnarray}\label{lambdal_minus21}
\lambda_{l,-2} = \lambda^-_{l}-\lambda^-_2 =\lambda^-_{l2}
\end{eqnarray}
as the result of $\Delta^-_1+\Delta^-_2=\lambda^-_2$.

Keeping running this iterative algorithm to the $(-m)$-th level ($3 < m \le l$), we obtain the normally ordered eigen-equation
\begin{eqnarray}\label{eigen_equ_level_negative_m_after_rnmlz}
H^a_{-m} \Phi_{l,-m} = \lambda_{l,-m} \Phi_{l,-m},
\end{eqnarray}
where
\begin{eqnarray}
H^a_{-m} &\equiv& \mpzc{h}^\dagger_{-m} \mpzc{h}_{-m},\label{H_minus_m} \\  
 \mpzc{h}_{-m}&=&\mpzc{h}_0 + m \sqrt{p}'=-\sqrt{p}^{m+1} \od{}{x} \sqrt{p}^{-m} =-\mfk{h}^\dagger_m, \nonumber  \\ \label{lowering_ops_minus_mth}\\
 \mpzc{h}^\dagger_{-m} &=& \mpzc{h}^\dagger_0 + m \sqrt{p}'=\sqrt{p}^{-m} w^{-1} \od{}{x} \sqrt{p}^{m+1}w = -\mfk{h}_m, \nonumber \\  \label{raising_ops_minus_mth} 
 \Phi_{l,-m} &=& \frac{\mpzc{h}_{-m+1}}{\lambda^{\frac{1}{2}}_{l,-m+1}} \Phi_{l,-m+1},\label{lowering_recursion_relation_minus_mth} \\
 \Delta^-_n &\equiv & \mpzc{h}_{-n+1}\mpzc{h}^\dagger_{-n+1}-\mpzc{h}^\dagger_{-n}\mpzc{h}_{-n},\nonumber \\  
            &=&-q'-\smbr{n-1} p'',\label{SI_associated_decending} \\
            &=&\Delta^+_n, \nonumber  \\
 \lambda_{l,-m} &\equiv& \lambda_l-\sum^m_{n=1} \Delta^-_n=\lambda^-_l-\lambda^-_m=\lambda_{lm}. \label{eigenvalue_negative_level}
\end{eqnarray}

The combination of Eqs. \eqnm{H_plus_m}, \eqnm{H_minus_m}, \eqnm{lowering_ops_minus_mth} and \eqnm{raising_ops_minus_mth} shows that $H^a_{-m}=H^a_m$. On the other hand, \eq{H_m_expansion} says that $H^a_m$ is not even in $m$. Thus, $H^a_{-m}$ can not be viewed as the operator obtained simply from $H^a_m$ by replacing $m$ with $-m$. In \eq{eigenvalue_negative_level}, $\lambda_{l,-m}$ should be merely viewed as the notation $\lambda_{l,i} \equiv -(l+i)q'-[l(l-1)-i(i+1)]p''/2$ for $i =-m<0$, which is formally different from the other notation $\lambda_{lm}$ (with $m\ge 0$) given in \eq{lambda_lm}. That is, in \eq{eigenvalue_negative_level} the notation $\lambda_{lm}$ should not be misunderstood as an even function of $m$.

Consecutively making use of Eqs. \eqnm{lowering_recursion_relation_minus_mth} and \eqnm{lowering_ops_minus_mth} (with $m\rightarrow j$) for $0 < j \le m \le l$ ($m \le l$), we obtain
\begin{eqnarray}\label{Phi_lm_bottom_up_approach_negative}
 \Phi_{l,-m} &=& \frac{\mpzc{h}_{-m+1}}{\lambda^{\frac{1}{2}}_{l,-m+1}} \frac{\mpzc{h}_{-m+2}}{\lambda^{\frac{1}{2}}_{l,-m+2}} \cdots  \frac{\mpzc{h}_{-1}}{\lambda^{\frac{1}{2}}_{l,-1}} \frac{\mpzc{h}_0}{\lambda^{\frac{1}{2}}_{l0}} \Phi_l,\nonumber  \\
             &=&\smbr{-1}^m \displaystyle \prod^m_{j=1} \lambda^{-\frac{1}{2}}_{l,-j+1}  \sqrt{p}^m\smbr{\od{}{x}}^{m}\Phi_l, \label{generic_Phi_l_minusm} \\
             &=&\smbr{-1}^m \Phi_{lm}, \nonumber 
\end{eqnarray}
where to reach the second equality, the lumped form of $\mpzc{h}_{-j}$ (the second equality in \eq{lowering_ops_minus_mth}) for $j=0,\;1,\;\cdots,\; m-1$ has been applied to shorten the operator product, and \eq{Phi_lm_bottom_up_approach} has been used to reach the third equality.

According to \eq{eigenvalue_negative_level}, for the top association level $-m=-l$, $\lambda_{l,-l}=0$ and the  corresponding eigen-equation $\mpzc{h}^\dagger_{-l} \mpzc{h}_{-l}\Phi_{l,-l}=0$ can be interpreted as the result of the annihilation condition
\begin{eqnarray}
 \mpzc{h}_{-l}\Phi_{l,-l} = -\smbr{\sqrt{p}\od{}{x} - l\sqrt{p}'}\Phi_{l,-l}=0.
\end{eqnarray}
This equation has the solution
\begin{eqnarray}\label{Phi_l_minus_l}
 \Phi_{l,-l} = \sqbr{(-1)^l \Phi^{(l)}_l \displaystyle \prod^l_{j=1} \lambda^{-\frac{1}{2}}_{l,-j+1} }\sqrt{p}^l,
\end{eqnarray}
where the prefactor inside the square brackets is chosen to be consistent with the version of $\Phi_{l,-l}$ that follows from \eq{generic_Phi_l_minusm} in the case $m=l$. 

On the other hand,  making the substitution for $\mpzc{h}_{-m} \Phi_{l,-m}$ in \eq{eigen_equ_level_negative_m_after_rnmlz} through the $m$ version of the lowering recurrence relation \eqnm{lowering_recursion_relation_minus_mth} gives rise to the raising recurrence relation
\begin{eqnarray}\label{raising_recursion_relation_negative_association_level}
 \Phi_{l,-m} = \frac{\mpzc{h}^\dagger_{-m}}{\lambda^{\frac{1}{2}}_{l,-m} }   \Phi_{l,-(m+1)}.
\end{eqnarray}

Thus, starting with the solution \eqnm{Phi_l_minus_l} of top association level, all $\Phi_{l,-m}$ ($0\le m<l$) can be obtained by repeatedly applying \eq{raising_recursion_relation_negative_association_level} (with $m\rightarrow j$) for $j=l-1,\; l-2,\;\cdots, \; m$, i.e.
\begin{eqnarray}\label{Phi_l_minusm_top_down}
\Phi_{l,-m} &=& \frac{\mpzc{h}^\dagger_{-m}}{\lambda^{\frac{1}{2}}_{l,-m} } \frac{\mpzc{h}^\dagger_{-(m+1)}}{\lambda^{\frac{1}{2}}_{l,-(m+1)} } \cdots \frac{\mpzc{h}^\dagger_{-(l-2)}}{\lambda^{\frac{1}{2}}_{l,-(l-2)} } \frac{\mpzc{h}^\dagger_{-(l-1)}}{\lambda^{\frac{1}{2}}_{l,-(l-1)} }   \Phi_{l,-l}, \nonumber \\
            &=& \smbr{\Phi^{(l)}_l\displaystyle \prod^{l-1}_{j=m} \lambda^{-\frac{1}{2}}_{l,-j}} \frac{1}{\sqrt{p}^{m} w} \smbr{\od{}{x}}^{l-m} \smbr{-p}^l w, \nonumber \\
\end{eqnarray}
where $\Phi_{l,-l}$  has been substituted for via \eq{Phi_l_minus_l}, and the operator product has been shortened by making use of the lumped form of $\mpzc{h}^\dagger_{-j}$ given in the second equality in \eq{raising_ops_minus_mth}. \eq{Phi_l_minusm_top_down} is the analog of \eq{Phi_lm_top_down_approach}, which provides the top-down approach for calculating all the associated hypergeometric-like functions with negative association level as well as the base level.

\subsection{Passing to the Standard {\it z}-coordinate Representation of the SUSYQM Algorithm for Generic Association Level}\label{associated_standard_reps}
One can construct the associated hypergeometric-like differential equations of all higher association levels and their solutions \emph{directly in $z$-coordinate} representation, based on the background knowledge prepared in Subsection \ref{2nd_type_momentum_op}. and following the similar procedure illustrated in Subsection \ref{positive_association_levels}. This algorithm is almost the same as that has been demonstrated in Subsection \ref{positive_association_levels} or \ref{negative_association_levels} for the $x$-coordinate representation. For instance, in order to build the associated hypergeometric-like equations and their eigen-solutions of all positive association levels, one starts with \eq{standardized_factorization_base_level} -- the supersymmetrically factorized eigen-equation of association level $m=0$, and introduces the ladder operators of association level $m=1$ in the form
\begin{equation}\label{raising_op_association_level_1_z_coord}
  \begin{aligned}
  \mathsf{h}^\dagger_1 (z) &\equiv \mathsf{h}^\dagger_0(z)-\sqrt{p}'|_{x=x(z)},\\
  \mathsf{h}_1 (z) &\equiv \mathsf{h}_0(z)-\sqrt{p}'|_{x=x(z)},
  \end{aligned}
\end{equation}
and the pattern invariance condition
\begin{eqnarray}\label{quasi_commutator_positive_first_z_coord}
\Delta^+_1=\mathsf{h}^\dagger_0(z)\mathsf{h}_0(z)- \mathsf{h}_1(z) \mathsf{h}^\dagger_1(z) =- q'.
\end{eqnarray}
The condition \eqnm{quasi_commutator_positive_first_z_coord} then allows us to renormalize the superpartner eigen-equation 
\begin{eqnarray}\label{superparnter_eigen_equ_base_level}
\mathsf{h}^\dagger_0(z)\mathsf{h}_0(z)  \mathsf{\Phi}_{l1}(z) = \lambda_l \mathsf{\Phi}_{l1}(z)
\end{eqnarray}
of \eq{Schrodinger_equ_base_level} into the following one
\begin{eqnarray}
 \mathsf{h}_1(z) \mathsf{h}^\dagger_1(z)  \mathsf{\Phi}_{l1}(z) = \lambda_{l1} \mathsf{\Phi}_{l1}(z),
\end{eqnarray}
where $\mathsf{\Phi}_{l1}(z) = \smbr{\mathsf{h}^\dagger_0(z)/\sqrt{\lambda_l}}\mathsf{\Phi}_l(z)$ and $\lambda_{l1}=\lambda_l-\Delta^+_1$.
% for generating the eigen-equation of the first positive association level ($m=1$) and its solution.
One can keep going this way to construct the supersymmetrically factorized associated hypergeometric-like equation and its eigen-solution of generic association level $m>1$.

Although the above suggested procedure is quite straightforward, we will not follow it here. Instead, we will make use of the isomorphism between the $x$ and $z$-coordinate representations of the \emph{same} SUSYQM algorithm, some instances of this isomorphism, established by the supersymmetrization ${\cal S}$ defined in \eq{S_transformation} and the inverse ${\cal S}^{-1}$ in \eq{S_inverse_transformation}, and the momentum operator map \eqnm{momentum_map_type2}, have already been revealed at the initial level ($m=0$) in Subsection \ref{2nd_type_momentum_op}. We will see that these agents also establish the isomorphism between the two coordinate representations \emph{at all higher association levels}. This implies that, we can obtain all main results, such as the supersymmetrically factorized eigen-equations, eigen-solutions, the recurrence relations and the pattern invariance conditions in the $z$-coordinate representation, by simply applying the supersymmetrization transformation $S$ to those obtained already in the $x$-coordinate representation, that have already been achieved in in Subsections \ref{positive_association_levels} and \ref{negative_association_levels}. For instance, it is quite obvious that, the two operators in \eq{raising_op_association_level_1_z_coord} can be obtained by applying $S$ to those two in Eqs. \eqnm{raising_op_association_level_1} and \eqnm{lowering_op_association_level_1}, and the condition \eqnm{quasi_commutator_positive_first_z_coord} is reached by applying $S$ to the condition \eqnm{quasi_commutator_positive_first}.

Similarly, in view of Eqs. \eqnm{down_shift_S}, \eqnm{S_inverse_transformed_h} and \eqnm{S_transformed_h}, applying $S$ to $\mfk{h}_m(x)$ in \eq{lowering_ops_mth} and $\mfk{h}^\dagger_m(x)$ in \eq{raising_ops_mth} yields respectively
\begin{equation}\label{standard_h_m}
\begin{aligned}
  \mathsf{h}_m(z) &\equiv {\cal S}\smbr{\mfk{h}_m(x)}, \\
                  &= \mfk{h}_m(x) - W^a_0(x),\\
                  &= -\sqrt{p}\od{}{x} + W^a_m(x),\\
                  &= -\od{}{z}+ \mathsf{W}^a_m(z), \\ 
\end{aligned}
\end{equation}
\begin{equation}\label{standard_h_m_dagger}
\begin{aligned}
  \mathsf{h}^\dagger_m &\equiv {\cal S}\smbr{\mfk{h}^\dagger_m(x)},\\
                      &= \mfk{h}^\dagger_m(x) + W^a_0(x),\\
                      &= -\sqrt{p}\od{}{x} + W^a_m(x), \\
                      &=\od{}{z}+ \mathsf{W}^a_m(z), 
\end{aligned}
\end{equation}
where
\begin{eqnarray}\label{Wam_xcoord}
 W^a_m(x)&=& W^a_0(x) -m \sqrt{p}',\\
        &=&-\frac{1}{2\sqrt{p}}\sqbr{\smbr{m-\frac{1}{2}}p'+q} \nonumber 
\end{eqnarray}
is the superpotential of the supersymmetric factorization $H^a_m(x)=\mfk{h}_m(x)\mfk{h}^\dagger_m(x)$ (\eq{H_plus_m}) in $x$-coordinate, and $\mathsf{W}^a_m(z)\equiv W^a_m(x(z))$ is the counterpart in $z$-coordinate. The second line in \eq{standard_h_m} and that in  \eq{standard_h_m_dagger} are the analogs in \eq{xy_connection}.

In the second half of Subsection \ref{2nd_type_momentum_op}, we have demonstrated that, for the starting level ($m=0$), multiplying  \eq{eigen_equ_level_m_after_rnmlz} with the rescaling factor $\smbr{\sqrt{p} w}^{1/2}$ (the left factor of $\cal S$) and applying the supersymmetrization transformation ${\cal S}$ turn \eq{eigen_equ}, with $H_0$ being the asymmetric factorization \eqnm{H0_Hermitian_factorization} in $x$-coordinate, into the transparently supersymmetric factorization \eqnm{Schrodinger_equ_base_level} in $z$-coordinate. Now, for generic association level $m$, in view of the last equation in \eqnm{standard_h_m} and that in \eqnm{standard_h_m_dagger}, the same operations would supersymmetrize the asymmetric factorization \eqnm{eigen_equ_level_m_after_rnmlz} in $x$-coordinate into the following explicit supersymmetric factorization in $z$-coordinate
\begin{eqnarray}\label{S_equ_standard_association_m}
 \mathsf{H}^a_m(z) \mathsf{\Phi}_{lm}(z) &=&\mathsf{h}_m(z) \mathsf{h}^\dagger_m(z) \mathsf{\Phi}_{lm}(z) = \lambda_{lm} \mathsf{\Phi}_{lm}(z), \nonumber \\  
\end{eqnarray}
where
\begin{eqnarray}
\mathsf{H}^a_m(z)&\equiv& \left.{\cal S}\smbr{ H^a_m(x)}\right|_{x=x(z)}, \nonumber \\
                 &=& \sqbr{-\sqrt{p}\od{}{x} + W^a_m(x)} \sqbr{\sqrt{p}\od{}{x} + W^a_m(x)}, \nonumber \\ 
                 &=&\mathsf{h}_m(z) \mathsf{h}^\dagger_m(z),\label{H_m_factorized} \\
 \mathsf{\Phi}_{lm}(z)&\equiv& \left. p^{\frac{1}{4}}\smbr{x} w^{\frac{1}{2}}\smbr{x} \Phi_{lm}\smbr{x}\right|_{x=x(z)}. \label{rescaling_association} 
\end{eqnarray}

The two operator relations in the second lines of \eqnm{standard_h_m} and \eqnm{standard_h_m_dagger} show that, in order to quickly get the $z$-coordinate superpotential $\mathsf{W}^a_m(z)\equiv W^a_m(x(z))$ from the $x$-coordinate factorization $H^a_m(x)=\mfk{h}_m(x)\mfk{h}^\dagger_m(x)$, $W^a_0(x)$ has to be subtracted from the first factor $\mfk{h}_m(x)$ and to be added to the second factor $\mfk{h}^\dagger_m(x)$ to obtain the \emph{common} function $W^a_m(x)$ appearing in the third lines of Eqs. \eqnm{standard_h_m} and \eqnm{standard_h_m_dagger}. These two operations account for the effects of the supersymmetrization transformation $\cal S$ on $\mfk{h}_0(x)$ and $\mfk{h}^\dagger_0(x)$, i.e. shifting them respectively by $\mp W^a_0(x)$.

Applying the transformation ${\cal S}$ to condition \eqnm{SI_associated_ascending} yields the pattern invariance condition in $z$-coordinate 
\begin{eqnarray}\label{SI_associated_ascending_z_coord}
\Delta^+_n \equiv \mathsf{h}^\dagger_{n-1}(z) \mathsf{h}_{n-1}(z)- \mathsf{h}_n(z) \mathsf{h}^\dagger_n(z).
\end{eqnarray}

The above demonstrated isomorphism between the two representations of the same SUSYQM in $x$ and $z$-coordinates tells us that if the eigenfunction of Schr\"odinger equation \eq{S_equ_standard_association_m} can be expressed in terms of the associated hypergeometric-like functions $\Phi_{lm}\smbr{x}$ in the form \eqnm{rescaling_association}, its Hamiltonian $\mathsf{H}^a_m(z)$ always admits the supersymmetric factorization $\mathsf{H}^a_m(z)=\mathsf{h}_m(z) \mathsf{h}^\dagger_m(z)$ directly within $z$-coordinate representation. In many quantum mechanic problems one is faced with solving the bound-state solution $\mathsf{\Phi}_{lm}(z)$ to this Schr\"odinger equation for certain given analytical potentials $\mathsf{V}^a_m\smbr{z}$ that is related to through $\mathsf{W}^a_m(z)$
\begin{eqnarray}\label{V_a_m_potential}
  \mathsf{V}^a_m\smbr{z} &=&-\od{\mathsf{W}^a_m(z)}{z} + \sqbr{\mathsf{W}^a_m(z)}^2, \\
                         &=& \frac{k_m(x)}{4 p(x)}\sqbr{k_m(x)-p'(x)}+\frac{1}{2}k'_m(x) \nonumber 
\end{eqnarray}
for $k_m(x)\equiv \smbr{m-1/2} p'(x)+q(x)$. 
If the superpotential $\mathsf{W}^a_m(z)$ (or $\mathsf{W}^a_0(z)$) is easy to be identified through solving this equation, then the solution can be obtained quickly {\it within} this $z$-coordinate representation via the standard SUSYQM algorithm.

However, for many types of potentials, the correspondences between their \emph{raw} Schr\"odinger equations and  certain types of associated hypergeometric-like differential equations, if any, are not so straightforwardly established by such simple \emph{single} supersymmetrization transformation like ${\cal S}$. As in the case of converting the raw Schr\"odinger equations to certain principal hypergeometric-like differential equations, one may have to perform one or all of the operations such as (partial) supersymmetrization, inverse momentum map, and other algebraic manipulations at different stages of the conversion of a Schr\"odinger equation into  a certain associated hypergeometric-like differential equation. This issue has to be studied case by case.

\subsection{An Equivalent Eigen-equation of the Associated Hypergeometric-like Differential Equation}\label{equivalent_associated_eigen_equ}
Parallel to what we have achieved in Subsection \ref{equivalent_eigen_equ}, we can derive some eigen-equation that is equivalent to the standard associated hypergeometic-like differential equation \eqnm{eigen_equ_level_m_after_rnmlz} by a certain partial asymmetrization transformation.

Making use of the defintion for $\mfk{h}_0(x)$ in \eq{mfkh_00} and that for $\mfk{h}^\dagger_0(x)$ in \eq{mfkh0_dagger}, Eqs. \eqnm{lowering_ops_mth} and \eqnm{raising_ops_mth} can be respectively rewritten as
\begin{equation}\label{mfkhm_x}
\begin{aligned}
 \mfk{h}_m(x) &= \mfk{h}_0(x) + \Delta W^a_m(x) , \\
 \mfk{h}^\dagger_m(x) &= \mfk{h}^\dagger_0(x) + \Delta W^a_m(x),
\end{aligned}
\end{equation}
where $\Delta W^a_m(x) \equiv W^a_m(x) - W^a_0(x) = -m\sqrt{p}'$  and  $W^a_m(x)$
is the superpotential defined in \eq{Wam_xcoord} for the supersymmetric factorization $H_m \equiv \mfk{h}_m(x)\mfk{h}^\dagger_m(x)$ (\eq{H_plus_m}) 
The two equations in \eqnm{mfkhm_x} are the analogs of Eqs. \eqnm{Bl_B0_relation1} and \eqnm{Al_A0_relation1}.

Either by \emph{partially lumping} the common piece $\Delta W^a_m $ inside $\mfk{h}_m$  and inside $\mfk{h}^\dagger_m$ respectively into the bilateral transformations
\begin{eqnarray}
 {\cal S}^{-1}_\Delta\smbr{-} \equiv \sqrt{p}^{-m} \smbr{-} \sqrt{p}^m
\end{eqnarray}
and 
\begin{eqnarray}
 {\cal S}_\Delta\smbr{-}\equiv  \sqrt{p}^m \smbr{-} \sqrt{p}^{-m},
\end{eqnarray}
or by directly making use of Eqs. \eqnm{h_j_lumped} and \eqnm{h_j_dagger_lumped}, the two equations in \eqnm{mfkhm_x} are rewritten as
\begin{eqnarray}
 \mfk{h}_m ={\cal S}^{-1}_\Delta\smbr{\mfk{h}_0},\quad \mfk{h}^\dagger_m = {\cal S}_\Delta\smbr{\mfk{h}^\dagger_0}.
\end{eqnarray}
Consequently,
\begin{eqnarray}\label{asymmtrization_associated_hypergeometric_like_operator}
 {\cal S}^{-1}_\Delta \smbr{\mfk{h}_m \mfk{h}^\dagger_m }&=&\mfk{h}_{2m}\mfk{h}^\dagger_0= \smbr{\mfk{h}_0 - 2m \sqrt{p}'}\mfk{h}^\dagger_0,
\end{eqnarray}
where $\mfk{h}_{2m}\equiv \sqrt{p}^{-2m} \mfk{h}_0 \sqrt{p}^{2m}$. That is, ${\cal S}^{-1}_\Delta$  is a \emph{partial asymmetrization} transformation of the factorization $\mfk{h}_m \mfk{h}^\dagger_m$, in that effectively it splits the supersymmetric factorization $\mfk{h}_m \mfk{h}^\dagger_m$ into the asymmetric first-order differential term $-2m \sqrt{p}' \mfk{h}^\dagger_0$ plus the remaining supersymmetric factorization $\mfk{h}_0\mfk{h}^\dagger_0=H^a_0$. Its inverse ${\cal S}_\Delta$ is a \emph{partial supersymmetrization} transformation, that turns the partially supersymmetic factorization $\mfk{h}_{2m}\mfk{h}^\dagger_0$ into the full supersymmetric factorization $H^a_m=\mfk{h}_m\mfk{h}^\dagger_m$ (\eq{H_plus_m}), by evenly distributing $- 2 m \sqrt{p}'=2\Delta W^a_m(x)$ inside the first factor in the factorization \eqnm{asymmtrization_associated_hypergeometric_like_operator} over the two factors. 

Accordingly, the supersymmetrically factorized \emph{associated} hypergeometric-like differential  equation \eqnm{eigen_equ_level_m_after_rnmlz} can be \emph{partially asymmetrized} into the following equivalent one
\begin{eqnarray}\label{asymmetrized_associated_equ}
 \mfk{h}_{2m}\mfk{h}^\dagger_0 \varphi_{lm} =\lambda_{lm} \varphi_{lm},
\end{eqnarray}
simply by multiplying \eq{eigen_equ_level_m_after_rnmlz} with $\sqrt{p}^{-m}$ and making use of \eq{asymmtrization_associated_hypergeometric_like_operator}. Here 
\begin{eqnarray}\label{asymmetrized_Phi_lm}
 \varphi_{lm}\equiv \sqrt{p}^{-m} \Phi_{lm} \propto \smbr{\od{}{x}}^m \Phi_l,
\end{eqnarray}
where the proportionality has been reached by making use of the compact expression \eqnm{Phi_lm_bottom_up_approach} for $\Phi_{lm}$. This result shows that \eq{asymmetrized_associated_equ} can be derived from the \emph{principal} hypergeometric-like differential equation  \eqnm{eigen_equ} for $\Phi_l$, \emph{merely} by performing $m$-fold $x$-differentiations -- the first operation conducted in the traditional differential way of deriving the \emph{associated} hypergeometric-like differential equation of $m$-th association level from \eq{eigen_equ}, but \emph{without} performing the second and third operations -- multiplying with $\sqrt{p}^m$ and moving it next to $\smbr{d/dx}^m$. 

Making use of the definition \eqnm{mfkh_00} for $\mfk{h}_0$ and \eqnm{mfkh0_dagger} for $\mfk{h}^\dagger_0$, the operator product in \eq{asymmtrization_associated_hypergeometric_like_operator} can be expanded as
\begin{eqnarray}\label{h2m_h0dagger}
 \mfk{h}_{2m}\mfk{h}^\dagger_0 
 &=&-p\frac{d^2}{dx^2} -\smbr{q + m\,p'}\od{}{x}, 
\end{eqnarray}
which is surprisingly a \emph{principal} hypergeometric-type differential operator that is similar to $H_0$ and can be directly obtained from $H_0$ via the substitution $q \rightarrow q_m \equiv q + m\, p'$. Or, this $\mfk{h}_{2m}\mfk{h}^\dagger_0 $ is a nontrivial generalization of $H_0=\mfk{h}_0\mfk{h}^\dagger_0$ in that the former reduces to the latter in the case $m=0$.  Accordingly, \eq{asymmetrized_associated_equ} -- a variant of the \emph{associated} hypergeometric-like \eq{eigen_equ_level_m_after_rnmlz} -- is of the \emph{same} type of \emph{principal} hypergeometric-like differential equation as the startup eign-equation \eqnm{eigen_equ_base_level} or \eqnm{eigen_equ}.
In other words, the traditional associated hypergeometric-like equation \eqnm{asymmetrized_associated_equ} and its relevant principal counterpart \eqnm{eigen_equ} are unified into the same type of principal hypergeometric-like equations \emph{with similar outlooks}.  Because of such similarity,  one can solve \eq{asymmetrized_associated_equ} by running the SUSYQM algorithm developed in Subsection \ref{non_standard_factorization} for solving the principal hypergeometric-like differential equation \eqnm{eigen_equ}, by regarding $q_m$ as $q$ to obtain both $\varphi_{lm}$ and $\lambda_{lm}$ and then apply the supersymmetrization  to reach the supersymmetrically factorized \eq{H_plus_m} and the eigen-function $\Phi_{lm}=\sqrt{p}^m \varphi_{lm}$. In contrast, the associated hypergeometric-like differential operator $H^a_m$ given by \eq{H_m_expansion} looks quite different from $H_0$ (except for $m=0$) due to the existence of the last several pieces, which eventually makes the function type of the eigen-functions of $H^a_m$ different from those of $H_0$, recalling that the former that we are interested in are usually not polynomials in $x$, as shown in \eq{Phi_lm_bottom_up_approach},  while the latter are polynomials in $x$. 

The operator $\mfk{h}_{2m}\mfk{h}^\dagger_0$ given in \eq{h2m_h0dagger} or \eq{H_plus_m} looks much simpler than $H^a_m$ in \eq{H_m_expansion}. This reminds one that, if one tries to solve some differential equation by converting it into the associated hypergeometic-like equation of $H^a_m$, he/she can target at converting it into the equivalent though simpler \eq{asymmetrized_associated_equ}. 

Now, beside the iterative SUSYQM algorithm demonstrated in Subsections \ref{positive_association_levels} and \ref{negative_association_levels}, another quick route of deriving the associated hypergeometric-like differential equation that comes in its supersymmetric factorized form from the base hypergeometric equation \eqnm{eigen_equ} surfaces out. Firstly, take the $m$-fold differentiation of \eq{eigen_equ} to reach the asymmetrically factorized \eq{asymmetrized_Phi_lm}. This is formally accomplished by replacing $q$ with $q+m\,p'$ in the operator $H_0$ to get $\mfk{h}_{2m}\mfk{h}^\dagger_0$, $\Phi_l$ with $\varphi_{lm}$ and $\lambda_l$ with $\lambda_{lm}$. Secondly, construct a supersymmetrization transformation to supersymmetrize \eq{asymmetrized_associated_equ} into \eq{S_equ_standard_association_m} \emph{in one step}.
With Eqs. \eqnm{mfkh_00} for $\mfk{h}_0$ and \eqnm{mfkh0_dagger} for $\mfk{h}^\dagger_0$, the second equality in \eq{asymmtrization_associated_hypergeometric_like_operator} can be rewritten as
\begin{eqnarray}\label{asymmtrization_associated_hypergeometric_like_operator1}
 \mfk{h}_{2m}\mfk{h}^\dagger_0 &=& \sqbr{-\sqrt{p}\od{}{x} + 2 W^a_m} \sqbr{\sqrt{p}\od{}{x}}, \\
                               &=& \smbr{\sqrt{w}\sqrt{p}^{m+\frac{1}{2}}}^{-2} \sqbr{-\sqrt{p}\od{}{x}} \smbr{\sqrt{w}\sqrt{p}^{m+\frac{1}{2}}}^2 \nonumber \\
                               &&\cdot \sqbr{\sqrt{p}\od{}{x}}.\nonumber 
\end{eqnarray}
This reminds us to introduce the overall supersymmetrization transformation
\begin{eqnarray}
 {\cal S}_m (-) \equiv \smbr{\sqrt{w}\sqrt{p}^{m+\frac{1}{2}}}^{-1}(-)\smbr{\sqrt{w}\sqrt{p}^{m+\frac{1}{2}}},
\end{eqnarray}
and use it to accomplish the following supersymmetrization
\begin{eqnarray}\label{supersymmtrization_h2m-h0dagger}
 {\cal S}_m \smbr{\mfk{h}_{2m}\mfk{h}^\dagger_0} &=& \sqbr{-\sqrt{p}\od{}{x} + W^a_m}\sqbr{\sqrt{p}\od{}{x} + W^a_m}, \nonumber \\
                                                 &=& \mathsf{H}^a_m(z).
\end{eqnarray}
 Accordingly, ${\cal S}_m$ \emph{completely} supersymmetrizes \eq{asymmetrized_associated_equ} into \eq{S_equ_standard_association_m}.

Obviously, ${\cal S}_m$ reduces to ${\cal S}$  given in \eq{S_transformation} for $m=0$, and, as a composite supersymmetrization, ${\cal S}_m\smbr{-}={\cal S}^{-1}_\Delta\smbr{{\cal S}\smbr{-}}$.

\subsection{Another Type of Factorization of the Associated Hypergeometric-like Differential Equation}\label{another_factorization_associated_eigen_equ}
From the perspective of factorization, what we have achieved in Subsection \ref{positive_association_levels} amounts to providing a light way of reaching the SUSYQM factorization \eqnm{H_plus_m} of the associated hypergeometric-like differential operator $H^a_m$ in its expanded form \eqnm{H_m_expansion} with $m$ a positive integer, initiated by preparing the kinetic energy operator $-\smbr{\sqrt{p}d/dx}^2$, as demonstrated in Subsection \ref{2nd_type_momentum_op}. It is worthwhile to point out that,  the factorization of this $H^a_m$ is {\it not} unique and we will see that $H^a_m$ also admits another type of supersymmetric factorization. The argument for this can go as follows. In view of the specific structure of the last two rational fractions in $H^a_m$ in \eq{H_m_expansion}, multiplying $p$ would turn the two $m$-dependent pieces into a binomial in $x$, i.e.
\begin{eqnarray}\label{pHm_expanded}
 p H^a_m(x)= p H_0(x) +  \frac{m}{2}\sqbr{ p''p +\smbr{q-p'}p'}+ \smbr{\frac{m}{2}}^2 p'^2. \nonumber \\
\end{eqnarray}
This $p H^a_m(x)$ can be directly motivated by preparing the kinetic energy operator $-\smbr{p d/dx}^2$ from $H^a_m(x)$ in \eq{H_m_expansion}, just as we have constructed $p H_0(x)$ from $H_0(x)$ in \ref{base_kinetic_energy_operator}, which is the special $m=0$ case of $H^a_m(x)$. Therefore,  applying the same active supersymmetrization ${\cal T}\smbr{-}$ to $p H^a_m(x)$ would partially supersymmetrize it through completely supersymmetrizing $p H_0(x)$. This reminds us to pursue a supersymmetic factorization of the equivalent eigen-equation 
\begin{eqnarray}\label{pHm_equ}
\sqbr{p H^a_m(x) -\lambda_{lm}p + E_{lm}}\Phi_{lm}(x)=E_{lm} \Phi_{lm}(x)  
\end{eqnarray}
that is obtained from the associated hypergeometric-like  \eq{eigen_equ_level_m_after_rnmlz} for $H^a_m(x)$ by multiplying both sides with $p$ and adding the term proportional to the new eigenvalue $E_{lm}$ that is both $l$ and $m$-dependent, in the same way \eq{eigen_equ_with_multiplier} has been derived from \eq{eigen_equ}. Seeking for a supersymmetric factorization of the operator inside the square brackets in \eq{pHm_equ} is always feasible by the similar procedure demonstrated in Subsections \ref{Standard_Hermitian_factorization}-\ref{non_standard_factorization} for factorizing the operator $p H_0(x)-\lambda_l p + E_l$ in \eq{eigen_equ_with_multiplier}, since the structures of these two operators are similar, in that each of them consists of $p H_0(x)$ and another overall binomial in $x$, and admits a certain supersymmetric factorization that has linear superpotential  in $x$. Notice that, by design, \eq{pHm_equ} is expected to reduce to \eq{eigen_equ_with_multiplier} in the case $m=0$. Thus, we expect to reach the factorization  
\begin{eqnarray}\label{pHm_generalized}
p H^a_m - \lambda_{lm} p + E_{lm}=\mathsf{B}^-_{lm} \mathsf{A}^-_{lm}
\end{eqnarray}
with 
\begin{eqnarray}\label{BAlm}
\mathsf{B}^-_{lm}=\mathsf{B}^-_l+ C^-_{lm},\quad \mathsf{A}^-_{lm}=\mathsf{A}^-_l + C^-_{lm}.
\end{eqnarray}
Here $\mathsf{A}^-_l$ and $\mathsf{B}^-_l$ have already been given in Eqs.\eqnm{raising_ops_NHF} and \eqnm{lowering_ops_NHF} as the factorization of the $m=0$ case of \eq{pHm_equ}, and $C^-_{lm}$ is a constant as yet to be determined that is both $l$ and $m$-dependent; in particular, $C^-_{lm}$ is required to be proportional to $m$, so that this factorization for generic $m$ returns to the $m=0$ case, i.e. $\mathsf{B}^-_{l0}=\mathsf{B}^-_l$ and $\mathsf{A}^-_{l0}=\mathsf{A}^-_l$. 

With $p H^a_m(x)$ given in \eq{pHm_expanded} and $\lambda_{lm}$ in \eq{lambda_lm0} and $p H_0-p\lambda^-_l =\mathsf{B}^-_l \mathsf{A}^-_l - E^-_l$ (\eq{eigen_equ_NHF}), we have
\begin{eqnarray}\label{pHm_singled_out}
 p H^a_m(x)-\lambda_{lm} p &=& \mathsf{B}^-_l \mathsf{A}^-_l-E^-_l +  \lambda^-_m p \\
 &&+ \frac{m}{2}\sqbr{ p''p +\smbr{q-p'}p'}+ \smbr{\frac{m}{2}}^2 p'^2. \nonumber
\end{eqnarray}

On the right hand side of \eq{pHm_generalized} substituting for $\mathsf{B}^-_{lm}$ and $\mathsf{A}^-_{lm}$ via \eq{BAlm},  and on the left hand side substituting for $p H^a_m(x)-\lambda_{lm} p$ via \eq{pHm_singled_out}, we obtain
\begin{eqnarray}\label{coeffs_balance}
 E_{lm}-E^-_l &=& 2 C^-_{lm} W^-_l(x) +\smbr{C^-_{lm}}^2 - \lambda^-_m p  \\
 &&- \frac{m}{2}\sqbr{ p''p +\smbr{q-p'}p'} - \smbr{\frac{m}{2}}^2 p'^2,\nonumber 
\end{eqnarray}
where $\mathsf{B}^-_l + \mathsf{A}^-_l=2 W^-_l(x)$, resulting from the combination of  Eqs. \eqnm{raising_ops_NHF} and \eqnm{lowering_ops_NHF}, has been made use of. On the right hand side of \eq{coeffs_balance}, further making the substitutions $W^-_l(x)=\alpha^-_l x+\beta^-_l$ (\eq{Wl_superpotential}), $p(x)=(p''/2)x^2+ p'_0 x+ p_0$, $q(x)=q' x +q_0$ and the expression of $\lambda^-_m$ in \eq{lambda_minu_eigenvalue}, it is easy to check that the overall coefficient of $x^2$, that collects those only from the last three terms on the right hand side, automatically vanishes. Balancing the coefficients of $x$ that come from all terms but the second one on the right hand side yields
\begin{eqnarray}
 C^-_{lm} = \frac{m}{4}\frac{p''q_0 - p'_0 q'}{\alpha^-_l} =\frac{m}{4} \frac{p'_0 q' - p''q_0}{c_{l-1}},  
\end{eqnarray}
where \eq{alpha_l_minus_recursion} has been made use of to reach the second equality. Balancing the coefficients of $x^0$ on both sides of \eq{coeffs_balance} gives us
\begin{eqnarray}
 E_{lm} &=&E^-_l + C^-_{lm} \smbr{C^-_{lm} + 2 \beta^-_l} \\
        &+& m\sqbr{q'+\frac{m-2}{2} p''}p_0 -\frac{m}{2}\sqbr{q_0+ \frac{m-2}{2}p'_0}p'_0.\nonumber 
\end{eqnarray}

In the case of negative association level, a highly similar factorization scheme for $p H^a_{-m}$ with $H^a_{-m}$ given in \eq{H_minus_m} can be detailed.

Here the factorization \eqnm{pHm_generalized} has been determined by the awkward {\it two-step} method in that the associated hypergeometric-like differential equation (the expanded form of \eqnm{eigen_equ_level_m_after_rnmlz}) has already been derived by the approach other than this factorization algorithm itself. An iterative SUSYQM algorithm of deriving the factorization $\mathsf{B}^-_{lm} \mathsf{A}^-_{lm}$ without $H^a_m$ or $\lambda_{lm}$ being given beforehand, by treating the base factorization \eqnm{eigen_equ_NHF} as the initial (the level $m=0$) factorization, is expected to be more natural and illuminative. However, we will not continue to address this issue since it is out of the main stream of this paper.

\subsection{Some Remarks}\label{some_remarks_associated_eigen_equ}
At this stage, it is worthwhile to point out some key differences, between our iterative SUSYQM algorithms of deriving the associated hypergeometric-like differential equation and its solution, presented in Subsections \ref{2nd_type_momentum_op}-\ref{associated_standard_reps}, and the factorization methods of this type of equation achieved by Jafarizadeh {\it et al} \cite{jafarizadeh_SUSY_and_diff_equs_1997} and by Cotfas \cite{Cotfas_SI_raising_lowering_hypergeometric_equations_2002}.

From the principal hypergeometric-like differential equation \eqnm{eigen_equ}, Jafarizadeh {\it et al} \cite{jafarizadeh_SUSY_and_diff_equs_1997}  and Cotfas \cite{Cotfas_SI_raising_lowering_hypergeometric_equations_2002} \emph{first} derived the associated hypergeometric-like differential equation for {\it positive} association level $m$ in the form ${\tilde H}^a_{lm}(x)\Phi_{lm}(x)=0$ and  the relation $\Phi_{lm}\propto \sqrt{p}^m (d/dx)^m \Phi_l$, with ${\tilde H}^a_{lm}(x)$ essentially being $H^a_m(x)-\lambda_{lm}$ in our notations, in the same traditional differential manner the associated Legendre differential equation is derived from the (principal) Legendre differential equation, which can be found in many middle-level texts on classical electrodynamics, quantum mechanics and methods of mathematical physics. Their ${\tilde H}^a_{lm}(x)$ take rather messy forms, as they come out as the reshuffle of the expanded form \eqnm{H_m_expansion} of $H^a_m(x)$  and $\lambda_{lm}$ in the form \eqnm{lambda_lm}. 
Jafarizadeh and Fakhri \cite{jafarizadeh_SUSY_and_diff_equs_1997} pursued a direct factorization of $H^a_{lm}(x)$ in the same way they factorized $H_0(x)$, without invoking the concept of superpotential or explicitly demonstrating the SISY of their factorization. Cotfas \cite{Cotfas_SI_raising_lowering_hypergeometric_equations_2002} found a factorization of the associated hypergeometric-like operator that amounts to $H^a_m(x)+\lambda^-_m$ in our notations. In his factorization, both the raising and lowering operators were derived from some two-term and \emph{three-term} recurrence relations, which were further derived from the relation $\Phi_{lm}\propto \sqrt{p}^m (d/dx)^m \Phi_l$ that is assumed to be already known. Also at the stage of factorizing $H^a_m+\lambda^-_m$, the superpotentials were not introduced. 
Cotfas did check the SISY of the ladder operators between two neighboring association levels \emph{after} the factorization was achieved, 

A common feature of these authors' work is that their factorization schemes were achieved in {\it clear-cut two separate stages}. Firstly, they derived the associated hypergeometric-like differential equations from the base hypergeometric-like counterpart by the traditional differential means, so that the explicit forms of the associated hypergeometric-like differential operator and the relation $\Phi_{lm}\propto \sqrt{p}^m (d/dx)^m \Phi_l$ were prepared before starting their factorizations. Without such inputs from this traditional derivation, their ensuing factorizations could not proceed. Secondly, they further pursued the {\it direct} factorizations of the derived associated hypergeometric-like differential operators of \emph{generic} association level that do not appear in such well-organized forms that SUSY and SISY are self-evident, making the task of factorizing them somewhat heavy.

In comparison,  the \emph{iterative} SUSYQM algorithm developed in this section does comes with the four key features, which we have promised in the introduction section and rephrase here in more concrete fashions.

Rather than directly factorizing the lengthy and messy associated hypegeometric-like differential operator of generic association level in one single step, we start with factorizing the shorter hence easier principal hypegeometric-like operator $H_0$ as the initial ($m=0$) association level, and approaches the factorization of the operator of generic association level \emph{in multiple steps} (iteratively). As shown in Subsection \ref{positive_association_levels} for positive association levels, the initial supersymmetric factorization \eqnm{eigen_equ_base_level}, which is actually asymmetrical in $x$-coordinate, has been trivially prepared from the base eigen-equation \eqnm{eigen_equ}, once one recognizes the simple fact that the kinetic energy operator $-(\sqrt{p}d/dx)^2$ and the superpotential $W_0^a(x)$ are hidden in $H_0(x)$. The SUSY of the factorization \eqnm{eigen_equ_base_level} becomes more manifest when going to $z$-coordinate through the active supersymmetrization transformation $\cal S$ \eqnm{S_transformation} and the momentum operator map \eqnm{momentum_map_type2}.  The superpartner eigen-equation of each {\it present-level} eigen-equation, which is also the pre-normally-ordered one of the next level, is reached by performing the \emph{routine intertwining operation} on the normally ordered eigen-equation of present association level, this way the SUSY of this superpartner eigen-equation is automatically ensured. The raising and lowering operators of the next level are carefully designed in such a fashion that together with those of the present level they form the generalized commutator relation, which is an incarnation of the SISY between the potential and its superpartner potential, hence the SISY from present level to the next one is then guaranteed. This commutator relation is then used to renormalize the superpartner eigen-equation of present level into that of the next level. In doing so, the SUSY and SISY indeed not only become self-evident  but also are augmented to the {\it governing principles} that guide us to march level by level for consecutively generating the eigen-equations in their supersymmetrically factorized forms and their eigen-solutions \emph{simultaneously}. Since the SUSY and SISY are already manifest inside this algorithm, there is need to do the final check of these two symmetries in the ultimately achieved factorization.

In this algorithm, the breakthrough step is the design of the raising operator $\mfk{h}^\dagger_1$ of the first association level in the form \eqnm{raising_op_association_level_1}, which is motivated by the raising role of $\mfk{h}^\dagger_0$, where the latter is a natural consequence of intertwining action -- a primary though trivial operation in SUSYQM. The same SUSYQM algorithm has also been constructed in Subsection \ref{negative_association_levels} for negative association levels.

We see that, rather than as the inputs that are out of some other traditional methods for constructing the algorithm of factorization, the concrete form of associated hypergeometric-like differential equation of generic association level, the relation $\Phi_{lm}\propto \sqrt{p}^m (d/dx)^m \Phi_l(x)$, some three-term recurrence relations and the generalized Rodriguez' formula all come out naturally as the \emph{outputs} or \emph{byproducts} of this iterative algorithm \emph{itself}. In other words, the construction of this algorithm does not receive any input from any traditional method. This  makes our algorithm  as logically independent one of solving the discrete eigenvalue problem of this type of associated hypergeometric-like operators.

We also see that the key algebraic operations involved in building our iterative algorithm, such as constructing the active supersymmetrization transformation ${\cal S}$, introducing of momentum operator map, designing the raising and lowering operators, 
intertwining action and renormalization,  are indeed quite natural and \emph{elementary}. In contrast, the traditional one big stride of passing from the base level hypergeometric-like eigen-equation \eqnm{eigen_equ} to the associated one of generic association level and pinning down its eigen-solution in the form \eqnm{Phi_lm_bottom_up_approach} require much higher mathematical intuition and wisdom.
Although the algebraic steps of our SUSYQM algorithm man not be  necessarily fewer than the traditional differential method, it seems to be fair to claim that, in terms of the algebraic and conceptual simplicities and the logical independence, our SUSYQM algorithm is a strong candidate for supplanting the traditional Frobenius method of solving the eigen-value problem of the associated hypergeometric-like eigen-equation.

\section{The Degenerate Cases}\label{degenerate_cases}
\subsection{The Coincidence of the Two Types of SUSYQM Algebras}\label{operator_algebra_dg}
A natural question that arises is under what condition the two types of hypergeometric-like functions coincide, or the associated functions of generic association levels collapse into that of the base level. The immediate intuitive answer is that this will happen when the two types of momentum operators $-i p d/dx$ and $-i\sqrt{p} d/dx$, or the two types of kinetic energy operator $-\smbr{p d/dx}^2$ and $-\smbr{\sqrt{p} d/dx}^2$, in $x$-coordinate, become identical, which then means that $p(x)=1$.  Under this condition, the three coordinates $x$, $y$ defined in \eq{y_x_coord_trans} and $z$ defined in \eq{z_coord_transformation} also become identical, and the two supersymmetrization transformations ${\cal T}$ in \eq{T_transformation} and $\cal S$ in \eq{S_transformation} coincide as well. $H_0(x)$ has the natural factorization
\begin{eqnarray}\label{H0_degenerate}
 H_0(x) = -\frac{d^2}{d x^2} -q(x)\od{}{x}=\sqbr{-\od{}{x}-q(x)}\sqbr{\od{}{x}}.
\end{eqnarray}
The actual degeneracy condition can be relaxed to  $p(x)=p_0$ for a non-vanishing positive constant $p_0$.  By performing the rescalings $p_0^{-1/2} x \rightarrow x$ (assuming $p_0>0$) and $p_0^{-1/2} q(x) \rightarrow q(x)$ in \eq{eigen_equ}, we see that $p(x)=p_0$ is equivalent to $p(x)=1$. More straightforwardly, $H_m(x)$ defined in \eq{H_m_expansion} reduces to $H_0(x)$ when $p(x)=p_0$.

Now, let us check in the special case $p(x)=1$ how the minus route factorization of the principal hypergeometric-like differential equation, demonstrated in Subsections \ref{minus_route_iteration} and \ref{non_standard_factorization}, reduces. 

The upper equation in \eqnm{W0_minus} reduces to  
\begin{eqnarray}\label{W0_reduced}
 W_0(x)=W^-_0(x)=-\frac{1}{2} q(x).
\end{eqnarray}
According to the upper equation in \eqnm{alpha_0_beta_0_minus} and \eq{coefficient_x^2},
\begin{eqnarray}
\alpha^-_l \longrightarrow\alpha^-_0=-\frac{q'}{2},
\end{eqnarray}
and to the lower equation in \eqnm{alpha_0_beta_0_minus} and \eqnm{coefficient_x},
\begin{eqnarray}
 \beta^-_l \longrightarrow \beta^-_0 =-\frac{q_0}{2},
\end{eqnarray}
hence the collapse of the superpotential goes like
\begin{eqnarray}
 W^-_l(x)\longrightarrow W^-_0(x).
\end{eqnarray}
Consequently, according to Eqs. \eqnm{raising_ops_NHF} and \eqnm{lowering_ops_NHF}, there come the two collapses 
\begin{equation}\label{BAl_collapse}
\begin{aligned}
 \mathsf{B}_l &\longrightarrow \mathsf{B}_0 =-\od{}{x}-q, \\
 \mathsf{A}_l &\longrightarrow \mathsf{A}_0 =\od{}{x}.
 \end{aligned}
\end{equation}
This factorization $H_0=\mathsf{B}_0\mathsf{A}_0$ is obviously the same as  the trivial factorization of $H_0$ in \eq{H0_degenerate}. \eq{coefficient_x^0} gives $\delta^-_l \rightarrow -q'$. In consequence,  $E^-_l \rightarrow -l q'$ by combining the uper equation in \eqnm{E0_minus} and the lower one in \eqnm{alpha_l_minus}. Furthermore, $\lambda^-_l \rightarrow -l q'$ according to \eq{lambda_minu_eigenvalue}. Thus, we see the collapse of the prefactorized operator ${\mathbb{H}}_l$ goes as
\begin{eqnarray}\label{degenerate_factorization_base}
{\mathbb{H}}_l=p H_0-\lambda^-_l p + E^-_l \longrightarrow H_0,
\end{eqnarray}
which is the same as the collapse of the factorization ${\mathbb{H}}_l=\mathsf{B}_l\mathsf{A}_l$ given in \eq{BAl_collapse}. The generalized commutator relation \eqnm{SIC_minus_NHF} now degenerates into the {\it genuine} commutator relation
\begin{eqnarray}\label{degenerated_commutator}
 \sqbr{\mathsf{A}_0,\, \mathsf{B}_0}=-q',
\end{eqnarray}
which is \emph{$l$-independent}.

Next let us check how the minus route factorization of the associated hypergeometric-like differential equation for positive association levels, demonstrated in Subsections \ref{2nd_type_momentum_op} and \ref{positive_association_levels}, collapses. 

$W_0^a(x)$ in \eq{superpotential_W0a} reduces to
\begin{eqnarray}
 W_0^a(x)=-\frac{1}{2}q = W^-_0(x).
\end{eqnarray}
Conseqently,  $\mfk{h}_0$ in \eq{mfkh_00} and $\mfk{h}^\dagger_0$ in \eq{mfkh0_dagger} reduce respectively to
\begin{equation}
\begin{aligned}
 \mfk{h}_0&=-\od{}{x}-q=\mathsf{B}_0, \\
 \mfk{h}^\dagger_0&=\od{}{x}=\mathsf{A}_0.
\end{aligned}
\end{equation}
According to \eq{Wam_xcoord}, $W_m^a(x)$ undergoes the collapse
\begin{eqnarray}
 W_m^a(x) \longrightarrow W^a_0(x)=W^-_0(x)
\end{eqnarray}
for all $0<m\le l$. Consequently, 
\begin{equation}
\begin{aligned}
\mfk{h}_m &\longrightarrow \mfk{h}_0, \\
\mfk{h}^\dagger_m &\longrightarrow \mfk{h}^\dagger_0,
\end{aligned}
\end{equation}
referring to Eqs. \eqnm{lowering_ops_mth} and \eqnm{raising_ops_mth}. That is, the factorization $H^a_m=\mfk{h}_m \mfk{h}^\dagger_m$ in \eq{eigen_equ_level_m_after_rnmlz} for all higher association levels 
$m$ collapses into the factorization $H_0=\mfk{h}_0\mfk{h}^\dagger_0$ of the departure level in \eq{H0_Hermitian_factorization}, which is also the degenerate factorization of the Hamiltonian ${\mathbb{H}}_l$ in \eq{degenerate_factorization_base}.

In \eq{Delta_plus_n}, $\Delta^+_n=-q'$. In consequence, the generalized commutator relation \eqnm{SI_associated_ascending} also collapse to the commutator relation \eqnm{degenerated_commutator}, which means that we use the \emph{same} commutator relation to conduct the renormalization for generating the  eigen-functions and eigenvalues of higher levels in the case of finding the principal eigen-solutions as well as in the case of finding the associated eigen-solutions by running the same iterative SUSYQM algorithm. Therefore, the algorithm demonstrated in \ref{positive_association_levels} for generating the {\it associated} hypergeometric-like eigen-equations of higher association levels and their eigen-solutions and the algorithm demonstrated in \ref{proliferation_minus_route} for generating the principal hypergeometric-like eigen-equations and their eigen-solutions, will result in the same eigen-equations and eigen-solutions of higher levels. In other words, the associated hypergeometric-like functions and the principal ones coincide.

Let us dive into some details of the collapses of the associated eigen-solutions. The eigen-equation \eqnm{eigen_equ_level_m_after_rnmlz} reduces to
\begin{eqnarray}\label{eigen_equ_collapsed}
 H_0 \Phi_{lm} =\lambda_{lm} \Phi_{lm}.
\end{eqnarray}
On the one hand, as a result of $q''=0$, the combination of Eqs. \eqnm{lambda_minu_eigenvalue} and \eqnm{lambda_lm0} gives the collapse of the associated eigenvalue in the fashion
\begin{eqnarray}\label{eigenvalue_coincidence}
 \lambda_{lm} = -\smbr{l-m}q'=\lambda_{l-m}.
\end{eqnarray}
On the other,  comparing $\Phi_{lm}$ in \eq{Phi_lm_top_down_approach} with $\Phi_l$ in \eq{Rodriguez_formula},  we see that the associated eigenfunction $\Phi_{lm}$ collapses in the following special way
\begin{eqnarray}\label{eigenfunction_coincidence}
 \Phi_{lm} = \Phi_{l-m}.
\end{eqnarray}
Thus, \eq{eigen_equ_collapsed}  has no more solutions than the base level eigen-equation $H_0 \Phi_l =\lambda^-_l \Phi_l$.
 
\subsection{The Eigen-solution via the SUSYQM Algorithm}\label{eigen_solution_dg}
In the case $p(x)=1$, although the principal and the associated hypergeometric-like differential operators collapse into the same one and their SUSYQM algebras collapse into the same one as well, there still exist a hierarchy of eigen-solutions for this reduced hypergeometric-type operator. In this case,  \eq{eigen_equ} degrades to 
\begin{eqnarray}\label{Coincidence_equ}
\sqbr{-\frac{d^2}{dx^2}-\smbr{q' x+q_0} \od{}{x}}\Phi_l= \lambda_l\Phi_l. 
\end{eqnarray}
Since this equation takes such a simple form, we do not need to strictly follow the steps of the SUSYQM algorithm demonstrated in Section \ref{susyqm_H_like_functions} or Subsection \ref{positive_association_levels} to get its solution. Instead, we can solve it directly in the spirit of the generic iterative SUSYQM algorithm.

For $q'<0$, by virtue of the linear coordinate transformation $x=\sqrt{-2/q'} \xi -q_0/q'$, the above equation is turned into the familiar Hermite equation
\begin{eqnarray}\label{Hermite_equ}
L\smbr{\xi}\varphi_l(\xi) =B(\xi)A\smbr{\xi} \varphi_l(\xi)= 2 {\varLambda}_l \varphi_l(\xi),
\end{eqnarray}
where 
\begin{eqnarray}\label{factorized_Hermite_operator}
L\smbr{\xi}&\equiv& -\frac{d^2}{d \xi^2}+ 2\xi \od{}{\xi}, \\
B(\xi) &\equiv& -\od{}{\xi} + 2\xi =- \frac{1}{w\smbr{\xi}} \od{}{\xi} w\smbr{\xi}, \label{A_B_operators} \\
A(\xi) &\equiv& \od{}{\xi},
\end{eqnarray}
$w(\xi) \equiv \exp\smbr{-\xi^2}$ being the weight function, $\varphi_l(\xi)\equiv \Phi_l\sqbr{x(\xi)}$ and ${\varLambda}_l\equiv - \lambda_l/q'$. The  factorization \eqnm{factorized_Hermite_operator} is just the self-adjoint splitting of Hermite differential operator $L\smbr{\xi}$, with respect to $w(\xi)$-weighted inner product. Both $A$ and $B$ are independent of the level index $l$ and there exists the commutator relation
 \begin{eqnarray}\label{Commutator_hermite}
 \sqbr{B, A}\equiv BA-AB=-2,
 \end{eqnarray}
which is just the $\xi$-coordinate representation of the commutator relation \eqnm{degenerated_commutator}.

For $l=0$, set $\varLambda_0=0$, \eq{Hermite_equ} reduces to 
\begin{eqnarray}\label{ground_level_dg}
B A\varphi_0=0, 
\end{eqnarray}
which holds when 
\begin{eqnarray}
A \varphi_0=0,
\end{eqnarray}
that is, the ground state eigenfunction can be taken as $\varphi_0=1$ and $A$ works as the lowering operator to annihilate it.  Making use of the relation \eqnm{Commutator_hermite} to move the raising operator $B$ to the right of $A$, \eq{ground_level_dg} is rewritten as 
\begin{eqnarray}
A B \varphi_0=2\varphi_0. 
\end{eqnarray}
Intertwiningly acting this equation with $B$ and defining $\varLambda_1=1$ and $\varphi_1\equiv \smbr{B/\sqrt{2\varLambda_1}}\varphi_0$, we obtain the renormalized eigen-equation 
\begin{eqnarray}\label{first_level_dg}
BA \varphi_1 =2 \varphi_1, 
\end{eqnarray}
which is just the $l=1$ case of \eq{Hermite_equ}. Making use of the relation \eqnm{Commutator_hermite} again, \eq{first_level_dg} is rewritten as 
\begin{eqnarray}
A B \varphi_1 =2\varLambda_2 \varphi_1.
\end{eqnarray}
Intertwiningly acting this equation with $B$ and defining $\varLambda_2=2$ and $\varphi_2\equiv \smbr{B/\sqrt{2\varLambda_2}} \varphi_1$, we obtain the renormalized eigen-equation
\begin{eqnarray}
B A \varphi_2 =2\varLambda_2 \varphi_2,  
\end{eqnarray}
which is the $l=2$ case of \eq{Hermite_equ}. Keeping running this SUSYQM algorithm to the $l$-th level ($l>2$), we obtain the eigenvalue $\varLambda_l\equiv l$ and the eigen-solution as the normalized Hermite polynomials
\begin{eqnarray}\label{varphi_l}
\varphi_l(\xi) &\equiv& \frac{B}{\sqrt{2 l}} \varphi_{l-1}(\xi)=\frac{B^l}{\sqrt{2}^l \sqrt{l!}} \varphi_0(\xi) \nonumber \\
&=&\frac{(-1)^l}{\sqrt{2^l l!} }  \exp\smbr{\xi^2} \smbr{\od{}{\xi}}^l \exp\smbr{-\xi^2},
\end{eqnarray}
Where $B^l$ is shortened by the lumped form of $B$ given in \eq{A_B_operators}. Here, once again, we see  that the generic Hermite equation and its polynomial solution are efficiently built {\it simultaneously} through running the iterative SUSYQM algorithm, starting with $B A\varphi_0=0$. Hence, it is fair to claim that this start-up factorization and the shape invariance condition \eqnm{Commutator_hermite} together completely determine the generic eigen-equation \eq{Hermite_equ} and its solution via this SUSYQM algorithm. 

The standard representation of the SUSYQM of this Hermite operator is readily obtained. Rewrite the self-adjoint factorization $L=BA$ as 
\begin{eqnarray}
L&=&\frac{1}{\sqrt{w}}\cubr{\sqbr{\frac{1}{\sqrt{w}}\smbr{- \od{}{\xi}}\sqrt{w}}\sqbr{\sqrt{w}\smbr{\od{}{\xi}}\frac{1}{\sqrt{w}}}}\sqrt{w}, \nonumber \\
 &=&{\cal T}^{-1}\cubr{\sqbr{{\cal T}^{-1}\smbr{- \od{}{\xi}}}\sqbr{{\cal T}\smbr{\od{}{\xi}}}},
\end{eqnarray}
where the active supersymmetrization transformation is defined as ${\cal T}\smbr{-} \equiv \sqrt{w}\smbr{-}  \sqrt{w}^{-1}$. Applying $\cal T$ to the above equation, the asymmetric factorization $L=BA$ is then supersymmetrized into 
\begin{eqnarray}
\widehat{L}\equiv {\cal T} \smbr{L(\xi)}= a^\dagger\smbr{\xi}a\smbr{\xi} 
\end{eqnarray}
with the ladder operators 
\begin{eqnarray}
a^\dagger\smbr{\xi} &\equiv& {\cal T} \smbr{B(\xi)}=-\od{}{\xi}+ W(\xi)={\cal T}^{-1}\smbr{- \od{}{\xi}},\nonumber \\ 
a\smbr{\xi} &\equiv& {\cal T}\smbr{A(\xi)}=\od{}{\xi}+ W(\xi)\nonumber
\end{eqnarray}
and the superpotential $W(\xi)\equiv \xi$. Accordingly, upon multiplying $\sqrt{w}=\exp\smbr{-\xi^2/2}$ and defining the normalized wave function $\widehat{\varphi_l}(\xi)\equiv \sqrt{w} \varphi_l(\xi)/\sqrt[4]{\pi}$, \eq{Hermite_equ} is turned into the supersymmetrically factorized Schr\"odinger equation 
\begin{eqnarray}
 \widehat{L} \widehat{\varphi_l}=a^\dagger\smbr{\xi} a\smbr{\xi}\widehat{\varphi_l}= 2 l\widehat{\varphi_l},
\end{eqnarray}
 for the familiar harmonic oscillator characterized by the shifted potential $V(\xi) \equiv -dW(\xi)/d\xi +W^2(\xi)=\xi^2-1$. Multiplying \eq{varphi_l} with $\sqrt{w}$ then yields
\begin{eqnarray}
 \widehat{\varphi_l}(\xi)&=& \smbr{a^\dagger}^l \widehat{\varphi}_0, \nonumber \\
                         &=& \frac{(-1)^l}{\sqrt{\sqrt{\pi} 2^l l!} }  \exp\smbr{\frac{\xi^2}{2}} \smbr{\od{}{\xi}}^l \exp\smbr{-\xi^2}\nonumber 
\end{eqnarray}
with $\widehat{\varphi}_0(\xi)\equiv \sqrt{w}/\sqrt[4]{\pi} = \exp\smbr{-\xi^2/2}/\sqrt[4]{\pi}$ being the ground-state wave function. This algebraic approach to the solution of quantum harmonic oscillator is well-known in quantum physics community.

For $q'>0$, utilizing the transformation $x=\sqrt{2/q'}\zeta-q_0/q'$, \eq{Coincidence_equ} is recast into the \emph{quasi}-Hermite equation
\begin{eqnarray}\label{Quasi_Hermite_equ}
 \sqbr{-\frac{d^2}{d \zeta^2} - 2\zeta \od{}{\zeta}} \tilde{\varphi}_l(\zeta) = D(\zeta) C(\zeta) \tilde{\varphi}_l(\zeta) =-2 \varLambda_l \tilde{\varphi}_l(\zeta),\nonumber \\
\end{eqnarray}
where $\tilde{\varphi}_l(\zeta)\equiv {\varphi}_l(x(\zeta))$, 
\begin{eqnarray}
D(\zeta) &\equiv& -\od{}{\zeta}-2 \zeta= - {\mathfrak w}^{-1}\smbr{\zeta} \od{}{\zeta} {\mathfrak w} \smbr{\zeta}, \nonumber \\
C(\zeta) &\equiv& \od{}{\zeta}, 
\end{eqnarray}
$\mathfrak{w}\smbr{\zeta}\equiv \exp\smbr{\zeta^2}$ and $\varLambda_l\equiv -\lambda_l/q'$. There exists the commutator relation 
\begin{eqnarray}\label{Commutator_quasi_hermite}
 \sqbr{D,\, C}\equiv D C- C D=2.
 \end{eqnarray}
For $l=0$, we can take $\tilde{\varphi}_0(\zeta)=1$ and $\varLambda_0=0$, which ensures that $C \tilde{\varphi}_0(\zeta)=0$ and $D\, C \tilde{\varphi}_0(\zeta)=0$. Similar to the case for $q'<0$, running the iterative SUSYQM algorithm will yield $\varLambda_l = l $ for any integer $l>1$ as well as the raising recurrence relation
\begin{eqnarray}
\tilde{\varphi}_l(\zeta)\equiv \frac{D(\zeta)}{\sqrt{-2l}} \tilde{\varphi}_{l-1}(\zeta).
\end{eqnarray}

Consequently, the  eigenfunctions are obtained as the quasi-Hermite polynomials
\begin{eqnarray}\label{Quasi_Hermite_polynomial}
 \tilde{\varphi}_l(\zeta)&=&\frac{D^l(\zeta)}{\sqrt{(-2)^l l!}} \tilde{\varphi}_0(\zeta), \nonumber \\
                        &=&\frac{i^l}{\sqrt{2^l l!}}\exp\smbr{-\zeta^2} \smbr{\od{}{\zeta}}^l \exp\smbr{\zeta^2},
\end{eqnarray}
where $i$ is the imaginary unit, and a specific normalization has been adopted that allows $\tilde{\varphi}_l(\zeta)$ to carry a pure imaginary phase factor for odd $l$. Formally, this solution can be obtained from \eq{varphi_l} by the substitution $\xi=i\zeta$. Indeed, under this transformation, \eq{Quasi_Hermite_equ} goes back to \eq{Hermite_equ}.

Multiplying with ${\mathfrak w}^{1/2}\smbr{\zeta}$, \eq{Quasi_Hermite_equ} is supersymmetrized into
\begin{eqnarray}\label{Quasi_Hermite_equ1}
 b^+ b \widetilde{\varphi}_l = -2 l \widetilde{\varphi}_l, 
\end{eqnarray}
where $b^+\equiv \sqrt{\mathfrak w}^{-1} D \sqrt{\mathfrak w}=-d/d\zeta + W(\zeta)$, $b \equiv \sqrt{\mathfrak w} C \sqrt{\mathfrak w}^{-1}=d/d\zeta + W\smbr{\zeta}$, $W\smbr{\zeta}\equiv -\zeta$ and $\widetilde{\varphi}_l\equiv \sqrt{\mathfrak w} \tilde{\varphi}_l$. \eq{Quasi_Hermite_polynomial} then yields
\begin{eqnarray}
\widetilde{\varphi}_l &=& \frac{\smbr{b^+}^l}{\sqrt{(-2)^l l!}} \widetilde{\varphi}_0(\zeta),\nonumber \\
                      &=&\frac{i^l}{\sqrt{2^l l!}}\exp\smbr{-\frac{\zeta^2}{2}}\smbr{\od{}{\zeta}}^l\exp\smbr{\zeta^2}, 
\end{eqnarray}
where $\widetilde{\varphi}_0(\zeta)=\exp\smbr{\zeta^2/2}$. Obviously, these $\widetilde{\varphi}_l$ ($l\ge 0$) are \emph{not normalizable} over the domain $\smbr{-\infty, \infty}$, so they can not work as bound-state wave functions.

\eq{Quasi_Hermite_equ1} can be rewritten as
\begin{eqnarray}
 \smbr{-\frac{1}{2}\frac{d^2}{d\zeta^2}+\frac{1}{2}\zeta^2}\widetilde{\varphi}_l=-\smbr{l+\frac{1}{2}}\widetilde{\varphi}_l,
\end{eqnarray}
which is the static Schr\"odinger equation for the harmonic oscillator with {\it negative} energy levels. Mathematically, the potential energy minus the total energy is \emph{positively} definite for any non-negative integer $l$, so the second derivative of $\widetilde{\varphi}_l$ never vanishes and $\widetilde{\varphi}_l$ does not have any turning point so that it keeps increasing with $\abs{\zeta}$. Graphically, the horizontal line of a negative total energy level never intersects with the curve of the non-negative hyperbola potential $\zeta^2/2$. Thus, this type of solutions diverge as $\zeta\rightarrow \pm \infty$. In fact, both the weight function ${\mathfrak w}(\zeta)=\exp\smbr{\zeta^2}$ and the ground-state eigen-function $\widetilde{\varphi}_0(\zeta)=\exp\smbr{\zeta^2/2}$ diverge as $\zeta\rightarrow \pm\infty$.  This explains why  the quantum harmonic oscillator does not have bound-state wave function with negative energy level, as it is well-known that the bound states come with positive energy levels.

For $q'(x)=0$, \eq{eigen_equ} further degrades to
\begin{eqnarray}
 \frac{d^2\Phi_l}{dx^2}+q_0 \od{\Phi_l}{x}=\lambda^-_l \Phi_l.
\end{eqnarray}
If $q_0\ne 0$, this equation has the constant solution $\Phi_l(x)=c_1$ -- the simplest “polynomial". Note that the other solution $\propto e^{-q_0 x}$ is not a polynomial. If $q_0=0$, this equation has the monomial solution $\Phi=c_2+c_3x$ for constants $c_2$ and $c_3$. In both cases, the eigenvalue $\lambda^-_l=0$, which is consistent with the formula \eqnm{lambda_minu_eigenvalue} for $\lambda^-_l$ in the case $ p(x)=1$ and $q'(x)=0$.

It has been noticed for a while that the \emph{associated} Hermite differential equation is the same as the Hermite differential equation \emph{itself} (e.g., see \cite{jafarizadeh_SUSY_and_diff_equs_1997}), which is encoded here in the fact that all three shape invariance conditions \eqnm{SIC_minus_NHF},  \eqnm{SI_operator_product_minus} and \eqnm{SI_associated_ascending} for the SUSYQM of the generic hypergeometric-like operator $H_0$ collapse into the commutator relation \eqnm{degenerated_commutator} or \eqnm{Commutator_hermite}. However, along the line of the reasoning provided by us in this section, among all the hypergeometric-like operators the Hermite and quasi-Hermite operators (see \eq{Quasi_Hermite_equ}) are the {\it only nontrivial hypergeometric-like operators} that have such degenerate SUSYQM algebras, so that their associated hypergeometric-like functions coincide with their principal counterparts.

\section{Summary and Conclusions}\label{summary}
Systematic iterative SUSYQM algorithms for solving the principal  as well as the associated hypergeometric-like differential equations have been developed in this paper. These algorithms are based on the SUSY and SISY themselves, completely independent of the traditional methods or the results of the latter.  This type of algorithms consist of four key operations: 1) preparing the supersymmetric factorization of the initial-level eigen-equation; 2) obtaining the superpartner eigen-equation by taking the intertwining action on the present-level eigen-equation; 3) designing appropriate generalized commutator relation between the ladder operators of present level and those of next neighboring level, which ar equivalent to the shape invariance symmetry owned by a potential and its superparnter potential; 4) renormalizing the superpartner eigen-equation of present level into the normally ordered one of the next neighboring level, by making use of the generalized commutator relation designed or taking the intertwining action.  The operations 2)-4) may need to be performed repeatedly for many turns to complete this type of iterations.

\begin{table*}[t]
\begin{tabularx}{0.986\textwidth}{|p{0.17\textwidth}|p{0.13\textwidth}|p{0.21\textwidth}||p{0.09\textwidth}|p{0.12\textwidth}|p{0.09\textwidth}|p{0.11\textwidth}|}
 \multicolumn{7}{c}{} \\
 \hline
 \scalebox{0.8}{initial} \scalebox{0.8}{asymmetric} \scalebox{0.8}{factorization}& \scalebox{0.8}{supersymetrization} \scalebox{0.8}{transformation} &\scalebox{0.8}{initial} \scalebox{0.8}{manifestly} \scalebox{0.8}{supersymetric} \scalebox{0.8}{factorization}& \scalebox{0.8}{momentum} \scalebox{0.8}{operator} \scalebox{0.8}{map} &\scalebox{0.8}{kinetic} \scalebox{0.8}{energy} \scalebox{0.8}{operator}  \scalebox{0.8}{map}&\scalebox{0.8}{superpotential}& \scalebox{0.8}{rescaling} \scalebox{0.8}{factor}\\
 \hline
 \scalebox{0.8}{$p H_0(x)$} \scalebox{0.8}{$=\sqbr{-p \od{}{x}+2 W_0}\sqbr{p \od{}{x}}$}   & \multirow{2}{4em}{\scalebox{0.8}{${\cal T}(-)=r_t\smbr{-} r_t^{-1}$}}    &\scalebox{0.8}{${\cal T}\smbr{p H_0(x)}$} \scalebox{0.8}{$=\sqbr{-p \od{}{x}+ W_0}\sqbr{p \od{}{x}+ W_0}$} &\multirow{2}{4em}{ \scalebox{0.8}{$p\, \od{}{x} \equiv \od{}{y}$} } & \multirow{2}{4em}{\scalebox{0.7}{$-\smbr{p\,\frac{d}{dx}}^2=-\frac{d^2}{dy^2}$}} & \multirow{2}{4em}{\scalebox{0.7}{$W_0(x)$}} & \multirow{2}{4em}{\scalebox{0.7}{$r_t\equiv\sqrt{w}$}}\\
   & & & & & & \\
 \scalebox{0.8}{$H_0(x)$} \scalebox{0.8}{$= \sqbr{-\sqrt{p}\od{}{x}+2 W^a_0} \sqbr{\sqrt{p}\od{}{x}}$} & \multirow{2}{4em}{\scalebox{0.8}{ ${\cal S}(-)=r_s\smbr{-}r^{-1}_s$} } & \scalebox{0.8}{${\cal S}\smbr{H_0(x)}$} \scalebox{0.8}{$= \sqbr{-\sqrt{p}\od{}{x}+ W^a_0} \sqbr{\sqrt{p}\od{}{x}+ W^a_0}$} & \multirow{2}{4em}{\scalebox{0.8}{$ \sqrt{p}\, \od{}{x}\equiv \od{}{z}$}} & \multirow{2}{4em}{\scalebox{0.8}{$- \smbr{\sqrt{p}\, \od{}{x}}^2=-\frac{d^2}{d z^2}$} }& \scalebox{0.8}{$W_0^a(x)$} & \scalebox{0.8}{$r_s\equiv\smbr{w\sqrt{p}}^{\frac{1}{2}}$} \\
 \hline 
\end{tabularx}
\setcounter{table}{0}
\caption{The roles of two active supersymmetrization transformations $\cal T$ and $\cal S$ on the two initial asymmetric factorizations of $ pH_0(x)$ and $H_0(x)$ respectively and the resultant maps of momentum and kinetic energy operators, superpotentials and eigen-function rescaling factors.} \label{two_initial_supersymmetrizations}
\end{table*}

\begin{table*}[t]
\begin{tabularx}{0.99\textwidth}{|p{0.175\textwidth}|p{0.13\textwidth}|p{0.21\textwidth}||p{0.09\textwidth}|p{0.12\textwidth}|p{0.09\textwidth}|p{0.11\textwidth}|}
 \multicolumn{7}{c}{} \\
 \hline
\scalebox{0.8}{asymmetric factorization} & \scalebox{0.8}{supersymetrization} \scalebox{0.8}{transformation} &\scalebox{0.8}{manifestly} \scalebox{0.8}{supersymetric} \scalebox{0.8}{factorization}& \scalebox{0.8}{momentum} \scalebox{0.8}{operator} \scalebox{0.8}{map} &\scalebox{0.8}{kinetic} \scalebox{0.8}{energy} \scalebox{0.8}{operator}  \scalebox{0.8}{map}&\scalebox{0.8}{superpotential}& \scalebox{0.8}{rescaling} \scalebox{0.8}{factor}\\
 \hline
 \scalebox{0.8}{$p H_l(x)$} \scalebox{0.8}{$=\sqbr{-p \od{}{x}+2 W^-_l}\sqbr{p \od{}{x}}$}   & \multirow{2}{4em}{\scalebox{0.8}{${\cal T}_l(-)=r_l\smbr{-} r_l^{-1}$}}    &\scalebox{0.8}{${\cal T}_l\smbr{p H_l(x)}$} \scalebox{0.8}{$=\sqbr{-p \od{}{x}+ W^-_l}\sqbr{p \od{}{x}+ W^-_l}$} &\multirow{2}{4em}{ \scalebox{0.8}{$p\, \od{}{x} \equiv \od{}{y}$} } & \multirow{2}{4em}{\scalebox{0.7}{$-\smbr{p\,\frac{d}{dx}}^2=-\frac{d^2}{dy^2}$}} & \multirow{2}{4em}{\scalebox{0.7}{$W^-_l(x)$}} & \multirow{2}{4em}{\scalebox{0.7}{$r_l\equiv\sqrt{w_l}$}}\\
    & & & & & & \\
 \scalebox{0.8}{${\mathfrak h}_{2m} {\mathfrak h}^\dagger_0$} \scalebox{0.8}{$= \sqbr{-\sqrt{p}\od{}{x}+2 W^a_m} \sqbr{\sqrt{p}\od{}{x}}$} & \multirow{2}{4em}{\scalebox{0.8}{ ${\cal S}_m(-)=r_m\smbr{-}r^{-1}_m$} } & \scalebox{0.8}{${\cal S}_m\smbr{{\mathfrak h}_{2m} {\mathfrak h}^\dagger_0}=\mathsf{H}^a_m(z)$} \scalebox{0.8}{$= \sqbr{-\sqrt{p}\od{}{x}+ W^a_m} \sqbr{\sqrt{p}\od{}{x}+ W^a_m}$} & \multirow{2}{4em}{\scalebox{0.8}{$ \sqrt{p}\, \od{}{x}\equiv \od{}{z}$}} & \multirow{2}{4em}{\scalebox{0.8}{$- \smbr{\sqrt{p}\, \od{}{x}}^2=-\frac{d^2}{d z^2}$} }& \scalebox{0.8}{$W_m^a(x)$} & \scalebox{0.8}{$r_m\equiv\sqrt{w}\sqrt{p}^{m+\frac{1}{2}}$} \\
 \hline
\end{tabularx}
\setcounter{table}{1}
\caption{The roles of the active supersymmetrization transformation ${\cal T}_l$ on $ p H_l(x)$ and the active supersymmetrization transformation ${\cal S}_m$ on ${\mathfrak h}_{2m} {\mathfrak h}^\dagger_0$ involving the higher level superpotentials $W^-_l(x)$ and $W^a_m(x)$, respectively.}
\setcounter{table}{1}
\label{two_generic_supersymmetrizations}
\end{table*}

In order to illustrate that \eq{eigen_equ} of initial level ($l=0$), viewed as a \emph{principal} hypergeometric-like differential equation, admits a certain type of manifest supersymmetric factorization, the operator ${\mathbb H}_0(x)\equiv pH_0(x)$ is chosen as the $x$-coordinate version of the Hamiltoian, which contains the special squared form $-\smbr{pd/dx}^2$ as its kinetic energy operator. For this ${\mathbb H}_0(x)$, the active supersymmetrization transformation ${\cal T}$ in the form \eqnm{T_transformation}, that has been devised by some simple algebraic manipulations, is utilized to \emph{supersymmetrize} the trivial asymmetric factorization \eqnm{simple_factorization} of ${\mathbb H}_0(x)$ into the supersymmetric factorized form \eqnm{pH0_supersymmetrized}, in the sense that the unevenly distributed $2W_0$ in ${\mathbb H}_0(x)$ is evenly partitioned over the two factorial operators, so that $W_0$ works as the initial-level superpotential of this supersymmetric factorization. After such supersymmetrization, the momentum operator map $\hat{p}_y\equiv-i\,d/dy = -i\,pd/dx$ has been naturally introduced, which identifies $p\,d/dx$ with $d/dy$ and defines the coordinate transformation $y(x)$. The $x$-coordinate supersymmetric factorization \eqnm{pH0_supersymmetrized} then becomes the $y$-coordinate incarnation \eqnm{cal_H_0} -- the standard supersymmetric factorization of the Hamiltonian ${\cal H}_0(y)$. 

In order to illustrate that the \emph{same}  \eq{eigen_equ}, now viewed as the \emph{associated} hypergeometric-like differential equation of initial level, admits \emph{another} type of transparent supersymmetric factorization, \emph{another} squared form $-\smbr{\sqrt{p}d/dx}^2$, which comes directly with $H_0(x)$ itself, is singled out and regarded as \emph{another} kinetic-energy operator in $x$-coordinate. This $H_0(x)$ has the trival asymmetric factorization \eqnm{assymetric_factorization_H0II}. It is then turned into the supersymmetric factorization \eqnm{supersymmetrized_H0} with $W^a_0$ as its superpotential, by \emph{another} algebraic active supersymmetrization transformation ${\cal S}$ in the form \eqnm{S_transformation}. Naturally, by introducing \emph{another} type of momentum operator map $\hat{p}_z\equiv-i\,d/dz = -i\,\sqrt{p}d/dx$, which identifies $\sqrt{p} d/dx$ with $d/dz$ and defines the coordinate transformation $z(x)$, the $x$-coordinate supersymmetric factorization \eqnm{supersymmetrized_H0} is mapped to its $z$-coordinate incarnation -- the standard supersymmetric factorization of the Hamiltonian ${\cal H}_0(z)$. 

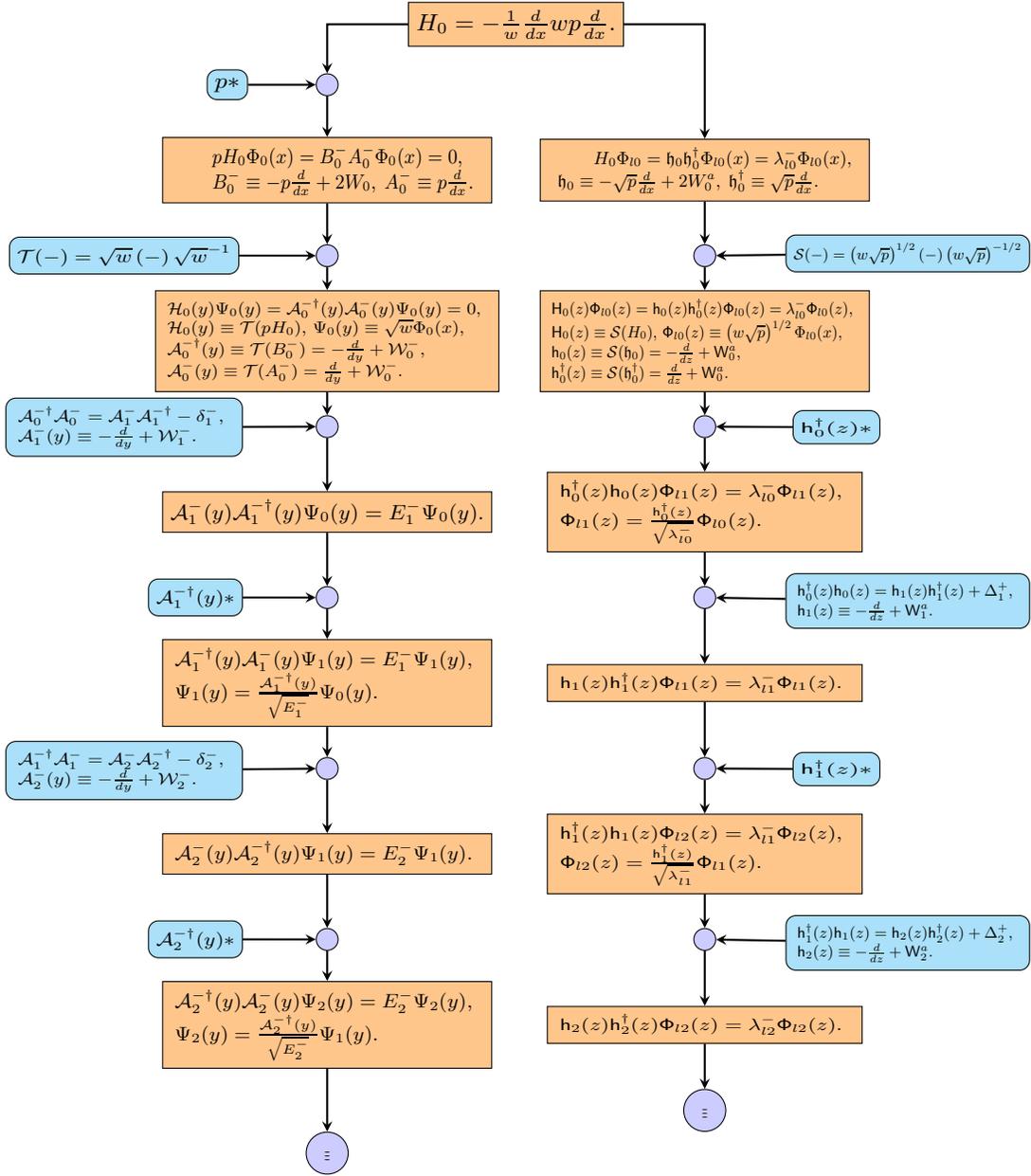
\begin{figure*}[t]
\renewcommand\figurename{Diagram}
\begin{tikzpicture}[node distance=1.2cm]

\node (start) [process] {\resizebox*{2.8cm}{0.4cm}{$H_0=-\frac{1}{w}\od{}{x} w  p \od{}{x}$.}};

\node (inpntl1) [inpoint, below left of= start, xshift=-1.8cm, yshift=-0.0cm] {};
\node (p_multiplication) [input, below left of=start,xshift=-3.2cm, yshift=-0.0cm] {{\resizebox*{0.35cm}{0.2cm}{$p*$}}};
\node (pro1a) [process, below of =inpntl1, xshift=-0.0cm, yshift=-0.0cm] {\resizebox*{4.4cm}{0.7cm}{\begin{tabular}{l}
                                                                    $\qquad p H_0 \Phi_0(x)= B^-_0 A^-_0\Phi_0(x)=0$,\\                                                                 
                                                                 $\qquad B^-_0 \equiv -p \frac{d}{dx}+2W_0,\; A^-_0\equiv p\frac{d}{dx}$.
                                                                \end{tabular} } };
\node (inpntl2) [inpoint, below of= pro1a, xshift=-0.0cm, yshift=-0.0cm] {};
\node (T_trans) [input, below left of=pro1a,xshift=-2.0cm, yshift=-0.35cm] {\resizebox*{3.0cm}{0.3cm}{{${\cal T}(-)=\sqrt{w}\smbr{-}\sqrt{w}^{-1}$}}};

\node (pro2a) [process, below of = inpntl2, xshift=-0.0cm, yshift=-0.0cm] {\resizebox*{4.6cm}{1.2cm}{\begin{tabular}{l}
                                                           ${\cal H}_0(y)\Psi_0(y)={\cal A}^{-\dagger}_0(y) {\cal A}^-_0(y)\Psi_0(y)=0$,\\
                                                           ${\cal H}_0(y)\equiv {\cal T}(p H_0)$, $\Psi_0(y)\equiv \sqrt{w} \Phi_0(x)$,\\
                                                           ${\cal A}^{-\dagger}_0(y) \equiv {\cal T}(B^-_0)= -\od{}{y}+{\cal W}^-_0 $,\\
                                                           ${\cal A}^-_0(y)\equiv {\cal T}(A^-_0)= \od{}{y}+{\cal W}^-_0$.
                                                          \end{tabular} } };
\node (inpntl3) [inpoint, below of= pro2a, xshift=-0.0cm, yshift=-0.0cm] {};
\node (SIl1) [input, below left of=pro2a,xshift=-2.0cm, yshift=-0.36cm] {\resizebox*{3.1cm}{0.6cm}{\begin{tabular}{l}
                                                                                                    ${\cal A}^{-\dagger}_0 {\cal A}^-_0={\cal A}^-_1 {\cal A}^{-\dagger}_1 -\delta^-_1$,\\
                                                                                                    $ {\cal A}^-_1(y)\equiv -\od{}{y}+{\cal W}^-_1$.
                                                                                                   \end{tabular} }};                                                          
                                                                                                   
\node (pro3a) [process, below of = inpntl3, xshift=-0.0cm, yshift=-0.0cm] {\resizebox*{4.4cm}{0.35cm}{${\cal A}^-_1(y) {\cal A}^{-\dagger}_1(y)\Psi_0(y)=E^-_1 \Psi_0(y)$.} };

\node (inpntl4) [inpoint, below of= pro3a, xshift=-0.0cm, yshift=-0.0cm] {};
\node (A_1_multiplication) [input, below left of = pro3a, xshift=-1.0cm, yshift=-0.35cm] {\resizebox*{1.1cm}{0.3cm}{${\cal A}^{-\dagger}_1(y)*$} };

\node (pro4a) [process, below of = inpntl4, xshift=-0.0cm, yshift=-0.0cm] {\resizebox*{4.4cm}{1.0cm}{\begin{tabular}{l}
                                                                                                    ${\cal A}^{-\dagger}_1(y) {\cal A}^-_1(y) \Psi_1(y)=E^-_1 \Psi_1(y)$,\\
                                                                                                    $ \Psi_1(y) = \frac{{\cal A}^{-\dagger}_1(y)}{\sqrt{E^-_1}}\Psi_0(y)$.
                                                                                                   \end{tabular} } };
\node (inpntl5) [inpoint, below of= pro4a, xshift=-0.0cm, yshift=-0.0cm] {};
\node (SIl2) [input, below left of=pro4a,xshift=-2.0cm, yshift=-0.38cm] {\resizebox*{3.1cm}{0.6cm}{\begin{tabular}{l}
                                                                                                    ${\cal A}^{-\dagger}_1 {\cal A}^-_1 = {\cal A}^-_2 {\cal A}^{-\dagger}_2 -\delta^-_2$,\\
                                                                                                    $ {\cal A}^-_2(y)\equiv -\od{}{y}+{\cal W}^-_2$.
                                                                                                   \end{tabular} }};                                                          
\node (pro5a) [process, below of = inpntl5, xshift=-0.0cm, yshift=-0.0cm] {\resizebox*{4.4cm}{0.35cm}{\begin{tabular}{l}
                                                                                                    ${\cal A}^-_2(y) {\cal A}^{-\dagger}_2(y) \Psi_1(y)=E^-_2 \Psi_1(y)$.
                                                                                                   \end{tabular} } };
\node (inpntl6) [inpoint, below of= pro5a, xshift=-0.0cm, yshift=-0.0cm] {};
\node (A_2_multiplication) [input, below left of = pro5a, xshift=-1.0cm, yshift=-0.35cm] {\resizebox*{1.1cm}{0.3cm}{${\cal A}^{-\dagger}_2(y)*$} };
\node (pro6a) [process, below of = inpntl6, xshift=-0.0cm, yshift=-0.0cm] {\resizebox*{4.4cm}{1.0cm}{\begin{tabular}{l}
                                                                                                    ${\cal A}^{-\dagger}_2(y) {\cal A}^-_2(y) \Psi_2(y)=E^-_2 \Psi_2(y)$,\\
                                                                                                    $ \Psi_2(y) = \frac{{\cal A}^{-\dagger}_2(y)}{\sqrt{E^-_2}}\Psi_1(y)$.
                                                                                                   \end{tabular} } };
\node (inpntl7) [inpoint, below of= pro6a, xshift=-0.0cm, yshift=-0.6cm] {\resizebox*{0.2cm}{0.2cm}{$\vdots$}};

\node (pro1b) [process, below right of = start, xshift=1.8cm,yshift=-1.2cm] {\resizebox*{4.4cm}{0.65cm}{ \begin{tabular}{l}
                                                                 $\qquad H_0 \Phi_{l0} =\mfk{h}_0 \mfk{h}^\dagger_0\Phi_{l0}(x)=\lambda^-_{l0}\Phi_{l0}(x)$,\\
                                                                 $\mfk{h}_0 \equiv -\sqrt{p} \od{}{x}+2W^a_0,\;\mfk{h}^\dagger_0 \equiv \sqrt{p}\od{}{x}$.
                                                                \end{tabular} }};                                                          
                                                          
\node (inpntr1) [inpoint, below of= pro1b, xshift=-0.0cm, yshift=-0.0cm] {};
\node (S_trans) [input, below right of=pro1b,xshift=2.0cm, yshift=-0.32cm] {\resizebox*{3.2cm}{0.3cm}{${\cal S}(-)=\smbr{w\sqrt{p}}^{1/2}\smbr{-}\smbr{w\sqrt{p}}^{-1/2}$}};
\node (pro2b) [process, below of = inpntr1, xshift=0.0cm, yshift=-0.0cm] {\resizebox*{4.4cm}{1.2cm}{\begin{tabular}{l}
                                                           $\mathsf{H}_0(z)\mathsf{\Phi}_{l0}(z)=\mathsf{ h}_0(z) \mathsf{ h}^\dagger_0(z)\mathsf{\Phi}_{l0}(z)=\lambda^-_{l0} \mathsf{\Phi}_{l0}(z)$,\\
                                                           $\mathsf{H}_0(z)\equiv  {\cal S}(H_0)$, $\mathsf{\Phi}_{l0}(z)\equiv \smbr{w\sqrt{p}}^{1/2}\Phi_{l0}(x)$,\\
                                                           $\mathsf{ h}_0(z)\equiv {\cal S}(\mfk{h}_0) =-\od{}{z}+\mathsf{W}^a_0$,\\
                                                           $\mathsf{ h}^\dagger_0(z)\equiv {\cal S}(\mfk{h}^\dagger_0) = \od{}{z}+\mathsf{W}^a_0$.
                                                           \end{tabular} }};                                                   
\node (inpntr2) [inpoint, below of= pro2b, xshift=-0.0cm, yshift=-0.0cm] {};
\node (h0_multiplication) [input, below right of=pro2b,xshift=1.cm, yshift=-0.36cm] {\resizebox*{1.cm}{0.25cm}{$\mathsf{ h}^\dagger_0(z)*$}};
\node (pro3b) [process, below of = inpntr2, xshift=0.0cm, yshift=-0.0cm] {\resizebox*{4.2cm}{0.9cm}{\begin{tabular}{l}
                                                           $\mathsf{ h}^\dagger_0 (z) \mathsf{ h}_0 (z) \mathsf{\Phi}_{l1}(z)=\lambda^-_{l0} \mathsf{\Phi}_{l1}(z)$,\\
                                                           $\mathsf{\Phi}_{l1}(z)= \frac{\mathsf{ h}^\dagger_0(z)}{\sqrt{\lambda^-_{l0}}} \mathsf{\Phi}_{l0}(z)$.
                                                           \end{tabular} }};                                                   
\node (inpntr3) [inpoint, below of= pro3b, xshift=-0.0cm, yshift=-0.0cm] {};
\node (SIr1) [input, below right of=pro3b,xshift=2.0cm, yshift=-0.4cm] {\resizebox*{3.2cm}{0.55cm}{\begin{tabular}{l}
                                                                                                    $\mathsf{ h}^\dagger_0 (z) \mathsf{ h}_0 (z)=\mathsf{ h}_1 (z) \mathsf{ h}^\dagger_1 (z) + \Delta^+_1$,\\
                                                                                                    $ \mathsf{ h}_1 (z)\equiv -\od{}{z} + \mathsf{ W}^a_1$.
                                                                                                   \end{tabular} }};           
\node (pro4b) [process, below of = inpntr3, xshift=0.0cm, yshift=-0.0cm] {\resizebox*{4.2cm}{0.3cm}{\begin{tabular}{l}
                                                           $\mathsf{ h}_1 (z) \mathsf{ h}^\dagger_1 (z) \mathsf{\Phi}_{l1}(z)=\lambda^-_{l1} \mathsf{\Phi}_{l1}(z)$.
                                                           \end{tabular} }};     
\node (inpntr4) [inpoint, below of= pro4b, xshift=-0.0cm, yshift=-0.0cm] {};                                                           
\node (h1_multiplication) [input, below right of=pro4b,xshift=1.cm, yshift=-0.36cm] {\resizebox*{1.0cm}{0.25cm}{$\mathsf{ h}^\dagger_1(z)*$}};                                                           
\node (pro5b) [process, below of = inpntr4, xshift=0.0cm, yshift=-0.0cm] {\resizebox*{4.2cm}{0.9cm}{\begin{tabular}{l}
                                                           $\mathsf{ h}^\dagger_1 (z) \mathsf{ h}_1 (z) \mathsf{\Phi}_{l2}(z)=\lambda^-_{l1} \mathsf{\Phi}_{l2}(z)$,\\
                                                           $\mathsf{\Phi}_{l2}(z)= \frac{\mathsf{ h}^\dagger_1(z)}{\sqrt{\lambda^-_{l1}}} \mathsf{\Phi}_{l1}(z)$.
                                                           \end{tabular} }};                                                   
\node (inpntr5) [inpoint, below of= pro5b, xshift=-0.0cm, yshift=-0.0cm] {};
\node (SIr2) [input, below right of=pro5b,xshift=2.0cm, yshift=-0.4cm] {\resizebox*{3.2cm}{0.55cm}{\begin{tabular}{l}
                                                                                                    $\mathsf{ h}^\dagger_1 (z) \mathsf{ h}_1 (z)=\mathsf{ h}_2 (z) \mathsf{ h}^\dagger_2 (z) + \Delta^+_2$,\\
                                                                                                    $ \mathsf{ h}_2 (z)\equiv -\od{}{z} + \mathsf{ W}^a_2$.
                                                                                                   \end{tabular} }};                                                                      
\node (pro6b) [process, below of = inpntr5, xshift=0.0cm, yshift=-0.0cm] {\resizebox*{4.2cm}{0.3cm}{\begin{tabular}{l}
                                                           $\mathsf{ h}_2 (z) \mathsf{ h}^\dagger_2 (z) \mathsf{\Phi}_{l2}(z)=\lambda^-_{l2} \mathsf{\Phi}_{l2}(z)$.
                                                           \end{tabular} }};     
\node (inpntr6) [inpoint, below of= pro6b, xshift=-0.0cm, yshift=-0.0cm] {\resizebox*{0.2cm}{0.2cm}{$\vdots$}};                                                           
                                                           % \node (out1) [io, below of=pro2a] {Output};                                                           

\draw [arrow] (start) -|  (inpntl1);

\draw [arrow] (p_multiplication) -- (inpntl1);
\draw [arrow] (inpntl1) --  (pro1a);
\draw [arrow] (pro1a) --  (inpntl2);
\draw [arrow] (T_trans) -- (inpntl2);
\draw [arrow] (inpntl2) -- (pro2a);
\draw [arrow] (pro2a) -- (inpntl3); 
\draw [arrow] (SIl1) -- (inpntl3); 
\draw [arrow] (inpntl3) -- (pro3a); 
\draw [arrow] (pro3a) -- (inpntl4); 
\draw [arrow] (A_1_multiplication) -- (inpntl4); 
\draw [arrow] (inpntl4) -- (pro4a); 
\draw [arrow] (pro4a) -- (inpntl5); 
\draw [arrow] (SIl2) -- (inpntl5); 
\draw [arrow] (inpntl5) -- (pro5a); 
\draw [arrow] (pro5a) -- (inpntl6); 
\draw [arrow] (A_2_multiplication) -- (inpntl6); 
\draw [arrow] (inpntl6) -- (pro6a);
\draw [arrow] (pro6a) -- (inpntl7);

\draw [arrow] (start) -|  (pro1b);
\draw [arrow] (pro1b) -- (inpntr1);
\draw [arrow] (S_trans) -- (inpntr1);
\draw [arrow] (inpntr1) -- (pro2b);
\draw [arrow] (pro2b) -- (inpntr2);
\draw [arrow] (h0_multiplication) -- (inpntr2);
\draw [arrow] (inpntr2) -- (pro3b);
\draw [arrow] (pro3b) -- (inpntr3);
\draw [arrow] (SIr1) -- (inpntr3);
\draw [arrow] (inpntr3) -- (pro4b);
\draw [arrow] (pro4b) -- (inpntr4);
\draw [arrow] (h1_multiplication) -- (inpntr4);
\draw [arrow] (inpntr4) -- (pro5b);
\draw [arrow] (pro5b) -- (inpntr5);
\draw [arrow] (SIr2) -- (inpntr5);
\draw [arrow] (inpntr5) -- (pro6b);
\draw [arrow] (pro6b) -- (inpntr6);

\end{tikzpicture}
\caption{Side by side comparison between the two flow charts of running the same SUSYQM algorithm for iteratively generating a hierarchy of the principal (left) hypergeometric-like differential equations and their eigen-solutions as well as the associated (right) counterparts, stemming from the same hypergeometric-like operator $H_0$.}
\label{diagram1}
\end{figure*}

We see that  there exist \emph{two} distinct types of supersymmetric factorizations that are closely related to $H_0(x)$. These can be reached by supersymmetrizing the two different asymmetric factorizations that can be constructed essentially from the \emph{same} self-adjoint form \eqnm{self_adjoint_form} of $H_0(x)$ via the two active supersymmetrization transformations, i.e. ${\cal T}$ in the form \eqnm{T_transformation} and ${\cal S}$ in the form \eqnm{S_transformation}, which have been designed aiming at respectively preparing the squared forms  $-\smbr{p\, d/dx}^2$ and $-\smbr{\sqrt{p}\,d/dx}^2$.
These two active supersymmetrizations and their spin-offs that can be immediately read off from them, such as the momentum and kinetic energy operator maps, the superpotentials and the rescaling factors of the eigen-functions for passing to the corresponding standard Schr\"odinger equations, are summarized in Table \ref{two_initial_supersymmetrizations}. The two processes of these active supersymmetrizations and their yields are also shown  from the second layer to the third one in the two flows in Diagram \ref{diagram1}.

In parallel, the roles of the two more general active supersymmetrizations ${\cal T}_l$ and ${\cal S}_m$, in the sense that ${\cal T}\equiv {\cal T}_0$ and ${\cal S} \equiv {\cal S}_0$, are summarized in Table \ref{two_generic_supersymmetrizations}. ${\cal T}_l$ supersymmtrizes the asymmetric form \eqnm{partially_asymmetrized_Hl} of $p H_l(x)$, where $H_l(x)$ is a generalization of $H_0(x)$ in that it involves the superpotential $W^-_l$ of generic principal level $l$, into the supersymmetric operator product in \eq{Schr_equ_x_coord_factorized}. That is, ${\cal T}_l$ turns \eq{asymmetrized_eigen_equ_general_l0}, an equivalent of \eq{eigen_equ_with_multiplier}, into the supersymmetrically factorized \eq{Schr_equ_x_coord_factorized}. ${\cal S}_m$ supersymmetrizes the asymmetric factorization  \eqnm{asymmtrization_associated_hypergeometric_like_operator1} of ${\mathfrak h}_{2m} {\mathfrak h}^\dagger_0$, which is a principal hypergeometric-like incarnation of the associated hypergeometric-like operator $H^a_m(x)$ and involves the superpotential $W^a_m$ of generic association level $m$,  into the supersymmetric factorization \eqnm{supersymmtrization_h2m-h0dagger}. In other words, ${\cal S}_m$ turns \eq{asymmetrized_associated_equ}, an equivalent of the associated hypergeomtric-like differential \eq{eigen_equ_level_m_after_rnmlz}, into the supersymmetrically factorized Schr\"odinger \eq{S_equ_standard_association_m}.

It is obvious that, for the four supersymmetrizations that are listed in Tables \ref{two_initial_supersymmetrizations} and \ref{two_generic_supersymmetrizations}, the superpotentials $W_0(x)$, $W^a_0(x)$, $W^-_l(x)$ and $W^a_m$ actually pop out and are easily read off directly from the left factors of the asymmetric factorizations of the operators $P\, H_0(x)$, $H_0(x)$, $P\, H_l(x)$ and $\mathfrak{h}_{2m} \mathfrak{h}^\dagger_0$,  even \emph{before} these supersymmetric factorizations are achieved.

We see that the constructions of the active supersymmetrizations $\cal T$ in \eq{T_transformation} and $\cal S$ in \eq{S_transformation} involve only elementary algebraic manipulations, and the coordinate transformations $y(x)$ and $z(x)$ are naturally and \emph{actively} reached by the two momentum operator maps \eqnm{momentum_map_type1} and \eqnm{momentum_map_type2}, the initial superpotentials $W_0(x)$ in \eq{W_0} and $W^a_0(x)$ in \eq{superpotential_W0a} and the rescaling factors of the eigenfunctions $r_t$ and $r_s$ also come out naturally and simultaneously  along with these two active supersymmetrizations. The two trivially different asymmetric factorizations of $p H_0(x)$ and $H_0(x)$ are then actively turned into the two initial level Hamiltonians ${\cal H}_0(y)\equiv {\cal T}\smbr{p H_0(x)}$ and $\mathsf{H}_0(z)\equiv {\cal S}\smbr{H_0(x)}$ in their supersymmetrized forms. In such a pure algebraic approach, the unpleasant first-order derivative in \eq{eigen_equ} is {\it not} eliminated in the `hard' way, but simply disguised or absorbed algebraically into the squared forms of some two-term  differential operators of first order, as shown in \eq{pH0_symmetrized}. Eventually,  the obtained supersymmetric factorization of ${\cal H}_0(y)$ or $\mathsf{H}_0(z)$ guarantees that the first-order derivatives  automatically cancel, hence the corresponding standard Schr\"odinger equations are reached in such a lighter algebraic way. 

This approach of active supersymmetrization is in sharp contrast to the traditional Courant-Hilbert \emph{differential} approach that is utilized to pass from a linear differential equation of second order in $x$-coordinate such as \eq{eigen_equ} to a certain standard Schr\"odinger type equation. In the latter approach, a unknown coordinate transformation and a unknown rescaling factor of the eigenfunction are invoked to transform the original differential equation into a new one. By forcing the overall coefficient of the first-order derivative in this new equation to vanish, another differential equation of second order for the coordinate transformation and rescaling factor is obtained and solved. We see that, in such a traditional \emph{differential} approach that is full of mathematical flavor, the first-order differential operator (derivative) is eliminated {\it bona fide}, and the coordinate transformation and rescaling factor are obtained \emph{passively} by solving some resulting differential equation of second order. In most cases, the resultant Schr\"odinger type equation does not come in a supersymmetrically factorized form. In order to reach such a supersymmetric factorization, a Riccarti equation of the resultant potential still has to be solved for the corresponding superpotential. 

In contrast, in our active supersymmetrization approach, all these are accomplished {\it in one go} without solving  any intermediate differential equation of second order for the coordinate transformation or the first order Ricatti equation for the superpotential. In this sense, the active supersymmetrization  is a method that is full of the spirit of {\it doing calculus with almost no calculus}, which well fits in with the spirit of algebraic SUSYQMs, and more importantly the algebras involved in this active supersymmetrization are almost trivial and this method is more efficient than the widely used Courant-Hilbert differential method, particularly for the purpose of developing SUSYQM algorithms. We think that this method of active supersymmetrization deserves more amplification.

In Appendix \ref{generalized_DO}, the same idea of active supersymmetrization has been applied to more general Sturm-Liouville-type operator. In fact, we have devised two types of such transformations to supersymmetrize the sum of the differential term of second order and that of first order inside a Sturm-Liouville-type operator, hence the eigen-equation of this type of operator is the transformed into {\it two} distinct standard Schr\"odinger equations.   Inside each of these two resultant Schr\"odinger equations, if the non-differential term is of the same type function as any of the terms of the potential of the primary supersymmetrizations of the two differential terms, or if the non-differential term can absorbed into the first-order differential term via a certain partial asymmetrization -- the inverse of some \emph{partial supersymmetrization}, the {\it complete}  supersymmetric factorizations of the Sturm-Liouville eigen-equation can be reached. In some special cases, the resultant overall potentials further bear shape invariance symmetry, thus the SUSYQM algorithms for solving this type of eigenvalue problems, similar to those of $H_0$, can be constructed. 

The active supersymmetrization transformation ${\cal T}$ not only supersymmetrizes the \emph{initial level} eigen-equation \eqnm{initial_eigen_equ_assymmetric} into \eq{initial_eigen_equ_minus} but also supersymmetrizes the \emph{generic level} eigen-equation \eqnm{eigen_equ_NHF} for the principal hypergeometric-like function  into \eq{Schr_equ_x_coord_factorized}. By virtue of the first type of momentum operator map \eqnm{momentum_map_type1}, this $x$-coordinate supersymmetric factorization \eqnm{Schr_equ_x_coord_factorized} is then turned into that of standard Schr\"odinger equation \eqnm{Schr_equ_y_coord} in $y$-coordinate, with the factorized Hamiltonian ${\cal H}_l(y)$ given in \eq{Hamiltonian_SHF}, as demonstrated in Subsection \ref{Standard_Hermitian_factorization}.
Similarly, combining with the second type of momentum operator map \eqnm{momentum_map_type2}, the active supersymmetrization transformation ${\cal S}$, not only supersymmetrizes the asymmetric factorization  \eqnm{asymmetric_eigen_equ_initial_association_level} in $x$-coordinate  of \eq{eigen_equ} of \emph{initial association level $m=0$} into the supersymmetrically factorized Schr\"odinger \eq{Schrodinger_equ_base_level} in $z$-coordinate, but also supersymmetrizes the asymmetrically factorized  eigen-equation \eqnm{eigen_equ_level_m_after_rnmlz} of the \emph{generic association level}  for the associated hypergeometric-like function into the supersymmetrically factorized Schr\"odinger \eq{S_equ_standard_association_m}, as demonstrated in Subsection \ref{associated_standard_reps}. 
In addition, both these two supersymmetrization transformations and the two momentum operator maps have their inverses. For instance, as demonstrated in Subsection \ref{non_standard_factorization}, the inverse transformation ${\cal T}^{-1}$ and the inverse momentum operator map of the first type  have been utilized to guide us to establish the nonstandard ($x$-coordinate) SUSYQM algorithm for obtaining the eigen-solutions of the principal hypergeometric-like differential equation from the standard one in $y$-coordinate.  Thus, these active supersymmetrization transformations and the accompanying momentum operator maps, together with their inverses, establish \emph{a one-to-one correspondence or an isomorphism} between the SUSYQMs represented in $x$-coordinate and those in $y$ or $z$-coordinate. Under such isomorphisms, the SUSYQM algebraic structures, such as the supersymmetrically factorized eigen-equations, the generalized commutator relations (shape invariance conditions), the raising/lowering recurrence relations, the three-term recurrence relations of eigen-functions, etc., 
 are faithfully preserved when passing from one coordinate representation to the other around. In this sense, rather than regarding the standard $y$ and $z$-coordinate supersymmetric factorizations as the true SUSYQMs of the principal and associated hypergeometric-like differential operators, it is better to think of the $x$-coordinate realizations and the $y$ and $z$-coordinate realizations as just \emph{two different coordinate representations of the  same SUSYQMs}, even though in the $x$-coordinate representation, the factorizations \eqnm{eigen_equ_NHF} and \eqnm{eigen_equ_level_m_after_rnmlz} do not look manifestly supersymmetric. Those coordinate representations in which the Hamiltonians appear in manifest supersymmetric forms are called the standard ones. 

The isomorphisms between the original $x$-coordinate representation and the new $y$ or $z$-coordinate representation of the same SUSY and SISY owned by the hypergeometric-like differential operator $H_0(x)$ and its associate tell us that one can implement this type of SUSYQM algorithms in either coordinate representation at one's disposal. For example, in Subsections \ref{Standard_Hermitian_factorization}-\ref{upward_SIC}, we have demonstrated in detail that the SUSYQM algorithm for the principal hypergeometric-like operator $H_0(x)$ can be constructed  in the {\it nonstandard} $x$-coordinate representation in Subsection \ref{non_standard_factorization} as well as in the standard $y$-coordinate representation, \emph{with almost equal efficiency}. Also in Subsections \ref{positive_association_levels} and \ref{negative_association_levels}, we have shown in detail that the SUSYQM of the associated hypergeometric-like operator $H^a_m(x)$ can be efficiently constructed in the nonstandard $x$-coordinate. In practice, for developing the SUSYQM algorithms, one should take the advantages of both the non-standard and standard representations, and switch smoothly between them to meet one's needs. For instance, since in the standard coordinate representations the supersymmetric factorizations are manifestly defined, it is more instructive to resort to these coordinate representations for quickly pinning down the concrete forms of the active supersymmetrization transformations, the superpotentials and the momentum operator maps, as demonstrated in Subsections \ref{base_kinetic_energy_operator} and \ref{2nd_type_momentum_op}.  From these standard representations, one can even obtain some guidance for how to design the ladder operators in nonstandard $x$-coordinate representations.  However, once these identifications are done, one should take the innate advantage of the $x$-coordinate representations -- the two functions $p(x)$ and $q(x)$ in $H_0(x)$ take extremely simple forms. As a result, there exists  the simple linear superpotential $W_l(x)=\alpha_l x+\beta_l$ for establishing the SUSYQM of the principal hypergeometric-like operator $H_0(x)$, and the factorization parameters have also been conveniently  determined in this $x$-coordinate representation in Subsections \ref{Standard_Hermitian_factorization}-\ref{upward_SIC}, even though the operator algebras are mainly represented in the standard $y$-coordinate there. A large class of analytical soluble potentials that appear in the original standard coordinate representations of the Hamiltonians like drastically different from the binomial potentials exhibited in Eqs. \eqnm{Vl_potential} and \eqnm{pHm_equ}. Traditionally, people like to go conversely from the standard representation to the nonstandard one, i.e. to try to convert their standard Schr\"odinger equations into the base or associated differential equations of $H_0$ in $x$-coordinate and then seek their eigen-solutions. This is mainly because their analytical solutions do exist and in term of which the original solutions can be expressed analytically after all, as we have shown in this paper. From the perspective of establishing SUSYQM, there is also another reason to make this jump of coordinate representation, i.e. in $x$-coordinate the \emph{binomial} nature of the potentials in Eqs. \eqnm{Vl_potential} and \eqnm{pHm_equ} allows us to easily finding out their \emph{linear} superpotentials.  

Now, a conspicuous picture emerges in front of us, for why there exist simultaneously {\it two} types of -- the principal and associated -- special functions that are associated with the {\it same} hypergeometric-like differential operator $H_0$, and for how these two types of functions and the related eigenvalues are iteratively built through essentially the {\it same} SUSYQM algorithm.  This picture become extremely clear when we look at the two iteration flow charts juxtaposed side by side in Diagram \ref{diagram1}, the left one illustrates the process of proliferating a hierarchy of the \emph{principal} hypergeometric-like eigen-equations and their solutions, that has been demonstrated in detail in \ref{proliferation_minus_route}, and the right one illustrates the process of proliferating a hierarchy of the \emph{associated} hypergeometric-like eigen-equations and their solutions, that has been demonstrated in detail in \ref{positive_association_levels}. 

Seated inside the orange box in the top layer is the hypergeometric-like operator $H_0$ in its natural self-adjoint form in $x$-coordinate, which consists of two asymmetric layers of first-order differential operators. As already listed in the first column of Table I, this asymmetric factorization of $H_0(x)$ can bifurcate into two distinct asymmetric factorizations, one is for $p\, H_0$ (indicated also in Diagram \ref{diagram1} as the result of $p$-multiplication inserted between the first two layers along the left flow), initiated by preparing the kinetic energy operator $-\smbr{pd/dx}^2$, and the other is for $H_0$ itself, initiated by preparing another type of kinetic energy operator $-\smbr{\sqrt{p} d/dx}^2$. Consequently, there exist two distinct initial-level eigen-equations \eqnm{initial_eigen_equ_assymmetric} and \eqnm{eigen_equ_base_level}  in their asymmetrically factorized forms for these two operators, as seated inside the two orange boxes in the second layers of Diagram \ref{diagram1}.

Going from layer 2 to layer 3 of the two main (orange) flows, the two boxes with light blue are inserted as the left and right wings of these two flow charts. The left box contains the active supersymmetrizations $\cal T$ (defined in \eq{T_transformation}) and the right one contains $\cal S$ (defined in \eq{S_transformation}). These two are utilized to turn those two asymmetric factorized eigen-equations of initial level, listed in layer 2, into the standard supersymmetrically factorized Schr\"odinger equations in the two new coordinates $y$ and $z$, which are defined by the two accompanying momentum operator maps \eqnm{momentum_map_type1} and \eqnm{momentum_map_type2}, respectively. These operations have also been summarized in Table 1.

From the third layer on, along the two main orange flows, the pivotal parts of the iterative SUSYQM algorithms are illustrated, i.e. the supersymmetrically factorized eigen-equations and their eigen-solutions of next higher level are repeatedly proliferated from those of the present level, by incorporating the generalized commutator relations (shape invariance conditions) and intertwiningly acting the eigen-equation of the present level with some raising ladder operators. The left column of the flow summarizes the minus route of the bottom-up iterative SUSYQM algorithm of proliferating the \emph{principal} hypergeometric-like eigen-equations and their solutions detailed in \ref{proliferation_minus_route}, while the right column summarizes the similar algorithm of proliferating the \emph{associated} hypergeometric-like eigen-equations and their solutions detailed in the first half part of \ref{positive_association_levels} but represented in the standard $z$-coordinate, or the same algorithm directly demonstrated in $z$-coordinate in Subsection \ref{associated_standard_reps}. However, the order of performing these two operations differs in these two particular flows. For instance,  from layer 4 to layer 5, in the left flow, firstly, the generalized commutator relation \eqnm{SI_operator_product_minus_l1} is incorporated to substitute the ladder operators with subscript $1$ for those with subscript $0$, in consequence, the raising ladder operator ${\cal A}^{-\dagger}_1(y)$ (for the minus route of factorization)  sits immediately next to $\Psi_0(y)$ (as shown inside the left third orange box) for replacing the \emph{incorrectly} situated lowering operator ${\cal A}^{-}_0(y)$, and the eigenvalue $E^-_0=0$ is renormalized to $E^-_1$; secondly, by intertwiningly act the resultant eigen-equation (inside the third orange box) with the same raising operator ${\cal A}^{-\dagger}_1(y)$, the superpartner eigen-equation in its normally ordered form is generated (shown as the first line inside the left fourth orange box), which works as the new startup equation for the next turn of iteration, as a consequence, the eigen-function $\Psi_1(y)$ of level $1$ is generated. In contrast, in the right flow, from layer 4 to layer 5, since the raising operator $\mathsf{ h}^\dagger_0(z)$ has been already situated in the correct position, firstly, intertwiningly acts the eigen-equation of level $0$ (the first line inside the second orange box) with this operator results in its superpartner eigen-equation (shown as the first line inside the third orange box), as a consequence, the level-$1$ eigenfunction $\mathsf{\Phi}_{l1}(z)$ is created; secondly, by incorporating the generalized commutator relation \eqnm{quasi_commutator_positive_first_z_coord} to substitute $\mathsf{ h}^\dagger_1(z)\mathsf{ h}^\dagger_1(z)$ for $\mathsf{ h}^\dagger_0(z)\mathsf{ h}_0(z)$, this superpartner eigen-equation is then renormalized into the normally ordered one (shown inside the third orange box), which works as the new startup eigen-equation for the next turn of iteration.

In the left column of the iteration flow, the initial eigenvalue is set to be vanishing, so the annihilation condition \eqnm{ground_state_minus_route} reduced from the initial eigen-equation gives birth to the simplest initial eigen-function $\Phi_0(x)$ -- a constant, and the iteration can keep going {\it ad infinitum}. In contrast, in the right column of the iteration flow the iteration terminates at the highest association level $m=l$. The corresponding annihilation condition \eqnm{annihilation_condition_top_level} actually gives birth to the simple eigen-function of the \emph{top iteration level} (see \eq{Phi_ll}). With such eigen-function in hand, one can obtain the eigen-equations and eigen-solutions of all lower $m$ levels by running the SUSYQM algorithm backwards. This latter top-down approach actually generates the associated hypergeometric-like eigen-functions of all association levels in the compact form \eqnm{Phi_lm_top_down_approach}, without resorting to the left column algorithm for providing the concrete expression of the principal hypergeometric-like function $\Phi_l(x)$. 

Despite of the detailed differences mentioned in the above two paragraphs between the left and right column of iteration flows in Diagram \ref{diagram1}, we see that essentially the {\it same} iterative SUSYQM algorithm has been run {\it twice}, once for building the {\it base} hypergeometric-like differential equations and their eigen-solutions, the other for building the {\it associated} counterparts. In other words, it is not the SUSYQM algorithm itself but the two pivotal distinct startup supersymmetric factorizations of the same $H_0$, listed in the two orange boxes in the second layer and in the first column of Table 1, that initiate the bifurcation of the flows of running of the same algorithm. Thus, from the perspectives of SUSYQM, there are two key facts that are directly responsible for the existence of two distinct types of -- the base and associated -- hypergeometric-like functions in general for the same $H_0$: 1) there are \emph{two  distinct quantum kinetic energy operators} $- (p d/dx)^2$ and $-(\sqrt{p}d/dx)^2$), rooted in the {\it same} $H_0(x)$, that directly lead to \emph{two distinct initial supersymmetric factorizations}; 2) the two startup potentials/superpotentials $V_0(x)$/$W_0(x)$ in Eq. \eqnm{Riccati_equ_V0}/\eqnm{W_0} and $V^a_0(x)$/$W_0^a(x)$ in Eq. \eqnm{initial_potential_association}/\eqnm{superpotential_W0a} can proliferate into \emph{two series of descendent potentials/superpotentials of higher levels in shape-invariant fashions}. 

It is noteworthy that,  the SUSYQM algorithm in the right column of Diagram \ref{diagram1} actually \emph{accomplishes three tasks in one go} without any inputs from that in the left column or from any traditional methods. Not only is the differential equation for the associated hypergeometric-like operator $H^a_m$ derived from \eq{eigen_equ} -- that of the principal hypergeometirc-like differential operator $H_0$, but also is the eigen-solutions of $H^a_m$ generated, in a simple algebraic way in which the SUSY and SISY are manifest. In addition, the eigen-solutions of \eq{eigen_equ} \emph{itself} also come out along with this SUSYQM algorithm, i.e.  the eigen-function comes in the compact form of the generalized Rodriguez' formula \eqnm{Rodriguez_formula}, and the eigenvalue $\lambda^-_l$, given in \eq{lambda_m_def}, comes out as the accumulation of all the constants $\Delta^+_m$ appearing in a series of generalized commutators relations, such as \eqnm{SI_associated_ascending}, for the supersymmetric factorization of $H^a_m$. If we just limit our concern to looking for a quick approach of producing the eigen-solutions of \eq{eigen_equ}, this latter approach seems to be much more efficient than the SUSYQM algorithm shown in the left column of Diagram \ref{diagram1} or in Subsection \ref{susyqm_H_like_functions}, because there is no need for determining those factorization parameters, as done in \ref{Minus_route_factorization} and \ref{Plus_route_factorization}. 

Inside $H_0$, when $p(x)$ is just a non-vanishing constant, the associated hypergeometric-like operator $H_m(x)$ degenerates into the base hypergeometric-like operator $H_0(x)$. In other word, the two flows in Diagram \ref{diagram1} will produce the same eigen-equation, eigenfunction and eigenvalue of generic level $l$. In particular, as clearly revealed in Eqs \eqnm{eigenvalue_coincidence} and \eqnm{eigenfunction_coincidence}, their is a level-to-level coincidence between bottom-up flow in the right column and the top-down flow in the right column; for instance, the eigen-equation of the top association level $m=l$ on the right is the same as that of the initial base level $l=0$ on the left. We find that for the hypergeometric-like operator $H_0(x)$, except for further degenerate cases, the Hermite-type operator is the {\it only} type of operator whose generalized commutator relations degrade to the same level-independent one and whose base eigen-functions and associated ones degenerate into the same class of bounded functions -- Hermite polynomials.

For the potential $V_l(x)$ in \eq{Vl_potential} of the base Hamiltonian $\mathbb{H}_l(x)$,  its superpartner potential $V^s_l(x)$ in \eq{potential_x_coord} can be either downward connected to the potential of its nearest neighboring $(l-1)$-th level in the form of the shape invariance condition \eqnm{SI_potential_minus} or upward connected to the potential of its nearest neighboring $(l+1)$-th level in the form of the shape invariance condition \eqnm{SI_potential_plus}.  This entails two different yet closely related routes/branches of factorizations of $\mathbb{H}_l(x)$ for solving the eigenvalue problem of $H_0(x)$ via  essentially the same iterative SUSYQM algorithm.
The intimate connections between the two sets of factorization parameters for these two routes of factorizations are listed in Subsection \ref{Plus_route_factorization}. The origin of such two routes of factorizations of the same operator $\mathbb{H}_l(x)$ and their affinities can be ultimately traced back to the existence of the `superpartner' eigen-equation \eqnm{eigenequ_lambda_plusl} of the original hypergeometric-like equation \eqnm{eigenequ_lambda_minusl} (or \eqnm{eigen_equ}). These two equations have common eigenfunctions while different eigenvalues.

The generic SUSYQM algorithms provided here for solving the discrete eigenvalue problems of several important cases of $H_0$, such as the Legendre and Laguerre operators, and even some Sturm-Liouville-type operators like Bessel operator, can be further concretized, by incorporating the specific characters of these operators. This type of work  will be published separately.

In our iterative SUSYQM algorithm designed for solving the discrete eigenvalue problems of the hypergeometric-like differential operator $H_0(x)$ and the $H_0(x)$-born operator $H^a_m$,  not only the underlying SUSY and SISY of these two operators are manifestly brought to the fore at each level or from one level to the next neighboring level,  but also these two symmetries are actually augmented to the {\it governing principles} that guide us to build this algorithm, and they indeed dictate the very existence of a hierarchy of the base hypergeometric-like functions as well as the associated counterparts in their canonical forms.  In particular, through our algorithm, the tasks of deriving the associated hypergeometric-like eigen-equation itself, that comes out automatically in its supersymmetrically factorized form, and that of seeking for its eigen-solutions  are accomplished neatly \emph{in one go} and in an elementary algebraic way, starting from the trivial initial factorization \eqnm{assymetric_factorization_H0II} of $H_0(x)$. 

The SUSYQM algorithms developed by us for solving the eigenvalue problems of the hypergeometric-like differential operators $H_0$ and $H^a_m$  and the pure algebraic supersymmetrization transformations, on the one hand, do not rest on any result coming out the traditional advanced methods such as the Frobenius and the generating function methods. In fact, several results, such the three-term recurrence relations, the generalized Rodriguez' formula, that have been made use of by others as  the inputs for factorizing the operators $H_0$ and $H^a_m$, just come out as the byproducts of our algorithm.  Therefore our algorithms, that are based on the two symmetries (SUSY and SISY) alone,  are  \emph{logically completely independent of those traditional methods}. On the other hand, our algorithm merely involve some elementary algebraic manipulations and basic calculus  that can  even be accessible by high school students or freshmen. Such conceptual simplicity and algebraic elementariness make the hypergeometric-like eigen-functions and their associated counterparts, traditionally classified as special functions due to the complicated ways of defining and determining them,  {\it not so special} as people used to think of them to be.   These two aspects of our SUSYQM algorithms make them strong competitive candidates for replacing the conventional methods of approaching the base and the associated hypergeometric-like functions.

An immediate reward of our SUSYQM algorithms is that to some extent they might relieve the burdens people experience from the traditional ways of tackling the eigenvalue problems of $H_0$ and $H^a_m$, especially when first learning or teaching those solving  techniques. In traditional mathematic, physical and engineering texts, those materials involving solving the eigenvalue problems of $H_0$ and $H^a_m$ are often considered as hard and advanced (high-tech) ones and usually be postponed to teach in classes only for seniors and even graduates. A famous example is that the rigorous derivation of Bohr formula and the eigenfunction for Coulomb potential can not be given in the course of elementary atomic physics  because of the two stumbling stones -- the radial Schr\"odinger equation and the associated Legendre differential equation, until later in the quantum mechanics courses of medium or advanced level when these two eigen-equations are solved in the traditional Frobenius method. Now our algebraic SUSYQM algorithms raise the possibility of teaching these materials to freshmen or even to high school students, by solving the confluent hypergeometric equation (as a principle hypergeometric type equation), which is derived from the radial Schr\"odinger equation, either via the algorithm presented in Subsections \ref{minus_route_iteration} and \ref{non_standard_factorization}, or via the algorithm in Subsections \ref{positive_association_levels}; meanwhile, the associated Legendre differential equation (as an associated hypergeometric-type differential equation) and its solution can be generated via the algorithm in Subsections \ref{positive_association_levels} and \ref{negative_association_levels}, once these students are quickly exposed to the elementary background knowledges involved in these SUSYQM algorithms. In this resect, we also note that the radial Schr\"odinger equation for Coulomb potential can even be solved directly in another economic SUSYQM way \cite{Valance_Coulumb_Hamiltonian_SUSY_1990} without turning it into a confluent hypergeometric differential equation.

In Eqs. \eqnm{eigen_equ} and \eqnm{eigen_equ_level_m_after_rnmlz}, the coefficients of the binomial $p(x)$ and monomial $q(x)$ and the eigenvalues $\lambda$ and $\lambda_{lm}$ have been assumed to be constants.  In fact, these constants can be think of as functions of some common parameter like $\omega$, so that Eqs. \eqnm{eigen_equ} and \eqnm{eigen_equ_level_m_after_rnmlz} become \emph{generalized} eigen-equations for the eigenvalue $\omega$. The eigen-solutions of Eqs. \eqnm{eigen_equ} and \eqnm{eigen_equ_level_m_after_rnmlz} in this case are still provided by the SUSYQM algorithms constructed in this paper. For instance, the discrete eigenvalue $\omega_l$ of \eq{eigen_equ} is the solution of the following algebraic equation
\begin{eqnarray}
 \lambda(\omega_l)=\lambda^-_l(\omega_l),
\end{eqnarray}
where $\lambda^-_l(\omega)$ is given by \eq{lambda_minu_eigenvalue} and the $\omega$-dependencies in it are brought around through $q'(\omega)$ and $p''(\omega)$. Of course, in general this algebraic equation for $\omega_l$ holds only for a few integers $l$.  These seemingly man-made generalized eigen-equations could find some practical applications.  For instance, these equations could be related to the differential equations satisfied by some standing waves existing in magnetized plasmas of high temperature, in which $\omega$ represents the constant eigen-frequency.  In the following we will give a concrete example.

In \cite{Wang_GGAM_2022}, the authors are faced with solving the eigen-solutions of a type of generalized Schr\"odinger equation
\begin{eqnarray}\label{Schr_equ_shape_varying}
 -\rho^2_{0i} \frac{d^2 \hat{\Psi}}{d r^2} + D\smbr{\omega, r} \hat{\Psi}=0,
\end{eqnarray}
which describes a type of global modes existing in magnetized plasmas of of high temperature, where $\rho_{0i}$ is a constant parameter, $\hat{\Psi}$ is proportional to the mode amplitude, and $D(\omega, r)$ is the function that has both the frequency $\omega$ dependency and  the spatial coordinate $r$ dependency. There is an outstanding feature of this equation that makes it different from the ordinary  Schr\"odinger equation \eqnm{Schr_equ_y_coord}. In \eq{Schr_equ_y_coord} the potential ${\cal V}_l(y)$ and $E_l$ can also be trivially lumped into the one with two terms by introducing the obviously \emph{separable} function ${\cal V}_l(y)-E_l$. The $y$-dependency graph of this function just shifts vertically from that of ${\cal V}_l(y)$ but the shapes of these two graphs do not vary with $E_l$. In contrast, in \eq{Schr_equ_shape_varying}, $D(\omega, r)$ takes such a form that the $\omega$ dependency and $r$-dependency are not separable, and moreover, due to its complicated $\omega$-dependency, the shapes of the $r$-dependency graphs of $D(\omega, r)$ strongly vary with $\omega$. This feature means that, if we still want to obtain the approximate analytical eigen-solutions of \eq{Schr_equ_shape_varying} by treating it as an ordinary Schr\"odinger equation, in different bands of frequency $\omega$, in the propagating regions of the modes, we have to approximate $D\smbr{\omega, r}$ by \emph{different types} of $r$-dependent functions ${\cal D}_\omega(r)$, and $\omega$ still enters the characteristic parameters that determine these functions. \eq{Schr_equ_shape_varying} is then approximated by
\begin{eqnarray}\label{Schr_equ_shape_varying1}
 -\rho^2_{0i} \frac{d^2 \hat{\Psi}}{d r^2} + {\cal D}_\omega \smbr{r} \hat{\Psi}={\cal E_\omega}\hat{\Psi},
\end{eqnarray}
where the $\omega$-dependent eigenvalue ${\cal E_\omega}$ is chosen in such a way that the graph of ${\cal D}_\omega-{\cal E_\omega}$ almost coincides with that of $D\smbr{\omega, r}$, e.g., ${\cal E_\omega}$ can be the negative of the minimum of $D\smbr{\omega, r}$ with respect to its $r$-dependency. Of course, the favorable types of potentials ${\cal D}_\omega \smbr{r}$ to be considered are those for which \eq{Schr_equ_shape_varying1} has analytical bounded solutions, or for which \eq{Schr_equ_shape_varying1} can be converted into \eq{eigen_equ} for principal hypergeomtric-like operator $H_0(x)$ or \eq{eigen_equ_level_m_after_rnmlz} for associated hypergeomtric-like operator $H^a_m(x)$, in view that the Schr\"odinger equations for a large class of potentials can be transformed into these two types of equations. In the latter case the coefficients of $p(x)$ and $q(x)$ and the eigenvalues $\lambda_l$ and $\lambda_{lm}$ are naturally $\omega$-dependent.  In fact, for relatively low eigen-frequency, ${\cal D}_\omega \smbr{r}$ was taken as a quadratic function of $r$, and the eigen-solutions of \eq{Schr_equ_shape_varying1} are picked out as a subset of those of quantum harmonic oscillator, which are closely related to those of Hermite operator, as shown in Subsection \ref{degenerate_cases}.  For higher bands of eigen-frequencies, we have to go beyond the quadratic approximation of $D\smbr{\omega, r}$ for more accurate analytical eigen-solutions.  Actually, this is the original driving force that motivates the author to write this paper.

The author would like to devote this paper to the memory of the late Professor Herbert L. Berk.

\section*{Acknowledgments}
The author is grateful to Dr. Yu Wang for his patiently checking the spellings of this draft. This work is supported by National Natural Science Foundation of China under Grant No. 12075070. 
\appendix
\section{Calculation of \mathinhead{\alpha^\pm_l}{}, \mathinhead{\beta^\pm_l}{} and \mathinhead{E^\pm_l}{}--by Directly Matching Two Forms of \mathinhead{V_l}{}}\label{solution_Wl_and_El}

In Subsections \ref{Minus_route_factorization} and \ref{Plus_route_factorization}, we have demonstrated that $\alpha^\pm_l$, $\beta^\pm_l$ and $E^\pm_l$ are expressed in terms of the elements in series $\{c_l\}$ and $\{d_l\}$ in an iterative approach. Here we give another approach to arrive at these results. In the two expressions of $V_l(x)$ (a binomial of $x$) in Eqs. \eqnm{Vl_potential} and \eqnm{potential_x_coord}, the coefficients of either one are expressed in terms of $\alpha^\pm_l$, $\beta^\pm_l$ and $E^\pm_l$, while the coefficients of the other directly in terms of the elements in $\{c_l\}$ and $\{d_l\}$. Then the direct match of these two sets of coefficients will give us the same answer. 

Now  we have to resort to an \emph{external} way of fixing $\lambda^-_l$, as done in \cite{jafarizadeh_SUSY_and_diff_equs_1997}. Assuming that $\Phi_l$ is a polynomial of $l$-th order in $x$ in view that $p(x)$ inside the operator $H_0(x)$ is a binomial in $x$ and $q(x)$ is linear in $x$, which is confirmed \emph{posteriori} by the explicit form \eqnm{Rodriguez_formula} of $\Phi_l$,  then $\lambda^-_{l}$ in the form \eqnm{lambda_minu_eigenvalue} can be obtained by balancing the coefficients of the highest order on both sides of \eq{eigen_equ_base_level}. 

From the definitions $c_l\equiv \smbr{l p''+q'}/2$ and $d_l\equiv l p'_0+q_0$, we can make the substitutions for $p''$ via $p''=2\smbr{c_l-c_{l-1}}$, for $q'$  via $q'=2 \sqbr{l c_{l-1}-\smbr{l-1}c_l }$, for $p'_0$ via $p'_0=d_l- d_{l-1}$ and for $q_0$ via $q_0=d_0$ in the equation wherever $p''$, $p'_0$, $q'$ and $q_0$ appear, in order to ultimately express $\alpha^\pm_l$, $\beta^\pm_l$ and $E^\pm_l$ in terms of the elements in $\{c_l\}$ and $\{d_l\}$. For instance,
\begin{eqnarray}\label{lambda_plus_minus2}
\lambda^-_l &=& -lq'(x)-\frac{1}{2}l(l-1) p''(x),\nonumber \\
            &=& l\sqbr{\smbr{l-1}c_l-\smbr{l+1}c_{l-1}}.
\end{eqnarray}
Although $l$ originally is restricted to a non-negative integer, it can be extended to taking $-1$ in order to define $c_{-1}\equiv (q'-p'')/2$, consistent with the definition of $c_l$.

With $p(x)=(1/2)p'' x^2+ p'_0 x+ p_0$ , $q(x)=q'_0 x + q_0$, $\lambda_l=\lambda^-_l$ given in \eq{lambda_plus_minus2}, $c_l\equiv \smbr{l p''+q'}/2$ and $d_l\equiv l p'_0+q_0$, the binomial form of $V_l(x)$ given in \eq{Vl_potential} reads
\begin{eqnarray}
 V_l(x)&=&c_l c_{l-1} x^2 + \sqbr{\smbr{l+1}c_{l-1}d_l-l c_l d_{l-1}} x \\
       && +\frac{1}{4}\sqbr{\smbr{l+1}d_{l-1}-l d_l}^2-p_0\sqbr{l^2c_l -(l+1)^2 c_{l-1}} \nonumber \\
       && +E_l.\nonumber 
\end{eqnarray}

On the other hand, upon substituting for $p(x)= \smbr{p''/2} x^2+ p'_0 x+ p_0$ and $W_l(x)=\alpha_l x+\beta_l$, the binomial form of $V_l(x)$ given \eq{potential_x_coord} reads
\begin{eqnarray}
V_l(x) =\alpha_l \smbr{\alpha_l-\frac{p''}{2}} x^2+\alpha_l\smbr{2\beta_l-p'_0}x+ \beta^2_l-\alpha_l p_0, \nonumber \\
\end{eqnarray}
where $\alpha_l$ and $\beta_l$ are retained in the coefficients.

Now to express $\alpha_l$, $\beta_l$ and $E_l$ in terms of the elements in series $\{c_l\}$ and $\{d_l\}$, we equate the coefficients in front of $x^2$, $x$ and $x^0$ in these two forms of $V_l(x)$ and obtain
\begin{eqnarray}\label{alha_quadratic_0}
 \alpha^2_l+\smbr{c_{l-1}-c_l} \alpha_l - c_{l-1} c_l =\smbr{\alpha_l-c_l}\smbr{\alpha_l+c_{l-1}}=0,\nonumber \\
\end{eqnarray}
\begin{eqnarray}\label{beta_equ_0}
 \beta_l = -\frac{d_{l-1}}{2}\smbr{1+ l \frac{c_l}{\alpha_l}}  +\frac{d_l}{2} \sqbr{1+ \smbr{l+1} \frac{c_{l-1}}{\alpha_l}},
\end{eqnarray}
and 
\begin{eqnarray}\label{E_l_equ_0}
 E_l &=& \beta^2_l-\frac{1}{4}\sqbr{\smbr{l+1}d_{l-1}-l d_l}^2 \nonumber \\
        &&+ p_0\sqbr{l^2c_l -(l+1)^2 c_{l-1} + \alpha_l },\nonumber \\
     &=&\frac{1}{4 \alpha^2_l}\cubr{\sqbr{(l+1)c_{l-1} -(l-1)\alpha_l} d_l + l \smbr{\alpha_l-c_l}  d_{l-1}}  \nonumber\\
                                        &&\cdot\cubr{(l+1)\smbr{c_{l-1} +\alpha_l} d_l-\sqbr{l c_l+(l+2)\alpha_l}d_{l-1}}\nonumber \\
                                        &&+p_0\sqbr{l^2c_l -(l+1)^2 c_{l-1}-\alpha_l}.
\end{eqnarray}

\eq{alha_quadratic_0} gives the two routes of solutions for $\alpha_l$ in the forms
\begin{equation}\label{alpha_l_plus_minus_third_way}
 \begin{aligned}
  \alpha^+_l&=c_l,\\
  \alpha^-_l&=-c_{l-1}=-\alpha^+_{l-1}, 
 \end{aligned}
\end{equation}
where the solution $\alpha^+_l$ is the same as that given by the upper equation in \eqnm{alpha_l_plus} and $\alpha^-_l$ the same as that given by the upper equation of \eqnm{alpha_l_minus}.

For these $\alpha^\pm_l$, \eq{beta_equ_0} has two solutions
\begin{eqnarray}\label{beta_l_plus}
 \beta^+_l &=& \frac{1}{2c_l} \cubr{-(l+1) c_l d_{l-1} + \sqbr{ (l+1) c_{l-1} + c_l } d_l}, \nonumber \\ 
\end{eqnarray}
\begin{eqnarray}\label{beta_l_minus_other_way}
\beta^-_l = \frac{1}{2c_{l-1}}\sqbr{\smbr{l c_l -c_{l-1}} d_{l-1} - l c_{l-1} d_l },
\end{eqnarray}
and \eq{E_l_equ_0} has two solutions
\begin{eqnarray}
E^+_l &=&\smbr{l+1} \cubr{\frac{d_l}{4 c^2_l}\sqbr{ \smbr{c_{l-1} + c_l} d_l -  2c_l d_{l-1}}  - p_0} \nonumber \\
        &&\cdot \sqbr{\smbr{l+1} c_{l-1}-\smbr{l-1}c_l}, \label {E_l_plus}\\
 E^-_l &=&l\cubr{\frac{d_{l-1}}{4 c^2_{l-1}}  \sqbr{2 c_{l-1} d_l - \smbr{c_{l-1}+ c_l } d_{l-1}} -  p_0} \nonumber \\
        &&\cdot \sqbr{(l+2) c_{l-1} - l c_l }. \label{E_l_minus}
\end{eqnarray}

Notice that both $\cubr{c_l}$ and $\cubr{d_l}$ are simple arithmetical series, i.e. $c_{l+1}-c_l=c_l-c_{l-1}$ and $d_{l+1}-d_l=d_l-d_{l-1}$. Then it is easy to check that $\beta^-_{l+1}=-\beta^+_l$ and $E^-_{l+1}=E^+_l$.

Although formally both Eqs. \eqnm{E_l_plus} and \eqnm{E_l_minus} have solutions for $l<0$, such as $E^+_{-1} = 0$, these equations should start with $l=0$, as the lowest level of polynomial solution is for $l=0$ by definition. For $l=0$, \eq{E_l_minus} gives $E^-_0=0$, while \eq{E_l_plus} shows that $E^+_0$ is not vanishing in general ($E^+_0 = E^-_1$).

Note that in this direct match, $\lambda^-_l$ in \eq{lambda_plus_minus2} has to be given beforehand by other approaches or arguments.

\section{Supersymmetrizing the Same Sturm-Liouville-Type Eigen-equation into Two Types of Schr\"odinger Equations} \label{generalized_DO}
Note that in Subsections \ref{base_kinetic_energy_operator}, \ref{Standard_Hermitian_factorization} and \ref{2nd_type_momentum_op}, we have made use of the method of active supersymmetrization to quickly pass from the eigen-equation \eqnm{eigen_equ} to the two Schr\"odinger equations \eqnm{Schr_equ_y_coord} and \eqnm{Schrodinger_equ_base_level} in their standard forms,  respectively via combining the momentum operator maps \eqnm{momentum_map_type1} and \eqnm{momentum_map_type2} and the unitary transformations ${\cal T}$ and ${\cal S}$, the concrete quadratic nature of $p(x)$ and the linear nature of $q(x)$ are not invoked. In addition, adding another zeroth derivative term, a function of $x$, to the two-term-like differential operator $H_0(x)$ will not affect the viability of this passage. These two facts immediately imply that the active supersymmetrization can be applied to quickly transforming the more generic Sturm-Liouville-type eigen-equation into certain standard Schr\"odinger equations, i.e. 
\begin{eqnarray}\label{SL_eigen_equ0}
 SL(x)\psi(x)=\Lambda \psi(x)
\end{eqnarray}
for constant $\Lambda$ and the Sturm-Liouville-type linear differential operator
\begin{eqnarray}\label{operator_L}
 SL(x)\equiv -P(x)\frac{d^2}{d x^2} -Q(x)\od{}{x} - R(x)
\end{eqnarray}
where $P(x)$, $Q(x)$ and $R(x)$ are general functions in $x$.

By pulling $d/dx$ out from the right, the first two terms in $SL(x)$ has the factorized form $-(Pd/dx+Q)d/dx$. In terms of the integration factor $I(x)\equiv e^{\int dx Q/P}$ for the left linear differential operator, the first terms are lumped into the following asymmetric double-layer structure
\begin{eqnarray}\label{first_two_term0}
 - P\frac{d^2}{dx^2} - Q \od{}{x}= I^{-1}(x)\smbr{-P\od{}{x} I(x) } \od{}{x}.
\end{eqnarray}
Furthermore, if we introduce the weight function $\rho(x)$ such that the one under the integral symbol in the expression of $I(x)$ is an exact 1-form, i.e. $dx Q/P= d \log \smbr{\rho P}$, then 
$I(x)=\rho(x)P(x)$ and
\eq{first_two_term0} is rewritten as the familiar self-adjoint form
\begin{eqnarray}\label{first_two_term}
 - \smbr{P\od{}{x} + Q} \od{}{x} &=& - \frac{1}{\rho}\smbr{\od{}{x} \rho P } \od{}{x}
\end{eqnarray}
 with $\rho(x) =P^{-1}(x) e^{\int dx Q/P} $. 

\subsection{Aiming at Preparing Momentum Operator \mathinhead{-i P d/dx}{}}\label{1st_type_momentum_operator}
From the asymmetric double-layer structure \eqnm{first_two_term}, in order to construct a Hamiltonian  that has the squared form $-\smbr{P d/dx}^2$ as its kinetic energy operator (respectively, $-i P d/dx$ as its momentum operator in $x$-coordinate), $P$ is multiplied from the left and it is illuminating to turn the factorization \eqnm{first_two_term} into the more symmetric form
\begin{eqnarray}
 &&- \frac{1}{\rho}\smbr{P \od{}{x}} \rho  \smbr{P  \od{}{x}} \nonumber \\
 &=& \frac{1}{\sqrt{\rho}}\sqbr{\frac{1}{\sqrt{\rho}}\smbr{-P\od{}{x}} \sqrt{\rho}} \sqbr{\sqrt{\rho} \smbr{P  \od{}{x}}\frac{1}{\sqrt{\rho}}}\sqrt{\rho},\nonumber\\
 &\equiv& {\cal T}^{-1}_\rho\cubr{\sqbr{{\cal T}^{-1}_\rho \smbr{-P \od{}{x}}} \sqbr{{\cal T}_\rho\smbr{P \od{}{x}}}},
\end{eqnarray}
where in the second equality, the square root of the middle factor $\rho$ in the first line, has been equally partitioned to the two neighboring operators $\pm P d/dx$, and then by some identical manipulations these two operators are wrapped up with the transformation ${\cal T}_\rho \smbr{-} \equiv \sqrt{\rho}\smbr{-}\sqrt{\rho}^{-1}$ and its inverse ${\cal T}^{-1}_\rho$, respectively.

Now it is evident that the combined transformation $P{\cal T}_\rho\smbr{-}$ {\it actively} supersymmetrizes the first two terms of $SL(x)$ in \eqnm{operator_L} into the following supersymmetric factorization
\begin{eqnarray}\label{supersymmetrization_SL_type_I}
 &&-P{\cal T}_\rho\smbr{ P\frac{d^2}{dx^2} + Q \od{}{x}} \nonumber \\
 &=& \sqbr{{\cal T}^{-1}_\rho \smbr{-P \od{}{x}}} \sqbr{{\cal T}_\rho\smbr{P \od{}{x}}}, \\
&=&\sqbr{-P \od{}{x}+G(x)}\sqbr{P \od{}{x}+G(x)},\nonumber
\end{eqnarray}
where $G(x)\equiv -P(x)\sqbr{\log \rho(x)}'/2$, which is obtained by expanding each factor inside the square brackets in the first equation above, can be regarded as the {\it partial } superpotential associated with $SL(x)$. 

Multiplying $P\sqrt{\rho}$ from the left of \eq{SL_eigen_equ0} and inserting the unit operator $\sqrt{\rho}^{-1}\sqrt{\rho}$ between $SL(x)$ and $\psi(x)$ in this equation for preparing the transformation $P{\cal T}_\rho\smbr{-}$, and adding the term $ E \psi(x)$ ($E$ is viewed as the new eigenvalue) to both sides,  then, in view of  \eq{supersymmetrization_SL_type_I}, \eq{SL_eigen_equ0} is turned into the Schr\"odinger equation
\begin{eqnarray}\label{symmetrized_eigen_equ2}
 \sqbr{\smbr{-i P\od{}{x}}^2 +U(x)}\rho^{\frac{1}{2}}\psi(x)= E\; \rho^{\frac{1}{2}}\psi(x), \nonumber \\
\end{eqnarray}
where $-i P d/dx$ is viewed as the momentum operator and
\begin{eqnarray}\label{U_potential}
 U(x) \equiv - P(x)\od{G(x)}{x}+ G^2(x)- R(x)-\Lambda P(x)+E \nonumber \\
\end{eqnarray}
as the overall potential energy.

If $R(x)$, $-\Lambda P(x)$ and $E$ can find the same types of functions among those terms appearing in $-\sqrt{P} d G/dx+G^2(x)$, these similar terms can be merged and an overall superpotential $\widetilde{G}(x)$  can be defined for $U(x)$ in the form
\begin{eqnarray}
 U(x) = -P\od{\widetilde{G}}{x} +\widetilde{G}^2(x)
\end{eqnarray}
and \eq{symmetrized_eigen_equ2} then admits the full SUSY factorization 
\begin{eqnarray}\label{symmetrized_eigen_SUSYF}
 \sqbr{- P\od{}{x} + \widetilde{G}}\sqbr{P\od{}{x} + \widetilde{G}}\rho^{\frac{1}{2}}\psi(x)= E\; \rho^{\frac{1}{2}}\psi(x).\nonumber \\
\end{eqnarray}

Under the momentum operator map 
\begin{eqnarray}\label{momentum_map_type1_u}
 -iP(x)\od{}{x}\equiv -i \od{}{u},
\end{eqnarray}
which defines the coordinate transformation
\begin{eqnarray}
  u(x)=\int^x  \frac{d {\bar x}}{P(\bar x)},
\end{eqnarray}
$(-i P d/dx)^2$ is identified as the kinetic energy operator $- d^2/d u^2$ in $u$-coordinate, and \eq{symmetrized_eigen_equ2} becomes the standard Schr\"odinger equation
\begin{eqnarray}\label{standardized_SL_equ}
 \sqbr{-\frac{d^2}{du^2} + {\cal U}(u)} \Psi(u)= E  \Psi(u)
\end{eqnarray}
for $\Psi(u)\equiv \left.\rho^{1/2}(x)\psi(x)\right|_{x=x(u)}$ and ${\cal U}(u)\equiv U(x(u))$.

\subsection{Aiming at Preparing Momentum Operator \mathinhead{-i \sqrt{P} d/dx}{}}\label{2nd_type_momentum_operator}
In the following, we will show that there exists another factorization of \eqnm{first_two_term} that allows us to regard $-i\sqrt{P}d/dx$ as the momentum operator of another Hamiltonian associated with the same operator $SL(x)$. Aiming at some factorization that has the basic building block $\sqrt{P}d/dx$, the factorization \eqnm{first_two_term} can be identically transformed into the following one
\begin{eqnarray}\label{asymmetric_double_layer0}
&-& \frac{1}{\rho}\smbr{\od{}{x} \rho P } \od{}{x} \nonumber \\
&=& \frac{1}{\rho\sqrt{P}}\smbr{-\sqrt{P}\od{}{x}} \rho \sqrt{P}  \smbr{\sqrt{P}\od{}{x}} \nonumber \\
&=& \smbr{\rho \sqrt{P}}^{-\frac{1}{2}} \sqbr{\smbr{\rho \sqrt{P}}^{-\frac{1}{2}} \smbr{-\sqrt{P}\od{}{x}} \smbr{\rho \sqrt{P}}^{\frac{1}{2}}} \nonumber \\
&\cdot &\sqbr{\smbr{\rho \sqrt{P}}^{\frac{1}{2}} \smbr{\sqrt{P}\od{}{x}}\smbr{\rho \sqrt{P}}^{-\frac{1}{2}}}\smbr{\rho \sqrt{P}}^{\frac{1}{2}}.
\end{eqnarray}
This reminds us to introduce the wrapper transformation
\begin{equation*}
S_\rho \smbr{-} \equiv \smbr{\rho \sqrt{P}}^{\frac{1}{2}}\smbr{-}\smbr{\rho \sqrt{P}}^{-\frac{1}{2}} 
\end{equation*}
and its inverse 
\begin{equation*}
S^{-1}_\rho\smbr{-}\equiv \smbr{\rho \sqrt{P}}^{-\frac{1}{2}}\smbr{-}\smbr{\rho \sqrt{P}}^{\frac{1}{2}}
\end{equation*}
and denote \eq{asymmetric_double_layer0} as
\begin{eqnarray}\label{asymmetric_double_layer}
- \frac{1}{\rho}\smbr{\od{}{x} \rho P } \od{}{x} &\equiv& S^{-1}_\rho \left\{  \sqbr{S^{-1}_\rho\smbr{-\sqrt{P}\od{}{x}}} \right. \\
                                                &&\left.\cdot \sqbr{S_\rho\smbr{\sqrt{P}\od{}{x}}} \right\}, \nonumber 
\end{eqnarray}

Obviously, \eq{asymmetric_double_layer} implies that applying the transformation $S_\rho \smbr{-}$ to the first two terms of $SL(x)$ in \eqnm{operator_L} yields the following supersymmetric factorization
\begin{eqnarray}\label{supersymmetrization_SL_type_II}
&& S_\rho \smbr{-P\frac{d^2}{d x^2} - Q\od{}{x}}  \nonumber \\
&=&\sqbr{S^{-1}_\rho\smbr{-\sqrt{P}\od{}{x}}} \sqbr{S_\rho\smbr{\sqrt{P}\od{}{x}}},\nonumber\\
&=&\sqbr{-\sqrt{P}\od{}{x}+W_\rho(x)}\sqbr{\sqrt{P}\od{}{x}+W_\rho(x)},
\end{eqnarray}
where 
\begin{eqnarray}\label{Wrho}
W_\rho(x) \equiv -\sqrt{P}\smbr{\log \sqrt{\rho \sqrt{P}}}' 
\end{eqnarray}
Both \eqnm{asymmetric_double_layer} and \eqnm{supersymmetrization_SL_type_II} contain the squared form $-(\sqrt{P}d/dx)^2$. 

Therefore, by multiplying with $\smbr{\rho \sqrt{P}}^{1/2}$ (the left factor of $S_\rho$) from the left side of $SL(x)$ for preparing the transformation  $S_\rho$ and making use of \eq{supersymmetrization_SL_type_II}, \eq{SL_eigen_equ0} is then transformed into 
\begin{eqnarray}\label{nonstandard_SCH_equ2}
 &&\sqbr{-\smbr{\sqrt{P}\od{}{x}}^2 + V_\rho(x)} \sqbr{\smbr{\rho \sqrt{P}}^{\frac{1}{2}}\psi(x)} \nonumber \\
 &=&\Lambda \sqbr{\smbr{\rho \sqrt{P}}^{\frac{1}{2}}\psi(x)}
\end{eqnarray}
with
\begin{eqnarray}\label{Vrho_potential}
 V_\rho(x) \equiv -\sqrt{P}\od{W_\rho}{x} +W^2_\rho - R(x).
\end{eqnarray}

In general, $-R(x)$ may not be incorporated into $W_\rho$ to define the overall superpotential $\widetilde{W}_\rho(x)$ such that
\eq{Vrho_potential} can be rewritten as
\begin{eqnarray}\label{Vrho_x}
 V_\rho(x)\equiv -\sqrt{P}\od{\widetilde{W}_\rho}{x} +\widetilde{W}^2_\rho.
\end{eqnarray}
However, if $-R(x)$ is of the same type term as any one appearing in $-\sqrt{P} d W_\rho/dx+W^2_\rho(x)$, it can be merged into the latter and such $\widetilde{W}_\rho$ can be introduced. In this case, \eq{nonstandard_SCH_equ2} admits the full SUSY factorization
\begin{eqnarray}\label{nonstandard_SCH_SUSYF}
 &&\sqbr{-\sqrt{P}\od{}{x} + \widetilde{W}_\rho} \sqbr{\sqrt{P}\od{}{x} + \widetilde{W}_\rho} \sqbr{\smbr{\rho \sqrt{P}}^{\frac{1}{2}}\psi(x)} \nonumber \\
 &=&\tilde{\Lambda} \sqbr{\smbr{\rho \sqrt{P}}^{\frac{1}{2}}\psi(x)},
\end{eqnarray}
where $\tilde{\Lambda}$ counts in the additional constant possibly brought around along the introduction of $\widetilde{W}_\rho$.

Under the second type of momentum operator map 
\begin{eqnarray}\label{v_momentum_map}
 -i\sqrt{P}\od{}{x} \equiv -i \od{}{v},
\end{eqnarray}
which defines the new coordinate $v$ as
\begin{eqnarray}
 v(x) =\int^x \frac{d{\bar x} }{P^{\frac{1}{2}}(\bar x)}, \label{v_coordinate}
\end{eqnarray}
\eq{nonstandard_SCH_equ2} becomes another standard Schr\"odinger equation
\begin{eqnarray}\label{standard_SCH_equ_2nd}
 \sqbr{-\frac{d^2}{d v^2} + {\cal V}(v)}\Phi(v)=\Lambda \Phi(v)
\end{eqnarray}
in $v$-coordinate, where
\begin{eqnarray}
 \Phi(v) \equiv \left.\rho^{\frac{1}{2}}(x) P^{\frac{1}{4}}(x)\psi(x)\right|_{x=x(v)}, \nonumber \\ \label{eigenfunction2_rescaling}
\end{eqnarray}
and ${\cal V}(v) \equiv V_\rho(x(v))$. Unlike the standardized Schr\"odinger equation \eqnm{standardized_SL_equ} which has the eigenvalue that is different from the original eigen-equation \eqnm{SL_eigen_equ0}, this standardized Schr\"odinger equation {\it shares} the eigenvalue $\Lambda$ with the original equation \eqnm{SL_eigen_equ0}.

A simple example of the type of differential operator $P(x) d^2/dx^2+Q(x) d/dx =\rho^{-1}(x) (d/dx) \rho(x) P(x) (d/dx)$ that is worthy of mentioning is the radial part of a D-dimensional Laplacian operator $r^{-(D-1)} (d/dr) r^{D-1} (d/dr)$ in cylindrical or spherical coordinate system with $x=r$ being the radial coordinate, which comes naturally in the factorized form with $\rho(r)= r^{D-1}$ and $P(r)=1$. As results, the rescaling factor reduces $\sqbr{\rho(r) P^{1/2}(r)}^{1/2}=r^{(D-1)/2}$ and the superpoential in \eq{Wrho} to $W_\rho(r) =(1-D)/(2 r)$. This $W_\rho (r)$ contributes to the potential $V_\rho (r)$ in \eq{Vrho_x} the term $\smbr{1-D}\smbr{3-D}/\smbr{4 r^2}$. An interesting and important case is $D=3$. In this case, $W_\rho(r)$ has no contribution to $V_\rho(r)$ at all and $V_\rho(r)=-R(r)$. Moreover, the overall superpotential $\widetilde{W}_\rho(r)$ in \eq{Vrho_x}, if any, is merely determined by the function $R(r)$ through $-d \widetilde{W}_\rho(r)/dr + \widetilde{W}^2_\rho(r)=-R(r)$. In 3-D quantum mechanic problems, $R(r)$ often represents the effective potential consisting of the centrifugal potential and interaction potential. In the case $D=2$, it is the contribution $-1/(4 r^2)$ to $V_\rho(r)$ from $W_\rho(x)$ and $R(r)$ {\it together} that determine the analytical overall superpotential $\widetilde{W}_\rho(r)$. Thus, in order to have an overall    analytical superpotential $\widetilde{W}_\rho(r)$, the functional form of $R(r)$ in $D=2$ case is more restrictive than in $D=3$ case. As the consequence of $P(r)=1$, \eq{v_momentum_map} says that $v(r)=r$. That is, the eigen-equation $SL(r)\psi(r)=\Lambda \psi(r)$ can be transformed into the standard Schr\"odinger equation {\it within the original $r$-coordinate} merely by rescaling the eigen-function $\psi(r)$ into $\phi(r)\equiv r^{(D-1)/2} \psi(r)$. In particular, the rescaling factor is just $r$ in 3-D case, and $\sqrt{r}$ in 2-D case. The SUSYQMs of cylindrical and spherical Bessel functions can be readily established by this second type of active supersymmetrization transformation, which just introduces such simple rescaling but no radial coordinate change. In 3-D case, this transformation does not result in any additional non-derivative term to the overall potential.

In order to pass from the Sturm-Liouville-type equation \eqnm{SL_eigen_equ0} to a standard Schr\"odinger equation \eqnm{standard_SCH_equ_2nd}, traditionally the coordinate transformation \eqnm{v_coordinate} and the rescaling factor $\smbr{\rho \sqrt{P}}^{1/2}$ are  obtained by solving some differential equations of second order satisfied by $v(x)$ and this rescaling factor, which result from aiming at eliminating {\it bona fide} the first order derivative in \eq{SL_eigen_equ0} (see this, e.g. on page 292 in \cite{Courant_and_Hilbert_1931, *Courant_and_Hilbert_1989}). In contrast,  by aiming at directly constructing the kinetic energy operator $-\smbr{\sqrt{P}d/dx}^2$ or the momentum operator $-i \sqrt{P}d/dx$, we implement this passage by applying the pure algebraic {\it active supersymmetrization} transformation $S_\rho$, which just disguises the first order derive and gives rise to the partial superpotential, the coordinate transformation and the rescaling factor $\smbr{\rho \sqrt{P}}^{1/2}$ simultaneously merely by elementary algebraic manipulations. Not only is this active supersymmetrization much simpler, but also full of quantum mechanic flavor.

\subsection{Another Sufficient Condition for Fully Supersymmetrizing \mathinhead{SL(x)}{}}
Around Eqs. \eqnm{symmetrized_eigen_SUSYF} and \eqnm{nonstandard_SCH_SUSYF}, we have discussed the possible situations in which $SL(x)$ can be completely supersymmetrized. Now, we address another sufficient condition for fully supersymmetrizing $SL(x)$. Similar to the partial supersymmetrization that results in a non-differential term in the operator in \eq{2nd_equivalent_eigen_equ}, we can think of the non-differential term $-R$ inside the operator $SL(x)$ as a result of a certain partial supersymmetrization transformation when it is applied to the operator $-P(x)\,d^2/dx^2 -\sqbr{Q(x)+Q_1(x)}d/dx$ for eliminating the part $-Q_1d/x$. 
\begin{eqnarray}
 -P \frac{d^2}{d x^2} -Q_1 \od{}{x} &=& P\sqbr{-\od{}{x} - 2 W_{Q_1}}\od{}{x},\nonumber \\
                                    &=& P \eta^{-1}\smbr{-\od{}{x}} \eta \od{}{x}
\end{eqnarray}
where $W_{Q_1}\equiv Q_1/2 P$ and $\eta(x)\equiv  e^{2 \int^x d \tilde{x}  W_{Q_1}\smbr{\tilde{x}} }$ is the integration factor. Consequently, by introducing the transformation ${\cal S}_\eta \equiv \sqrt{\eta}(-)\sqrt{\eta}^{-1}$, we have the partial supersymmetrization
\begin{eqnarray}\label{partial_supersymmetrization}
 {\cal S}_{\eta}\smbr{-P \frac{d^2}{d x^2} -Q_1 \od{}{x}} &=& P\, {\cal S}^{-1}_{\eta}\smbr{-\od{}{x}} {\cal S}_{\eta}\smbr{\od{}{x}}, \nonumber \\
                                                          &=&-P\,\frac{d^2}{d x^2} + P\sqbr{W'_{Q_1} + W^2_{Q_1}}. \nonumber \\
\end{eqnarray}

On the other hand, ${\cal S}_{\eta}$ shifts $-Q\, d/dx$ according to
\begin{eqnarray}\label{partial_shift}
   {\cal S}_{\eta}\smbr{ -Q \od{}{x}} = - Q \od{}{x} + Q\; W_{Q_1}.
\end{eqnarray}

Adding Eqs. \eqnm{partial_supersymmetrization} and \eqnm{partial_shift} up then yields
\begin{eqnarray}
 &&{\cal S}_{\eta}\sqbr{-P \frac{d^2}{d x^2} -\smbr{Q+Q_1} \od{}{x}}  \\
 &=&-P\,\frac{d^2}{d x^2} - Q \od{}{x} + P\sqbr{W'_{Q_1} + W^2_{Q_1}} + Q\; W_{Q_1}.\nonumber
\end{eqnarray}
Comparing the right hand side of the above equation with $SL(x)$ in \eq{operator_L}, if the two non-differential terms happen to be of the same type up to a constant $\Lambda_1$, i.e.
\begin{eqnarray}
 P\sqbr{W'_{Q_1} + W^2_{Q_1}} + Q\;W_{Q_1}  =- R - \Lambda_1,
\end{eqnarray}
which is a nonlinear differential equation of first order for $Q_1$ for given $R$, then
\begin{eqnarray}
 SL(x) ={\cal S}_{\eta}\sqbr{-P \frac{d^2}{d x^2} -\tilde{Q} \od{}{x}} + \Lambda_1
\end{eqnarray}
for $\tilde{Q}\equiv Q+Q_1$. Consequently, \eq{SL_eigen_equ0} gives rise to the eigen-equation
\begin{eqnarray}\label{SL_eigen_equ_equivalent}
 \sqbr{-P \frac{d^2}{d x^2} -\tilde{Q} \od{}{x}} \tilde{\psi}(x)=\tilde{\Lambda} \tilde{\psi}(x)
\end{eqnarray} 
for $\tilde{\psi}(x) \equiv \sqrt{\eta}^{-1}\psi(x)$ and $\tilde{\Lambda}=\Lambda-\Lambda_1$. 

Note that the two-term differential operator $-P d^2/dx^2 - \tilde{Q}d/dx$ in \eq{SL_eigen_equ_equivalent} is of the same type as $-P d^2/dx^2 - Q d/dx$ in \eq{first_two_term}. Thus, by simply replacing $Q$ with $\tilde{Q}$  and following the procedure in Subsection in \ref{2nd_type_momentum_operator}, aiming at preparing the kinetic energy operator $-\smbr{\sqrt{P}d/dx}$, the operator $-P d^2/dx^2 - \tilde{Q}d/dx$  is then easy to be supersymmetrically factorized. When following the procedure in Subsection \ref{1st_type_momentum_operator}, aiming at preparing the kinetic energy operator $-\smbr{P d/dx}^2$, if the term $\tilde{\Lambda} P+ \tilde{E}$, added into \eq{SL_eigen_equ_equivalent} for constructing new eigen-equation, can be incorporated into a certain partial superpotential, an analog of $G(x)$ in \eq{supersymmetrization_SL_type_I}, to define a new overall superpotential, then the operator $-P d^2/dx^2 - \tilde{Q}d/dx$ can be fully supersymmetrically factorized.
In these two ways, the operator $SL(x)$ is fully supersymmetrized.

\bibliographystyle{unsrt}
\bibliography{SUSYQM_of_hypergeometric-like_differential_operators_arxiv_upload_v1.bib}

\end{document}